\documentclass[rmp,aps,twocolumn,amssymb,amsfonts,nofootinbib,showkeys]{revtex4}
\tighten
\usepackage{graphicx}
\usepackage{mathrsfs}
\usepackage{bm}
\usepackage{chapterbib}
\usepackage{amsmath}
\usepackage{upgreek}
\usepackage{wasysym}
\usepackage{mathbbol}

\newcommand{\be}{\begin{equation}}
\newcommand{\ee}{\end{equation}}
\newcommand{\bea}{\begin{eqnarray}}
\newcommand{\eea}{\end{eqnarray}}

\newcommand{\br}{{\bf r}}

\newcommand{\wc}{\omega_{\rm c}}

\def\eac{\epsilon_{\mathrm{ac}}}
\def\edc{\epsilon_{\mathrm{dc}}}
\def\eph{\epsilon_{\mathrm{ph}}}
\def\pac{\varphi_{\mathrm{ac}}}

\def\ro{\rho_\omega}

\def\os{\omega_\pi}
\def\olo{\omega_{\mbox{\scriptsize {LO}}}}

\def\rc{R_{\mbox{\scriptsize {c}}}}

\def\tpi{\tau_{\pi}}

\def\tq{\tau_{\mbox{\scriptsize {q}}}}
\def\ttr{\tau}

\def\tsh{\tau_{\mbox{\scriptsize {sh}}}}
\def\tsm{\tau_{\mbox{\scriptsize {q,sm}}}}
\def\tin{\tau_{\mbox{\scriptsize {in}}}}

\def\tst{\tau_\star}

\def\lb{\lambda_B}

\newcommand{\req}[1]{Eq.\,(\ref{#1})}

\newcommand{\reqs}[2]{Eqs.\,(\ref{#1}),\,(\ref{#2})}
\newcommand{\rfig}[1]{Fig.\,\ref{#1}}

\newcommand{\rsec}[1]{Sec.\,\ref{#1}}

\newcommand{\rsecs}[2]{Secs.~\ref{#1},\,\ref{#2}}

\newcommand\bn{{\bm n}}

\newcommand\bE{{\bm E}}
\newcommand\bR{{\bm R}}

\newcommand\ve{\varepsilon}

\newcommand\w{\omega}
\newcommand\St{{\rm St}}
\renewcommand\j{{\bm j}}

\newcommand\q{\bm{q}}
\newcommand\bzeta{\bm{\xi}}

\begin{document}
\title{Nonequilibrium phenomena in high Landau levels}
\author{I.~A. Dmitriev}
\affiliation{Institut f\"ur Theorie der Kondensierten Materie, Karlsruhe Institute of Technology, 76128 Karlsruhe, Germany}
\affiliation{Institut f\"ur Nanotechnologie, Karlsruhe Institute of Technology,  76021 Karlsruhe, Germany}
\affiliation{Ioffe Physical Technical Institute, 194021 St.~Petersburg, Russia}
\author{A.~D. Mirlin}
\affiliation{Institut f\"ur Nanotechnologie, Karlsruhe Institute of Technology,  76021 Karlsruhe, Germany}
\affiliation{Institut f\"ur Theorie der Kondensierten Materie, Karlsruhe Institute of Technology, 76128 Karlsruhe, Germany}
\affiliation{Petersburg Nuclear Physics Institute, 188300 St. Petersburg, Russia}
\author{D.~G.~Polyakov}
\affiliation{Institut f\"ur Nanotechnologie, Karlsruhe Institute of Technology, 76021 Karlsruhe, Germany}
\author{M.~A. Zudov}
\affiliation{School of Physics and Astronomy, University of Minnesota, Minneapolis, Minnesota 55455, USA}

\date{\today}

\begin{abstract}
Developments in the physics of 2D electron systems during the last decade revealed a new class
of nonequilibrium phenomena in the presence of a moderately strong magnetic field.
The hallmark of these phenomena is magnetoresistance oscillations generated by the external
forces that drive the electron system out of equilibrium. The rich set of dramatic phenomena
of this kind, discovered in high mobility semiconductor nanostructures, includes, in particular,
microwave radiation-induced resistance oscillations and zero-resistance states, as well as
Hall field-induced resistance oscillations and associated zero-differential resistance states.
The experimental manifestations of these phenomena and the unified theoretical framework for
describing them in terms of a quantum kinetic equation are reviewed. This survey also contains
a thorough discussion of the magnetotransport properties of 2D electrons in the linear-response
regime, as well as an outlook on future directions, including related nonequilibrium phenomena
in other 2D electron systems.
\end{abstract}

%\pacs{PACS numbers: 73.50.-h, 73.43.Qt, 73.63.Hs, 78.67.De}


\maketitle

\tableofcontents

\section{Introduction}
\label{s1}

Fundamental research on two-dimensional electron gases (2DEGs), as well as on
quantum wires and quantum dots made on the base of 2D structures, has largely
determined the development of condensed matter physics in the last half-century
\cite{beenakker:1991,ferry:1997}.
Continuous advances in the fabrication of semiconductor structures and the
development of experimental techniques and theoretical approaches keep bringing
up fascinating new areas of research. On the side of applications, these
structures, serving as a basis for the planar semiconductor technology, are of
enormous importance for modern nano- and optoelectronics
\cite{alferov:2001,kroemer:2001}.

A large body of research findings concerning the character of electronic states,
mechanisms of scattering, activated transport in the insulating regime, the
cyclotron resonance, quantum magnetooscillations, and many other properties of a
2DEG has been summarized by \citet{ando:1982a}. The discovery of the integer and
fractional quantum Hall (QH)
effects in 1980--1982 largely shifted the focus of research towards the range of
strong magnetic fields $B$ where the QH physics is observed
\cite{klitzing:1986,tsui:1999,stormer:1999,laughlin:1999}. More recently, the
behavior of very-low-density 2DEGs at zero $B$ attracted a great deal of
attention: the physics of these systems is governed by the interplay of quantum
localization and strong interactions, which leads to a metal-insulator
transition \cite{abrahams:2001}.

During the last two decades evidence has accumulated that the range of
moderately strong $B$, where the cyclotron dynamics already sets in whereas the
quantum localization effects and thus the Hall quantization are not developed
yet, also reveals a rich variety of important physical phenomena. These
include geometric resonances in periodically modulated structures
\cite{weiss:1989,weiss:1991b} and magnetoresistance due to
quasiclassical memory effects and to interaction-induced quantum corrections to
the conductivity [for a review, see \cite{dmitriev:2008a}].

A particularly prominent research arena was opened by the experimental discovery
of the microwave-induced resistance oscillations (MIRO) \cite{zudov:2001a,ye:2001}
and the zero-resistance states (ZRS) \cite{mani:2002,zudov:2003}. It became clear
soon after these discoveries that a new field of nonequilibrium physics in high
Landau levels (LLs) emerged. Most importantly, the observed phenomena demonstrated
that the combined effect of the weak Landau quantization and the relatively weak
microwave radiation can give rise to very strong changes of the transport
properties of a 2DEG. These novel nonequilibrium phenomena revealed unexpected
and conceptually interesting physics which has proven to be of interest in many
areas of condensed matter physics. In addition to MIRO and ZRS, a number of
other closely related phenomena have been observed, such as the Hall field-induced
resistance oscillations (HIRO) \cite{yang:2002}, the phonon-induced resistance
oscillations (PIRO) \cite{zudov:2001b}, in particular, under nonequilibrium
conditions created by a dc field \cite{zhang:2008}, the microwave-induced
$B$-periodic oscillations in stronger magnetic fields \cite{kukushkin:2004},
the zero-differential resistance states (ZdRS) \cite{bykov:2007,zhang:2008}, as well
as the fractional MIRO \cite{dorozhkin:2003,willett:2004,zudov:2003}.

In the decade that has passed since the above discoveries, intensive
experimental and theoretical work by many researchers has greatly
advanced the understanding of the physics of nonequilibrium phenomena in high
LLs. These advances have motivated us to write this review. While
several short reviews are available (mainly written on a much earlier stage of
development of the research field)
\cite{fitzgerald:2003,durst:2004,durst:2006,dmitriev:2008a,vitkalov:2009,lyapilin:2004,lei:2010},
there is a clear need for a full-scale review article on the subject.
In addition to summarizing past achievements, we also review important open
questions as well as perspectives for further developments (including related
nonequilibrium phenomena in other 2D systems). Particularly interesting in this
respect is the recent discovery of photoinduced dissipationless transport in a
2D electron system on liquid He \cite{konstantinov:2009b,konstantinov:2010}.

\section{Linear transport of 2D electrons in a moderately strong magnetic field}
\label{s2}

We begin by reviewing the transport properties of a 2DEG in the linear-response
regime, with emphasis on the effects that are qualitatively sensitive to the
specific nature of disorder.

\subsection{2D electron gas: Types of disorder}
\label{s2.1}

Almost all work on the novel nonequilibrium magnetotransport phenomena at low
magnetic field $B$ has relied on ultra-high mobility selectively doped
GaAs/AlGaAs heterostructures---with the electron mobility $\mu$ at the level of
$\mu\sim 10^7\,{\rm cm}^2/{\rm V\,s}$ (the mean free path $l\sim 10^2\,\mu{\rm
m}$). Therefore, we focus below on the scattering mechanisms that are specific
to these realizations of a 2DEG. Note that it is on the structures of this very
design and this sample quality that the most spectacular advances in the study
of the fractional QH effect have been made in the opposite limit of
strong $B$. Typically, the 2DEG in high mobility structures is confined to a
single quantum well (QW) in undoped (``pure") GaAs. In the single-interface setup, a
triangular-shaped QW is created by the conduction band offset at the
interface between AlGaAs and GaAs on one side
and by the electric field produced by the charge of dopants, and of the 2DEG
itself, on the other. The stability of the 2DEG charge confinement near the
interface is supported by the chemical potential fixed by the overall
distribution of charges. The double-interface design with a GaAs QW
squeezed between AlGaAs layers is also frequently employed.

\subsubsection{Remote donors}
\label{s2.1.1}

The key idea behind the suppression of impurity scattering in the
heterostructures is a spatial separation of the layer to which the 2DEG is
confined
and the ionized donor impurities by an undoped ``spacer" with a typical width
$d\sim 10^2\,{\rm nm}$. Since the first experiments on the modulation-doped
heterostructures \cite{dingle:1978}, the optimization of the structure design
has led to the growth of achieved mobilities from about $10^4\,{\rm cm}^2/{\rm
V\,s}$ to the values in excess of $3\times 10^7\,{\rm cm}^2/{\rm V\,s}$
\cite{umansky:2009,pfeiffer:2003}. Much of this progress has been related to the
strong dependence of the momentum relaxation rate $1/\tau=e/m\mu$ on $d$ for
scattering off remote charged impurities (here and below $\hbar=1$):
\be
{1\over\tau}={\pi n_i\over 8m(k_F d)^3}~,
\label{II.1}
\ee
where $m$ is the electron mass, $k_F$ is the Fermi wavevector of the 2DEG,
$n_i=\int\!dz\,c_R(z)(d/|z|)^3$, and $c_R(z)$ is the three-dimensional density
of remote charged impurities located at $|z|>d$ ($z$ is counted from the
interface).

The transport scattering rate at zero temperature
% RR $T$
is obtained as an integral
over the scattering angle $\phi$:
\be
{1\over \tau}=m\int_0^{2\pi}{d\phi\over
2\pi}\,(1-\cos\phi)W_{q=2k_F\sin(\phi/2)}~,
\label{II.2a}
\ee
where $W_q$ is the Fourier component of the correlation function of the screened
impurity potential in the plane of the 2DEG
\be
W(r)=\left<V(0)V({\bf r})\right>~,\quad {\bf r}=(x,y)
\label{II.2b}
\ee
at the transferred in-plane momentum $q$. In the limit $k_Fd\gg 1$ (typically
$k_Fd\sim 10$), scattering is predominantly on small angles and
Eq.~(\ref{II.2a}) reduces to
\be
{1\over \tau}={m\over k_F^3}\int_0^\infty\!\!{dq\over 2\pi}\,q^2W_q~.
\label{II.2}
\ee
The function $W_q$ that describes remote charged impurities and upon
substitution in Eq.~(\ref{II.2}) yields Eq.~(\ref{II.1}) is
\be
W_q=\left(\pi\over m\right)^2n_ie^{-2qd}~.
\label{II.3}
\ee
Equation (\ref{II.3}) assumes that (i) $q^{-1}$ is much larger than the 2DEG
thickness $w$ (typically $w\sim 10\,{\rm nm}$); (ii) the only source of
screening of the impurity potential is the 2DEG and the dielectric function
$\epsilon(q)$ is given by the 2D random phase approximation (RPA):
$\epsilon(q)=\epsilon_0(1+2/qa_B)$, where $\epsilon_0$ is the lattice dielectric
constant at the interface and $a_B=\epsilon_0/me^2$ is the Bohr radius; and (iii)
$q^{-1}$ is much larger than $a_B\simeq 10\,{\rm nm}$. Conditions (i) and (iii)
for the characteristic values of $q^{-1}\sim d$ are reasonably well satisfied in
the high-mobility structures. The use of the RPA in (ii) relies on the smallness
of the Wigner-Seitz parameter $r_s=(\pi n_ea_B^2)^{-1/2}$, where
$n_e=k_F^2/2\pi$ is the density of the 2DEG. For the typical value of
$n_e=3\times 10^{11}\,{\rm cm}^{-2}$, $r_s\simeq 1$ and the conventional use of
the RPA is only marginally justified.
%Fluctuations of the random potential with
%the correlation function (\ref{II.3}) can be accurately described by Gaussian
%statistics in view of the typically large parameter $n_id^2\gg 1$.
% RR
However, deviations from the RPA in Eq. (5) lead only to the parameter $m/\pi$
being replaced by the exact compressibility, so that at $r_s\sim 1$ Eq. (5)
remains valid up to a factor of order unity. We also note that fluctuations of
the random potential created by remote impurities are accurately described by
Gaussian statistics (i.e., higher cumulants can be neglected) in view of the
typically large parameter $n_i d^2 \gg 1$, independently of the value of $r_s$.

Note that the total scattering rate
\be
{1\over\tau_{\rm q}}=m\int_0^{2\pi}{d\phi\over 2\pi}\,W_{q=2k_F\sin(\phi/2)}
\label{II.3a}
\ee
(which determines, e.g., the LL broadening induced by disorder,
see Sec.~\ref{s2.3}) is, in the case of remote impurities, much larger than
$1/\tau$. In the limit $k_Fd\gg 1$, $1/\tau_{\rm q}$ is rewritten, similarly to
Eq.~(\ref{II.2}), as
\be
{1\over \tau_{\rm q}}={2m\over k_F}\int_0^\infty\!\!{dq\over 2\pi}\,W_q~,
\label{II.3b}
\ee
which for $W_q$ from Eq.~(\ref{II.3}) gives $\tau/\tau_{\rm q}=(2k_Fd)^2$.
% RR
The parametrically large, for $k_Fd\gg 1$, difference between $\tau$ and
$\tau_{\rm q}$ reflects a diffusive character of electron dynamics on the Fermi
surface: the ratio $\tau/\tau_{\rm q}$ is a characteristic number of small-angle
scattering events that is needed to change the direction of momentum by an angle
of order $\pi$.

In an important case of ``delta-doping," charged impurities are concentrated in
a thin layer whose thickness is much smaller than $d$ (down to a few lattice
constants), so that $c_R(z)$ can be approximated as $n_i\delta(z+d)$ with $n_i$
the sheet density of these impurities. The $\delta$-layer is grown
either by directly adding impurities in \mbox{AlGaAs} or by doping into a
short-period AlAs/GaAs superlattice embedded in the alloy
\cite{friedland:1996,umansky:2009}. In the latter case, X-valley electrons in
AlAs yield additional screening of disorder and a further reduction of $1/\tau$.
In modern high-mobility GaAs/AlGaAs structures, the growth sequence design
typically includes one or two Si-doped $\delta$-layers (these may be followed by
other layers with Si dopants which compensate for surface states at the boundary
with the vacuum), which are the main supply of electrons to the 2DEG and are
thought to be the main source of scattering in the 2DEG as far as the
intentionally doped layers are concerned. A setup with two $\delta$-doped
layers, one on each side of the 2DEG, has the advantage that it yields larger
$n_e$ for given $d$. The effect of the increase of $n_e$ on $1/\tau$ is stronger
than the enhancement of the strength of disorder. In the case of remote
impurities with $n_e=n_i$, the time $\tau$ for a QW symmetrically
doped from both sides is, according to Eq.~(\ref{II.1}), a factor of $\sqrt{2}$
larger than for the single-side doping. Moreover, the double-side doping has
been instrumental in attaining the highest reported mobilities for which the
scattering off remote impurities is likely to be of little importance, see
Sec.~\ref{s2.1.2}.

Equation (\ref{II.1}) implies that charged impurities are randomly distributed
in the 2D plane. This assumption appears to overestimate $1/\tau$
\cite{heiblum:1984,buks:1994,buks:1994a,coleridge:1997} and $1/\tau_{\rm q}$
\cite{coleridge:1991,coleridge:1997} measured in some heterostructures with
$d\alt 50\,{\rm nm}$. The reason why the measured mobility is higher than
expected was argued \cite{buks:1994,buks:1994a} to be related to the fact that
the Si-dopants in AlGaAs (in a range of Al content which includes the typical
values of 30-40$\%$) may exist in two configurations: as shallow donors (which
become positively charged after having supplied electrons for the 2DEG) and as
DX centers (whose energies lie deeper in the forbidden gap and which can be
negatively charged). \citet{buks:1994,buks:1994a} interpreted their experimental
results to prove that the DX centers {\it are} negatively charged in the ground
state and demonstrated that correlations in the spatial distribution of the
positively and negatively charged impurities may lead to a strong reduction of
the scattering rate, similar to the correlations considered earlier
\cite{efros:1990} for the case when not all donor impurities are ionized and
some remain neutral. Control of the ratio of the impurity densities in the
shallow and DX configurations indeed provided experimental evidence
\cite{buks:1994,buks:1994a,coleridge:1997} for the correlations capable of
substantially reducing $1/\tau$ (by a factor of up to about 6), depending on the
parameters of the technological process, and a similar behavior was also
demonstrated in the measurements of $1/\tau_{\rm q}$
\cite{coleridge:1997,shikler:1997}.

\subsubsection{Background impurities}
\label{s2.1.2}

Electrostatics of the heterostructure \cite{stern:1983} dictates that $n_e$ is
lowered with increasing $d$ \cite{heiblum:1984,umansky:1997}. Details of the
relation between $n_e$ and $d$ depend on the concrete doping design of the
heterostructure; however, as long as the electric field in the spacer layer is
mainly determined by the 2DEG charge, $n_e$ scales as $1/d$, which gives for
$n_e=n_i$, according to Eq.~(\ref{II.1}), $\mu\propto d^{5/2}$ \cite{lee:1983}.
The growth of $\mu$ with increasing spacer thickness, observed at smaller $d$,
typically stops in the ultra-high mobility samples at $d\simeq\,$60--70$\,{\rm
nm}$ \cite{umansky:1997}. This behavior is usually argued to be associated with
a competition between scattering off remote charged impurities
[Eq.~(\ref{II.1})] and scattering off ``background impurities" which are present
in a small concentration in the spacer layer and also in the GaAs layer. Being
distributed in close vicinity of the 2DEG, the background impurities can lead to
large-angle scattering, thus giving the main contribution to $1/\tau$ at
sufficiently large $d$.

The momentum relaxation rate due to scattering off background impurities
uniformly distributed with the three-dimensional densities $c_B^<$ on one side
of the interface and $c_B^>$ on the other is obtained as
\be
{1\over\tau}={\pi c_B\over mk_F^3a_B^2}\,\ln\left(\min
\{k_F,w^{-1}\}a_B\right)~,
\label{II.4}
\ee
where $c_B=c_B^<+c_B^>$ and the corresponding correlation function $W_q$ for
$qw\ll 1$ is given by
\be
W_q=(c_B/2q)[\,2\pi e^2/\epsilon (q)q\,]^2~.
\label{II.4b}
\ee
Equation (\ref{II.4}) is accurate provided both $k_F^{-1}$ and $w$ are much
smaller than $a_B$. In reality, both spatial scales in high-mobility GaAs/AlGaAs
structures are about $a_B\simeq 10\,{\rm nm}$, so that $1/\tau$ due to
scattering off background impurities should then be sensitive to the exact shape
of the electron density profile in the direction across the 2DEG plane. A
comparison of Eqs.~(\ref{II.1}) and (\ref{II.4}) shows that at $n_i/d^3\sim
c_B/a_B^2$ (up to the logarithmic factor) the two mechanisms of scattering give
equal contributions to the total scattering rate and at larger $d$ the
scattering on background impurities dominates.

The dependence of $\mu$ on $n_e$ varied {\it in situ} by a controlled
illumination in structures with $d$ larger than 85 and up to 200\,nm,
exhibited---when fitted to a power law over about one decade in $n_e$ within the
interval between $10^{10}$ and $2\times 10^{11}\ {\rm
cm}^{-2}$---the relation $\mu\propto n_e^x$ with $x$ in the range 0.6--0.7
\cite{shayegan:1988,pfeiffer:1989,umansky:1997}. \citet{pfeiffer:2003} also
emphasized that the power law with $x\simeq 0.7$ holds over two decades in $n_e$
for a series of samples with different $n_e$ for $\mu > 10^6\,{\rm cm^2/V\,s}$.
These values of $x$ are substantially smaller than 3/2 expected for scattering
off remote impurities [Eq.~(\ref{II.1})], which is commonly regarded as evidence
pointing toward the dominant role of background impurities. Note that
Eq.~(\ref{II.4}) for background impurities also predicts the exponent 3/2 for
sufficiently large $n_e$. However, in the low-density case $n_e\alt
10^{11}\,{\rm cm}^{-2}$, the dependence of $\mu$ on $n_e$---according to the
numerical calculations  \cite{ando:1982,gold:1989} using variational electron
wavefunctions to treat the effect of a finite thickness of the 2DEG---is indeed
substantially weaker for background impurities than $\mu\propto n_e^{3/2}$ [the
results by \citet{ando:1982} for background impurities are obtainable from those
for remote impurities by putting $d\to 0$]. On the other hand, the numerical
variational calculation by \citet{walukiewicz:1984} for $n_e>10^{11}\,{\rm
cm}^{-2}$ gave similar dependences of $\mu$ on $n_e$ for remote and
background impurities.

A systematic experimental study \cite{umansky:1997} of the relative
contributions of the scattering mechanisms was performed in single-interface
heterostructures grown in a wide range of $d$ and $n_i$ and mobilities as high
as $\mu\simeq 1.4\times 10^7\,{\rm cm^2/ V\,s}$. By measuring the mobility at
given $d$ and $n_e$ as a function of $n_i$ \citep[see also][]{umansky:2009},
experimental evidence was presented that for $d$ larger than $70\,{\rm nm}$ the
contribution of remote impurities to the total $1/\tau$ was smaller (in the
``best samples"---about 10--15\%) than that of background charges. These results
indicate that $\mu$ in the presently available ultra-high mobility structures of
various designs is likely to be mainly, or to a large extent, limited by
background impurities.

In both Eqs.~(\ref{II.1}) and (\ref{II.4}), $1/\tau$ was obtained  within the
Born approximation. While this level of approximation is well justified for
remote impurities, for $k_Fa_B\sim 1$ background impurities that are within the
distance $a_B$ from the 2DEG and give the main contribution (as far as the
background charges are concerned) to the scattering rate are not really weak
scatterers. The momentum relaxation time $\tau$ for the background impurities
sitting effectively right at the interface is given simply by the time of
flight $1/v_Fc_Ba_B^2$ (where $v_F$ is the Fermi velocity) between two of them
along the straight line. The presence of the strong scatterers at the interface
will have important ramifications in the subsequent sections.

\subsubsection{Surface roughness}
\label{s2.1.3}

Interface roughness is also an important source of disorder in high-mobility
GaAs/AlGaAs structures and is argued \cite{markus:1994,saku:1996,umansky:1997}
to be probed by the anisotropy displayed by $\mu$ when the current is measured
along different directions in the 2D plane. The anisotropy of interface
roughness is viewed as an inherent property of the kinetics of growth of
GaAs-based structures. Grown, as is usual in high-mobility structures, on the
(001) plane, they exhibit atomic-scale terracing at the interface, with randomly
shaped terraces being typically elongated in the $[\bar{1}10]$ direction.
Anisotropy of $\mu$ was reported to be in some high-mobility samples as large as
40\% \cite{markus:1994,umansky:1997,saku:1996,tokura:1992}, increasing with
electron density. The experimental results indicate that the strength of
interface roughness varies depending on the growth conditions: for two
ultra-high mobility samples in \citet{saku:1996,tokura:1992}, both having almost
the same $\mu\simeq 10^7\,{\rm cm^2/V\,s}$ along the $[\bar{1}10]$ direction,
the anisotropy was about 40\% in one of them and about 10\% in the other. In the
``best structures" studied by \citet{umansky:1997} the anisotropy was also
reported to be reduced to 5--10\%. The experiments suggest that the
characteristic correlation radius of the interface inhomogeneities\footnote{The
issue of anisotropic interface roughness in ultra-high mobility GaAs/AlGaAs
structures has attracted attention also in the context of cooperative effects in
electron transport at half-filling of high LLs \cite{willett:2001,cooper:2001}.
These experiments probed surface roughness at the interface with the vacuum and
were focused on long-range statistically anisotropic fluctuations of the
surface height with a much larger correlation radius on the scale of 1\,$\mu$m.}
that are relevant to the mobility is 10\,$\rm nm$
\cite{markus:1994,saku:1996,tokura:1992}.

Modeling the interface as a hard wall, the problem of scattering by surface
roughness reduces to that of a perturbation imposed on the boundary condition.
Spatial fluctuations of the height of the surface $h({\bf r})$ along the $z$
axis create then a 2D random potential whose correlation function at the
transferred momentum $\bf q$ is \cite{prange:1968,ando:1982a}
\be
W_{\bf q}=F^2\left<hh\right>_{\bf q}~,
\label{II.4a}
\ee
where $F=\int\!dz\, (\partial U/\partial z)\rho_{\rm D}
(z)=(\partial\chi/\partial z)^2\vert_{z=+0}/2m$, $U(z)$ is the unperturbed
confining potential, $\rho_{\rm D}(z)$, nonzero for $z>0$, is the unperturbed
electron density profile (normalized to unity) in the direction normal to the
2DEG plane, $\chi (z)=\rho_{\rm D}^{1/2}(z)$, and $\left<hh\right>_{\bf q}$ is
the correlation function of the surface corrugations. Equation (\ref{II.4a}) is
valid for the characteristic amplitude of $h({\bf r})$ much smaller than the
2DEG thickness. Estimates based on Eq.~(\ref{II.4a})
\cite{tokura:1992,saku:1996,ando:1982,markus:1994} show that the experimentally
observed strong anisotropy can be explained in terms of scattering by interface
roughness for realistic parameters of the latter. If the charge of the depletion
layer can be neglected compared to the 2DEG charge, then $F\propto n_e$ and the
dependence of the mobility on $n_e$, as it follows from Eq.~(\ref{II.4a}), is
$\mu\propto n_e^{-2}$, i.e., with increasing $n_e$ this mechanism of scattering
becomes stronger---in contrast to scattering off charged impurities. The
calculation by \citet{markus:1994} shows that the $n_e^{-2}$ scaling of $\mu$
(for $F\propto n_e$) changes to $n_e^{-1/2}$ if scattering by interface
roughness is predominately on small angles.

Although the term ``interface roughness" commonly refers to the long-range
fluctuations of $h({\bf r})$, there are also inherent imperfections in the
interface related to alloy disorder in the AlGaAs part of the heterostructure.
These are correlated on the scale of the lattice constant $a_0$ and are
described for $qa_0\ll 1$ by
\be
W_q=\Delta_c^2a_0^3I,
\label{II.4c}
\ee
where $\Delta_c$ is the energy scale of the fluctuations and the integral
$I=\int\!dz\,\rho^2(z)$, with $\rho(z)$ being the electron density profile
(normalized to unity) across the 2DEG, is taken over the region where alloy
scattering is present \cite{ando:1982}. The integral is estimated as
$I\sim\xi^5/w^6$, where $\xi\sim a_0$ is the tunneling length under the barrier
on the AlGaAs side of the interface and $w$ is the total thickness of the 2DEG.
Although the values of $\Delta_c$ and $I$ can be calculated only numerically,
early estimates \cite{ando:1982} show that for $n_e=3\times 10^{11}\,{\rm
cm}^{-2}$ and the Al fraction in the range 30--40\% alloy scattering at the
interface establishes the mobility limit at the level of about $10^7\,{\rm
cm^2/V\,s}$. This suggests that alloy scattering, commonly deemed irrelevant at
the GaAs/AlGaAs interface, may in fact be important in the modern ultra-high
mobility heterostructures.

\subsection{Magnetotransport beyond the Drude theory}
\label{s2.2}

At the most basic level, dc transport in a magnetic field $B$ is described by
the Drude formula according to which the resistivity tensor $\hat\rho(B)$ in two
dimensions reads
\be
\hat\rho(B)=\frac{m}{e^2n_e}
\left(\begin{array}{cc} 1/\tau & \omega_c \\
-\omega_c & 1/\tau  \end{array}\right)~,
\label{II.5}
\ee
where $\omega_c=eB/mc$ is the cyclotron frequency. Equation (\ref{II.5}) can
also be derived within the formalism of the Boltzmann kinetic equation which has
been very successful in explaining the magnetotransport properties of metals
\cite{lifshits:1973,pippard:1989}, the behavior of $\hat\rho(B)$ in which is
known to depend in an essential way on the shape and topology of the Fermi
surface. For an isotropic 2DEG, which is a virtually perfect approximation for
the 2DEG in GaAs as far as the band structure is concerned (note, however, the
possible importance of anisotropic disorder discussed at the end of
Sec.~\ref{s2.1}), this theory predicts the simple Drude result (\ref{II.5}). A
striking feature of Eq.~(\ref{II.5}) is that the longitudinal resistivity
$\rho_{xx}(B)=\rho_{\rm D}\equiv m/e^2n_e\tau$ is independent of $B$, so that
the deviation
\be
\Delta\rho_{xx}=\rho_{xx}(B)-\rho_{xx}(0)~,
\label{II.6}
\ee
termed a positive or negative magnetoresistance (MR), depending on the sign of
$\Delta\rho_{xx}$, is exactly zero.

\subsubsection{Quantum magnetoresistance}
\label{s2.2.1a}

The MR in 2D systems has been studied extensively---in the last three decades
particularly in regard to $T$-dependent contributions coming from the influence
of a magnetic field on the quantum corrections, small in the parameter
$1/k_Fl\alt 1$, to the transport coefficients \cite{altshuler:1985}. One quantum
contribution to the MR is related to the suppression of the weak-localization
correction to $\rho_{\rm D}$, which occurs in a ``classically weak" (in the
sense of $\omega_c\tau\ll 1$) magnetic field and results in a small sharp spike
of $\rho_{xx}(B)$ at $B=0$. Another quantum contribution is due to the
interaction between degenerate electrons multiply scattered by disorder. In the diffusive regime (where the
temperature $T\ll 1/\tau$),
% RR In the diffusive regime ($T\tau\ll 1$),
the Coulomb interaction-induced MR obeys (for
$r_s\ll 1$ and $\omega_c\tau\gg 1$) $\Delta\rho_{xx}/\rho_{\rm
D}=-(\omega_c^2\tau^2/\pi k_Fl)\ln (1/T\tau)$, irrespective of the nature of
disorder \cite{altshuler:1985,houghton:1982,girvin:1982}. The theory of this MR
was extended to arbitrary $T\tau$ by \citet{gornyi:2003,gornyi:2004}. In the
ballistic regime ($T\tau\gg 1)$, particularly relevant to high-mobility
structures, the MR caused by electron-electron interactions depends in an
essential way on the type of disorder, especially for $\omega_c\ll T$, where it
is strongly suppressed for smooth disorder. For $\omega_c\gg T$, if $1/\tau$ is
limited by smooth disorder, the Coulomb interaction-induced MR for $r_s\ll 1$
reads \cite{gornyi:2003,gornyi:2004}
\be
{\Delta\rho_{xx}\over\rho_{\rm D}}=-{3\zeta(3/2)\over 32\pi^{3/2}}{1\over
k_Fl}{(\omega_c\tau)^2\over (T\tau)^{1/2}}
\label{II.7}
\ee
for $T\tau\ll r_s^{-2}$, while for $T\tau\gg r_s^{-2}$ it retains the same
amplitude as in Eq.~(\ref{II.7}) but changes sign. If, by contrast, $1/\tau$ is
limited by white-noise disorder but the smooth component of disorder is still
strong enough to produce a contribution to the momentum relaxation rate
$1/\tau_L\gg \omega_c^3/k_F^2v_F^2$, the MR is enhanced compared to
Eq.~(\ref{II.7}) by a factor $4(\tau_L/\tau)^{1/2}$ \cite{gornyi:2004}.

A crossover in the temperature behavior of the quadratic-in-$B$ MR between the
diffusive and ballistic regimes was observed in the intermediate range of
$T\tau\sim 1$ in moderate-mobility GaAs/AlGaAs structures by \citet{li:2003}
[see also earlier works \cite{paalanen:1983,choi:1986}], in close agreement with
the theory for smooth disorder. The behavior of the MR in the crossover region
was also studied in Si/SiGe structures by \citet{olshanetsky:2006}. The
suppression of the MR as $T$ increases, observed by \citet{galaktionov:2006} in
samples similar to those in \cite{li:2003} in a wider range of $T\tau\agt 1$,
agrees favorably with the $T^{-1/2}$ scaling characteristic of the ballistic
regime. The enhancement of the MR compared to Eq.~(\ref{II.7}), which can be
largely attributed to the interplay of the short- and long-range components of
disorder, was reported in a number of experiments
\cite{olshanetsky:2006,galaktionov:2006,bockhorn:2011}, with a particularly
strong enhancement observed by \citet{bockhorn:2011} in ultra-high mobility
GaAs/AlGaAs structures.

\subsubsection{Classical magnetoresistance}
\label{s2.2.1}

Quite apart from the quantum MR, it was half a century ago that it was realized
that long-range inhomogeneities can have a profound effect on magnetotransport;
specifically, that even weak inhomogeneities with a spatial scale larger than
the mean free path $l$ can yield a strong MR
\cite{herring:1960,dreizin:1973,isichenko:1992}. In fact, even in
macroscopically homogeneous (on the scale of $l$) electron systems, disorder
with the correlation radius $d\gg k_F^{-1}$ can produce a strong quasiclassical
MR (QCMR). The QCMR was discussed in a variety of systems: in the 3D Coulomb
plasma \cite{polyakov:1986,murzin:1984}, in the 2D Lorentz-gas model
\cite{baskin:1978,bobylev:1995}, and in the 2DEG with smooth \cite{mirlin:1999}
and ``mixed" \cite{mirlin:2001,polyakov:2001} disorder. The essence of this
phenomenon is that transport retains signatures of the underlying quasiclassical
dynamics of electrons, which are not captured by the Boltzmann-Drude kinetic
theory. Specifically, the QCMR is due to correlations in the otherwise random
multiple scattering process at the points where the quasiclassical paths
self-intersect. The strength of these ``memory effects"---neglected in the
collision-integral formalism---grows as a power of $d/l$. Below, we briefly
discuss the results for the QCMR in a 2DEG.

The QCMR depends in an essential manner (even its sign does) on the correlation
properties of disorder. Let us first review the case of Gaussian (in the sense
of fluctuation statistics) disorder with the correlation radius $d\gg k_F^{-1}$.
The MR is in fact nonzero \cite{khveshchenko:1996,mirlin:1998b} within the
collision-integral formalism because of the cyclotron bending of electron paths
on the scale of $d$. This gives a small negative MR,
$\Delta\rho_{xx}/\rho_{\rm D}\sim -(d/R_c)^2$, where $R_c=v_F/\omega_c$ is the
cyclotron radius. Correlations of diffusive paths at the points of
self-intersection give rise to a much stronger positive MR: for $W_q$ from
Eq.~(\ref{II.3}) and $\omega_c\tau\gg 1$ \cite{mirlin:1999}
\be
{\Delta\rho_{xx}\over \rho_{\rm D}}=
{2\zeta(3/2)\over\pi}\left({d\over l}\right)^3 (\omega_c\tau)^{9/2}~.
\label{II.11}
\ee
The characteristic momentum transfer $q$ in the scattering processes leading to
Eq.~(\ref{II.11}) is $q\sim 1/\Lambda\ll 1/d$, where
\be
\Lambda=2\pi^{1/2}v_F\tau/(\omega_c\tau)^{3/2}
\label{II.12}
\ee
is the mean-square shift of the guiding center of a cyclotron orbit after one
revolution. The ratio $\Delta\rho_{xx}/\rho_{\rm D}$ in Eq.~(\ref{II.11}) is
much larger than that related to the effect of $B$ on the collision integral for
$\omega_c\tau\gg (l/d)^{2/5}$ (note that $l/d\sim 10^3$ in ultra-high mobility
structures) and becomes of order unity at $\omega_c\tau\sim (l/d)^{2/3}$. The
latter condition corresponds to $\Lambda\sim d$. As $B$ is increased further,
the strong positive MR is followed by an exponential falloff of $\rho_{xx}$
\cite{fogler:1997}:
\be
\ln (\rho_{xx}/\rho_{\rm D})\sim -(d/\Lambda)^{2/3}~,
\label{II.13}
\ee
which is due to the increasing adiabaticity of the electron dynamics and the
related classical localization. The MR in Eq.~(\ref{II.11}), induced by rare
($\Lambda\gg d$) self-intersections of diffusive trajectories, may be considered
as a precursor of the strong (adiabatic) localization.
Numerical simulations \cite{mirlin:1999} confirm the strong QCMR for the case of
smooth disorder. The QCMR of composite fermions scattered by effective random
magnetic
field was argued \cite{evers:1999} to explain the dependence of $\rho_{xx}(B)$
around a half-filling of LLs \citep[e.g.,][]{smet:1998}.

Rare strong scatterers randomly distributed on top of smooth weak potential
fluctuations can profoundly change the magnetotransport properties of a 2DEG
\cite{mirlin:2001,polyakov:2001}. As discussed in Sec.~\ref{s2.1}, the
combination of these two types of disorder---of comparable strength in what
concerns the momentum relaxation rate---is likely to be an adequate description
of a random potential in high mobility structures. The Drude formula
(\ref{II.5}) totally fails at $B\neq 0$ when only strong short-range scatterers
are present: in an important early work by \citet{baskin:1978}, a strong
negative MR was shown in the classical 2D Lorentz-gas model. In this model,
electrons are scattered by impenetrable hard disks (``voids"). In the limit of
the density of the voids $n_S\to\infty$ with the momentum relaxation time
$\tau_S$ held fixed, the model is exactly solvable \cite{bobylev:1995} for the
resistivity tensor $\hat\rho (B)$; in particular, $\rho_{xx}/\rho_{\rm D}\simeq
9\pi/8\omega_c\tau_S$ for $\omega_c\tau_S\gg 1$ and the MR is exponentially
small in the opposite limit. At finite $n_S$, the model shows a classical
metal-insulator transition at a critical value of $R_c\sim n_S^{-1/2}$
\cite{baskin:1978,bobylev:1995} and a quadratic MR in the low-$B$ limit
\cite{cheianov:2004}.

When both types of disorder are present, the momentum relaxation rate at zero
$B$ is $1/\tau\simeq 1/\tau_L+1/\tau_S$, where $1/\tau_{L,S}$ stands for the
contributions of smooth disorder $(L)$ and strong scatterers $(S)$. Consider the
relevant case of $\tau_S\ll\tau_L$ but smooth disorder being strong enough to
produce $\Lambda$ [Eq.~(\ref{II.12}) with $\tau\to\tau_L$] larger than both
correlation radii of disorder $d$ and $a$, where $a$ is the effective radius of
strong scatterers. Then \cite{mirlin:2001}
\be
\Delta\rho_{xx}/\rho_{\rm D}=-(\omega_c/\omega_0)^2
\label{II.15}
\ee
for $\omega_c\ll\omega_0$, where
\be
\omega_0=(2\pi n_S)^{1/2}v_F(2\tau_S^2/\tau_L\tau_{0S})^{1/4}~,
\label{II.16}
\ee
with $\tau_{0S}\sim \tau_S$ being the total scattering time for strong
scatterers. The ratio $(\omega_c/\omega_0)^2$ gives the fraction of the area
``explored" twice because of the self-intersections of electron paths, which
reduces the exploration rate and leads to a longer time between collisions with
{\it different} strong scatterers, hence the negative sign of the MR. This
should be contrasted with the positive MR (\ref{II.11}) for one-scale smooth
disorder, where the passages through the same area increase the scattering rate.
For $\omega_c\gg\omega_0$, the scattering rate is strongly suppressed by the
memory effects, which gives the $B^{-4}$ falloff as $B$ increases for
$n_SR_c^2\gg 1$ \cite{mirlin:2001,polyakov:2001},
\be
\rho_{xx}/\rho_{\rm D}\sim (\tau_S/\tau_L)(n_SR_c^2)^2~,
\label{II.17}
\ee
and a plateau of $\rho_{xx}$ for $n_SR_c^2\ll 1$ with
\be
\rho_{xx}/\rho_{\rm D}=\tau_S/\tau_L~,
\label{II.18}
\ee
i.e., the magnetic field ``switches off" short-range scatterers which give the
main contribution to $1/\tau$ at $B=0$. This behavior of the QCMR has been
observed in numerical simulations \cite{mirlin:2001}.

%%%%%%%%%%%%%%%%%%%%%%%%%%%%%%%%%%%%%%%%%%

\begin{figure}
\includegraphics[width=\columnwidth]{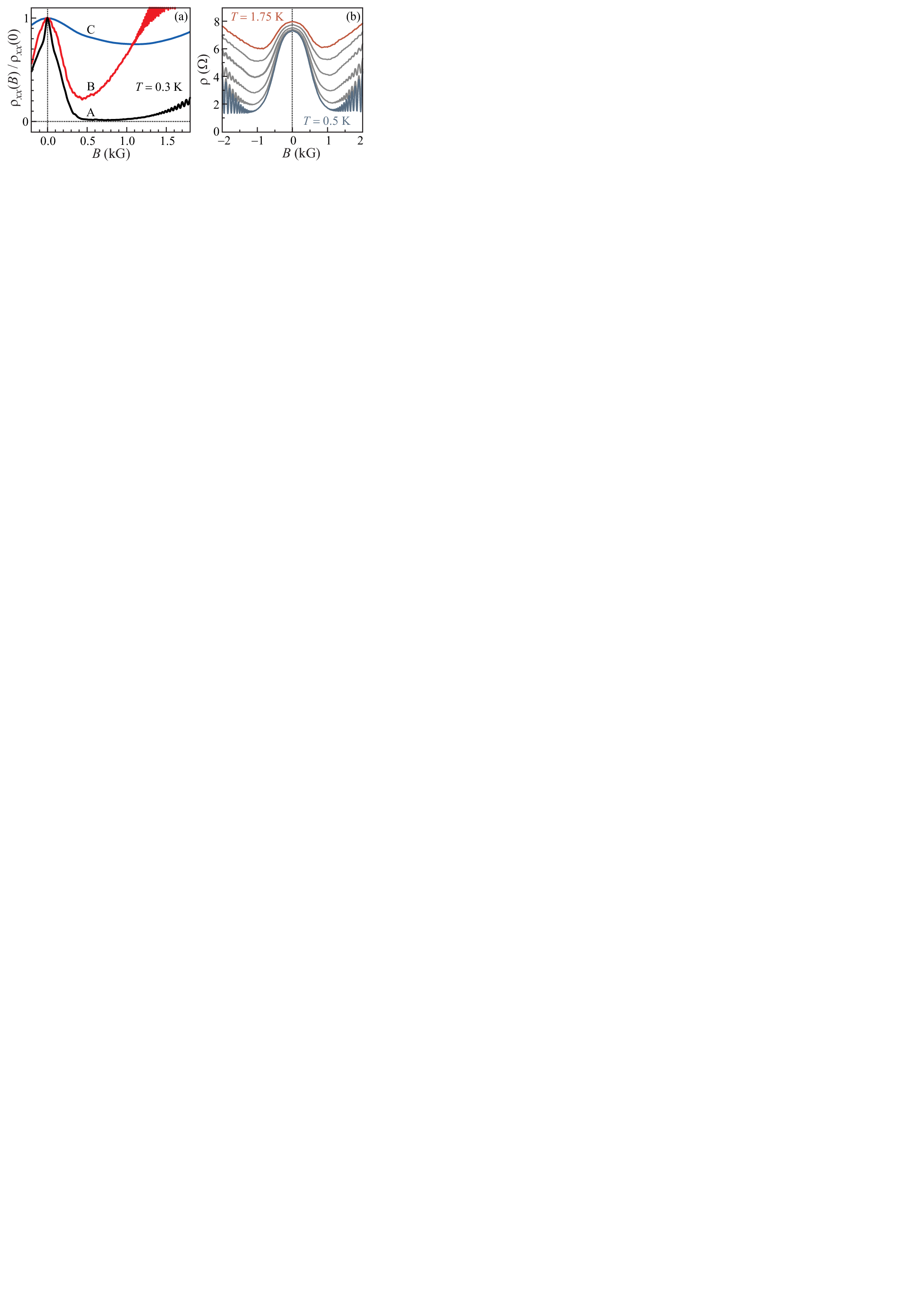}
\caption{(a) Magnetic-field dependence of the resistivity $\rho_{xx}$ (in units of
$\rho_{xx}$ at $B=0$) of three different samples A, B, and C with the mobility of (A)
$3\times 10^7\,{\rm cm}^2/{\rm V\,s}$, (B) $1.2\times 10^7\,{\rm cm}^2/{\rm
V\,s}$, and (C) $8.6\times 10^6\,{\rm cm}^2/{\rm V\,s}$ at $T=0.3\,{\rm K}$ and the electron density of (A) $2.9\times 10^{11}\,{\rm cm}^{-2}$
(B) $4.5\times 10^{11}\,{\rm cm}^{-2}$, and (C) $6\times 10^{11}\,{\rm cm}^{-2}$. Adapted from \citet{dai:2010}.
% RR
(b) Magnetic-field dependence of the resistivity
for different temperatures from 0.5 to $1.75 \,{\rm K}$ in steps of
$0.25 \,{\rm K}$ in a sample with $\mu\simeq 5.4 \times 10^6\,{\rm
cm^2/V\,s}$ and $n_e\simeq 1.6\times 10^{11}$ cm$^{-2}$. Adapted from \citet{hatke:2012a}.
}
\label{II.f0}
\end{figure}

%%%%%%%%%%%%%%%%%%%%%%%%%%%%%%%%%%%%%%%%%%%%

A negative MR with a pronounced resistance minimum at about $0.5\,{\rm kG}$ was
observed in early works on ultra-high mobility structures
\cite{umansky:1997,smet:1997}. Recently, a strong negative MR in the same range
of $B$ was reported in similar structures by \citet{dai:2010} and
\citet{hatke:2011b,hatke:2012a}.
% RR
In particular, in one of the samples studied by
\citet{dai:2010}, namely sample A
in Fig.~\ref{II.f0}a, the resistivity was shown
to decrease by a factor of about 50 between $B=0$ and $0.5\,{\rm kG}$ and
exhibit a wide plateau for larger $B$ before the onset of magnetooscillations,
in qualitative agreement with the picture of the QCMR for the model of
two-component disorder. With decreasing mobility (samples B and C in
Fig.~\ref{II.f0}a), the plateau disappears but the negative MR persists. Analysis
of the measured dependence of $\rho_{xx}(B)$ led \citet{dai:2010} to conclude
that the observed phenomenon can be consistently described in terms of the QCMR
theorized by \citet{mirlin:2001,polyakov:2001}.
% RR
A huge negative MR, with $\rho_{xx}(B)$ at the minimum of the $B$
dependence being at $T\simeq 0.3-0.5\,{\rm K}$ (for samples with different
mobilities) several times smaller than at $B=0$, was reported by \citet{hatke:2012a}
 to be suppressed as temperature increases and disappear at about $2\,{\rm
K}$. The $T$ dependence for one of the samples is shown in Fig.~\ref{II.f0}b.
% RR
According to the analysis performed by \citet{hatke:2012a}, the observed
$T$ dependence is too strong to be described in terms of the theory of the
quantum MR discussed in Sec.~\ref{s2.2.1a}.

The behavior of $\rho_{xx}(B)$ in the model of two-component disorder proves to
be substantially more intricate when, as $B$ increases, $\Lambda$ becomes
smaller than $2d$ (in high-mobility structures $d\gg a$) and scattering by the
smooth random potential acquires the character of an adiabatic drift
\cite{polyakov:2001}. Although the adiabaticity of scattering suppresses
$\rho_{xx}$ when only smooth disorder is present [Eq.~(\ref{II.13})], it can
also give rise to the growth of $\rho_{xx}$ with increasing $B$ in the presence
of short-range disorder. This behavior can be most clearly seen in the
``hydrodynamic" model where short-range disorder is characterized by white-noise
correlations and the problem is mapped onto that of advection-diffusion
transport \cite{isichenko:1992}. In the limit of large $B$, the hydrodynamic
model predicts $\rho_{xx}/\rho_{\rm D}\sim (\tau_S^2v_Fd/\tau_L R_c^2)^{5/13}$.
Note that for realistic parameters the full-fledged QCMR in the adiabatic regime
competes with the effects of Landau quantization (Sec.~\ref{s2.3}). A growth of
$\rho_{xx}$ with increasing $B$ in the experiments by \citet{dai:2010}
and \citet{hatke:2011b} was observed in a range of $B$ well
before the onset of oscillations [see also the earlier results
\citep[e.g.,][]{galaktionov:2006} on a positive MR observed at sufficiently high
$T$ in moderate-mobility structures]. The huge negative MR observed by
\citet{dai:2010} and \citet{hatke:2011b,hatke:2012a},
% RR
as well as the growth of $\rho_{xx}$
at higher $B$, warrants further study. In this respect, it is interesting to
note that, experimentally, there appears to be a connection between the huge
negative MR and a pronounced anomaly in the photoresponse; for details see
Sec.~\ref{s7.2.1prim}.

\subsubsection{Classical magnetooscillations of the ac conductivity}
\label{s2.2.3}

\begin{figure}
\includegraphics[width=\columnwidth]{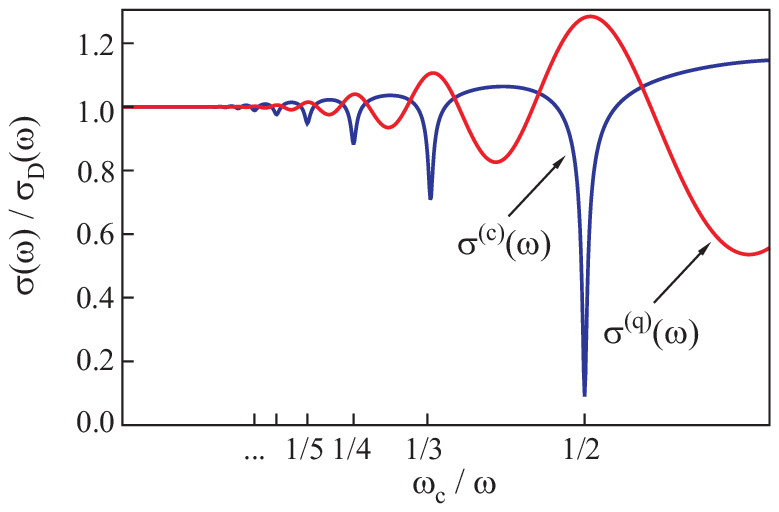}
\caption{
Classical (c) and quantum (q) oscillatory ac conductivity [normalized to the
Drude
conductivity $\sigma_{\rm D}(\omega)$] vs $\omega_c/\omega$ for
$\omega/2\pi=100\,{\rm GHz}$, $\tau_L=0.6\,{\rm ns}$, $\tau_L/\tau_{\rm q}=50$,
$\tau_S/\tau_L=0.1$, and $a/\Lambda=0.25$ at $\omega_c/\omega=1/2$. From
\citet{dmitriev:2004}.
}
\label{II.f1}
\end{figure}

Within the Boltzmann kinetic theory, the dissipative diagonal ac conductivity
$\sigma(\omega)=\sigma_+(\omega)+\sigma_-(\omega)$, which determines the
absorption rate for linearly polarized electromagnetic waves, is given by the
Drude formula
\be
\sigma_{\rm D,\pm}(\omega) = {e^2n_e\tau\over 2m}{1\over
1+(\omega_c\pm\omega)^2\tau^2}~.
\label{II.19}
\ee
The quasiclassical correlations in the dynamics of multiple scattering, which
lead to the MR discussed in Sec.~\ref{s2.2.1}, can also strongly modify the ac
response of a 2DEG at $B\neq 0$. In particular, they yield
periodic-in-$\omega/\omega_c$ oscillations in $\sigma(\omega)$ which are of
essentially classical nature. For the model of two-component disorder
(Sec.~\ref{s2.2.1}), the oscillatory classical correction
$\Delta\sigma^{(c)}(\omega)=\sigma^{(c)}(\omega)\!-\!\sigma_{\rm D}(\omega)$ in
units of the Drude conductivity $\sigma_{\rm D}(\omega)$ takes, in the large-$B$
limit, the form of sharp resonant features at the harmonics of the cyclotron
resonance (CR) at $\omega=M\omega_c$ with $|M|=1,2,\ldots$, the amplitude of
which is given \cite{dmitriev:2004} by
\be
{\Delta\sigma^{(c)}(M\omega_c)\over \sigma_{\rm
D}(M\omega_c)}=-{a(\omega_c\tau_L)^{1/2}\over \sqrt{3\pi}|M|\Lambda}~,
\label{II.19a}
\ee
where $a=1/n_Sv_F\tau_S$, and the width $\Gamma_M=3M^2/2\tau_L+1/\tau_{0S}$.
Equation (\ref{II.19a}) is valid for $\Gamma_M\ll\omega_c$. In the
opposite limit of strongly damped oscillations, $\Delta\sigma^{(c)}(\omega)$
reads \cite{dmitriev:2004}
\be
{\Delta\sigma^{(c)}(\omega)\over\sigma_{\rm D}(\omega)}
=-{a\over \pi^{1/2}\Lambda}
\,\cos{2\pi\omega\over\omega_c}
\,\exp \left(-{3\pi\omega^2\over\omega_c^3\tau_L}\right)~.
\label{II.19b}
\ee
The behavior of $\sigma^{(c)}(\w)$ is illustrated in Fig.~\ref{II.f1}, where
also the quantum
oscillations related to Landau quantization (Sec.~\ref{s2.3.2}), are shown. One
sees that the
classical oscillations can be much more pronounced than the quantum ones for
$\tau_{\rm q}\ll\tau_L$ (which is the case in high-mobility structures).
Resonant features, of similar origin, at the harmonics of the CR were also
discussed for the case of a random antidot array \cite{polyakov:2002}. The CR in
the opposite limit of a smooth random potential was considered by
\citet{fogler:1998} who found, in particular, a sharp jump, as $B$ is increased,
in the CR broadening from the Drude width $1/\tau$ to the width of separated LLs
$(\omega_c/\tau)^{1/2}$ at the crossover to the adiabatic localization regime.

It is also worth noting that the ac conductivity exhibits magnetooscillations
that are not associated with either the long-time correlations in disorder-induced
scattering [Eq.~(\ref{II.19b})] or Landau quantization but are entirely due to
electron-electron interactions \cite{sedrakyan:2008b}. Namely, they are related
to the sensitivity of screening of the impurity potential to the cyclotron
bending of quasiclassical electron paths. However, in contrast to the
oscillations in Fig.~\ref{II.f1}, the period of the interaction-induced
oscillations in $1/\omega_c$ is much larger (by a factor of the order of
$(\epsilon_F/|\omega|)^{1/2}\gg 1$) than $1/|\omega|$, so that for given $\omega$
these oscillations might only develop for much smaller $B$ and much smaller $T$.

%%%%%%%%%%%%%%%%%%%%%%%%%%%%%%%%%%%%%%

%%%%%%%%%%%%%%%%%%%%%%%%%%%%%%%%%%%%%%%

\subsection{Quantum magnetooscillations}
\label{s2.3}

In our discussion so far, we neglected the effects related to Landau
quantization by assuming that the electron DOS $\nu(\varepsilon)$ at
$\varepsilon$ equal to the Fermi energy $\varepsilon_F$ does not depend on $B$.
In this section, we review the oscillatory behavior of the transport
coefficients of a 2DEG with varying $\varepsilon_F/\omega_c$ and
$\omega/\omega_c$---which arises from Landau quantization and is collectively
known as ``quantum magnetooscillations"---in the linear-response regime [see
\citet{shoenberg:1984} for a review of the basic concepts and of the results for
magnetooscillations in 3D metals, and \citet{ando:1982a} for an early review of
magnetooscillations in a 2DEG].

\subsubsection{Magnetooscillations of the density of states}
\label{s2.3.1a}

In a clean noninteracting 2DEG the DOS per spin is a sum of delta-functions at
energies equal to those of LLs:
\be
\nu(\varepsilon)=(2\pi\lb^2)^{-1}\sum_N\delta (\varepsilon-\varepsilon_N)~,
\label{II.22}
\ee
where $\lb=(m\omega_c)^{-1/2}$ is the magnetic length and
$\varepsilon_N=(N+1/2)\omega_c$. As throughout the review, we concentrate on the
limit of moderately strong magnetic fields in which the number of occupied LLs
is large. For $N\gg 1$, the broadening of the $N$th peak in Eq.~(\ref{II.22}) by
disorder changes only weakly when $N$ changes by unity. The DOS of the $N$th LL
$\nu^{(N)}(\varepsilon)$ is then written as
\be
\nu^{(N)}(\varepsilon)=-{1\over 2\pi^2\lb^2}\,{\rm Im}\,{1\over
\varepsilon-\varepsilon_N-\Sigma(\varepsilon)}
\label{II.23}
\ee
where
$\Sigma (\varepsilon)$ is the self-energy of the disorder-averaged retarded
Green's function in the LL representation. The approximation is that $\Sigma
(\varepsilon)$ is treated as independent of $N$. The singular character of
Eq.~(\ref{II.22}) implies that $\Sigma (\varepsilon)$ cannot be calculated at
the level of the Born approximation when LLs are separated, i.e., when disorder
is so weak that $-{\rm Im}\,\Sigma(\varepsilon_N)\alt\omega_c$.

The mathematically simplest scheme to calculate $\Sigma (\varepsilon)$ is the
self-consistent Born approximation (SCBA)
\cite{ando:1974,raikh:1993,laikhtman:1994} in which $\Sigma (\varepsilon)$ obeys
\be
\Sigma(\varepsilon)={\omega_c\over 2\pi \tau_{\rm q}}\sum_N {1\over
\varepsilon-\varepsilon_N-\Sigma(\varepsilon)}~,
\label{II.24}
\ee
where $1/\tau_{\rm q}$ is the total (quantum) scattering rate
[Eq.~(\ref{II.3a})] at $B=0$. The SCBA gives in the limit $\omega_c\tau_{\rm
q}\gg 1$, where LLs are well separated:
\be
\nu^{(N)}(\varepsilon)=\nu_0\tau_{\rm q}\sqrt{{2\omega_c\over\pi\tau_{\rm
q}}-(\varepsilon-\varepsilon_N)^2}
\label{II.25}
\ee
if the argument of the square root is positive and zero otherwise, with
$\nu_0=m/2\pi$ being the DOS per spin at $B=0$. \citet{benedict:1986} and
\citet{carra:1989} showed that the SCBA---and thus the semicircle shape of the
broadened LLs in Eq.~(\ref{II.25})---are exact for white-noise disorder in the
limit $N\to\infty$. For a nonzero correlation length $d$ of Gaussian (in the
sense of fluctuation statistics) disorder, it was recognized by
\citet{raikh:1993} that the SCBA is a parametrically accurate approximation for
$N\gg 1$ provided $d\ll\lb$. The hard gaps in between are an artefact of the
SCBA, but the exponential tails of the DOS which fill the gaps smear only
slightly (for $N\gg 1$) the boundaries of the semicircles
\cite{efetov:1989,benedict:1987}. The gaps predicted by the SCBA close at
$\omega_c\tau_{\rm q}=\pi/2$ \cite{ando:1974b,laikhtman:1994}.

Apart from the conditions $N\gg 1$ and $d\ll\lb$, the SCBA also assumes that
disorder is weak; specifically, that at zero $B$ the conventional Born
approximation is valid, which for Gaussian disorder means\footnote{According to
Eq.~(\ref{II.3b}) with $W_q$ from Eq.~(\ref{II.3}) and $n_e=n_i$, this condition
can only be marginally satisfied for randomly distributed remote impurities.}
$d\ll v_F\tau_{\rm q}$. Moreover, the SCBA crucially assumes that the
non-Gaussian component of fluctuations of the random potential can be neglected.
The Gaussian character of disorder implies that individual impurities are weak
and their concentration $n_i$ obeys $n_i\max\{d^2,S\}\gg 1$, where $S$ is the
area over which disorder is averaged because of the quantum uncertainty of
electron paths. For separated LLs, the characteristic $S$ is given by the area
$2\pi\lb^2$ occupied by one electronic state, which is assumed within the
SCBA to be much larger than $d^2$. The DOS for disorder with non-Gaussian
statistics of fluctuations was considered beyond the SCBA by
\citet{ando:1974a,ando:1974b}, by \citet{brezin:1984} for $N=0$, and by
\citet{benedict:1986} for separated LLs in the limit $N\gg 1$. In particular, it
was found in these works that within a model of delta-function impurities there
develops a singularity in the DOS at $\varepsilon=\varepsilon_N$ for $2\pi
n_i\lb^2<1$, independently of the strength of the impurities.

If $\omega_c\tau_{\rm q}\ll 1$, Landau quantization only leads to a weak
modulation of the DOS and the SCBA predicts that $\nu^{(N)}(\varepsilon)$ in
this limit is described by a Lorentzian with ${\rm
Im}\,\Sigma(\varepsilon)=-1/2\tau_{\rm q}$. The DOS is then represented, by
means of the Poisson summation formula, as a sum over harmonics in the following
form:\footnote{Equation
(29) describes the asymptotic behavior of the amplitude of the $k$th harmonic in
the limit $\omega_c\tau_{\rm q}\ll 1$ (with exponential accuracy).
In the exact SCBA formula, valid for arbitrary
$\omega_c\tau_{\rm q}$ \cite{vavilov:2004},
each harmonic contains an extra factor $g_k=k^{-1} L_{k-1}^1(2\pi k/\wc\tq)$,
where $L_{k-1}^1$ is the generalized Laguerre polynomial.}
\be
\nu(\varepsilon)=\nu_0\left[1+2\sum_{k=1}^\infty (-\delta)^k\cos{2\pi
k\varepsilon\over\omega_c}\right]~,
\label{II.26}
\ee
where the Dingle factor \cite{shoenberg:1984}
\be
\delta=\exp (-\pi/\omega_c\tau_{\rm q})~.
\label{II.27}
\ee
The first harmonic---the least damped---yields the leading oscillatory
correction $\Delta\nu(\varepsilon)$ to the DOS:
\be
\Delta\nu(\varepsilon)=-2\nu_0\delta\cos(2\pi\varepsilon/\omega_c)~.
\label{II.28}
\ee

For $d\gg\lb$, the SCBA fails and instead the quasiclassical approximation
in which the DOS is represented as
\be
\nu^{(N)}(\varepsilon)=(2\pi\lb^2)^{-1}\left<\delta[
\varepsilon-\varepsilon_N-\bar{V}({\bf r})]\right>
\label{II.29}
\ee
is valid \cite{raikh:1993}. Here
\be
\bar{V}({\bf r})=\int_0^{2\pi}\!{d\theta\over
2\pi}\,V(x+R_c\cos\theta,y+R_c\sin\theta)
\label{II.30}
\ee
is the effective random potential acting on the guiding center of the cyclotron
orbit. For the case of Gaussian statistics of fluctuations of $V({\bf r})$, the
shape of $\nu^{(N)}(\varepsilon)$ is also Gaussian and is given
\cite{raikh:1993,mirlin:1996} by\footnote{The Gaussian shape of LLs was also obtained in an early work
by \cite{gerhardts:1975} within the cumulant expansion of the self-energy in the {\it
time} domain to first order in $W_q$. However, the control
parameters of the approximation were not correctly specified there. In
particular, Eq.~(34) was argued by \cite{gerhardts:1975} to be valid for disorder
with the correlation length $d\to 0$ (i.e., in the limit $d\ll\lambda$), whereas
the actual condition is $d\gg\lambda$.
% RR
%Equation (\ref{II.31}) was also
%obtained by \citet{gerhardts:1975} within a cumulant expansion of the
%self-energy (\ref{II.23}) in the {\it time} domain to first order in $W_q$.
%However, the control parameters of the approximation were not clearly specified;
%in fact, Eq.~(\ref{II.31}) was argued by \citet{gerhardts:1975} to describe
%short-range disorder with $d\ll\lb$, contrary to the actual condition which
%is the opposite.
}
\be
\nu^{(N)}(\varepsilon)={1\over 2\pi\lb^2}\left({\tau_{\rm
q}\over\omega_c}\right)^{1/2}\exp \left[-{\pi\tau_{\rm
q}\over\omega_c}(\varepsilon-\varepsilon_N)^2\right]
\label{II.31}
\ee
for $\lb\ll d\ll R_c$ and by
\be
\nu^{(N)}(\varepsilon)={1\over 2\pi\lb^2}[2\pi W(0)]^{-1/2}\exp
\left[-{(\varepsilon-\varepsilon_N)^2\over 2W(0)}\right]~.
\label{II.32}
\ee
for $d\gg R_c$ [in which limit $\bar{V}({\bf r})\simeq V({\bf r})$]. In
Eq.~(\ref{II.32}), $W(0)$ is the variance of $V({\bf r})$ [Eq.~(\ref{II.2b})].
Note that the broadening grows with increasing $B$ and saturates at $d\gg R_c$.
Equations (\ref{II.31}) and (\ref{II.32}) describe in effect inhomogeneous
broadening and are valid for both separated and overlapping LLs.

The sum of $\nu^{(N)}(\varepsilon)$ from Eq.~(\ref{II.31}) over $N$ gives the
sum over harmonics
\be
\nu(\varepsilon)=\nu_0\left[1+2\sum_{k=1}^\infty (-1)^k\delta^{k^2}\!\cos{2\pi
k\varepsilon\over\omega_c}\right]~.
\label{II.33}
\ee
Note that the only difference between Eqs.~(\ref{II.26}) and (\ref{II.33}) is in
the power of the factor $\delta$: it is $k$ in the former case and $k^2$ in the
latter. It follows that the damping of the leading $(k=1)$ oscillatory
correction in the limit of overlapping LLs is given by Eq.~(\ref{II.28}) in both
cases, despite the shape of LLs being quite different (Lorentzian vs Gaussian).
If $d\gg R_c$, the exponent of the damping factor for the $k$th harmonic is
proportional to $k^2$, similarly to Eq.~(\ref{II.33}), with the $k=1$ term in
$\nu(\varepsilon)$ given by
\be
\Delta\nu(\varepsilon)=-2\nu_0\exp[-2\pi^2W(0)/\omega_c^2]
\cos(2\pi\varepsilon/\omega_c)~.
\label{II.34}
\ee
The Dingle plot [the term conventionally used to describe the behavior of
$\ln\Delta\nu(\varepsilon)$ plotted in the low-$T$ limit as a function of $1/B$]
is seen to be quadratic in Eq.~(\ref{II.34}), in contrast to Eq.~(\ref{II.28})
where it is linear. The quadratic Dingle plot\footnote{If the damping of the
oscillations in the DOS is solely due to remote impurities, a finite range of
$B$ within which LLs are not separated and the Dingle plot is quadratic exists
provided disorder is sufficiently strong; specifically, if the ``out-scattering
length" $v_F\tau_{\rm q}\ll d$. In ultra-high mobility samples, $v_F\tau_{\rm
q}$ is typically of the order of $d$.} in the DOS can thus be used to probe the
presence of long-range (with the correlation radius $d\agt R_c$) inhomogeneities
in the sample.

\subsubsection{Shubnikov-de Haas oscillations}
\label{s2.3.1}

The magnetooscillations of the DOS give rise to the oscillatory behavior of the
dc transport coefficients as $\varepsilon_F/\omega_c$ is varied---the
Shubnikov-de Haas (SdH) effect. Within the SCBA \cite{ando:1974}, the
conductivity tensor for $N\gg 1$ was obtained for short-range $(d\ll k_F^{-1})$
disorder by \citet{ando:1974b} for the diagonal conductivity $\sigma_{xx}$ and
by \citet{ando:1975} for the Hall conductivity $\sigma_{xy}$. Note that the use
of the SCBA to treat also long-range disorder with $d\gg\lb$ in the early
works \cite{ando:1982a} is not justified since this condition violates the
applicability of the SCBA \cite{raikh:1993}. A generalization of the SCBA
approach which provides a framework for studying transport in the case of
disorder with an arbitrary correlation length $d\ll\lb$ was developed by
\citet{dmitriev:2003}. An essential ingredient of the theory describing
long-range disorder
% RR
($d\gtrsim k_F^{-1}$) is the inclusion of vertex corrections in averaging over
disorder. The result is that the theory
% RR
for all $d\ll \lambda$ can be formulated solely in terms of the
oscillatory DOS $\nu(\varepsilon)$ and the transport scattering rate
$\tau_B(\varepsilon)$ which is renormalized by Landau quantization as follows:
\be
{1\over\tau_B(\varepsilon)}={1\over\tau}{\nu(\varepsilon)\over\nu_0}~.
\label{II.35}
\ee
Expressed in terms of $\nu(\varepsilon)$ and $\tau_B(\varepsilon)$,
$\sigma_{xx}$ (neglecting the Zeeman splitting) reads \cite{dmitriev:2003}
\be
\sigma_{xx}=e^2v_F^2\int\!d\varepsilon\left(-{\partial
f_\ve^T\over\partial\varepsilon}\right){\nu(\varepsilon)\tau_B(\varepsilon)\over
1+\omega_c^2\tau_B^2(\varepsilon)}~,
\label{II.36}
\ee
where $f_\varepsilon^T$ is the thermal distribution function. Equation
(\ref{II.36}) describes both separated and overlapping LLs and is valid also in
the crossover in between. Note that, compared to the Drude formula $\sigma_{\rm
D}=(e^2n_e\tau/m)/(1+\omega_c^2\tau^2)$, Landau quantization manifests itself in
Eq.~(\ref{II.36}) in two ways: in the renormalization
$\tau\to\tau_B(\varepsilon)$ and in the appearance of the additional factor
$\nu(\varepsilon)/\nu_0$ in the numerator [where, in view of Eq.~(\ref{II.35}),
the DOS oscillations cancel out]. As $B$ is varied, $\sigma_{xx}$
at zero $T$ is represented in the limit of separated LLs as a series of peaks
the height of which is given by $\sigma_{xx}^{\rm max}\simeq\sigma_{\rm
D}(2\omega_c\tau_{\rm q}/\pi)\gg\sigma_{\rm D}$. For overlapping LLs, the
oscillatory correction $\Delta\sigma$ to $\sigma_{\rm D}$ reads
\be
{\Delta\sigma\over\sigma_{\rm D}}\simeq{2\omega_c^2\tau^2\over
1+\omega_c^2\tau^2}{\Delta\nu(\varepsilon_F)\over\nu_0}\,{\cal F}\!\left({2\pi^2
T\over\omega_c}\right)~,
\label{II.36a}
\ee
where $\Delta\nu(\varepsilon)$ is given by Eq.~(\ref{II.28}) and the factor
${\cal F}(x)=x/\sinh x$ describes the thermal averaging of the oscillations.
Importantly, in the case of long-range disorder $(k_Fd\gg 1$), the magnetic
field may be ``classically strong" in the sense of $\omega_c\tau\gg 1$ and at
the same time, if $\omega_c\tau_{\rm q}\ll 1$, lead to only weak SdH
oscillations even at $T=0$. Note also that for $T\gg\omega_c\gg\tau_{\rm
q}^{-1},\tau^{-1}$ the thermal averaging over the contributions of separated LLs
in Eq.~(\ref{II.36}) leads to a nonoscillatory contribution to the MR
\be
\rho_{xx}/\rho_{\rm D}=(8/3\pi)(2\omega_c\tau_{\rm q}/\pi)^{1/2}~.
\label{II.36b}
\ee

Extending Eq.~(\ref{II.35}) to the Hall conductivity yields
\be
\sigma_{xy}=-{en_ec\over B}+{e^2v_F^2\over
\omega_c}\int\!d\varepsilon\left(-{\partial
f_\ve^T\over\partial\varepsilon}\right){\nu(\varepsilon)\over
1+\omega_c^2\tau_B^2(\varepsilon)}~,
\label{II.37}
\ee
where the first term describes a collisionless drift of electrons in crossed
electric and magnetic fields. The oscillatory MR for overlapping LLs is then
given by\footnote{Inelastic electron-electron scattering (or, for that matter,
any scattering whose strength is proportional to $\varepsilon^2+\pi^2T^2$) does
not lead to an additional exponential damping of the SdH oscillations for
$T\gg\omega_c$ \cite{martin:2003,adamov:2006}. \label{fn_sdho}}
\be
{\Delta\rho\over\rho_{\rm D}}\simeq 2{\Delta\nu(\varepsilon_F)\over\nu_0}\,{\cal
F}\!\left({2\pi^2 T\over\omega_c}\right)
\label{II.38}
\ee
for arbitrary $\omega_c\tau\ll\tau/\tau_{\rm q}$. It is worth noting that the
main contribution to the oscillatory MR (\ref{II.38}) comes from the
oscillations of $\sigma_{xx}$ for $\omega_c\tau\gg 1$ and from the oscillations
of $\sigma_{xy}$ for $\omega_c\tau\ll 1$.
% RR
In the limit of white-noise disorder ($\tau=\tau_{\rm q}$),
%In the case of short-range disorder $(\tau=\tau_{\rm q})$,
 Eqs.~(\ref{II.36}) and (\ref{II.37}) agree with those
derived by \citet[Eq.~(2.12)]{ando:1974b} and by \citet[Eq.~(3.26)]{ando:1975},
respectively.\footnote{The expansion of $\sigma_{xy}$ to first order in $\delta$
obtained by \citet{ando:1975} was also derived by means of a different
representation of the Kubo formula for Hall transport by \citet{isihara:1986}.}
In the case of long-range disorder, they confirm the form of SdH oscillations
hypothesized by \citet{coleridge:1989}.

It is worth noting that the Fermi energy $\varepsilon_F$ for a fixed electron
density $n_e$ also exhibits magnetooscillations. These are qualitatively
important for the dependence of the MR on $B$ in the case of separated LLs at
zero $T$: it is because of the oscillations of $\varepsilon_F$ that $\rho_{xx}$
as a function of $B$ does not have gaps similar to those in the dependence of
$\nu(\varepsilon)$. For overlapping LLs, the oscillations of $\varepsilon_F$
yield only a subleading (in $\omega_c/\varepsilon_F\ll 1$) oscillatory term in
the MR. The thermal averaging for $T\gg\omega_c$ suppresses the oscillations of
$\varepsilon_F$ exponentially, similarly to Eq.~(\ref{II.36a}).

As discussed in Sec.~\ref{s2.1}, disorder in high-mobility heterostructures is a
mix of long- and short-range components characterized by vastly different
correlation radii. Because of the large ratio $\tau/\tau_{\rm q}$ for scattering
off the long-range component of disorder, it is possible that the mobility $\mu$
and the Dingle factor $\delta$ are determined by different sources of disorder.
Specifically, it is likely that, in ultra-high mobility samples, $\mu$ is
limited by background impurities and interface roughness, whereas the damping of
the magnetooscillations is mainly due to scattering off remote impurities.

Note also that, in real samples, there may exist ultra-long range
inhomogeneities with a correlation radius larger than $R_c$ in the regime of SdH
oscillations. These inhomogeneities may not affect the mobility but will provide
an additional damping of the oscillations of $\Delta\rho/\rho$ in
Eq.~(\ref{II.38}), which is described by the exponential factor in
Eq.~(\ref{II.34}). If $W(0)\gg \omega_c/\tau_q$, the damping of the SdH
oscillations will be mainly determined by these macroscopic inhomogeneities and
the Dingle plot for the MR will be quadratic in $1/B$, as was indeed reported
for some SdH measurements \citep[e.g.,][]{coleridge:1991,coleridge:1997}.

So far, the magnetooscillations of the MR and the DOS have been directly related
to each other, being characterized by the same Dingle factors. One important
reason why this is not true in general is the suppression of the transport
coefficients by localization effects. These
come in two varieties: classical and quantum. An example of classical
localization is the adiabatic drift along equipotential lines of a smooth random
potential, which results in the exponential suppression \cite{fogler:1997} of
$\sigma_{xx}$ with increasing $B$ [Eq.~(\ref{II.13})]. Oscillations of the MR
are then associated with the oscillatory DOS for only a subset of electron
paths; specifically, for those giving the main contribution to $\sigma_{xx}$.
The statistical properties of disorder present along the ``conducting" paths do
not necessarily coincide with those on average, which leads in general to
different Dingle factors for the DOS and the MR. The difference was shown
\cite{evers:1999} to be very pronounced for the case of a spatially random
magnetic field.

Quantum localization becomes strong when $\sigma_{xx}$ drops with increasing $B$
to a value of the order of the conductance quantum $e^2/2\pi$, at which field
the QH effect sets in.\footnote{Strictly speaking,
single-particle states of a 2DEG at $B\neq 0$ are localized except those at a
discrete set of critical energies---this localization constitutes the essence of
the QH effect. However, the temperature needed to observe localization is
exponentially small if $\sigma_{xx}\gg e^2/2\pi$ in the absence of localization,
so that in reality the QH effect becomes pronounced only when $\sigma_{xx}$ per
spin drops down to the values of a few conductance quanta.} Importantly, if
remote impurities, distributed randomly with the density $n_i=n_e$, are the only
source of disorder, the cyclotron frequency for the crossover field $B_{\rm
loc}$ obeys $\omega_c\tau_{\rm q}\sim (k_Fd)^{-1/2}\ll 1$, i.e., strong
localization develops when LLs are still well overlapped and the DOS does not
exhibit any gaps. If, however, the short-range component of disorder
substantially decreases $\tau$ while not affecting $\tau_{\rm q}$ (which is the
likely situation in ultra-high mobility structures, see above), $B_{\rm loc}$
shifts upward. It follows that it is only because of the admixture of
sufficiently strong short-range disorder that developed SdH oscillations can be
observed in high mobility samples before the crossover to the QH regime.

At this point, it is also worth emphasizing that quantum magnetooscillations of
the MR may not be related to those of the DOS at all. Even if the latter are
neglected, the quantum interference of diffusive waves for $\sigma_{xx}\gg
e^2/2\pi$ gives rise to the QH oscillations $\Delta\sigma_{xx}\propto\cos(2\pi
g_{xy})\exp (-2\pi g_{xx})$, where $g_{xx,xy}$ is $\sigma_{xx,xy}$ in units of
$e^2/2\pi$ \cite{pruisken:1990}. These are the oscillations into which the QH
effect transforms as $B$ is decreased---not into the SdH oscillations, contrary
to the common belief. Note that the period of the QH oscillations corresponds to
one flux quantum through the area $\min\{l^2,R_c^2\}$. That is, the
$B$-dependent interference occurs within the area the size of which is not
related to (and is smaller than) the localization length, so that the
observation of these oscillations for $\sigma_{xx}\gg e^2/2\pi$ does not require
exponentially low $T$---in contrast to the QH plateaus. See \citet{mirlin:1998b}
and \citet{evers:1999} for a discussion of the interplay of the SdH and QH
oscillations in the case of a random magnetic field.

\subsubsection{Quantum magnetooscillations of the ac conductivity}
\label{s2.3.2}

The oscillations of the DOS induced by Landau quantization have important
consequences also for ac transport. Specifically, the Drude formula in
Eq.~(\ref{II.19}) for $\sigma_\pm(\omega)$, which describes a single peak of the
CR, and Eq.~(\ref{II.36}) for the oscillatory $\sigma_{xx}$ at $\omega=0$
generalize\footnote{Note that Eq.~(\ref{II.39}) differs in an essential way from
the expression for $\sigma_\pm(\omega)$ obtained by \citet{ando:1975a} (for the
case of short-range disorder). In particular, the $\varepsilon$-dependent
factors in the integrand of Eq.~(\ref{II.39}) do not reduce near the CR to the
simple product
$(f_\varepsilon^T-f_{\varepsilon+\omega}^T)\,
\nu(\varepsilon)\nu(\varepsilon+\omega)$ in the limit of separated LLs, as
suggested by \citet{ando:1975a} for the CR lineshape.} to \cite{dmitriev:2003}
\bea
&&\hspace{-1cm}\sigma_\pm(\omega)={e^2v_F^2\over 2\omega}\nonumber\\
&&\hspace{-0.5cm}\times\int\!d\varepsilon\,{(f_\varepsilon^T-f_{
\varepsilon+\omega}^T)\nu(\varepsilon)\tau_B^{-1}
(\varepsilon+\omega)\over
[\tau_B^{-2}(\varepsilon)+\tau_B^{-2}(\varepsilon+\omega)]
/2+(\omega\pm\omega_c)^2}~.
\label{II.39}
\eea
At zero $T$, for the Fermi energy lying between two separated LLs,
Eq.~(\ref{II.39}) gives a CR peak whose height\footnote{The peak height found by
\citet{ando:1975a} for short-range disorder is larger than that given by
Eq.~(\ref{II.39a}) at $\tau=\tau_{\rm q}$ by a factor of $4/3$.}
\be
\sigma_-(\omega_c)=(e^2n_e\tau/m)(2/\pi\omega_c\tau_{\rm q})^{1/2}
\label{II.39a}
\ee
is much smaller than that following from the Drude theory  [Eq.~(\ref{II.19})].
In the case of long-range disorder $(\tau\gg\tau_{\rm q})$, the CR lineshape
from Eq.~(\ref{II.39}) is given by
\be
{\sigma_-(\omega)\over\sigma_-(\omega_c)}=f\left({\omega-\omega_c\over\Delta_{
\rm CR}}\right)~,
\label{II.40}
\ee
where
\be
f(x)=1-{x^2\over 2\sqrt{1+x^2}}\ln{\sqrt{1+x^2}+1\over \sqrt{1+x^2}-1}
\label{II.41}
\ee
and
\be
\Delta_{\rm CR}=(2\omega_c\tau_{\rm q}/\pi)^{1/2}/\tau~.
\label{II.42}
\ee
The width of the CR for separated LLs $\Delta_{\rm CR}$ is seen to be much
larger than $1/\tau$.  At the same time, in the case of long-range disorder, the
CR as a function of $\omega$ is much narrower than the peaks in the DOS.

Landau quantization in combination with disorder leads to the emergence of CR
harmonics at $\omega=M\omega_c$ with $|M|=2,3,\ldots$ \cite{ando:1975a}. For
separated LLs, Eq.~(\ref{II.39}) gives for the height of the $M$th peak
\be
\sigma (M\omega_c)={4\over 3\pi}{e^2n_e\over m\omega_c}\left({2\over
\pi\omega_c\tau_{\rm q}}\right)^{1/2}{\tau_{\rm q}\over\tau}{M^2+1\over
(M^2-1)^2}~,
\label{II.43}
\ee
which is seen to be smaller by the factor of $\tau_{\rm q}/\tau$ than the result
\cite{ando:1975a} for the case of short-range disorder. In contrast to the CR
peak, the width of the higher-harmonic peaks is given by the width of the peaks
in the DOS. Thus, if one imagines that the correlation radius of disorder is
increased at fixed $\tau$, the CR peak becomes higher and narrower, while the
tendency for the higher-harmonic peaks is the opposite.

For strongly overlapping LLs, the dynamic response of the 2DEG is given by the
CR, which is described for $\omega_c\tau_{\rm q}\ll 1$ by the Drude formula
(\ref{II.19}), and by the $\omega$-dependent analog of the SdH oscillations
(\ref{II.36a}). To first order in $\delta$, the oscillatory correction
$\Delta\sigma_\pm(\omega)$ to the Drude conductivity reads \cite{dmitriev:2003}
\be
{\Delta\sigma_\pm(\omega)\over\sigma_{\rm D,\pm}(\omega)}\simeq
{\Delta\nu(\varepsilon_F)\over\nu_0}{\cal F}\!\left({2\pi^2
T\over\omega_c}\right)\chi_1~,
\label{II.44}
\ee
where
\be
\chi_1={\Omega_\pm^2\over 1+\Omega_\pm^2}\,{\omega_c\over\pi\omega}\left(\sin
{2\pi\omega\over\omega_c}+{1+3\Omega_\pm^2\over
\Omega_\pm^3}\sin^2{\pi\omega\over\omega_c}\right)
\label{II.45}
\ee
and $\Omega_\pm=(\omega\pm\omega_c)\tau$. In the dc limit, Eq.~(\ref{II.44})
reduces to Eq.~(\ref{II.36a}). If $T$ is much larger than the Dingle temperature
$T_D=1/2\pi\tau_{\rm q}$, the main source of damping of the oscillations with
varying $\varepsilon_F/\omega_c$ is the thermal averaging. In ultra-high
mobility heterostructures, $T_D\sim 10^2\,{\rm mK}$ and for the typical
measurement temperature $T\sim 1\,{\rm K}$ the linear-in-$\delta$ oscillations
(\ref{II.44}) are completely washed out.

%%%%%%%%%%%%%%%%%%%%%%%%%%%%%%%%%%%%%%

\begin{figure}
\includegraphics[width=\columnwidth]{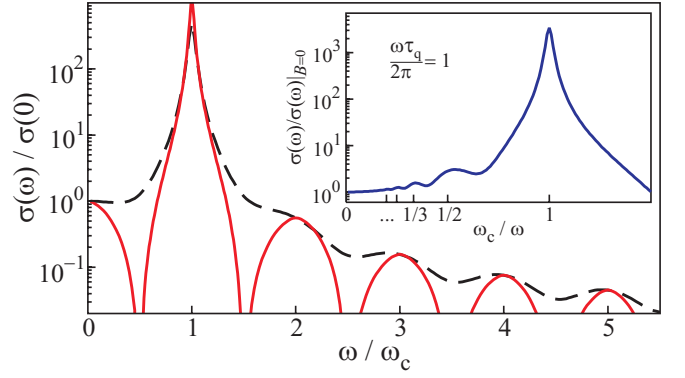}
\caption{Magnetooscillations of the ac conductivity of a 2DEG with smooth
disorder $(\tau/\tau_{\rm q}=10)$ in the limit $T\gg T_D$, in units of the dc
conductivity, as a function of $\omega/\omega_c$ for fixed $\omega_c\tau_{\rm
q}/\pi=3.25$ (solid line) and 1 (dashed). Inset: the conductivity vs
$\omega_c/\omega$ for fixed $\omega\tau_{\rm q}/2\pi=1$ in units of the ac
conductivity at $B=0$. From \citet{dmitriev:2003}.}
\label{II.f2}
\end{figure}

%%%%%%%%%%%%%%%%%%%%%%%%%%%%%%%%%%%%%%%%%

Unlike the oscillations with $\varepsilon_F/\omega_c$, which constitute the
essence of the SdH effect in the dc case, the oscillations with
$\omega/\omega_c$ survive the thermal averaging. Specifically, they survive in
the even-order terms in the expansion of $\sigma_\pm(\omega)$ in powers of
$\delta$, in which the intermodulation of the DOS oscillations at energies
separated by $\omega$ does not go away upon averaging over energy. For strongly
overlapping LLs, the oscillatory behavior of $\sigma_\pm(\omega)$ as
$\omega/\omega_c$ is varied at $T\gg T_D$ comes from the correction
$\Delta\sigma_\pm(\omega)$ of order $\delta^2$, which in the high-$T$ limit is
given by \cite{dmitriev:2003}
\be
\Delta\sigma_\pm(\omega)/\sigma_{\rm D,\pm}(\omega)\simeq 2\delta^2\chi_2
\label{II.46}
\ee
with
\bea
\hspace{-1cm}\chi_2\!&=&\!{\Omega_\pm^2\over (1+\Omega_\pm^2)^2}\nonumber\\
&\times&\!\left[\,(\Omega_\pm^2-3)\cos{2\pi\omega\over\omega_c}+{
3\Omega_\pm^2-1\over\Omega_\pm}\sin{2\pi\omega\over\omega_c}\,\right]~.
\label{II.47}
\eea
As noted already in the discussion of the SdH effect, in the case of long-range
disorder, for a classically strong magnetic field with $\omega_c\tau\gg 1$, the
modulation of the DOS is still small as long as $\omega_c\tau_{\rm q}\ll 1$.
Then, away from the CR (for $|\Omega_\pm|\gg 1$) Eq.~(\ref{II.46}) reduces to
the simple form
\be
\Delta\sigma_\pm(\omega)/\sigma_{\rm D,\pm}(\omega)\simeq 2\delta^2\cos
(2\pi\omega/\omega_c)~.
\label{II.48}
\ee
The overall behavior of the $T$-independent oscillations of $\sigma (\omega)$ is
illustrated in Fig.~\ref{II.f2}.

Intimately related to the independence of Eqs.~(\ref{II.46})--(\ref{II.48}) on
$T$ is the fact that the oscillations with $\omega/\omega_c$ are not sensitive
to the presence of macroscopic inhomogeneities (which may complicate the
determination of $\tau_{\rm q}$ from SdH experiments, as discussed at the end of
Sec.~\ref{s2.3.1}). The measurement of the damping of the
$\omega/\omega_c$--oscillations as a function of $B$ in the high-$T$ regime may
thus be the most reliable means of extracting $\tau_{\rm q}$ from the
magneto-oscillatory behavior of the transport coefficients. Note, however, that
the CR harmonics---and the related periodic modulation of $\sigma_\pm(\omega)$
with $\omega/\omega_c$---may be associated not only with Landau quantization but
also with the quasiclassical memory effects discussed in Sec.~\ref{s2.2.3}. In
fact, the latter can produce strong oscillations of $\sigma_\pm(\omega)$ with an
amplitude comparable to the Drude conductivity $\sigma_{\rm D,\pm}(\omega)$ [see
Fig.~\ref{II.f1}; for more details of the comparison between these two sources
of the magnetooscillations in the ac conductivity and the dc photoconductivity
see \citet{dmitriev:2004}].

The oscillations of $\sigma_\pm(\omega)$ as $\omega/\omega_c$ is varied on the
low-$B$ side of the CR, as well as the oscillations of the dynamic conductivity
with $\varepsilon_F/\omega_c$, have been observed in the early studies by
\citet{abstreiter:1976} in the far-infrared light transmission experiments on a
low-mobility 2DEG in Si inversion layers. However, only recently both these
types of magnetooscillations have also been observed in microwave absorption
experiments on high-mobility GaAs/AlGaAs heterostructures \cite{fedorych:2010}.
One reason why the measurements of $\sigma_\pm(\omega)$ in the
absorption/transmission experiments may be substantially complicated is that the
conductivity expresses the current as a response to the total (screened)
electric field whereas what is probed in this type of experiments is a response
to the (unscreened) field of the incident electromagnetic wave. As a result, the
behavior of $\sigma_\pm(\omega)$ in the absorption/transmission may be masked by
the excitation of magnetoplasmons
\cite{studenikin:2007,fedorych:2010,wirthmann:2007}.

%%%%%%%%%%%%%%%%%%%%%%%%%%%%%%%%%%%%%%%%%%%%%%%%%%%

\begin{figure}
\includegraphics[width=\columnwidth]{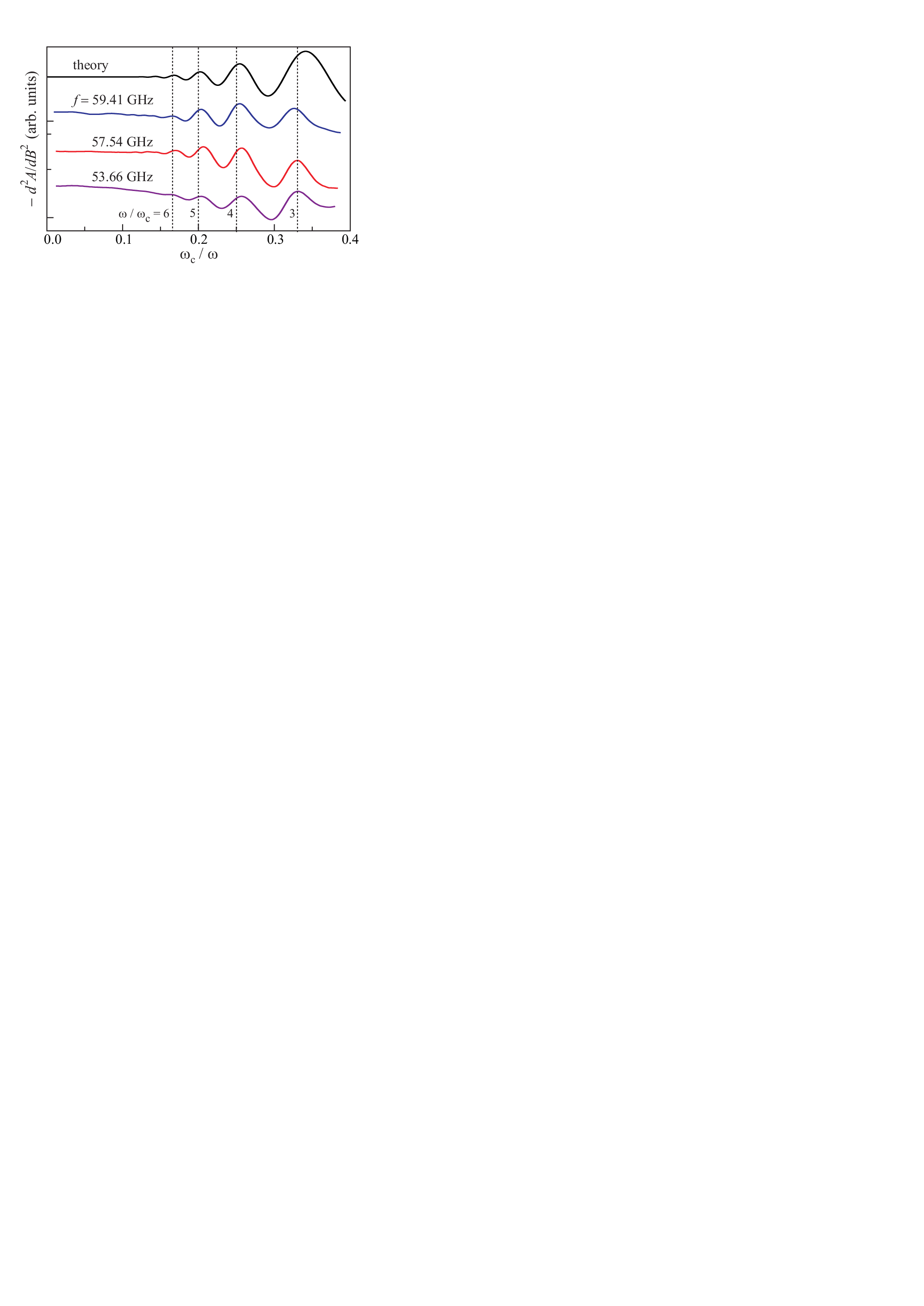}
\caption{
Absorption coefficient $\rm A$, differentiated twice with respect to $B$, as a
function of $\omega_c/\omega$ for fixed $\omega$. Three lower curves, marked by
different radiation frequencies $\omega/2\pi$, were taken at $T=2\,{\rm K}$ in a
sample with $n_e\simeq 3.6\times 10^{11}\,{\rm cm^{-2}}$ and $\mu\simeq 5\times
10^6\,{\rm cm^2/V\,s}$. The upper curve is the calculated dependence of
$-d^2{\rm A}/dB^2$ on $\omega_c/\omega$ at fixed $\omega$ according to
Eqs.~(\ref{II.48}) and (\ref{II.49}) with $\tau_{\rm q}=9.1\,{\rm ps}$
determined from the oscillatory-photoconductivity experiment. The curves are
offset for clarity. From \citet{fedorych:2010}.}
\label{II.f3}
\end{figure}

%%%%%%%%%%%%%%%%%%%%%%%%%%%%%%%%%%%%%%%%%%%%%%%%%%%%%%%

One more point---specific to a high-mobility 2DEG with $|\sigma_{xx}(\omega)\pm
i\sigma_{xy}(\omega)|\gg cn_r$, where $c$ is the speed of light in the vacuum
and $n_r$ is the refractive index of the medium on the sides of the 2DEG---is a
strong reflection of the electromagnetic wave caused by the dynamical screening
inside the 2DEG \cite{chiu:1976,falko:1989}. In a magnetic field, the enhanced
reflection leads to a suppression and an additional broadening of the CR
\cite{chiu:1976,mikhailov:2004} in the absorption/transmission coefficient. For
a linearly polarized wave normally incident on the 2DEG, the absorption
coefficient ${\rm A=A_++A_-}$ and the transmission coefficient $\rm T=T_++T_-$
are given by\footnote{Note that Eqs.~(\ref{II.49}) describe scattering by the
2DEG itself. To relate them to the absorption and transmission coefficients of
the electromagnetic wave incident on the sample, one should take into account
multiple Fresnel reflections from the boundaries of the latter. Changes of the
refractive index on scales smaller than the wavelength of the radiation are
averaged. If the wavelength inside the sample $2\pi c/\omega n_r$ is much larger
than the thickness of the sample, one can neglect the polarization of the medium
around the 2DEG and put $n_r\to 1$ in $g_\pm$ and $\gamma$.
When the distance from the 2DEG to the top surface of the heterostructure is
much smaller than the wavelength (which is a typical situation), scattering by
the 2DEG and the top surface \cite{fedorych:2010} is described by
Eqs.~(\ref{II.49}) with the change $n_r\to (1+n_r)/2$ in $g_\pm$ with a
simultaneous multiplication of $\rm A_\pm$ and $\rm T_\pm$ by $2n_r/(1+n_r)$ and
$4n_r/(1+n_r)^2$, respectively. \label{fn_screen}}
\be
{\rm A_\pm}={\rm Re}\,g_\pm/|1+g_\pm|^2~,\quad {\rm T}_\pm=1/2|1+g_\pm|^2~,
\label{II.49}
\ee
where $g_\pm=2\pi (\sigma_{xx}\pm i\sigma_{xy})/cn_r$ (neglecting the difference
in the dielectric properties of GaAs and AlGaAs).\footnote{To avoid notational
confusion: the abbreviation $\sigma_\pm$ is used in this review to denote ${\rm
Re}\,(\sigma_{xx}\pm i\sigma_{yx})/2$ \citep[as, e.g., in][]{dmitriev:2003},
whereas in a number of relevant papers \citep[e.g.,][]{fedorych:2010}
$\sigma_\pm$ denotes $\sigma_{xx}\pm i\sigma_{yx}$.} In the limit where the
Drude formula for $g_\pm$ can be applied,
\be
{\rm A_{D,\pm}}={2\gamma\tau\over 1+\Omega_\pm^2}{\rm
T_{D,\pm}}={\gamma\tau\over (1+\gamma\tau)^2+\Omega_\pm^2}
\label{II.50}
\ee
and the dynamical-screening-induced broadening
\be
\gamma=2\pi e^2n_e/mcn_r~.
\label{II.51}
\ee
In ultra-high mobility GaAs/AlGaAs heterostructures, the product $\gamma\tau$
may be as large as a few tens. A substantially enhanced---compared to the
transport scattering rate---broadening of the CR in the absorption (for magnetic
fields in which the LLs were not much separated) was reported in a high-mobility
structure by \citet{studenikin:2005}, similar to the very strong broadening of
the CR in the transmission through ultra-high mobility structures
\cite{tung:2009,smet:2005}. The oscillatory behavior of the absorption
coefficient as a function of $\omega/\omega_c$, observed by
\citet{fedorych:2010}, was related therein to the oscillations of the
conductivity tensor through Eq.~(\ref{II.49}). The low-$B$ portion of the data
shows good agreement with the simple asymptotic behavior resulting from
Eq.~(\ref{II.48}), see Fig.~\ref{II.f3}.

\subsubsection{Intersubband magnetooscillations}
\label{s2.3.3}

The mechanism in Sec.~\ref{s2.3.2} that relates the
$\omega/\omega_c$-oscillations of $\sigma (\omega)$---which survive the thermal
averaging [Eq.~(\ref{II.46})]---to the intermodulation of the DOS oscillations
at energies separated by $\omega$ can also lead to $T$-independent
magnetooscillations in dc transport. For that, one needs two (or more) parallel
2D electron gases, allowing for the exchange of electrons between them, with the
bottoms of their energy bands located at different energies. One system of this
type is a QW with two lowest subbands occupied by electrons, for which
\citet{polyanovsky:1988} and \citet{leadley:1992} pointed out that there should
exist oscillations of the MR periodic in $\Delta/\omega_c$, where $\Delta$ is
the subband spacing, not suppressed by the thermal averaging. The SCBA approach
to describe SdH oscillations in $\sigma_{xx}$ was generalized for short-range
disorder to the case of two subbands by \citet{raikh:1994} (see, however,
footnote 14 below). The SCBA theory of
SdH oscillations in the two-subband conductivity tensor for an arbitrary type of
disorder was developed by \citet{raichev:2008}.

The correction $\Delta\sigma_{\rm miso}$ to the dissipative Drude conductivity,
which describes the magnetointersubband oscillations (MISO), is a sum of two
terms $\Delta\sigma_{{\rm miso},\alpha}$ with $\alpha=1,2$, coming from
electrons in the $\alpha$th subband. For $\omega_c\tau_\alpha\ll 1$, the partial
contributions $\Delta\sigma_{{\rm miso},\alpha}$ are given for white-noise
disorder by \cite{raichev:2008,averkiev:2001}
\be
\Delta\sigma_{{\rm miso},\alpha}={2e^2n_\alpha\tau_\alpha^2\over
m\tau_{12}}\left(1-{2\tau_\alpha\over\tau_{12}}\right)\delta_1\delta_2\cos{
2\pi\Delta\over\omega_c}~,
\label{II.52}
\ee
where $n_\alpha$ and $1/\tau_\alpha$ are the electron density and the zero-$B$
scattering rate in the $\alpha$th subband, $1/\tau_\alpha$ includes both intra-
and intersubband scattering, $1/\tau_{12}$ is the zero-$B$ intersubband
scattering rate, and $\delta_\alpha$ is given by Eq.~(\ref{II.27}) with
$\tau_\alpha$ substituted for $\tau_{\rm q}$. Equation (\ref{II.52})
follows\footnote{The MISO amplitude resulting from Eq.~(\ref{II.36}) and the one
obtained by \citet{raichev:2008} agree with each other for arbitrary
$\omega_c\tau_\alpha$, whereas the one obtained by \citet{averkiev:2001}
coincides with them only for $\omega_c\tau_\alpha\to 0$.} directly from
Eq.~(\ref{II.36}). Importantly, Eq.~(\ref{II.35}) does not hold for the relation
between the oscillatory DOS $\nu_\alpha(\varepsilon)$ and the oscillatory
scattering rate $\tau_{B,\alpha}(\varepsilon)$ in the $\alpha$th subband, since
$1/\tau_{B,\alpha}(\varepsilon)$ contains the rate of intersubband scattering,
whereas $\nu_\alpha(\varepsilon)$ does not (in the leading approximation in
$1/\varepsilon_{F,\alpha}\tau_{12}$, where $\varepsilon_{F,\alpha}$ is the Fermi
energy in the $\alpha$th subband counted from its bottom).\footnote{At order
${\cal O}(\delta_1\delta_2)$, there are two contributions to MISO for
$\omega_c\tau_\alpha\ll 1$: one comes from the thermal averaging of the cross
term between the oscillations in $\nu_\alpha(\varepsilon)$ and those in
$\tau_{B,\alpha}(\varepsilon)$, the other---from the averaging of
$\tau_{B,\alpha}(\varepsilon)$ alone. In Eq.~(\ref{II.52}), the former brings
$-1$ to the expression in the brackets, the latter brings
$2(1-\tau_\alpha/\tau_{12})$. In the early work by \citet{raikh:1994}
%, where the
%SCBA approach to describe SdH oscillations in $\sigma_{xx}$ was generalized for
%short-range disorder to the case of two subbands
[see also an extension of the
theory to essentially similar oscillations in biased bilayer graphene
\cite{mkhitaryan:2011}], only the former contribution was taken into account.}
This is the reason why the preexponential factor in Eq.~(\ref{II.52}) does not
vanish at $B\to 0$, in contrast to Eq.~(\ref{II.46}) at $\omega=0$. As a result,
the MISO contribution to the MR $\Delta\rho_{\rm miso}/\rho_{\rm D}\simeq
-\Delta\sigma_{\rm miso}\rho_{\rm D}$ at $\omega_c\tau_\alpha\ll 1$ is mainly
given by the oscillatory {\it dissipative} component of the conductivity tensor
(in contrast to the oscillatory dc MR in the single-subband case, see
Sec.~\ref{s2.3.1}).

Within the SCBA, Eqs.~(\ref{II.36}) and (\ref{II.37}) allow one to describe
MISO for long-range disorder as well. Similar to Sec.~\ref{s2.3.1}, the
scattering rate in the exponent of $\delta_\alpha$ remains then the quantum
scattering rate, while $1/\tau_{B,\alpha}(\varepsilon)$ becomes the momentum
relaxation rate. However, in general, if scattering is not isotropic, the
zero-$B$ momentum relaxation rate in the $\alpha$th subband $1/\tau_\alpha$ is
related to the intra- and intersubband scattering rates in a nonlinear manner
% RR
\cite{ando:1982a,zaremba:1992,raichev:2008,mamani:2009a}. It is only in one simple case, when the intersubband
scattering is isotropic, that $1/\tau_\alpha$ is still given by a sum of the
momentum relaxation rate for (not necessarily isotropic) scattering within the
subband and the intersubband scattering rate $1/\tau_{12}$. In this
case,\footnote{The isotropy condition for intersubband transitions requires that
the Fermi wavelength in each of the subbands be much larger than $\min\{w,d\}$.
As discussed in Sec.~\ref{s2.1}, this condition is only marginally satisfied in
high-mobility GaAs/AlGaAs structures.} Eq.~(\ref{II.52}) for
$\omega_c\tau_\alpha\ll 1$ is reproduced also for long-range disorder.

The cumbersome expressions for the MISO amplitudes in $\sigma_{xx,xy}$ for the
general case of anisotropic scattering \cite{raichev:2008}
% RR
simplify significantly in the limit
 of a classically strong magnetic field
($\omega_c\tau_\alpha\gg 1$) but overlapping LLs ($\delta_\alpha\ll 1$)---which
conditions are only compatible in the case of long-range disorder. In this
limit \cite{mamani:2009a},
% RR
\be
\Delta\sigma_{\rm miso}={2e^2(n_1+n_2)\over
m\omega_c^2\bar{\tau}_{12}}\delta_1\delta_2\cos{2\pi\Delta\over\omega_c}
\label{II.53}
\ee
and, correspondingly, the MISO term\footnote{The expression for $\Delta\rho_{\rm
miso}$ in terms of the thermally averaged product
$\nu(\varepsilon)\nu(\varepsilon+\Delta)$ yielding Eq.~(\ref{II.54}) was
proposed by \citet{coleridge:1990}, although it was erroneously also suggested
there that MISO are suppressed by the thermal averaging if $T\gg\omega_c$.}
in the MR
\be
{\Delta\rho_{\rm miso}\over\rho_{\rm
D}}={2\over\bar{\tau}_{12}}\,{n_1\tau_1+n_2\tau_2\over
n_1+n_2}\,\delta_1\delta_2\cos{2\pi\Delta\over\omega_c}~,
\label{II.54}
\ee
where $1/\bar{\tau}_{12}$ is given by an integral over the scattering angle
$\phi$ [cf.\ Eq.~(\ref{II.2a})]:
\be
{1\over \bar{\tau}_{12}}=m\int_0^{2\pi}{d\phi\over
2\pi}\,\left(1-{2\sqrt{n_1n_2}\over n_1+n_2}\cos\phi\right)W^{(12)}_q~,
\label{II.55}
\ee
with $W^{(12)}_q$ being the Fourier component of the correlation function of the
intersubband matrix element of the random potential at the transferred in-plane
momentum
$q=[2m(\varepsilon_{F,1}+\varepsilon_{F,2}-2\sqrt{\varepsilon_{F,1}\varepsilon_{
F,2}}\cos\phi)]^{1/2}$.

Oscillations similar to MISO occur also in a double QW with the
lowest subband states split by tunneling or, for that matter, in any other
system whose energy spectrum exhibits two or more series of LLs, which are
offset with respect to each other, in the presence of scattering between the
series. Experimentally, MISO were studied in two-subband single QWs
\cite{leadely:1989,coleridge:1990,leadley:1992,sander:1998,rowe:2001,bykov:2010a,goran:2009}, in double \cite{mamani:2008,bykov:2008b} and triple
\cite{wiedmann:2009b} QWs, and in wide QWs split in two
layers electrostatically \cite{wiedmann:2010}.
% RR
MISO were also studied theoretically in 2D layers with electron states split by
spin-orbit interaction \cite{langenbuch:2004}.
The behavior of MISO in
nonequilibrium conditions is discussed in Secs.~\ref{s3.3.3} and \ref{s5.3}.

\section{Microwave-induced resistance oscillations (MIRO)}
\label{s3}

\subsection{MIRO: Experimental discovery and basic properties}
\label{s3.1}

%%%%%%%%%%%%%%%%%%%%%%%%%%%%%%%%%%%%%%%%%%%%%%%%%
\begin{figure}[t]
\includegraphics[width=\columnwidth]{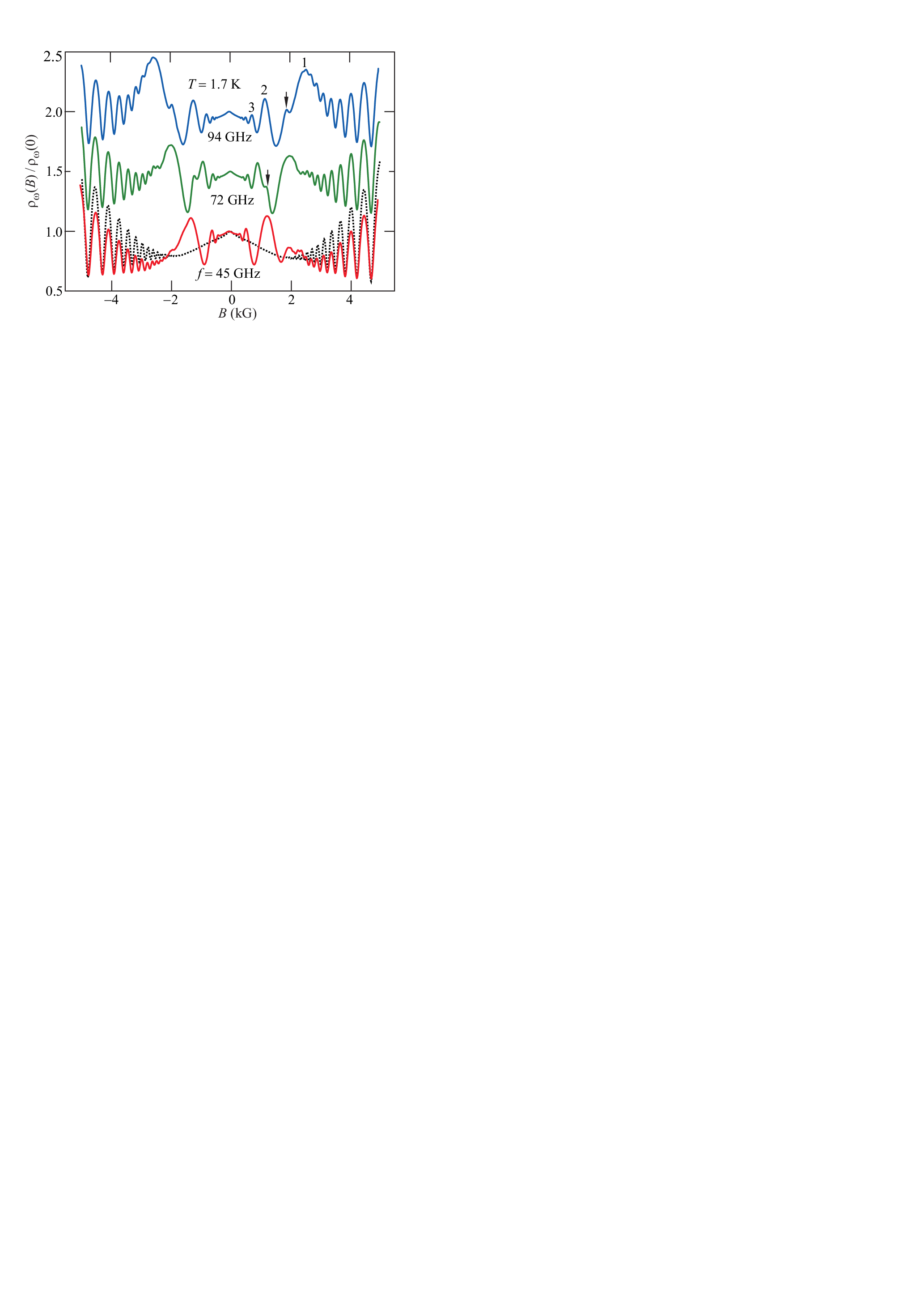}
\caption{Magnetoresistivity with (solid lines) and without (dashed)
microwave irradiation for different frequencies $f=45$, 72, and $94\,{\rm GHz}$,
normalized to its value at $B=0$.
The traces for different $f$ are vertically offset for clarity. Integers show
the order of MIRO peaks. The arrows mark the magnetoplasmon resonance. The
data were obtained at $T\simeq 1.7\,{\rm K}$ in a $200\,\mu{\rm m}$ wide Hall
bar sample with $n_e\simeq 2.0\times 10^{11}$ cm$^{-2}$ and $\mu\simeq 3.0
\times 10^6\,{\rm cm^2/V\,s}$. Adapted from \citet{zudov:2001a}.}
\label{fig.miro.1}
\end{figure}
%%%%%%%%%%%%%%%%%%%%%%%%%%%%%%%%%%%%%%%%%%%%%%%%%

%%%%%%%%%%%%%%%%%%%%%%%%%%%%%%%%%%%%%%%%%%%%%%%%%
\begin{figure}[t]
\includegraphics[width=\columnwidth]{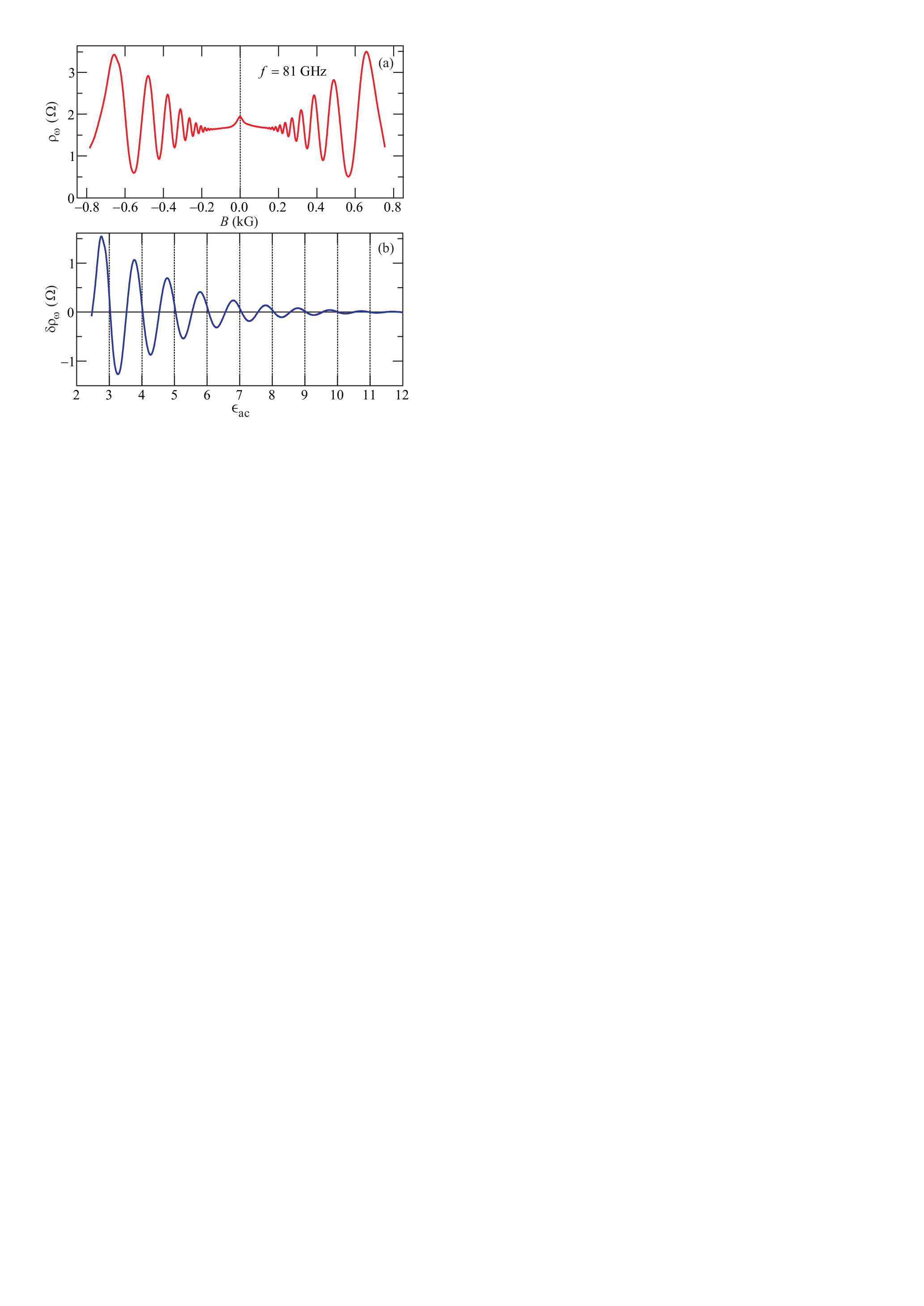}
\caption{
(a) Magnetoresistivity oscillations with varying $B$ at fixed $f=81\,{\rm GHz}$.
(b) Microwave-induced correction $\delta \ro$ to the resistivity as a function
of $\eac$, obtained by subtracting the slowly varying background. The data were
obtained at $T\simeq 1.5$ K in a $100\,\mu{\rm m}$ wide Hall bar sample with
$n_e\simeq 3.0\times 10^{11}$ cm$^{-2}$ and $\mu\simeq 1.2 \times 10^7\,{\rm
cm^2/V\,s}$.}
\label{fig.miro}
\end{figure}
%%%%%%%%%%%%%%%%%%%%%%%%%%%%%%%%%%%%%%%%%%%%%%%%%

When a sufficiently high-mobility 2DEG is subject to a weak magnetic field and
illuminated by microwave radiation, the longitudinal magnetoresistivity
$\rho(B)$ exhibits giant oscillations \cite{zudov:1997,zudov:2001a}, termed the
microwave-induced resistance oscillations (MIRO), see \rfig{fig.miro.1}.
In a Corbino disk-shaped 2DEG, experiments revealed corresponding
microwave-induced conductance oscillations \cite{yang:2003}. Most commonly,
MIRO are observed in dc measurements, with radiation being delivered to the 2DEG
via oversize waveguides [although other means, such as planar microwave
transmission lines patterned on top of the 2DEG \cite{ye:2001} and dipole
antennas \cite{willett:2004} have also been successfully implemented]. Recently,
they were also observed by \citet{bykov:2010d} and \citet{andreev:2011} in
contactless measurements in a capacitively coupled 2DEG. MIRO have been
studied not only in  high-mobility single-subband 2DEGs but in a variety of
other 2D systems, including 2DEGs patterned with a triangular antidot lattice
\cite{yuan:2006}, see \rsec{s3.3.4}; two-subband 2DEGs
\cite{bykov:2010e,wiedmann:2008}, see \rsec{s3.3.4}; and hole systems based on
C-doped GaAs/AlGaAs QWs \cite{du:2004a}. Phenomenologically similar
oscillations were recently discovered in a nondegenerate electron system on
surface of liquid $^3$He \cite{konstantinov:2009b}, see \rsec{s7.3.1}.

\emph{Period and phase.} The radiation-induced oscillatory part
$\delta\rho_\omega$ of the photoresistivity $\rho_\omega$ at a radiation
frequency $\omega$ oscillates with the ratio
\be
\eac\equiv \omega/\wc~,
\label{eac}
\ee
see \rfig{fig.miro}b. In contrast to PIRO and HIRO (\rsec{s5}), the MIRO
period is not sensitive to the carrier density $n_e$.
MIRO maxima ($\eac^{+}$) and minima ($\eac^{-}$) are roughly symmetrically
offset from the harmonics of the CR at $\eac=n$\footnote{Early experiments
\cite{zudov:1997,zudov:2001a,ye:2001} loosely associated the low-order
oscillation maxima with integer $\eac=n$. Now it is established that at $\eac =
n$ there are zero-response nodes where the photoresistivity
$\delta\rho_\omega=0$.} and occur at
\be
\eac^{\pm} \simeq n \mp \pac,\,\,n=1,2,3,...\,.
\label{eq.miro.max}
\ee
In accordance with theoretical predictions for the regime of overlapping LLs
(\rsec{phase}), the observations suggest that $\delta\rho_\omega\propto-\sin
2\pi\eac$ for $\eac\agt 2$, i.e., the phase $\pac$ in \req{eq.miro.max} is 1/4.
This value of the phase was reported as universal in many experiments
\cite{mani:2002,mani:2004a,mani:2004b,mani:2004c,mani:2004d,
mani:2004e,mani:2005,mani:2007,mani:2007b,mani:2008,mani:2009,mani:2010}.
Other experiments
\cite{zudov:2004,zudov:2006a,zudov:2006b,studenikin:2005,studenikin:2007,
dai:2010,zhang:2007b,hatke:2008a, hatke:2008b,hatke:2011b,hatke:2011c,hatke:2011d} found
that the lower-order maxima and minima are pushed toward the harmonics of the
CR and are best described by $\pac$ which is considerably smaller than 1/4. Such
a phase reduction is expected in the regime of separated LLs
\citep{zudov:2004,vavilov:2004,dmitriev:2005,studenikin:2005}, see \rsec{phase}.
Moreover, the phase decreases with increasing radiation power, as discussed next.

\begin{figure}[t]
\includegraphics[width=\columnwidth]{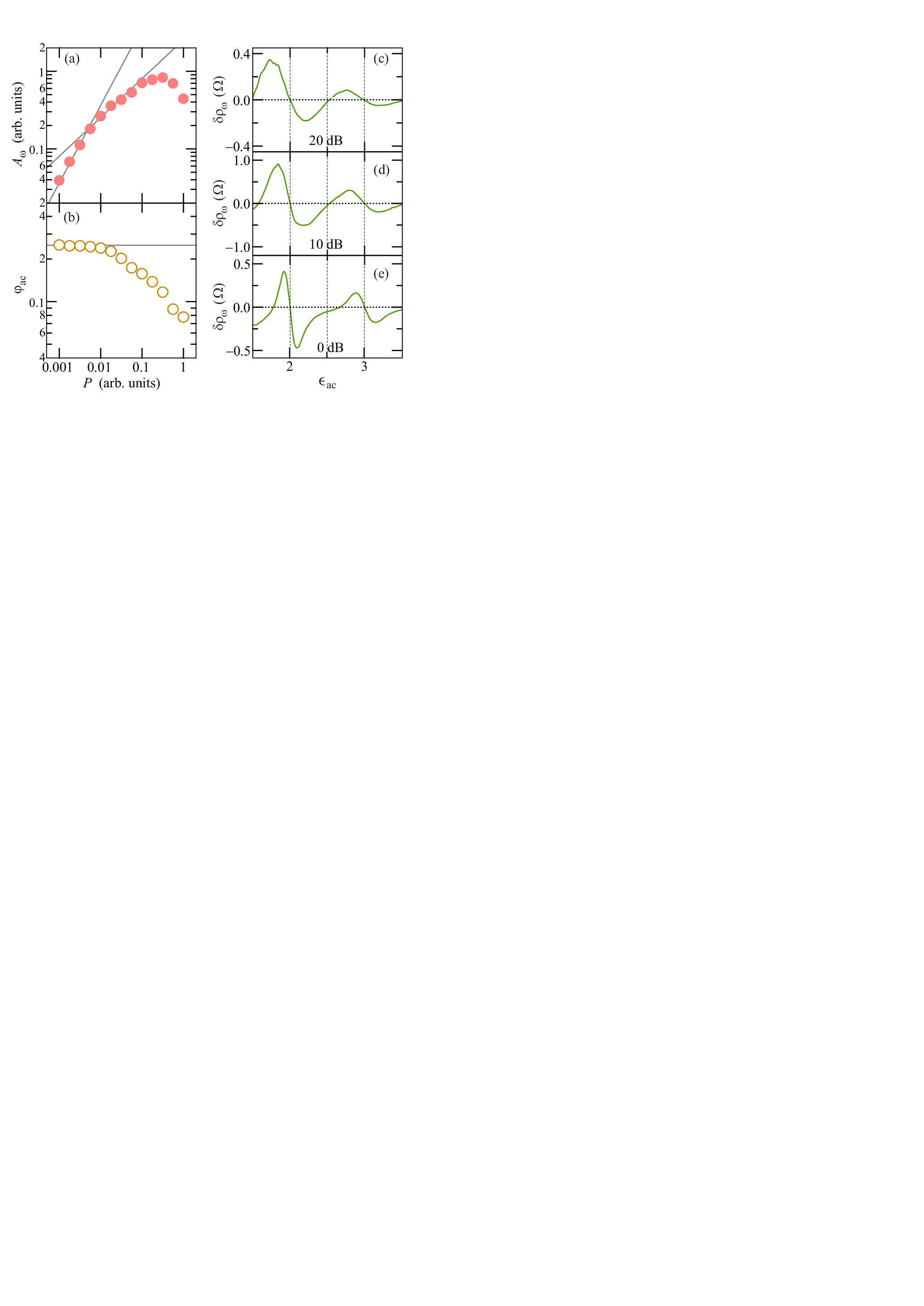}
\caption{(a) Amplitude $A_\w$ and (b) phase $\pac$ measured at the second MIRO
maximum vs microwave power $P$. Straight lines in part (a) represent, on the
log-log scale, the linear and square-root dependences on $P$. (c)-(e)
Photoresistivity $\delta\rho_\omega$ as a function of $\wc$ at fixed
$\w/2\pi=33\,{\rm GHz}$ for different $P$. The traces are labeled according to
the attenuation levels. The data were obtained at $T\simeq 1.5\,{\rm K}$ in a
$200\,\mu{\rm m}$ wide Hall bar sample with $n_e\simeq 2.9\times 10^{11}\,{\rm
cm}^{-2}$ and $\mu\simeq 2.4\times 10^7\,{\rm cm^2/V\,s}$. Adapted from
\citet{hatke:2011up1}.}
\label{fig.miro.power}
\end{figure}

\emph{Power dependence.}
At sufficiently low microwave power $P$, one expects that the amplitude $A_\omega$ [as
defined in Eq.~(\ref{phen_MIRO})] of the oscillatory photoresistivity
$\delta\rho_\omega$ is linear in $P$. While the linear dependence of $A_\omega$ on $P$
was clearly observed by \citet{zudov:2003,hatke:2011b}, a number of experiments
found a strongly sublinear  dependence
\cite{ye:2001,studenikin:2004,willett:2004,mani:2004a,mani:2010}. The apparently
conflicting experimental reports can be reconciled by the existence of two
distinct regimes of weak and strong power, characterized by the linear and
sublinear scaling of $A_\omega$ with $P$, respectively. The experimental data in
Fig.~\ref{fig.miro.power} show that the linear dependence of $A_\omega\propto P$
crosses over with increasing $P$ into $A_\omega\propto P^{1/2}$, accompanied by a
strong reduction of the phase $\pac$ compared to its value $\pac=1/4$ in the
linear regime. Such behavior is expected for the inelastic mechanism of MIRO,
see \rsec{s3.2.2}. For the displacement mechanism, a similar effect is expected
at the crossover to the multiphoton regime of the photoresponse, see \rsec{s6}.

%%%%%%%%%%%%%%%%%%%%%%%%%%%%%%%%%%%%%%%%%%%%%%%%%
%%%%%%%%%%%%%%%%%%%%%%%%%%%%%%%%%%%%%%%%%%%%%%%%%

%%%%%%%%%%%%%%%%%%%%%%%%%%%%%%%%%%%%%%%%%%%%%%%%%
\begin{figure}[t]
\includegraphics[width=\columnwidth]{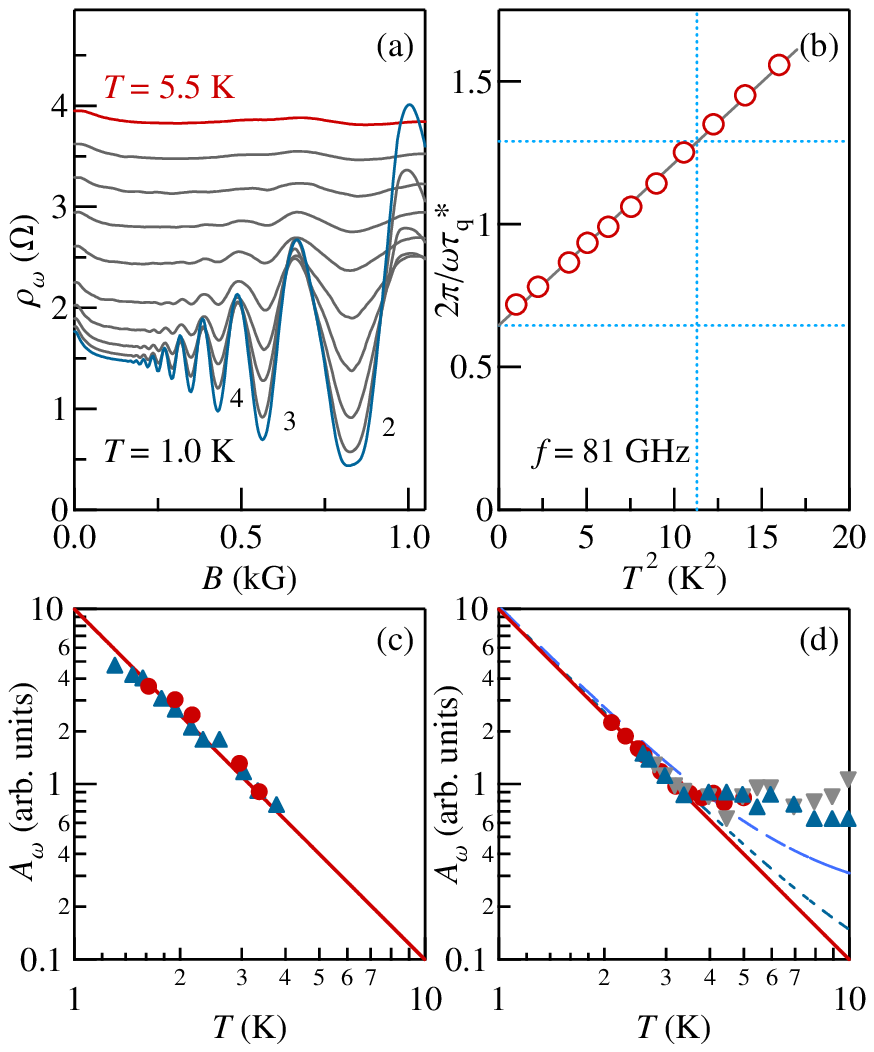}
\caption{(a) Evolution of MIRO with varying $T$ and (b) the $T$ dependence of
$2\pi/\w\tq^*$ extracted from these data by fitting them to \req{phen_MIRO}
\cite{hatke:2009a}. (c)-(d) The $T$ dependence of $A_\omega$ [\req{phen_MIRO}] obtained
by (c) \citet{studenikin:2007} and (d) \citet{wiedmann:2010a}.}
\label{fig.miro.temp}
\end{figure}
%%%%%%%%%%%%%%%%%%%%%%%%%%%%%%%%%%%%%%%%%%%%%%%%%

\emph{Temperature dependence.}
Most experiments show that MIRO are best observed at temperature $T\sim$
0.5--1$\,{\rm K}$, get strongly suppressed with increasing $T$, and become
almost invisible at $T\sim 4$--$7\,{\rm K}$. The overall behavior in the regime
of weak oscillations, illustrated in \rfig{fig.miro.temp}, is consistent with
the theoretical predictions, see \rfig{tau_star}b,c and the discussion in
\rsecs{temperature}{damping}, and \ref{in_vs_dis} [the $T$ dependence of MIRO in
the regime of ZRS is discussed in \rsec{s4.1}]. At low $B$ and $P$, the $B$
dependence of MIRO almost perfectly follows
\be\label{phen_MIRO}
\delta\rho_\omega=-A_\omega \eac\sin (2\pi\eac)\exp(-2\pi/\wc\tq^*)~,
\ee
where $A_\w$ is nearly $B$-independent for $\eac\agt 2$, see, e.g.,
\rfig{fig.miro.temp}a. Such fits of the $B$-traces of MIRO at different $T$
generally reveal that both $\tq^*$ in the factor $e^{-2\pi/\wc\tq^*}$ and $A_\omega$
depend on $T$. Specifically, the exponential factor in Eq.~(\ref{phen_MIRO}) is
well described by a quadratic-in-$T$ dependence of $1/\tq^*(T)=1/\tq+ \alpha
T^2$, see \rfig{fig.miro.temp}b, which is attributed to the LL
broadening induced by electron-electron interactions (\rsec{damping}). A similar
exponential dependence on $T$ was also reported for HIRO \cite{hatke:2009c},
PIRO \cite{hatke:2009b}, and MISO \cite{mamani:2008}.
\citet{studenikin:2005,studenikin:2007,wiedmann:2010a} found $A_\omega\propto T^{-2}$
at $T<3\,{\rm K}$, see \rfig{fig.miro.temp}c,d, consistent with the inelastic
mechanism of MIRO (\rsec{temperature}). The saturation of the dependence of $A_\omega$
on $T$ observed by \citet{wiedmann:2010a} at $T>3\,{\rm K}$, see
\rfig{fig.miro.temp}d, and the results of \citet{hatke:2009a}, where $A_\omega$ almost
independent of $T$ was reported, indicate that the displacement contribution to
MIRO [which produces $A_\omega={\rm const}(T)$ and strongly depends on the
correlation properties of disorder, see \rsec{mixed_disorder}] may remain
relevant down to low $T$.

\begin{figure}[ht]
\includegraphics[width=\columnwidth]{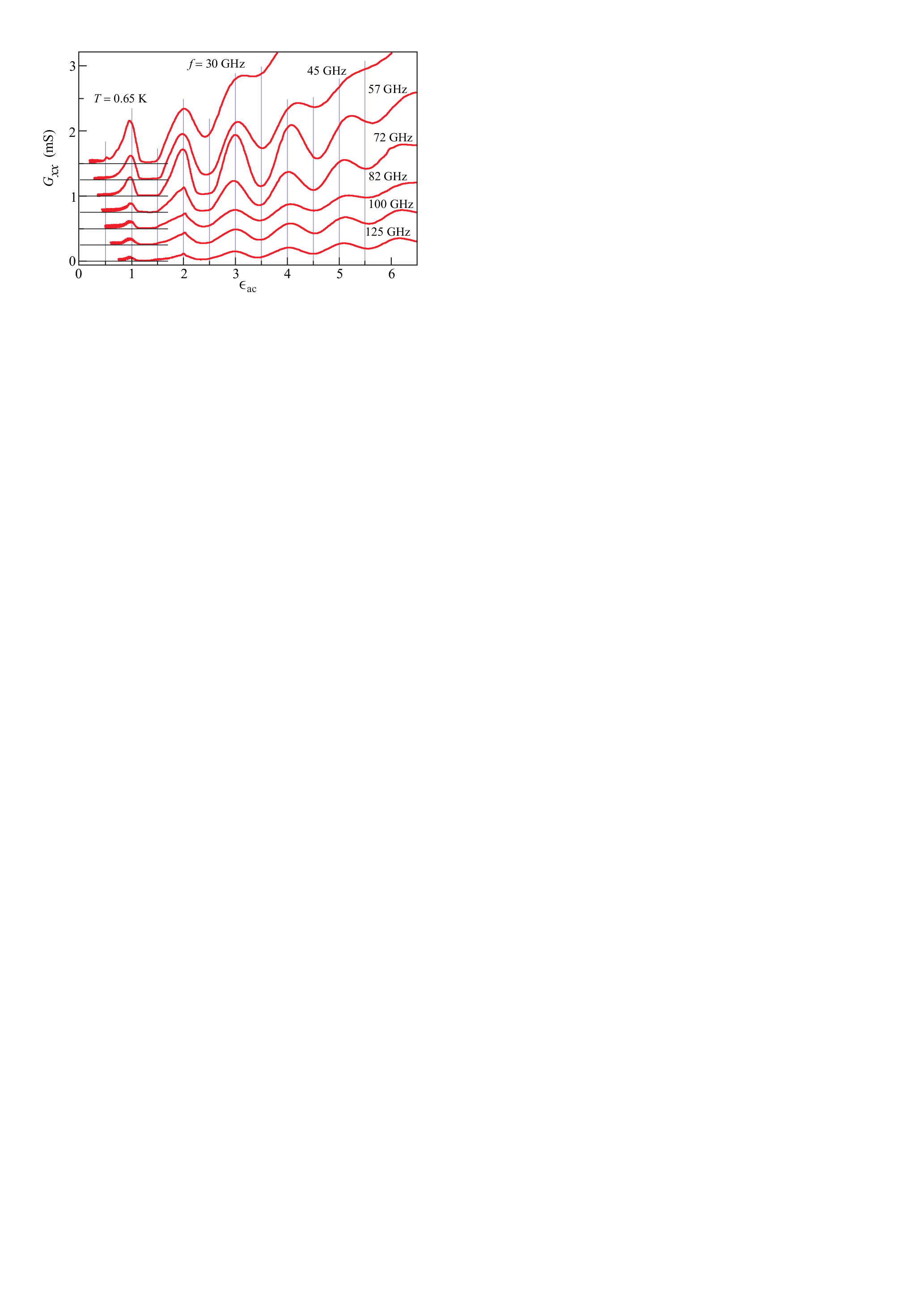}
\caption{Conductance $G_{xx}$ of a Corbino sample ($n_e= 3.55\times
10^{11}\,{\rm cm}^{-2}$, $\mu= 1.28 \times 10^{7}\,{\rm cm^2/V\,s}$) as a
function of $\w/\wc$ for different fixed $\w$. The traces, labeled according to
$f=2\pi\w$, are offset in steps of $0.25\,{\rm mS}$. Adapted from
\citet{yang:2003}.}
\label{fig.zcs.freq}
\end{figure}

\emph{Frequency dependence.}
Over the past decade MIRO have been observed in a wide range of radiation
frequencies, from 3 GHz \citep{willett:2004} to 1.5 THz \cite{wirthmann:2007}.
However, the majority of experiments employed frequencies from 30 to
150\,GHz, which appears to be the optimum range for a typical 2DEG. At lower
frequencies, MIRO are shifted towards weaker $B$ and are therefore
suppressed by the Dingle factor. At higher frequencies, the oscillation
amplitude decays as well \cite{yang:2003,studenikin:2007,tung:2009}, as
illustrated in \rfig{fig.zcs.freq} for the photoconductance measured in a
Corbino disk \cite{yang:2003}. At fixed both $\eac\geq 2$ and the microwave
intensity, the observed decay is consistent with
$\delta\rho_\omega\propto\omega^{-4}$,
in agreement with \reqs{P}{Epm},
(\ref{Epm1}), and (\ref{eq.miro}) below.

%%%%%%%%%%%%%%%%%%%%%%%%%%%%%%%%%%%%%%%%%%%%%%%%%
%%%%%%%%%%%%%%%%%%%%%%%%%%%%%%%%%%%%%%%%%%%%%%%%%

\subsection{Microscopic mechanisms of MIRO}
\label{s3.2}

Most experimental findings concerning MIRO can be explained as a combined effect
of Landau quantization and external fields either on the momentum relaxation
(displacement mechanism) or on the energy distribution of electrons within
disorder-broadened LLs (inelastic mechanism). Section \ref{s3.2.1} describes
both effects in terms of quasiclassical kinetics of the guiding centers of
cyclotron orbits. Section \ref{s3.2.2} deals with the main mechanism of MIRO
saturation which establishes the conditions for the observation of ZRS and fixes
the amplitude of a dc electric field in spontaneously formed domains. A quantum
kinetic equation formalism is described in \rsec{s3.2.3}. This formalism
provides a solid foundation for the approach used in Secs.~\ref{s3.2.1} and
\ref{s3.2.2} and serves as a basic tool for the theoretical description of other
nonequilibrium phenomena throughout the review. Additional quadrupole and
photovoltaic
contributions to MIRO (these govern, in particular, magnetooscillations of
the Hall part of the conductivity) are discussed in \rsec{s3.2.4}.  Alternative
mechanisms of MIRO, not related to the Landau quantization, are shortly
discussed in \rsec{s3.2.5}.

\subsubsection{Inelastic and displacement mechanisms}
\label{s3.2.1}

Initially MIRO were attributed \cite{durst:2003,vavilov:2004} to the
displacement mechanism which accounts for spatial displacements of
quasiclassical electron orbits due to radiation-assisted scattering off
disorder. Because of Landau quantization, which leads to a periodic modulation in
the DOS $\nu(\ve)\simeq\nu(\ve+\wc)$ (\rsec{s2.3.1a}), the
preferred direction of these displacements with respect to the symmetry-breaking
dc field oscillates with $\omega/\wc$. This results in MIRO with a phase and
a period which agree with those observed in experiment. Photoconductivity
oscillations governed by the displacement mechanism were in fact predicted long
ago \cite{ryzhii:1970,ryzhii:1986} in the limit of separated LLs and a strong dc
field. This mechanism was further studied using various approaches and
approximations in a number of theoretical works
\cite{anderson:2003,durst:2003,vavilov:2004,khodas:2008,dmitriev:2007,
dmitriev:2009b,%
lei:2003,park:2004,torres:2005,ryzhii:2004,shi:2003,volkov:2007,auerbach:2007,
kashuba:2006,kashuba:2006b,lee:2004,ryzhii:2003d}. Soon after the experimental
observations of MIRO, \citet{dmitriev:2003} proposed that the dominant
contribution to MIRO was due to the inelastic mechanism associated with
radiation-induced changes in the occupation numbers of electron states. Similar
ideas were discussed by \citet{dorozhkin:2003}; however, the calculation there
was not directly applicable to the experimentally relevant systems, in
particular, due to an unrealistic model of inelastic relaxation. Later studies
confirmed that the inelastic mechanism generally dominates the observed MIRO
\cite{dmitriev:2005} at low $T$, while the displacement mechanism can be
relevant at higher $T$ and only if a sufficient amount of short-range impurities
is present in the system \cite{khodas:2008,dmitriev:2009b}, or else, in the
limit of a strong dc field \cite{khodas:2008} or high microwave power
\cite{dmitriev:2007}.

\begin{figure}[ht]
\centerline{
\includegraphics[width=\columnwidth]{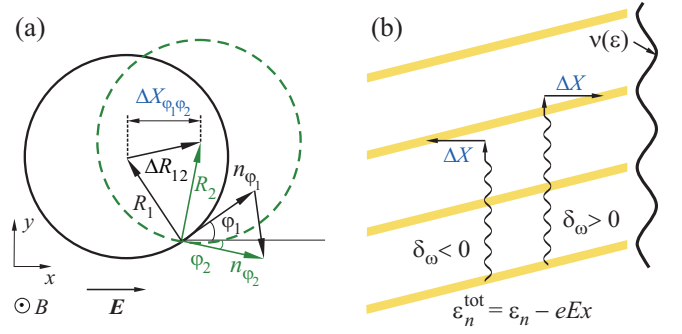}
}
\caption{(a) Shift of the guiding center of a cyclotron orbit due to
quasielastic scattering off disorder. (b) Schematics of the correlations between
the direction of the cyclotron-orbit shifts $\Delta X$ in the process of photon absorption
and the sign of the detuning $\delta_\w=\w/\wc-2$ for the second harmonic of the
CR. The yellow stripes mark the DOS maxima $\ve_n=(n+1/2)\wc$ in LLs tilted by a
dc field.}
\label{DeltaR}
\end{figure}

In this section, we formulate a description of both mechanisms in terms of
migration of the guiding centers of cyclotron orbits. A systematic way to obtain
the same results within the quantum kinetic approach is sketched in
\rsec{s3.2.3}. Quasiclassically, each scattering event leads to a shift of the
guiding center $\Delta\bR_{\varphi_1\varphi_2}=R_c {\bf
e}_z\times(\bn_{\varphi_1}-\bn_{\varphi_2})$, where
$\bn_{\varphi_k}=(\cos\varphi_k,\sin\varphi_k)$ with $k=1,2$ are the unit
vectors in the direction of motion before and after the collision, see
Fig.~\ref{DeltaR}.

%%%%%%%%%%%%%%%%%%%%%%%%%%%%%%%%%%%%%%%%%%%%%%%%%%%%%%%%%%%%%%%%%
%%%%%%%%%%%%%%%%%%%%%%%%%%%%%%%%%%%%%%%%%%%%%%%%%%%%%%%%%%%%%%%

For a macroscopically homogeneous 2DEG subjected to a dc electric field
$\bE={\bf e}_x E$, the dissipative current
\be\label{jx}
j_d=2\nu_0
e\!\int_{-\infty}^x\!dx_1\!\int_x^\infty\!dx_2\,(W_{x_1\to
x_2}-W_{x_2\to x_1})
\ee
is expressed in terms of the probabilities $W_{x_1\to x_2}$ of the
guiding-center shifts $x_1\to x_2$ along the ${\bf e}_x$ axis
$\Delta X_{\varphi_1\varphi_2}={\bf
e}_x\cdot\Delta\bR_{\varphi_1\varphi_2}=R_c(\sin\varphi_1-\sin\varphi_2)$. The probabilities are
given by the integrals over the initial and final energies in the scattering
event
\bea \nonumber
&& W_{x_1\to x_2}=\Big\langle\int\!\!d\ve_1\!\!\int\!\!d\ve_2\,{\cal M}_{\ve_1
\ve_2}\delta(x_1-x_2+\Delta X_{\varphi_1\varphi_2})
\\\label{W}
&&\times\left[\Gamma_{\varphi_1\varphi_2}^{(\rm el)}\delta(\Delta\ve^{\rm
tot}_{12})+\Gamma_{\varphi_1\varphi_2}^{(\rm ph)}\sum\nolimits_\pm
\delta(\Delta\ve^{\rm
tot}_{12}\pm\w)\right]\Big\rangle_{\varphi_1\varphi_2}~,\nonumber\\
\eea
where the angular brackets denote averaging over the initial and final angles
$\varphi_{1,2}$.
The delta functions in the square brackets express conservation of the total
electron energy $\Delta\ve^{\rm tot}_{12}=\ve_1-\ve_2+eE\Delta
X_{\varphi_1\varphi_2}=0$
in the elastic channel ($\propto\Gamma_{\varphi_1\varphi_2}^{(\rm el)}$) and its
change by $\pm \omega$
in the photon-assisted scattering channel
($\propto\Gamma_{\varphi_1\varphi_2}^{(\rm ph)}$).
Quantum magnetooscillations originate from the factor
\be\label{M}
{\cal M}_{\ve
\ve^\prime}=\tilde{\nu}_{\ve}\tilde{\nu}_{\ve^\prime}f_{\ve}\left(1-f_{
\ve^\prime}\right)~,
\ee
where $\tilde{\nu}_{\ve}=\nu(\ve)/\nu_0$ with $\nu_0=m/2\pi$ is the
dimensionless DOS in disorder-broadened LLs and $f_{\ve}$ is the nonequilibrium
distribution function in the steady state. In the homogeneous case, both
$\tilde{\nu}_{\ve}$ and $f_\ve$ are functions of the local kinetic energy $\ve$.
Since the disorder-induced broadening of LLs is determined by $\tq\ll\ttr$
(\rsec{s2.3.1a}), the effect of the external fields on $\tilde{\nu}_{\ve}$ is
negligible in the relevant range of the dc and microwave field strength. By
contrast, the modification of the distribution function is crucially important.
The nonequilibrium occupation of electron states is governed by the kinetic
equation
\bea\nonumber
&&\left\langle\tilde{\nu}_{\ve}^{-1}\Gamma_{\varphi\varphi^\prime}^{(\rm ph)}
\sum\nolimits_\pm(
{\cal M}_{\ve \ \ve^\prime\pm\w}-
{\cal M}_{\ve^\prime\pm\w \ \ve}
)\right\rangle_{\varphi\varphi^\prime}\\
&&+\left\langle\tilde{\nu}_{\ve}^{-1}\Gamma_{\varphi\varphi^\prime}^{(\rm el)}
({\cal M}_{\ve\ve^\prime}-
{\cal M}_{\ve^\prime\ve}
)\right\rangle_{\varphi\varphi^\prime}=\St_{\rm in}\{f_\ve\}~,
\label{f}\eea
where $\ve^\prime=\ve+eE\Delta X_{\varphi\varphi^\prime}$. To close the set of
equations, the type of inelastic scattering and the expressions for the rates
$\Gamma^{(\rm el)}$ and $\Gamma^{(\rm ph)}$ need to be specified. The
calculation in \rsec{temperature} below shows that the inelastic scattering
integral can be approximated in the form
\be\label{tin}
\St_{\rm in}\{f_\ve\}=\left(f^{T}_{\ve}-f_{\ve}\right)/\tin~,
\ee
which describes thermalization to the local Fermi distribution $f^{T}_{\ve}$
with the effective rate $1/\tin$. The scattering rates
$\Gamma_{\varphi\varphi^\prime}^{(\rm ph)}$ and
$\Gamma_{\varphi\varphi^\prime}^{(\rm el)}$ for transitions
$\varphi\to\varphi^\prime$ are given by \cite{khodas:2008}
\be\label{Gamma}
\Gamma_{\varphi\varphi^\prime}^{(\rm
ph)}=\frac{P_{\varphi+\varphi^\prime}}{2\tau_{\varphi-\varphi^\prime}}
\sin^2\frac{\varphi-\varphi^\prime}{2}\,,\quad
\Gamma_{\varphi\varphi^\prime}^{(\rm
el)}=\frac{1}{\tau_{\varphi-\varphi^\prime}}-2\Gamma_{\varphi\varphi^\prime}^{
(\rm ph)}\,.
\ee
The last term in $\Gamma_{\varphi\varphi^\prime}^{(\rm el)}$ describes the
microwave-induced modification of the elastic scattering. For the isotropic
2DEG, the disorder-induced scattering rate $\tau_{\varphi_1-\varphi_2}^{-1}$ is
expressible in the most general case as a series in angular harmonics
\be\label{tau_n}
\tau_{\varphi_1-\varphi_2}^{-1}=\sum_{n=-\infty}^\infty \tau_n^{-1}
e^{in(\varphi_1-\varphi_2)}, \ \ \ \tau_n=\tau_{-n}~.
\ee
For the two-component disorder model, the coefficients $\tau_n$ are discussed
in \rsec{mixed_disorder}. For the microwave field (screened by the 2DEG, see
\rsec{screening}) of the form
\be\label{Ew}
\bE_\w(t)=E_\omega \sum\nolimits_\pm {\rm Re}\left(s_\pm {\bf e}_\pm e^{i\w
t}\right)~,
\ee
where $2^{1/2}{\bf e}_\pm={\bf e}_x\pm i{\bf e}_y$ and $(s_+,s_-)$ is the
complex vector of unit length which determines the polarization of the field,
the dimensionless power $P_\theta$ in Eq.~(\ref{Gamma}) is written as
\bea\label{Ptheta}
&&P_\theta={\cal P}-2{\rm Re}\left({\cal E}_+{\cal E}_-^* e^{i\theta}\right)~,
\\\label{P}
&&{\cal P}=|{\cal E}_+|^2+|{\cal E}_-|^2~,
\\\label{Epm}
&&{\cal E}_\pm=s_\pm e v_F E_\w \w^{-1}(\w\pm\wc)^{-1}~.
\eea

The golden-rule approach formulated above describes a
rich variety of phenomena in strong dc and microwave fields. We first discuss it
in the case of MIRO and calculate the linear direct current at high temperature,
namely for $2\pi^2 T/\wc\gg 1$, to order $E E_\w^2$. The current reads
\be\label{jdisin}
j_d=\sigma_{\rm D} \left\langle\tilde{\nu}_\ve^2\right\rangle_\ve E+
\sigma^{\rm dis}E+\sigma^{\rm in}E~.
\ee
Here the first term---describing the linear dark conductivity---follows
directly from \req{II.36} and $\langle\ldots\rangle_\ve$ denotes the energy
averaging over the period $\wc$. The two other terms are microwave-induced
corrections, at this order in the external fields completely independent of each
other. The displacement contribution, proportional to $\sigma^{\rm dis}$,
originates from the photon-assisted displacements in Eqs.~(\ref{jx}), (\ref{W})
if one substitutes the equilibrium function $f^{(T)}_{\ve}$ for $f_{\ve}$. In
the limit $2\pi^2 T/\wc\gg 1$, a straightforward calculation yields
\bea\label{sigmadis}
\sigma^{\rm dis}&=&\sigma_{\rm D}\frac{\ttr}{4\tst}\left[\,{\cal P}-{\rm
Re}\left({\cal
E}_+{\cal E}_-^*\right)\,\right]({\cal R}_1-{\cal R}_3)~,
\\\label{R1}{\cal
R}_1&=&\w\partial_\w\left\langle\tilde{\nu}_\ve\tilde{\nu}_{\ve+\w}
\right\rangle_\ve~,
\\\label{R3}{\cal
R}_3&=&\left\langle\tilde{\nu}_\ve^2-\tilde{\nu}_\ve\tilde{\nu}_{\ve+\w}
\right\rangle_\ve~,
\eea
where $\tst^{-1}$ is expressed in terms of the partial contributions
(\ref{tau_n}) to the disorder-induced scattering rate as follows (the angle
brackets denote averaging over $\theta$):
\be\label{tau*}
\tst^{-1}=2\left\langle\tau_\theta^{-1}
(1-\cos\theta)^2\right\rangle_\theta=3\tau_0^{-1}-4\tau_1^{-1}+\tau_2^{-1}~.
\ee
The inelastic contribution to $j_d$, proportional to $\sigma^{\rm in}$,
accounts for the microwave-induced change $\delta\! f_{\ve}=f_\ve-f^{T}_\ve$ of
the distribution function. For $E\to 0$, the last term in Eq.~(\ref{f}) vanishes
and $\ve=\ve^\prime$. To first order in $E_\w^2$,
\bea\nonumber
\delta\! f_\ve&=&{\cal
P}\frac{\tin}{4\ttr}\sum\nolimits_\pm\left(f^{T}_{\ve\pm\w}-f^{T}
_\ve\right)\tilde{\nu}_{\ve\pm\w}\\
&\simeq& {\cal
P}\frac{\w\tin}{4\ttr}(\tilde{\nu}_{\ve+\w}-\tilde{\nu}_{\ve-\w})\partial_\ve
f^{T}_{\ve}~.
\label{f1}\eea
The approximation in the last line is valid for $T\gg\w$.

At order $EE_\w^2$ in $j_d$, the correction (\ref{f1}) of order $E_\w^2$ should
be substituted into Eqs.~(\ref{jx})-(\ref{M}) taken at first order in $E$, which
gives
\bea\label{sigmain}
&&\sigma^{\rm in}=-\sigma_{\rm D}\int \!d\ve\, \tilde{\nu}_{\ve}^2
\partial_\ve  \delta\! f_{\ve}
=\sigma_{\rm D}\frac{\tin}{4\ttr}{\cal P}{\cal R}_2,
\\\label{R2}&&{\cal
R}_2=\w\partial_\w\left\langle\tilde{\nu}^2_\ve(\tilde{\nu}_{\ve+\w}+\tilde{\nu}
_{\ve-\w})\right\rangle_\ve.
\eea

Next we summarize the most important properties of MIRO that follow from
the above results.
\paragraph{Period and phase.}\label{phase} These are determined by the factors
${\cal R}_i$ defined in Eqs.~(\ref{R1}), (\ref{R3}), and (\ref{R2}). In the
limit of overlapping LLs, the DOS is given by \req{II.28},
$\tilde{\nu}(\ve)=1-2\delta\cos(2\pi\ve/\omega_c)$ with
$\delta=\exp(-\pi/\omega_c\tau_{\rm q})\ll 1$, so that
\bea\nonumber
&&{\cal R}_1={\cal R}_2/4=-4\delta^2\frac{\pi\w}{\wc}\sin\frac{2\pi\w}{\wc},\\
&&{\cal R}_3=4\delta^2\sin^2\frac{\pi\w}{\wc}~.\label{RROvLLs}
\eea
We see that, for both the displacement and inelastic mechanisms, MIRO are
proportional to $\sin(2\pi\w/\wc)$ with a negative coefficient in front of it in
accord with the experimental findings
(for the displacement mechanism, the phase
of MIRO also agrees with the observed one for large $\pi\omega/\omega_c$, when
${\cal R}_3$ can be neglected compared to ${\cal R}_1$).

\begin{figure}[ht]
\includegraphics[width=\columnwidth]{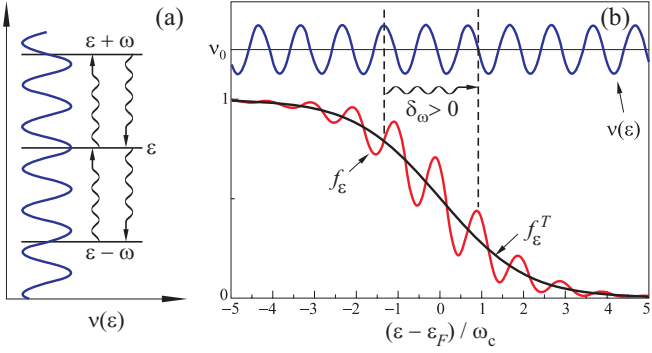}
\caption{(a) Emission/absorption of microwave quanta for the detuning
$\delta_\w=\w/\wc-2=1/4$ and (b) the DOS $\nu(\ve)$ and the resulting
oscillatory distribution function $f_\ve$ for $T=\wc$, compared to the thermal
distribution $f^T_\ve$. Adapted from \citet{dmitriev:2005}.}
\label{distribution}
\end{figure}

The phase of the oscillations can be qualitatively understood as follows. Owing
to the oscillatory behavior of $\tilde{\nu}_{\ve}\tilde{\nu}_{\ve\pm \omega+eE
\Delta X}$, which enters Eq.~(\ref{W}) via the factor (\ref{M}), the average
displacement of a cyclotron orbit $\Delta X$ (Fig.~\ref{DeltaR}b) is positive
(negative) for a small positive (negative) detuning $\delta_\w=\w/\wc-N$ from
the $N$th harmonic of the CR. The uphill drift for $\delta_\w>0$ produces a
contribution to the dissipative current (\ref{jx}) which is directed against the
electric field, i.e., $\sigma^{\rm dis}<0$. For the inelastic mechanism, a
small positive detuning $\delta_\w$ increases (decreases) the occupation of
states lying right above (below) the maxima of $\tilde{\nu}(\ve)$ compared to
equilibrium (Fig.~\ref{distribution}), while the negative detuning leads to the
opposite sign of the nonequlibirum correction to the distribution function. The
phase difference between the oscillations of $\nu(\varepsilon)$ and those of
$\partial_\ve\delta\! f_{\ve}$ in Eq.~(\ref{sigmain}) is such that the sign of
$-\partial_\ve\delta\! f_{\ve}$ at the maxima of $\tilde{\nu}(\ve)$ coincides
with the sign of $\sigma^{\rm in}$ (Fig.~\ref{distribution}), i.e.,
$\delta_\w>0$ yields $\sigma^{\rm in}<0$.

%%%%%%%%%%%%%%%%%%%%%%%%%%%%%%%%%%%%%%%%%%%%%%%%%%%%%%%%%%%%%%%%%
%%%%%%%%%%%%%%%%%%%%%%%%%%%%%%%%%%%%%%%%%%%%%%%%%%%%%%%%%%%%%%%

\begin{figure}[ht]
\centerline{
\includegraphics[width=\columnwidth]{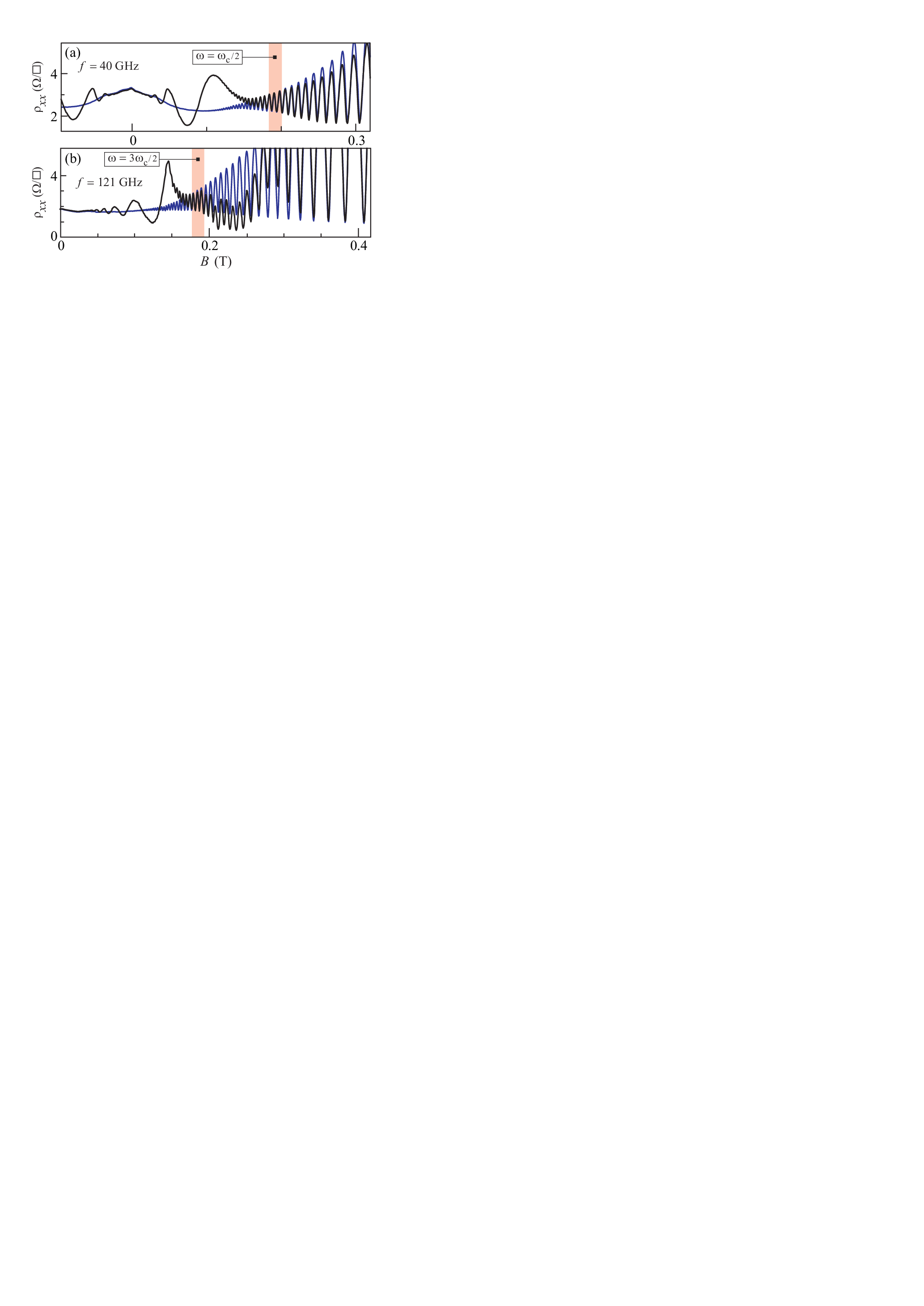}
}
\caption{Magnetoresistivity at a radiation frequency of (a) $40\,{\rm GHz}$ and
(b) $121\,{\rm GHz}$ as a function of $B$, compared to the magnetoresistivity in
the absence of radiation. The shaded boxes around the points at which (a)
$\w=\wc/2$ and (b) $\w=3\wc/2$ mark the range of $B$ within which the
magnetoresistivity with and without radiation coincide. Adapted from
\citet{dorozhkin:2005}.}
\label{windows}
\end{figure}

In the limit of separated LLs, $\wc\tau_{\rm q}\gg 1$, the DOS is a sequence of
semicircles (\ref{II.25}) of width $2\Gamma=2(2\wc/\pi\tau_{\rm
q})^{1/2}\ll\wc$, i.e., $\tilde{\nu}(\ve)=\tau_{\rm q}{\rm
Re}\sqrt{\Gamma^2-(\delta\ve)^2}$, where $\delta\ve$ is the detuning from the
center of the nearest LL. In this limit, one has
\bea
\label{R1sep}
&&{\cal R}_1={\cal R}_0\frac{\w}{\Gamma}\sum\limits_n{\rm sgn}(\Omega_n){\cal
H}_2(|\Omega_n|)\,,\\\label{R2sep}
&&{\cal R}_2=-{\cal R}_0\frac{4\w\wc}{\Gamma^2}
\sum\limits_n{\rm sgn}(\Omega_n) \Phi_2(|\Omega_n|)\,,\\
&&{\cal R}_3={\cal R}_0\left[1-\sum\limits_n{\cal
H}_1(|\Omega_n|)\right],\label{R3sep}
\eea
where ${\cal R}_0$ [which also enters \req{jdisin}] reads
\be
{\cal R}_0\equiv\langle\tilde{\nu}_\ve^2\rangle_\ve=16\wc/3\pi^2\Gamma~.
\label{R0sep}
\ee
The parameterless functions of $\Omega_n=(\w-n\wc)/\Gamma$ are nonzero at
$0\!<\!|\Omega_n|\!<\!2$, where they are given by
\bea\label{H1}
{\cal H}_1(x)\!&=&\!(2+x)\,[\,(4+x^2)E(Y)-4x K(Y)\,]/8\,,\qquad
\\\label{H2}
{\cal H}_2(x)\!&=&\!3x\,[\,(2+x)E(Y)-4K(Y)\,]/8\,,
\\\label{Phi2}
4\pi\,\Phi_2(x)\!&=&\! 3x\,{\rm arccos}(x-1)-x(1+x)\sqrt{x(2-x)}\,.
\eea
Here $Y=(2-x)^2/(2+x)^2$ and the functions $E$ and $K$ are the complete elliptic
integrals of the first and second kind, respectively [see \cite{dmitriev:2007b}
for a graphical representation of the functions (\ref{H1})-(\ref{Phi2})]. A
distinctive feature of the photoresponse in separated LLs is the presence of
windows in $\wc$ within which $|\w-n\wc|>2\Gamma$ for any integer $n$ and hence both
intra- and inter-LL single-photon transitions are impossible, i.e.,
$\tilde{\nu}_\ve\tilde{\nu}_{\ve+\w}=0$ for any $\ve$. The resulting gaps
in the dependence of the photoresponse on $B$ for a given radiation frequency
were observed by \textcite{dorozhkin:2005}, see \rfig{windows}. At sufficiently
high microwave power, the photoresponse in the gaps becomes visible due to
multiphoton effects (\rsec{s6}).

%%%%%%%%%%%%%%%%%%%%%%%%%%%%%%%%%%%%%%%%%%%%%%%%%%%%%%%%%%%%%%%%%
%%%%%%%%%%%%%%%%%%%%%%%%%%%%%%%%%%%%%%%%%%%%%%%%%%%%%%%%%%%%%%%

\paragraph{Polarization dependence.}
\label{polarization}
While the phase of MIRO in $\sigma^{\rm dis}$ and $\sigma^{\rm in}$ is
essentially the same, their polarization and temperature dependences are
qualitatively different. For linear polarization with $s_\pm=e^{\mp
i\psi}/\sqrt{2}$ characterized by the angle $\psi$ between the directions of the
microwave and dc fields, $\sigma^{\rm dis}$ contains the anisotropic term
${\rm Re}\left({\cal E}_+{\cal E}_-^*\right)\propto 2{\rm Re}\,s_+^2=\cos
2\psi$, while $\sigma^{\rm in}\propto {\cal P}$ does not depend on $\psi$. The
polarization dependence of $\sigma^{\rm dis}$ is illustrated in Fig.~13 in
\citet{vavilov:2004}. Experimental results on the polarization dependence are
discussed in \rsec{s7.2.4prim}.

\paragraph{Screening of the microwave field.}
\label{screening}
The dimensionless field ${\cal E}_\pm$ and the polarization parameters $s_\pm$
in \req{Epm} represent the strength and the polarization of the total (screened)
microwave field. As discussed in \rsec{s2.3.2} (see also \rsec{s7.2.4prim}), in
high mobility samples the active circular component of the incoming
electromagnetic wave is strongly suppressed near the CR due to a nearly complete
reflection. Expressed in terms of the field $E_\w^{(0)}$ and the polarization
$s^{(0)}_\pm$ of the incoming wave, \req{Epm} acquires the form
\be\label{Epm1}
{\cal E}_\pm=\frac{s^{(0)}_\pm e v_F E^{(0)}_\w }{\w(\w\pm\wc+i\gamma)}~,
\ee
where $\gamma$ [Eq.~\eqref{II.51}, see also footnote \ref{fn_screen}] is assumed
to be much larger than $\ttr^{-1}$. Since in typical experiments $\gamma$ is of
order $\w\gg\ttr^{-1}$, the CR in $\cal P$ [\req{P}] is strongly broadened by
the screening. Note also that the polarization of the screened field is
generally different from that of the field in the incident wave and depends on
$B$.

\begin{figure}[ht]
\centerline{
\includegraphics[width=\columnwidth]{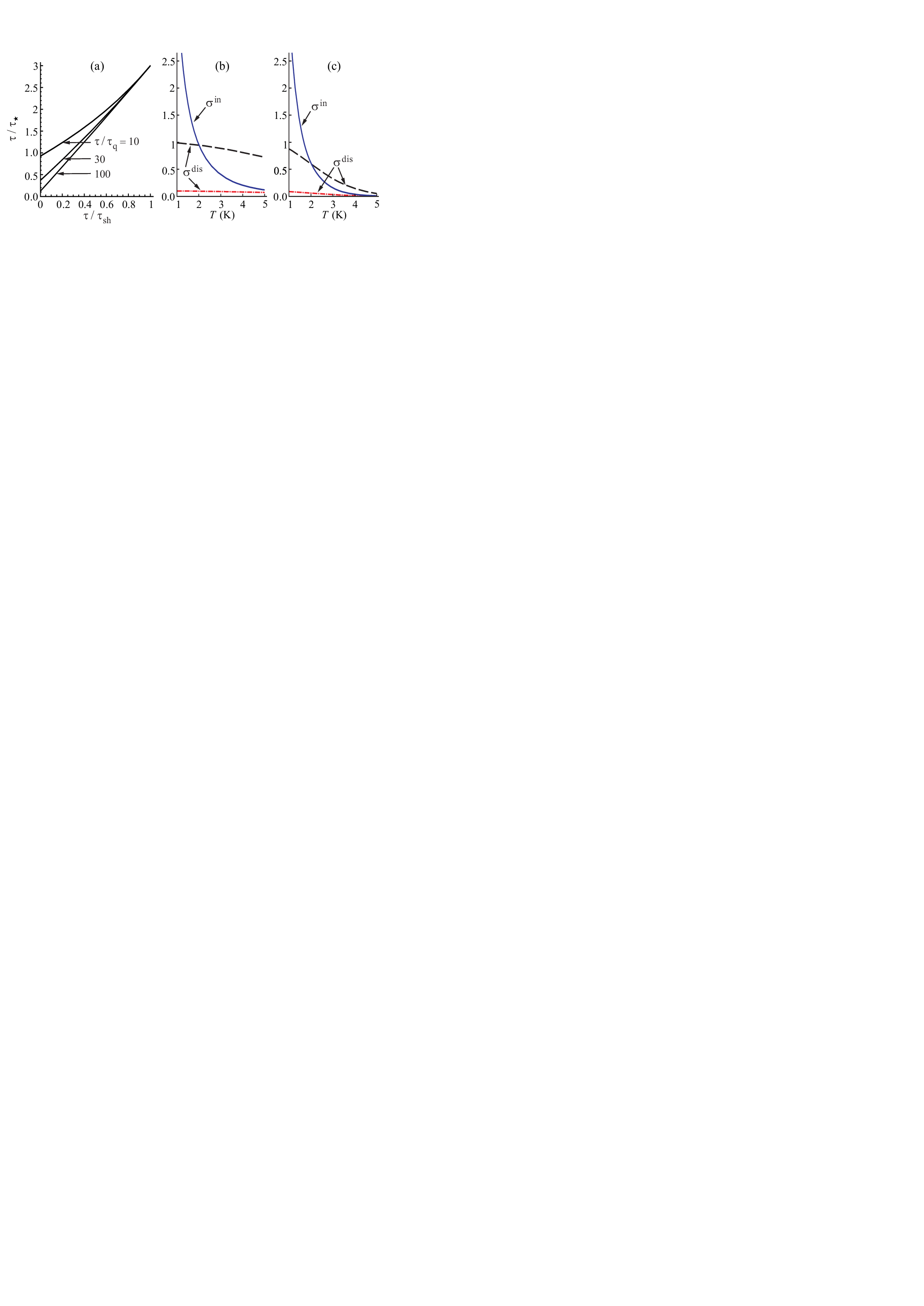}
}
\caption{(a) Dependence of $\ttr/\tst$ on the relative weight $\ttr/\tsh$ of the
short-range component of disorder in the transport scattering rate for fixed
$\ttr/\tq=100,\,30,$ and 10.
(b),(c) Dependence of $\sigma^{\rm in}$ and $\sigma^{\rm dis}$
% RR
(in arbitrary units) on $T$ for
(b) $T_{\rm q}=10\,{\rm K}$ and (c) $T_{\rm q}=3\,{\rm K}$.
The curves for $\sigma^{\rm in}$ are calculated for $\tin=2\ttr$ at $T=1\,{\rm
K}$; the dashed curves for $\sigma^{\rm dis}$---for strong short-range
disorder
 with $\ttr/2\tst=1$, $\ttr/\tsh=0.6$, and $T_\star=2\,{\rm K}$; the dash-dotted
curves for $\sigma^{\rm dis}$---for purely long-range disorder with
$\ttr/2\tst=0.1$, $\ttr/\tq=50$,
and $T_\star\simeq 6\,{\rm K}$.) Adapted from \citet{dmitriev:2009b}.}
\label{tau_star}
\end{figure}

\paragraph{Sensitivity to different types of disorder.}
\label{mixed_disorder}
The displacement contribution $\sigma^{\rm dis}$ is highly sensitive to the
details of the disorder potential \cite{khodas:2008,auerbach:2007}, which are
difficult to extract from standard transport measurements. In the mixed-disorder
model (Sec.~\ref{s2.1}), believed to provide an adequate description of high
mobility 2DEGs, the partial contributions (\ref{tau_n}) to the scattering rate
are given by
\begin{align} \label{mixed-disorder}
\frac{1}{\tau_n}=\frac{\delta_{n0}}{\tsh}+\frac{1}{\tsm}\frac{1}{1+\chi n^2}~,
\end{align}
where the first term describes scattering off the short-range component of the
random potential (modeled here as white-noise disorder) and the second term
describes small-angle scattering off the smooth random potential created by
remote donors. The characteristic scattering angle in the latter case is
$\sqrt{\chi}=(2 k_F d)^{-1}\ll 1$. Since in high mobility structures the quantum
scattering rate $\tau_{\rm q}^{-1}$, given by $\tau_0^{-1}$, is much larger than
the transport scattering rate $\tau^{-1}=\tau_0^{-1}-\tau_1^{-1}$, the quantum
scattering rate is dominated by the contribution of the long-range component:
$\tau_{\rm q,sm}^{-1}\gg\tau_{\rm sh}^{-1}$. That is, the LL broadening is largely
determined by the smooth disorder, whereas the relative weight of the short- and
long-range components in $1/\ttr\simeq{1}/{\tsh}+\chi/{\tsm}$ may be arbitrary.
Moreover, the displacement contribution $\sigma^{\rm dis}$
is proportional to the rate $\tst^{-1}$ given by Eq.~(\ref{tau*})
\cite{khodas:2008},
\be\label{t*}
\frac{ 1 }{ \tst }\simeq \frac{ 3 }{ \tsh } +\frac{12\chi^2}{ \tsm }~,
\ee
which contains one more power of the small parameter $\chi$ in front of
$\tau_{\rm q,sm}^{-1}$ compared to $\tau^{-1}$ \cite{vavilov:2004}. As a result,
even
a small amount of short-range scatterers may give the main contribution to
$\sigma^{\rm dis}$, as illustrated in Fig.~\ref{tau_star}a.

%%%%%%%%%%%%%%%%%%%%%%%%%%%%%%%%%%%%%%%%%%%%%%%%%%%%%%%%%%%%%%%
%%%%%%%%%%%%%%%%%%%%%%%%%%%%%%%%%%%%%%%%%%%%%%%%%%%%%%%%%%%%%%%

\paragraph{Temperature dependence.}
\label{temperature}
The displacement contribution [Eq.~\eqref{sigmadis}] is $T$-independent (apart
from the $T$ dependence of the DOS, which becomes relevant at high $T$ for both
mechanisms, see \rsec{damping}), while the inelastic contribution
[Eq.~\eqref{sigmain}], proportional to the inelastic scattering time $\tau_{\rm
in}$,
grows with lowering $T$. A detailed analysis of the inelastic relaxation was
performed by \citet{dmitriev:2005}. It was found that, for relevant $T$, the
amplitude of the oscillations in the distribution function is controlled by
electron-electron collisions. In overlapping LLs, the ansatz
$f=f_\ve^T+\varphi(\ve)\;\partial_\ve f_\ve^T$  with periodic
$\varphi(\ve)=\varphi(\ve+\wc)$ reduces the linearized collision integral
(\ref{f}) to
\bea
{\rm St}_{\rm in}\left\{f\right\}&=&-\frac{\pi^2 T^2+\ve^2}{2}\:\frac{\partial
f_\ve^T}{\partial\ve}\left\langle
A(E)\,[\,\varphi(\ve)-\varphi(\ve+E)\right.\nonumber\\
&+&\left.\!\varphi(\ve^\prime)-\varphi(\ve^\prime-E)\,]\,\right\rangle_{
\varepsilon^{\prime}E}~,
\label{St_tr}
\eea
where the angular brackets denote averaging over $\ve^\prime$ and $E$ within the
period $\omega_c$. For a harmonic modulation of the distribution function with
$\varphi(\ve)\propto \cos(2\pi\ve/\omega_c+\theta)$, the collision integral then
acquires the form ${\rm St}_{\rm in}\propto\langle A(E)[\,1-\cos(2\pi
E/\omega_c)\,]\rangle_E$, where the last factor strongly suppresses the
relevant energy-relaxation rate compared to the out-scattering rate. As shown by
\citet{dmitriev:2005}, the relaxation time approximation is justified for this
collision integral and the inelastic scattering time that enters
Eq.~(\ref{sigmain}) is given in the relevant domain $\wc\ll T\ll
\wc(\wc\ttr)^{1/2}$ by
\be
\tau_{\rm in}=\int\! \frac{d\ve}{2\omega}\,\tau_{\rm
ee}(\ve,T)\left(f^T_{\ve-\omega}-f^T_{\ve+\omega}\right)~,
\label{tau_gen}
\ee
where
\be
\tau_{ee}^{-1}(\ve,T)=\frac{\pi^2 T^2+\ve^2}{4 \pi\ve_F}\ln\frac{2
v_F/a_B}{\omega_c(\omega_c\ttr)^{1/2}}~.
\label{out}
\ee

For $T\gg\omega$, Eq.~(\ref{tau_gen}) gives $\tin\simeq 0.822
\tau_{ee}(0,T)\propto T^{-2}$ (\rfig{tau_star}b,c), while for $T\ll\omega$ the
result is $\tau_{\rm in}=(\pi^2 T/2\omega)\,\tau_{\rm ee}(0,T)$. It follows that
the $T^{-2}$ scaling of $\sigma^{\rm in}$ for $T\gg\omega$ crosses over to a
$T^{-1}$ scaling for $T\ll\omega$. In separated LLs, the relaxation time
approximation is not accurate parametrically; still, for an estimate---up to a
factor of order unity---one can use Eq.~\eqref{sigmain} with $\tin\sim
(\Gamma/\wc)\tau_{ee}(0,T)$.

\paragraph{Exponential $B$- and $T$-damping.}\label{damping}
According to Eqs.~(\ref{sigmadis}), (\ref{sigmain}), (\ref{RROvLLs}), the Dingle
factor squared, $\delta^2=\exp(-2\pi/\wc\tq)$, determines the damping of the
MIRO in the limit of overlapping LLs for both the displacement and inelastic
mechanisms. This same damping factor also describes the oscillations of the
absorption coefficient for $T\gg\omega_c$ (Sec.~\ref{s2.3.2}) and, as will be
seen in Sec.~\ref{s5}, HIRO and PIRO. Experimental data for these types
of oscillations measured in the same sample support this prediction. On the
other hand, the SdH measurements---if the data are fitted by the damping factor
$\delta$---systematically yield shorter $\tau_{\rm q}^{\rm (SdH)}<\tq$. As
discussed in \rsec{s2.3.1}, the reason for the enhanced damping is that the SdH
oscillations are sensitive to small, with an amplitude of the order of $\wc$,
macroscopic inhomogeneities of the chemical potential, whereas the
magnetooscillations that survive at $T\gg\wc$ are robust with respect to them.

The experimentally observed MIRO (\rsec{s3.1}) show at sufficiently high $T$ an
exponential suppression of the oscillation amplitude as $T$ is increased
\cite{hatke:2009a,wiedmann:2010a}. This effect can be explained in terms of a
$T$-dependent renormalization of the DOS by electron-electron interactions
\cite{chaplik:1971,ryzhii:2004,dmitriev:2009b}. Specifically, the quantum
scattering rate $\tau_{\rm q}^{-1}$ that enters the DOS
$\tilde\nu(\ve)=1-2\delta\cos(2\pi\ve/\wc)$ via the Dingle factor $\delta$
should be substituted by
\begin{equation}\label{delta_ee}
\tilde{\tau}_{\rm q}^{-1}=\tau_{\rm q}^{-1}+{\tau}_{\rm ee}^{-1}(\ve,T)~,
\end{equation}
where ${\tau}_{\rm ee}^{-1}$ is given by Eq.~(\ref{out}). Substitution of the
modified DOS into Eqs.~(\ref{W})-(\ref{f}) yields an additional
factor $\exp(-T^2/T_{\rm q}^2)$ in the amplitude of MIRO for $T$ above the
temperature $T_{\rm q}$ defined by $2\pi/\wc{\tau}_{\rm ee}(0,T_{\rm q})=1$
[which gives $T_{\rm q}\sim(\wc\ve_F)^{1/2}$ up to a logarithmic factor], see
Figs.~\ref{fig.miro.temp} and \ref{tau_star}b,c. It is worth recalling that
electron-electron scattering produces no additional exponential damping for the
SdH oscillations, see footnote \ref{fn_sdho}.
The difference in the manifestation of electron-electron interactions
in the damping factor for SdH oscillations and in the damping factor for MIRO is
related to the fact that the former oscillations emerge at linear order in
$\delta$, while the latter---at order $\delta^2$.

\paragraph{Relative weight of the inelastic and displacement
contributions.}\label{in_vs_dis}
The discussion in Secs.~III.B.1.d-f shows that the relative weight of
$\sigma^{\rm dis}$ and $\sigma^{\rm in}$ in the amplitude of MIRO
strongly depends on $T$ and the correlation properties of disorder. For an
estimate of their relative importance, we omit the anisotropic part
[Eq.~(\ref{sigmadis})] and the subleading for $\omega\gg\omega_c$ term
proportional to ${\cal R}_3$ [Eq.~(\ref{RROvLLs})] in $\sigma^{\rm dis}$, and
combine Eqs.~\eqref{sigmadis}--\eqref{R3} and \eqref{sigmain}--\eqref{RROvLLs}
into
\be
\frac {\sigma^{\rm in}+\sigma^{\rm dis}}{\sigma_{\rm D}} \!\sim
-\!\left(\frac{2\tin}{\ttr}+\frac{\ttr}{2\tst}\right) 4\delta^2 {\cal P}
\frac{\pi\w}{\wc} \sin \frac{2\pi\w}{\wc}~.
\label{eq.miro}
\ee
Since $\tin\sim \ve_F/T^2$ (\rsec{temperature}), \req{eq.miro} defines the
temperature $T_\star\sim(\ve_F\tst)^{1/2}/\ttr$ at which $\sigma^{\rm
dis}=\sigma^{\rm in}$. In the case of smooth disorder [when $\tst^{-1}$ is
given by the second term in \req{t*}], \req{eq.miro} yields
\be\label{BtoA}
\frac{\sigma^{\rm in}}{\sigma^{\rm
dis}}=\frac{4\tin(T)\tau_\star}{\ttr^2}\simeq\frac{\tin(T)}{3\tq},
\quad\frac{\ttr}{\tsh}< \frac{4\tq}{\ttr}~.
\ee
For overlapping LLs, \req{BtoA} gives $T_\star\sim(\ve_F/\tq)^{1/2}\gg T_{\rm
q}\sim(\wc\ve_F)^{1/2}$, i.e., in the case of smooth disorder, the inelastic
contribution dominates in the whole temperature range $T\alt T_{\rm q}$ where
MIRO can be observed (Fig.~\ref{tau_star}b,c). By contrast, for the case of a
strong short-range component of disorder ($\tau_{\rm sh}\sim\ttr\sim\tst$), one
has $T_\star\sim T_{\rm q}(\tst/\wc\ttr^2)^{1/2}$, much smaller than $T_{\rm q}$
for $\wc\ttr\gg 1$. As a result, the $\sigma^{\rm in}$-dominated $T^{-2}$
dependence of the MIRO amplitude for $T<T_\star$ crosses over as $T$ is
increased to a plateau (whose height is given by $\sigma^{\rm dis}$) in the
interval $T_\star<T<T_{\rm q}$ (Fig.~\ref{tau_star}b) before starting to fall
off exponentially for $T>T_{\rm q}$. In the intermediate case of $T_\star\sim
T_{\rm q}$, the range of $T$ in which MIRO are $T$-independent shrinks to
zero (Fig.~\ref{tau_star}c).

The experimental results, presented in \rfig{fig.miro.temp}, are in overall
agreement with the above theory. Apart from the $T$ dependence, the two
mechanisms are different in the photoresponse at the ``odd nodes''
$\w/\wc=n+1/2$, where $\sigma^{\rm in}=0$ while $\sigma^{(\rm
dis)}\propto{\cal R}_3\neq 0$, see \reqs{sigmadis}{R3}, and (\ref{RROvLLs}).
Moreover, their polarization dependence is different (\rsec{polarization}).
Experimental studies of these features [omitted in the estimate (\ref{eq.miro})]
can serve as an additional tool to quantify the relative magnitude of
$\sigma^{\rm dis}$ and $\sigma^{\rm in}$.

\subsubsection{Saturation of the inelastic contribution at high radiation power
and/or in a strong dc field}
\label{s3.2.2}
\begin{figure}[ht]
\centerline{
\includegraphics[width=\columnwidth]{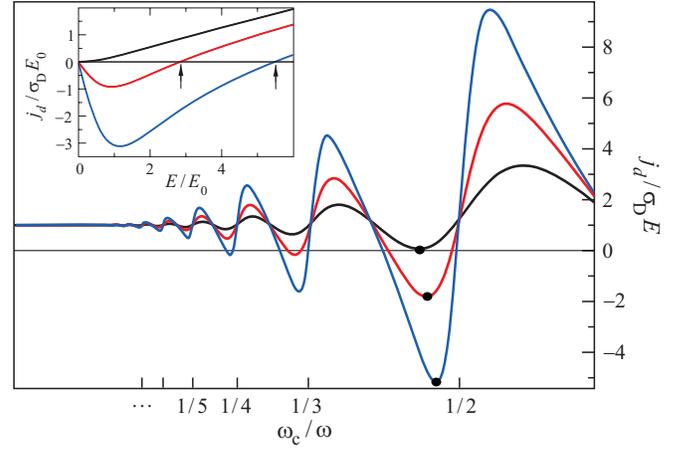}
}
\caption{Conductivity normalized to the dark Drude value (given, as we assume
$\wc\ttr\gg 1$, by $j_d/\sigma_{\rm D} E$) as a function of $\omega_c/\omega$ at
fixed $\omega\tau_{\rm q}=2\pi$ for different levels of microwave power
$(\tin/\ttr){\cal P}|_{\wc=0}=0.24\, ({\rm bottom}),\,0.8,\,2.4\, ({\rm top})$.
Inset: $I$-$V$ characteristics at the minima (marked by the dots) of the $\w/\w_c$
dependence of the resistivity. The arrows show the value of the dc field $E^*$
in domains in units of $E_0=\pi\wc\tau/2\tin e R_c$. Adapted from
\citet{dmitriev:2005}.}
 \label{MIROov}
 \end{figure}

\noindent
When the inelastic contribution dominates the oscillatory photoresponse
(\rsec{in_vs_dis}), the leading correction to $\sigma^{\rm in}$ linear in
${\cal P}$ and independent of $E$ [Eq.~(\ref{sigmain})] is given by the terms of
higher order in $\cal P$ and $E^2$ in the distribution function. These are
governed by the kinetic equation \eqref{f}, a generalized---to include higher
powers of $\cal P$ and $E^2$---solution of which in the regime of overlapping
LLs reads \cite{dmitriev:2005}
\bea
\label{fosc}
&&\delta f = \delta\,\frac{\omega_c}{2\pi}\: \frac{\partial f^T_\ve}{\partial
\ve}\: \sin \frac{2\pi\ve}{\omega_c}\:{\cal F}(\tin, {\cal P}, \zeta)\,,\\
\label{FF}
&&{\cal F}(\tin, {\cal P}, \zeta)=\frac{{\cal P}\frac{2\pi
\omega}{\omega_c}\sin\frac{2\pi\omega}{\omega_c}+2\zeta^2} {\ttr/\tin+{\cal P}
\sin^2\frac{\pi\omega}{\omega_c}+\zeta^2/2}~,
\eea
where the dimensionless parameter
\be\label{zeta1}
\zeta\equiv\pi\edc=2\pi e E R_c/\wc
\ee
characterizes the strength of the dc field. The current induced by the inelastic
mechanism is then given by
\be\label{inelasticMain}
j_d=\sigma_{\rm D} E\,[\,1+2\delta^2-2\delta^2 \,{\cal F}(\tin, {\cal P},
\zeta)\,]~.
\ee
The dependence of the resulting magnetoresistivity on ${\cal P}$ and $E$ is
illustrated in Fig.~\ref{MIROov}.

Let us consider first the limit $E\to 0$. At fixed $\w/\wc$, MIRO saturate
with increasing microwave power:
\be\label{sat}
{\cal F}(\tin, {\cal P}, 0)\to \frac{4\pi\omega}{\omega_c}\cot\frac{\pi
\omega}{\omega_c},\quad{\cal
P}\sin^2\frac{\pi\omega}{\omega_c}\gg\frac{\ttr}{\tin}~.
\ee
In the limit of high power ${\cal P}\gg\ttr/\tin$, the maxima and minima of
MIRO at $\eac=\eac^{\pm}$ [Eq.~(\ref{eq.miro.max})] shift towards the
nearest nodes at $\eac=n$, with the amplitude of MIRO at the extrema being
proportional to $(\tin{\cal P}/\ttr)^{1/2}$:
\be\label{highP}
\eac^{\pm}\!=n\mp\!{1\over\pi}\left(\frac{\ttr}{\tin{\cal P}}\right)^{1/2},\quad
{\cal F}|_{\eac=\eac^{\pm}}\!= \mp 2\pi\left(\frac{\tin{\cal
P}}{\ttr}\right)^{1/2}~.
\ee
Note that the $T$ dependence changes in the nonlinear regime.
As follows from Eq.~(\ref{FF}), there remains---for arbitrarily strong $\cal
P$---a range of $\eac$ around the nodes within which the photoresponse is
linear
in ${\cal P}$ (apart from heating by microwaves which can modify $\tin$);
however, this range shrinks with increasing $\cal P$. In a broader vicinity of
$\w/\w_c=n$ which includes
the nearest maxima and minima of MIRO, the result (\ref{inelasticMain})
applies---under the conditions $\tq\ll\wc^{-1}\ll\ttr\alt\tin$, which are met
in
typical experiments
\cite{dmitriev:2005,dmitriev:2007,khodas:2008,hatke:2011up1}---for arbitrary
microwave power and
for arbitrary relation between the components of mixed disorder
(\ref{mixed-disorder}). The resulting behavior with $\cal P$ reproduces the
observations in \rfig{fig.miro.power},
and is also in agreement with other experimental results
\cite{ye:2001,mani:2010}.

For $\tsh\sim\ttr$, i.e., in the presence of a strong short-range component of
disorder, one should for $\tin\alt\ttr$ take into account the displacement
contribution and also multiphoton processes (the latter are important at ${\cal
P}\sin^2\pi\w/\wc\agt 1$, see \rsec{sec.miro.dc.hp.exp}). This results in
\req{highP} with a factor of order
unity substituted for $\tin/\tau$ \cite{khodas:2008,hatke:2011up1}. If only the
smooth component of disorder is present, \reqs{inelasticMain}{FF} are
applicable
and give the main contribution to $j_d$ in a broader range ${\cal
P}\sin^2\pi\w/\wc<\ttr/\tq$; at higher ${\cal P}$, both the excitation of higher
angular and temporal harmonics of the
distribution function and the multiphoton processes become important, which
results in the emergence of a series of distinct strongly nonequilibrium
regimes
as $\w/\wc$ is varied [for more details see \cite{dmitriev:2007}]. The higher
harmonics of the distribution function also determine the Hall photoresponse
(\rsec{s3.2.4}).

The above consideration of the nonlinear-in-${\cal P}$ effects [\reqs{FF}{highP}]
shows that the linear-response conductivity [\req{inelasticMain} at $E\to0$] can
become negative despite $\delta\ll 1$, see Fig.~\ref{MIROov}. As discussed in
\rsec{s4.2}, this causes an electric instability leading to the formation of
domains. The electric field $E^*$ in the domains can be obtained from the
current-voltage characteristics (\ref{inelasticMain}), as illustrated in the
inset to Fig.~\ref{MIROov}, see \rsec{s4.3} for details. The behavior of the
magnetooscillations at high microwave power and in a strong dc field is further
discussed in \rsecs{s5}{s6}.

%%%%%%%%%%%%%%%%%%%%%%%%%%%%%%%%%%%%%%%%%%%%%%%%%%%%%%%%%%%%%
%%%%%%%%%%%%%%%%%%%%%%%%%%%%%%%%%%%%%%%%%%%%%%%%%%%%%%%%%%%%

\subsubsection{General approach: Quantum kinetic equation}
\label{s3.2.3}
A comprehensive approach to nonequilibrium transport in high LLs was initially
formulated for the case of smooth disorder
\cite{vavilov:2004,dmitriev:2005,dmitriev:2007} and later generalized to the
case of generic [Eq.~(\ref{tau_n})] disorder
\cite{vavilov:2007,khodas:2008,dmitriev:2009b}. The key steps of the derivation
are highlighted as follows.

\noindent(1) {\it Transformation $\br\rightarrow \br-\bzeta(t)$ to a moving
coordinate frame,} where $\bzeta(t)$ obeys
\be
\partial_t \bzeta(t)=\left(\frac{\partial_t -\w_c
\hat{\varepsilon}}{\partial_t^2+\w_c^2}\right)\frac{e}{m}[\bE_{\rm
dc}+\bE_\w(t)]
\label{bzeta}
\ee
with $\hat{\varepsilon}_{xy}=-\hat{\varepsilon}_{yx}=1$, is similar to the
transformation to Floquet states, used in a number of works
\cite{park:2004,lee:2004,lyapilin:2006,torres:2005,volkov:2007,kashuba:2006,
auerbach:2007} on the displacement mechanism of MIRO. This transformation
% RR
(unambiguously defined for $\omega \neq\omega_c$)
eliminates homogeneous electric fields in the Hamiltonian of a clean electron
system at the expense that the impurity potential, ``dressed by the electric
fields,'' becomes time-dependent. Scattering off the moving impurities can then
change the energy of electrons.

\noindent(2) {\it Keldysh equations within the SCBA.}
In the moving frame, the Green's functions $\hat{G}^{\alpha}$ and the
self-energies $\hat{\Sigma}^{\alpha}$ ($\alpha=R, A, K$ refer to the
retarded, advanced, and Keldysh components) are related within the SCBA
(Sec.~\ref{s2.3.1a}) as
\be
\hat{\Sigma}^{\alpha}_{21}=\int\!
\frac{d^2q}{(2\pi)^2}\,W_q\,e^{-i\q\bzeta_{21}}\left(e^{i\q\hat{\br}}\hat{G}^{
\alpha}
e^{-i\q\hat{\br}}\right)_{21}~,
\label{SCBAgen}
\ee
where $W_q$ is the correlation function of the bare disorder potential, the
subscript (21) denotes times $t_2$ and $t_1$ on the Keldysh contour, and
$\bzeta_{21}=\bzeta(t_2) - \bzeta(t_1)$.

\noindent(3) {\it Quasiclassical approximation.}
For a degenerate 2DEG in high LLs, Eq.~(\ref{SCBAgen}) can be reduced to a
simpler quasiclassical equation in the ``action-angle'' representation:
\bea
\label{SCBA}
&&\Sigma^{\alpha}_{21}(\varphi)=
-i\hat{\cal K}_{21}\,g^{\alpha}_{21}(\varphi)~,\\
\label{gg}
&& g^{\alpha}_{21}(\varphi)\equiv i\wc\sum_k
\hat{G}^{\alpha}_{21}(\hat{n}+k;\,\hat{\varphi})~,
\eea
where the operators $\hat{n}$ and $\hat{\varphi}$ are canonically conjugated,
$[\hat{n},\hat{\varphi}]=-i$. The eigenvalues of $\hat{n}$ and $\hat{\varphi}$
are the LL index and the angle coordinate of the momentum, respectively. The
effects of disorder and external fields are encoded in the integral operator
$\hat{\cal K}$,
\be \label{K}
\hat{\mathcal{K}}_{21} F(\varphi)=\int \frac{ d \varphi' }{ 2 \pi}\frac{ e^{ i
k_{\rm F}(\bn_{\varphi}-\bn_{\varphi'})
\bzeta_{21}}}{\tau_{\varphi - \varphi'} } F(\varphi')~.
\ee
The distribution function $\hat{f}$, defined by
$\hat{G}^R-\hat{G}^A-\hat{G}^K=2(\hat{G}^R\hat{f}-\hat{f}\hat{G}^A)$, commutes
with $\hat{\varphi}$. Accordingly, the operator $\hat{\varphi}$, which enters
Eq.~(\ref{SCBA}) and the impurity collision integral,
\be
\label{St}
i\,\St_{\rm im}\{ f \}=\hat{\Sigma}^R\hat{f}-\hat{f}\hat{\Sigma}^A+
(\hat{\Sigma}^K+\hat{\Sigma}^A-\hat{\Sigma}^R)/2~,
\ee
can be treated as a $c$-number. The result is the quantum kinetic equation
\bea
\label{kineq}
\left(\partial_t+\wc\partial_\varphi\right)f_{21}&-&
\St_{\rm in}\{f\}_{21}=\St_{\rm im}\{f\}_{21}~,\\
\St_{\rm im}\{f\}_{21}&=&\int\! dt_3\, \left[\,\hat{\cal K}_{21}\left(g_{23}^R
f_{31}-f_{23}g_{31}^A\right)\right.\nonumber \\
&-&\left. f_{31}\hat{\cal  K}_{23}g_{23}^R
+f_{23}\hat{\cal K}_{31}g_{31}^A\,\right]~.
\label{St_im}
\eea
Here $t=(t_1+t_2)/2$ is the ``center-of-mass'' time. The inelastic collision
integral $\St_{\rm in}\{f\}_{21}$ accounts for electron--electron scattering and
for the coupling to a thermal (phonon) bath.

Except for specific cases (\rsec{VI.A.3}) in the regime of separated LLs, the
effect of external fields on the spectrum can be neglected, meaning
$g_{t_1-t_2}^R$ in the moving frame remains the same as in the static frame and
does not depend on $t$ and $\phi$. It follows that the direct current
\be
\label{curgen}
{\j}=2ev_F\int d\ve \,\nu(\ve) \langle
\overline{{\bn}_\varphi f(\ve,\varphi,t)}
\rangle- 2 e \nu_0 \ve_F\overline{\partial_t \bzeta(t)}~,
\ee
is determined by the first angular harmonic of the Wigner-transformed
distribution function $f(\ve,\varphi,t)$. The bar denotes time averaging over
the period of the microwave field in the steady state.

The approach formulated above [Eqs.~(\ref{K}), (\ref{kineq})--(\ref{curgen})]
validates the description of the inelastic and displacement mechanisms of
MIRO in \rsec{s3.2.1}. It has been applied to a wealth of nonequilibrium
phenomena that we address in the remaining part of the review.

\subsubsection{Quadrupole and photovoltaic contributions to the
photoconductivity}
\label{s3.2.4}

\begin{figure}[ht]
\centerline{
\includegraphics[width=\columnwidth]{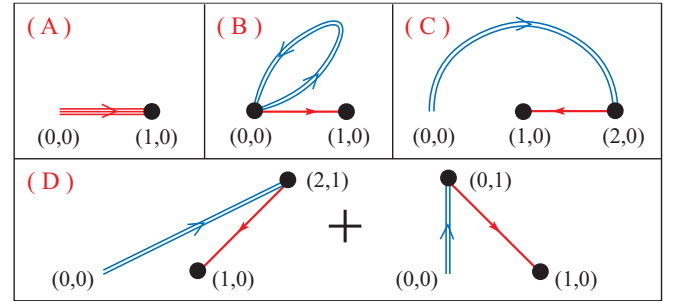}
}
\caption{Diagrams representing four distinct quantum contributions (A)-(D) to
the photocurrent at order $E E_\w^2$. Adapted from \citet{dmitriev:2007}.
}
 \label{diagQ}
 \end{figure}

Careful study of the quantum kinetic equation \eqref{kineq} shows
\cite{dmitriev:2007} that there are in total four different contributions to the
photocurrent at the leading order $E E_\w^2$. Diagrams (A)-(D) in
Fig.~\ref{diagQ} represent four ways to obtain perturbatively the first angular
harmonic $f_{10}$ of the distribution function $f(\ve,\varphi,t)\equiv\sum
f_{\nu n} e^{i\nu\varphi+i n \w t}$, which defines the current \eqref{curgen},
starting from the isotropic dark distribution $f^T_\ve$. In the displacement
contribution (A), $f_{10}\propto (\wc\partial_\varphi)^{-1} \St_{\rm
im}\{f^T_\ve\}$
results directly from the action of the collision operator $\St_{\rm im}\propto
E E_\w^2$. In the inelastic contribution (B), $f_{10}\propto
(\wc\partial_\varphi)^{-1} \St_{\rm im}\{\delta f_{00}\}$, where $\St_{\rm
im}\propto E$ and $\delta f_{00}\propto \tin E_\w^2$ is the microwave-induced
correction to the isotropic part of $f$. Apart from $\delta f_{00}$, the action
of $\St_{\rm im}\propto E_\w^2$ on $f^T_\ve$ results in excitation of the second
angular harmonics $f_{20}$. The linear dc response in the resulting state,
$f_{10}\propto (\wc\partial_\varphi)^{-1} \St_{\rm im}\{f_{20}\}$ with $\St_{\rm
im}\propto E$, produces the ``quadrupole'' contribution (C) to the direct
current. Finally, in the ``photovoltaic'' mechanism (D), a combined action of
the microwave and dc fields $\St_{\rm im}\propto E E_\w$ produces nonzero
temporal harmonics $f_{10}$ and $f_{12}$. The ac response in the resulting
state, $f_{10}\propto (\partial_t+\wc\partial_\varphi)^{-1} \St_{\rm
im}\{f_{10}, f_{12}\}$ with $\St_{\rm im}\propto E_\w$, also gives rise to the
direct current.

%%%%%%%%%%%%%%%%%%%%%%%%%%%%%%%%%%%%%%%%%%%%%%%%%%%%%%%%%%%%%
%%%%%%%%%%%%%%%%%%%%%%%%%%%%%%%%%%%%%%%%%%%%%%%%%%%%%%%%%%%%

Calculations performed in \cite{dmitriev:2007} (for smooth disorder) and
\cite{dmitriev:2009b} (for generic disorder) show that all four mechanisms
(A)-(D) produce microwave-induced corrections to the direct current ${\bf j-j}_{\rm
dark}=\hat{\sigma}_{ph}\bE$. The full photoconductivity tensor
$\hat{\sigma}^{\rm (ph)}\propto E_\w^2$ consists of components of different
symmetry
\be
\label{matrix} \hat{\sigma}^{\rm (ph)}=-2 \delta^2 \sigma_{\rm D}\left(\!
\begin{array}{cc}d_s+d_a & h_s+h_a \\ h_s-h_a & d_s-d_a \end{array}
\!\right)
\ee
oscillating with $\w/\wc$. The Hall part is solely due to mechanisms (C) and
(D). In particular, the quadrupole mechanism (C) yields an unusual symmetric
off-diagonal term $h_s$ violating the Onsager symmetry, while the antisymmetric
$h_a$ is produced by the photovoltaic mechanism (D). The photovoltaic mechanism
also yields corrections to both isotropic ($d_s$) and anisotropic ($d_a$)
diagonal components. It is worth mentioning that, unlike contributions (A)-(C),
the photovoltaic contribution (D) cannot be described in terms of the scattering
rates similar to \eqref{W}. In particular, it involves not only the DOS given by
the imaginary part of $G^{R}$ but also quantum corrections to the real
part of the Green's function which do not appear in the golden-rule
approximation.

In the case of smooth disorder, the largest contribution comes from the
inelastic mechanism (B), which contributes only to $d_s$. In this case, $d_s\gg
d_a\sim h_s \sim h_a$, meaning the contributions of subleading mechanisms (A),
(C), and (D) are comparable in magnitude. A sufficiently large amount of
short-range scatterers enhances contribution (A), see \rsec{mixed_disorder},
while the magnitude of contributions (C) and (D) remains essentially the same,
$d_s\sim d_a\gg h_s \sim h_a$. Thus, the description of MIRO in \rsec{s3.2.1},
which accounts only for the inelastic and displacement contributions, is
justified as long as the diagonal part of $\hat{\sigma}^{\rm (ph)}$ is
concerned. Mechanisms (C) and (D) yield oscillations
% RR
in the Hall component (in the terms $h_a$ and $h_s$). Upon tensor
inversion, the terms $d_a$ and $d_a$ also contribute to the Hall
components of the resistivity tensor. Although these contributions are
suppressed by an additional small factor $1/\omega_c\tau$, they may
compete with those originating from $h_{s,a}$, since $d_{s,a}$ may be much
larger than $h_{s,a}$.
Microwave-induced oscillations in the Hall resistivity were observed
experimentally
%in the Hall component (in
%the terms $h_a$ and $h_s$), which were also observed experimentally
\cite{mani:2004b,studenikin:2004,wiedmann:2011b}. While the amplitude of the
oscillations in the Hall component observed by \citet{wiedmann:2011b} is in good
agreement with theory, \citet{mani:2004b,studenikin:2004} reported considerably
stronger oscillations. In fact, a similar issue is known for SdH oscillations
whose measured amplitude in the Hall component of the resistivity often
considerably exceeds the theoretical predictions. It is likely that the physical
mechanism of the enhancement of the Hall component is the same in both cases.

\subsubsection{Classical mechanisms of MIRO}
\label{s3.2.5}
Most experimental results on MIRO reported so far have been consistently
explained within the quantum kinetic approach described above. These quantum
effects are directly linked \cite{dmitriev:2003,fedorych:2010} to quantum
magnetooscillations $\sigma^{\rm (q)}(\w)$ in the dynamic conductivity
(\rsec{s2.3.2}). On the other hand, the quasiclassical memory effects, discussed
in \rsec{s2.2.3}, also produce large $\w/\wc$--oscillations in the ac
conductivity, $\sigma^{\rm (c)}(\w)$ [\reqs{II.19a}{II.19b}], see the comparison
of the classical and quantum contributions in Fig.~\ref{II.f1}. In turn, the
oscillatory $\sigma^{\rm (c)}(\w)$ translates into  $\w/\wc$-oscillations of the
electronic temperature $T_e$, while the oscillating $T_e$ manifests itself in dc
transport via the renormalization of the elastic scattering rate by
electron-electron interactions. The resulting oscillatory contribution to the
photoconductivity has the form \cite{dmitriev:2004}
\be\label{ph-c}
\frac{\Delta\sigma_{\rm ph}^{(c)}(0)}{\sigma_{\rm D}}\sim -\frac{\tin^{\rm
e-ph}}{\ttr}{\cal P}
\frac{\w^2}{\ve_F T}\frac{\Delta\sigma^{(c)}(\omega)}{\sigma_{\rm D}(\omega)}~,
\ee
where $\tin^{\rm e-ph}$ is the electron-phonon inelastic relaxation time (which
controls the heating of the 2DEG) and $\Delta\sigma^{(c)}$ is given by
Eqs.~(\ref{II.19a}) and (\ref{II.19b}). The phase of the
magnetooscillations in
$\Delta\sigma_{\rm ph}^{(c)}(0)$ is opposite to the phase in the dynamical
conductivity $ \Delta\sigma^{(c)}(\omega)$ and is shifted by $\pi/4$ with
respect to the quantum oscillations [Eqs.~(\ref{sigmain}) and (\ref{sigmadis})].
At low $T$, the magnitude of the oscillations in Eq.~(\ref{ph-c}) may become
comparable to that in Eq.~(\ref{sigmain}) since the ratio
$\Delta\sigma^{(c)}(\omega)/\sigma_{\rm D}(\omega)$ can be of the order of unity
(Fig.~\ref{II.f1}), while the small parameter $\w/\ve_F\sim 10^{-1}-10^{-2}$ is
easily compensated by the large ratio $\tin^{\rm e-ph}/\tin$ of the
electron-phonon and electron-electron inelastic relaxation times
[electron-electron scattering provides for relaxation of the fast oscillations
in the energy distribution of electrons in Eq.~(\ref{sigmain}) but conserves the
total energy of the electron gas]. However, unlike the quantum oscillations of
the photoconductivity, the oscillations described by Eq.~(\ref{ph-c}) saturate
with increasing ${\cal P}$ well before ${\Delta\sigma_{\rm
ph}^{(c)}(0)}/{\sigma_{\rm D}}$ becomes of order unity, and therefore, cannot
cause ZRS.
Namely, in the regime of strong heating ($T_e-T\agt T$), ${\Delta\sigma_{\rm
ph}^{(c)}(0)}/{\sigma_{\rm D}}$ does not exceed $T_e/\ve_F\ll 1$.

It is important to mention that all classical corrections to the Drude formula
induced by the homogeneous external ac and dc fields vanish to zero in the
Boltzmann equation framework in the case of a parabolic dispersion relation and
$\ve$-independent $\ttr$ \cite{dmitriev:2004}. Apart from the
interaction-induced effects discussed above, finite nonlinear corrections may
result either from the energy dependence of $\ttr$ [which typically occurs on
the scale of $\ve_F$, thus yielding an additional small factor $T/\ve_F$
compared to Eq.~\eqref{ph-c} \cite{dmitriev:2004}] or from a weak
nonparabolicity
[which leads to the appearance of an even larger energy scale of the order of
the bandgap in the denominator \cite{koulakov:2003,joas:2004}]. By contrast, the
Landau quantization provides the much smaller scale $\wc$ in the dependence on
$\ve$ [Eqs.~(\ref{R1}) and (\ref{R2})], which explains the leading role of the
quantum mechanisms of MIRO.

Two recent works addressed classical effects close to the boundary of a 2DEG.
\citet{chepelianskii:2009} studied the influence of the microwave field on the
phase space portrait of electron dynamics near the sample edge. The conclusion
was that at $\w/\wc = n + 1/4$ the microwave radiation tends to trap electrons
in trajectories propagating near the edge, while at $\w/\wc = n$ the particles
are more efficiently kicked out into the bulk. \citet{mikhailov:2011}
discussed an effective electrostatic (ponderomotive) potential for electrons
(whose cyclotron dynamics is modeled using Newton's equation with a friction
term), created by a strong inhomogeneity \cite{mikhailov:2006} of the microwave
field in the near-contact region. Depending on the ratio $\w/\wc$, the
ponderomotive force repels electrons from or attracts them to the near-contact
area. Since the above works did not develop a systematic theoretical
analysis of the contributions of the boundary effects to the resistance of the
2D system, it is difficult to compare these contributions to those predicted by
the bulk theory (\rsec{s3.2.1}). In any case, in the systems in which MIRO
are observed transport is dominated by the bulk contribution, so that even a
strong modification of edge transport may modify the total current only weakly.
Indeed, experimentally, no dependence of the MIRO amplitude on the sample
dimension or geometry, characteristic of the edge effects, has been reported so
far. Moreover, edge transport obviously plays no role in Corbino geometry.
Furthermore, recent observations of MIRO using various contactless
techniques \cite{bykov:2010d,andreev:2011} have ruled out the contact-related
phenomena as a generic cause of MIRO.

\subsection{Related microwave-induced phenomena}
\label{s3.3}

\subsubsection{ Microwave-induced photovoltaic effects}
\label{s3.3.1}
\begin{figure}[ht]
\centerline{
\includegraphics[width=\columnwidth]{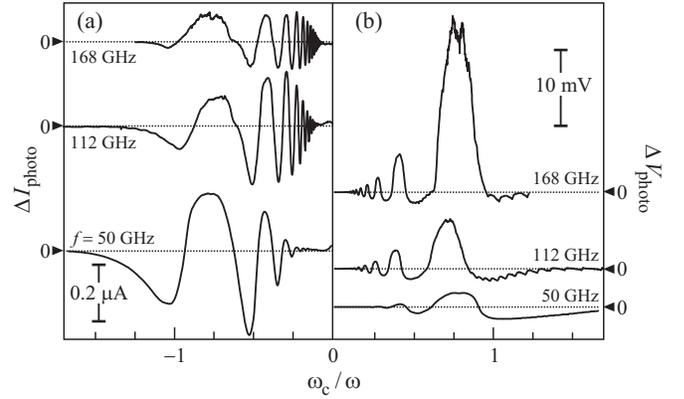}
}
\caption{(a) Photovoltaic current and (b) voltage at a fixed radiation frequency
$f=50,\,112$, and $168\,{\rm GHz}$ versus $\w_c/\w$ in the absence of dc bias
applied to the sample. The curves are offset for clarity. Adapted from
\citet{dorozhkin:2009}.
}
 \label{photovoltaics}
 \end{figure}

Recent experiments \cite{bykov:2008d,dorozhkin:2009}, performed using an
asymmetric contact configuration, discovered that microwaves can induce
oscillatory current and voltage signals very similar to MIRO even in the
absence of external dc driving, see Fig.~\ref{photovoltaics}. These photovoltaic
effects were attributed \cite{dmitriev:2009a} to the violation of the Einstein
relation between the conductivity and the diffusion coefficient, induced by the
microwave illumination. Consider an inhomogeneous system characterized by the
electrostatic potential $\phi(\br)$ and the electron density $n_e(\br)$,
both smoothly varying in space. In equilibrium, the electron system is characterized
by the constant electrochemical potential
$\eta=e\phi(r)+n_e(\br)/\chi^{\rm(dark)}$, while a generic weak perturbation
leads to the current $j^{\rm(dark)}=-\sigma^{\rm(dark)}\nabla\phi-e
D^{\rm(dark)}\nabla n_e\equiv-e D^{\rm(dark)}\chi^{\rm(dark)}\nabla\eta$. At
$2\pi^2 T/\wc\gg 1$, the dark compressibility $\chi^{\rm(dark)}=2\nu_0$, so that
the Einstein relation in equilibrium reads $\sigma^{\rm(dark)}=2e^2 \nu_0
D^{\rm(dark)}$, where $D^{\rm(dark)}=\langle\tilde{\nu}^2(\ve)\rangle_\ve
R_c^2/2\ttr$, see \req{jdisin}.

%%%%%%%%%%%%%%%%%%%%%%%%%%%%%%%%%%%%%%%%%%%%%%%%%%%%%%%%%%%%%
%%%%%%%%%%%%%%%%%%%%%%%%%%%%%%%%%%%%%%%%%%%%%%%%%%%%%%%%%%%%

Analysis of Eqs.~\eqref{jx}-\eqref{f}, adapted to the inhomogeneous conditions,
shows \cite{dmitriev:2009a} that the direct current under microwave
illumination,
\be\label{sigmaeta}
j=-\sigma_\eta\nabla\phi-2 e\nu_0 D\nabla\eta~,
\ee
necessarily contains an extra ``anomalous term" $-\sigma_\eta\nabla\phi$ which
violates the Einstein law. The term $\sigma_\eta=\sigma^{\rm in}+\sigma^{\rm
dis}_1$ includes the inelastic contribution to MIRO \eqref{sigmain} and the
most important part $\sigma^{\rm dis}_1\propto {\cal R}_1$ of the displacement
contribution \eqref{sigmadis}. The other part $\sigma^{\rm dis}_3\propto {\cal
R}_3$ enters the nonequilibrium diffusion coefficient,
$D=D^{\rm(dark)}+\sigma^{\rm dis}_3/2e^2 \nu_0$. The total conductivity
$\sigma$, which defines the direct current $j=\sigma E$ in a homogeneous
system, is given by $\sigma=\sigma_\eta+2e^2\nu_0 D$.

The current \eqref{sigmaeta} is nonzero even at $\nabla\eta=0$, provided there
is a nonzero built-in electric field (created, e.g. by an asymmetric contact
configuration or by local gates), as observed in
\cite{bykov:2008d,bykov:2010b,willett:2004,dorozhkin:2009,dorozhkin:2011}. The
simplest model, proposed by \textcite{dmitriev:2009a}, assumes that the built-in
field ${\cal U}_c/L$ is provided by two contacts with the difference of work
functions ${\cal U}_c$,
attached to a 2DEG stripe of width $L$. The $I$-$V$ characteristic for this
setup reads
\be\label{CVCstripe}
j=\sigma_\eta {\cal U}_c/L+ \sigma V/L~,
\ee
where the voltage between the contacts $eV\equiv\eta|_{x=0}-\eta|_{x=L}$. As
seen from Eq.~(\ref{CVCstripe}), there is a finite oscillatory photocurrent $j$
at zero bias voltage $V=0$, as well as a finite photovoltage $V$ (asymmetric
with respect to $V=0$) at $j=0$, as shown in \rfig{photovoltaics}. The corresponding field and density
distributions are calculated in \cite{dorozhkin:2011a}.

\subsubsection{Microwave-induced compressibility oscillations}
\label{s3.3.2}
The violation of the Einstein relation under microwave illumination has a number
of other important consequences. In particular, the generalized compressibility
$\chi^{(\phi)}_q\equiv n_q/(-e\phi_q)$, defined as a static density response
$n_q e^{i q r}$ to a weak local perturbation $\phi_q e^{i q r}$ of the screened
electrostatic potential, strongly deviates from its equilibrium value
$\chi^{\rm(dark)}= 2\nu_0$. Assuming a stationary nonequilibrium state with the
homogeneous component of the current $j=0$, one obtains from
Eq.~\eqref{sigmaeta}:
\be\label{chiq0}
\chi^{(\phi)}_{q\to0}=2\nu_0+\sigma_\eta/e^2 D=\sigma/e^2 D~.
\ee
Note that, in contrast, the quantity $\chi^{(\mu)}_q\equiv n_q/\mu_q$ is not
modified: $\chi^{(\mu)}_q=2\nu_0$. Using the generalized compressibility, one
can define the nonequlibrium screening length
% RR
\cite{dorozhkin:2011a}
\be\label{lambda}
\lambda=\frac{\epsilon D}{2\pi\sigma}\equiv\frac{\epsilon }{2\pi
e^2\chi^{(\phi)}_q}~,
\ee
which replaces the equilibrium Thomas-Fermi screening length
$\lambda_0=\epsilon/2\pi e^2\chi^{\rm (dark)}$ in all electrostatic problems.

\begin{figure}[ht]
\centerline{
\includegraphics[width=\columnwidth]{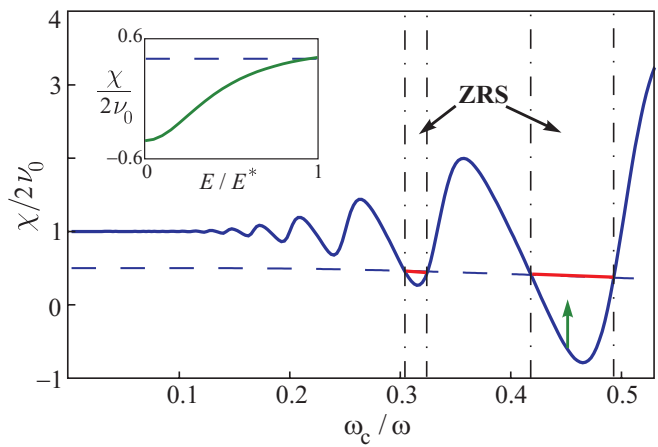}
}
\caption{Microwave-induced oscillations of the compressibility as a function of
$\wc/\w$ at fixed $\w\tau_{\rm q}=2\pi$ and $(\tau_{\rm in}/\tau){\cal
P}\vert_{\wc=0}=1$. In ZRS, the electric field inside domains $E^*$ fixes
the compressibility at the level shown by a dashed line. Inset: the dependence
of the compressibility on the electric field $E$ inside the domain wall.
Adapted from \citet{vavilov:2004b}.}
\label{comp}
\end{figure}

At $q^{-1}\gg l_{\rm in}\equiv(D\tin)^{1/2}$ but $q R_c\ll 1$, the local
approximation for the current \eqref{sigmaeta} is still justified. At the same
time, the effect of spatial variations of the electric field on the distribution
function becomes essential \cite{vavilov:2004b}. Summarizing the results of
\citet{dmitriev:2009a} and \citet{vavilov:2004b} for the leading inelastic
contribution, the generalized compressibility at $q R_c\ll 1$ is given by
\be\label{chiqin}
\frac{\chi^{(\phi)}_q}{2\nu_0}=1-\delta^2 \,{\cal F}(\tin, {\cal P},
\zeta)\,\frac{2+q^2 l_{\rm in}^2}{1+q^2 l_{\rm in}^2}~,
\ee
meaning the amplitude of the oscillations in $\chi^{(\phi)}_q$ reduces by a factor
of 2 at $ql_{\rm in}\gg 1$ compared to the limit $q\to 0$ [function ${\cal F}$ is defined in \eqref{FF}]. These results suggest
that ZRS corresponds to a plateau in the compressibility, as illustrated in
Fig.~\ref{comp} for the case $ql_{\rm in}\gg 1$. Therefore, local measurements
of the compressibility [e.g., using the techniques that utilize single-electron
transistors \cite{ilani:2000,ilani:2001}]
may provide a real space snapshot of the domain structure in ZRS.
Experimental work in this direction is currently underway.

%%%%%%%%%%%%%%%%%%%%%%%%%%%%%%%%%%%%%%%%%%%%%%%%%%%%%%%%%%%%%
%%%%%%%%%%%%%%%%%%%%%%%%%%%%%%%%%%%%%%%%%%%%%%%%%%%%%%%%%%%%

\subsubsection{Effects of a parallel magnetic field}
\label{s3.3.2prim}

\citet{yang:2006} observed the effect of a parallel magnetic field $B_\parallel$
on MIRO: the oscillations were suppressed as
$B_\parallel$ is increased and virtually disappeared at $B_\parallel \sim
0.5\:{\rm T}$. This observation came as a surprise because, according to the
theory presented above, MIRO is an orbital effect and should not be
essentially sensitive to the spin degree of freedom. Moreover, the transverse
motion of 2D electrons (that would be affected by the parallel field) is frozen.
In \cite{mani:2005}, MIRO were indeed practically independent of $B_\parallel$
up to about $1\;{\rm T}$, i.e., in the same range of $B_\parallel$ as in
\textcite{yang:2006}.
The most likely explanation of the experimental data of \citet{yang:2006} is the
effect of $B_\parallel$ on the quantum scattering rate $1/\tq$. Indeed, the
parallel magnetic field does affect the structure of the wave function across
the 2DEG plane, which can in general modify the impurity scattering rate. This
is a nonuniversal effect, as it depends on microscopic details of the
nanostructure in which the 2DEG is formed, as well as on the character of
imperfections. It is plausible that in the QW used by
\citet{yang:2006} this effect may be stronger than in a typical single-interface
structure. In fact, a closer inspection of the experimental data of
\citet{yang:2006} shows that the dark resistivity (and thus the transport
scattering rate $1/\tau$) was enhanced by the parallel field by a factor of
about 3. While in general the enhancement factors for $1/\tau$ and $1/\tq$
may be different (because $1/\tau$ and $1/\tau_{\rm q}$ may be controlled by
different components of disorder), this clearly shows that $B_\parallel$ did
strongly enhance disorder-induced scattering. The effects of $B_\parallel$ were
also studied by \citet{hatke:2011a} in the nonlinear (with respect to the dc
field) transport regime, see \rsec{s5.1}. It was demonstrated that $B_\parallel$
suppresses quantum oscillations via the enhancement of $1/\tq$.

\subsubsection{MIRO in multisubband structures}
\label{s3.3.3}

The microwave-induced effects in magnetotransport have also been intensively
studied in structures with several occupied subbands, both experimentally
\cite{wiedmann:2011b,gusev:2011,bykov:2008e,wiedmann:2008,wiedmann:2010b,
wiedmann:2010a,wiedmann:2009b,bykov:2010c} and theoretically
\cite{raichev:2008,wiedmann:2008,wiedmann:2009b,wiedmann:2010a}. Already the
first experiments in a double QW in the Hall-bar geometry
\cite{wiedmann:2008} as well as in van der Pauw and Corbino geometries
\cite{bykov:2008e} demonstrated strong interference between MIRO and MISO
(the latter are discussed in Sec.~\ref{s2.3.3}). Specifically, it was observed
that (i) MIRO are strongly enhanced (suppressed) in the maxima (minima) of
MISO and that (ii) MISO in the minima of MIRO are inverted by
sufficiently strong microwave radiation. Similar effects were recently observed
also in triple QWs \cite{wiedmann:2009b}.

\citet{wiedmann:2008} performed a direct comparison of the experimental results
with theory that combines the inelastic mechanism of MIRO
\cite{dmitriev:2005} and the intersubband scattering effects
\cite{raichev:2008}. The theory reproduced with a good accuracy the complicated
interference pattern, as well as the nontrivial dependence of the combined
oscillations on the microwave power, frequency, and temperature, observed in the
experiment. In all cases \cite{wiedmann:2008,wiedmann:2009b,wiedmann:2010a}, the
comparison confirmed the $T^{-2}$ dependence of MIRO amplitude
\cite{dmitriev:2005} up to $T\sim 4\,{\rm K}$. Similar $T^{-2}$ dependence of the
inelastic scattering time was extracted from the exponential high-$T$ damping of
MISO \cite{mamani:2008}. This damping, governed by the interaction-induced
broadening of LLs, has the same origin as the high-$T$ damping of MIRO
discussed in \rsec{temperature}.

\begin{figure}[ht]
\centerline{
\includegraphics[width=\columnwidth]{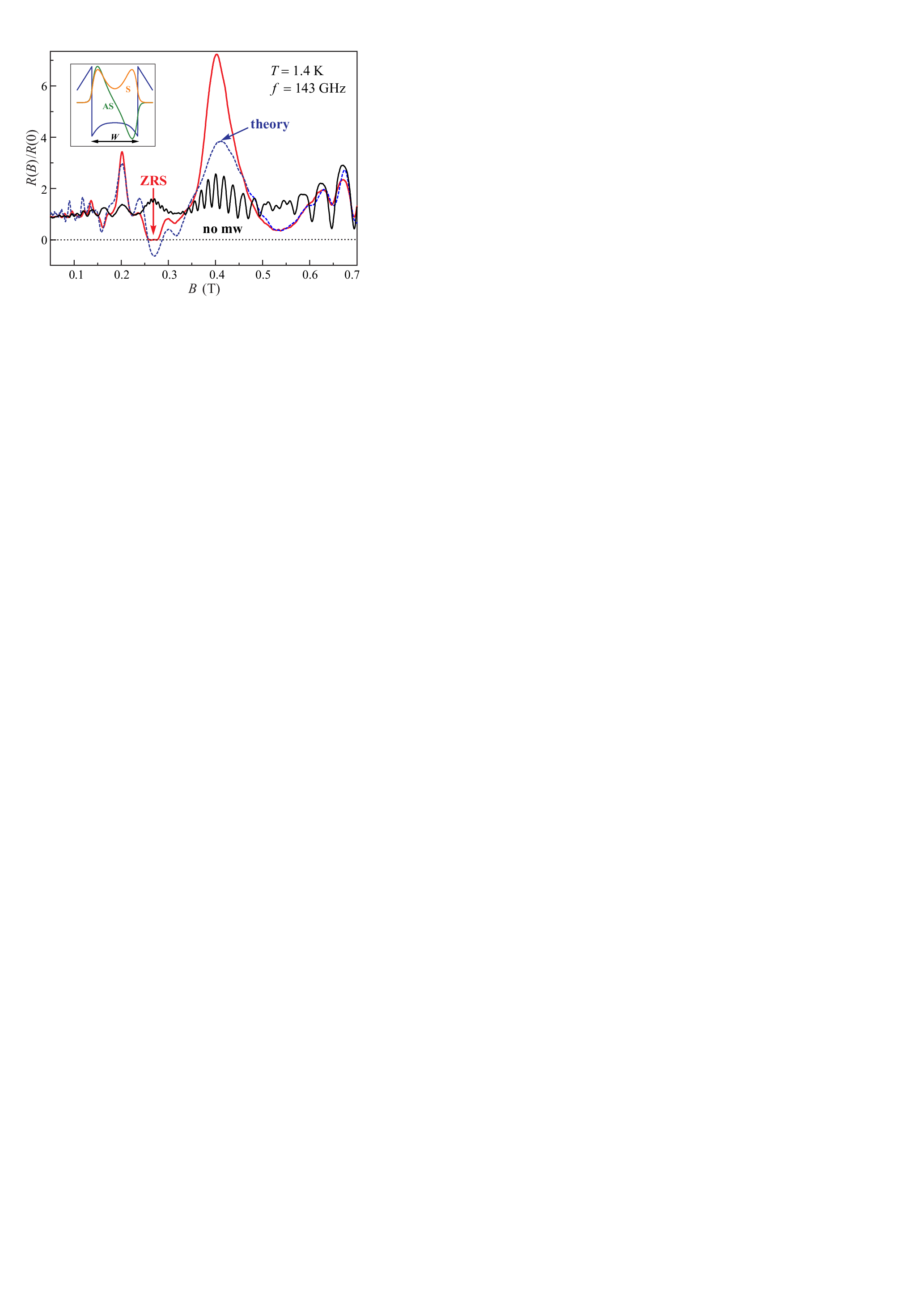}
}
\caption{Measured resistance (solid lines) of a two-subband electron system
without (no mw) and under microwave irradiation at $f=143\,{\rm GHz}$ as a
function of $B$. The inverted by microwaves MISO peak at $B\simeq 0.27\,{\rm T}$
exhibits vanishing resistance. Dashed line: theory. Inset: symmetric (S) and
antisymmetric (AS) wavefunctions for the two lowest subbands in a wide QW.
From \citet{wiedmann:2010b}.
}
 \label{ZRS_DQW}
 \end{figure}

In the case of a balanced double QW, with negligible interlayer
correlations of the scattering potential and identical
scattering rates, the theoretical result of \citet{raichev:2008} reads [cf.\
Eq.~\eqref{inelasticMain} and Eqs.~\eqref{II.52}-\eqref{II.54}]
\be\label{DQW}
\frac{j_d}{\sigma_{\rm D} E}-1=\delta^2[\,1-{\cal F}(\tin, {\cal P},
\zeta)\,]\left(1+\cos\frac{2\pi\Delta}{\wc}\right),
\ee
where ${\cal F}$ is given by Eq.~\eqref{FF}. This expression clearly shows
features (i) and (ii) mentioned above.
Indeed, MISO are inverted  for ${\cal F}>1$.
Further, if $2\delta^2({\cal F}-1)$ exceeds unity, the conductivity (and,
therefore, resistivity) at the MISO maximum becomes negative, which leads to the
emergence of ZRS (Sec.~\ref{s4}).
ZRS of this kind have been observed by \citet{wiedmann:2010b}, see
Fig.~\ref{ZRS_DQW}.

%%%%%%%%%%%%%%%%%%%%%%%%%%%%%%%%%%%%%%%%%%%%%%%%%%%%%%%%%%%%%
%%%%%%%%%%%%%%%%%%%%%%%%%%%%%%%%%%%%%%%%%%%%%%%%%%%%%%%%%%%%

\subsubsection{MIRO in spatially modulated systems}
\label{s3.3.4}
The discussion in Secs.~\ref{s3.3.1} and \ref{s3.3.2} shows that the violation
of the Einstein relation strongly modifies the transport and thermodynamic
properties of the microwave-illuminated 2DEG already in the presence of a weak
smooth inhomogeneity. The opposite limit of a strong short-period spatial
modulation was studied in
\cite{dietel:2005,joas:2005,dietel:2006,robinson:2004,kennett:2005,torres:2006}.

\citet{dietel:2005,joas:2005,dietel:2006} examined the transport properties of
the 2DEG with a unidirectional static modulation [this system has been
intensively studied before in the context of Weiss oscillations and transport
anisotropies
\cite{weiss:1989,gerhards:1989,beenakker:1989,zhang:1990,mirlin:1998}]. In the
presence of a periodic potential $V(x)=\tilde{V} \cos Q x$ the degeneracy of LLs
is lifted and LL bands appear with the dispersion relation
\be\label{band}
\ve_{n k}=\wc(n+1/2)+\tilde{V} J_0(Q R_c) \cos (Qkl_{\rm B}^2)~,
\ee
where the Bessel function $J_0$ is an approximation that is valid for high LLs.
The DOS of the clean modulated system has square-root singularities at the band
edges. In the $x$ direction, the photoconductivity can be calculated using the
golden-rule approach formulated in Sec.~\ref{s3.2.1}. The results for the
displacement and inelastic mechanisms are given by Eqs.~\eqref{sigmadis} and
\eqref{sigmain} with the oscillating factors \eqref{R1}, \eqref{R3}, and
\eqref{R2} calculated using the DOS corresponding to Eq.~\eqref{band}. The
oscillating factors acquire additional sign changes, related to singular DOS,
which can be detected in the experiment. In the perpendicular direction, the
linear (in the dc field) photoresponse diverges for the displacement mechanism.
The divergence is cut off by inelastic processes. As a result, both the
displacement and inelastic contributions to the transverse photoresponse are
proportional to $\tin$ and have a similar magnitude.

\begin{figure}[ht]
\centerline{
\includegraphics[width=\columnwidth]{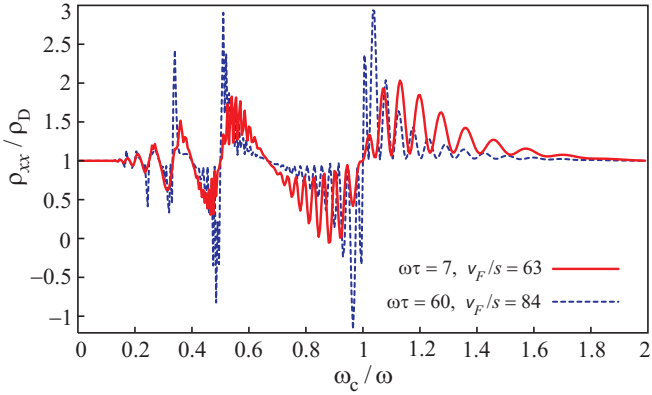}
}
\caption{Calculated resistivity magnetooscillations in the presence of a surface
acoustic wave for $\omega\tau=7$ and $v_F/s=63$ (solid line) and $\omega\tau=60$
and $v_F/s=84$ (dashed), both curves for $\omega\tau_{\rm q}=6$ and
$\omega\tau_{\rm in}=302$. From \citet{kennett:2005}.}
\label{SAW}
\end{figure}

\citet{robinson:2004,kennett:2005} studied the magnetoresistivity in the
presence of surface acoustic waves (SAW). This perturbation combines
\cite{levinson:1998} both the unidirectional spatial modulation with the
wavevector $Q$ and the ac excitation at frequency $\w=sQ$, where $s$ is the
sound velocity. Apart from an ac analog of the Weiss oscillations, which
results in an anisotropic positive MR oscillating with both $Q R_c$ and
$\w/\wc$, \citet{robinson:2004,kennett:2005} calculated the associated quantum
contribution for the inelastic mechanism of MIRO, which potentially gives
rise to SAW-induced ZRS. The quantum contribution is given by
Eqs.~\eqref{inelasticMain} and \eqref{FF}, where the microwave power ${\cal P}$
is replaced by properly normalized SAW power absorbed by the 2DEG. Since the
absorbed SAW power is proportional to the dynamic conductivity of the classical
system at the wavevector $Q$, $\sigma^{(c)}_{\w Q}$, which oscillates with $Q
R_c$, $\rho_{xx}$ (Fig.~\ref{SAW}) shows additional $Q R_c$-oscillations on top of
relatively slow $\w/\wc$-oscillations identical to MIRO.

%%%%%%%%%%%%%%%%%%%%%%%%%%%%%%%%%%%%%%%%%%%%%%%%%%%%%%%%%%%%%
%%%%%%%%%%%%%%%%%%%%%%%%%%%%%%%%%%%%%%%%%%%%%%%%%%%%%%%%%%%%

\citet{yuan:2006} investigated experimentally the microwave photoresistance in
the presence of a triangular antidot superlattice. This study
revealed the conventional MIRO with a superimposed magnetoplasmon peak on top of
a series of narrow geometrical resonances positioned at $2 R_c=\gamma_n a$,
where $a$ is the lattice period and $\gamma_n$ are numbers of order unity
corresponding to various commensurate orbits. Essentially no interference
between MIRO and the commensurability oscillations was observed. Note,
however, that a strong interplay of the two types of oscillations was predicted
\cite{torres:2006} for a short-period superlattice with $a$ of the order of
$l_{\rm B}\ll R_c$.

\subsubsection{Microwave-induced $B$-periodic oscillations}
\label{s3.3.5}

\begin{figure}[ht]
\centerline{
\includegraphics[width=\columnwidth]{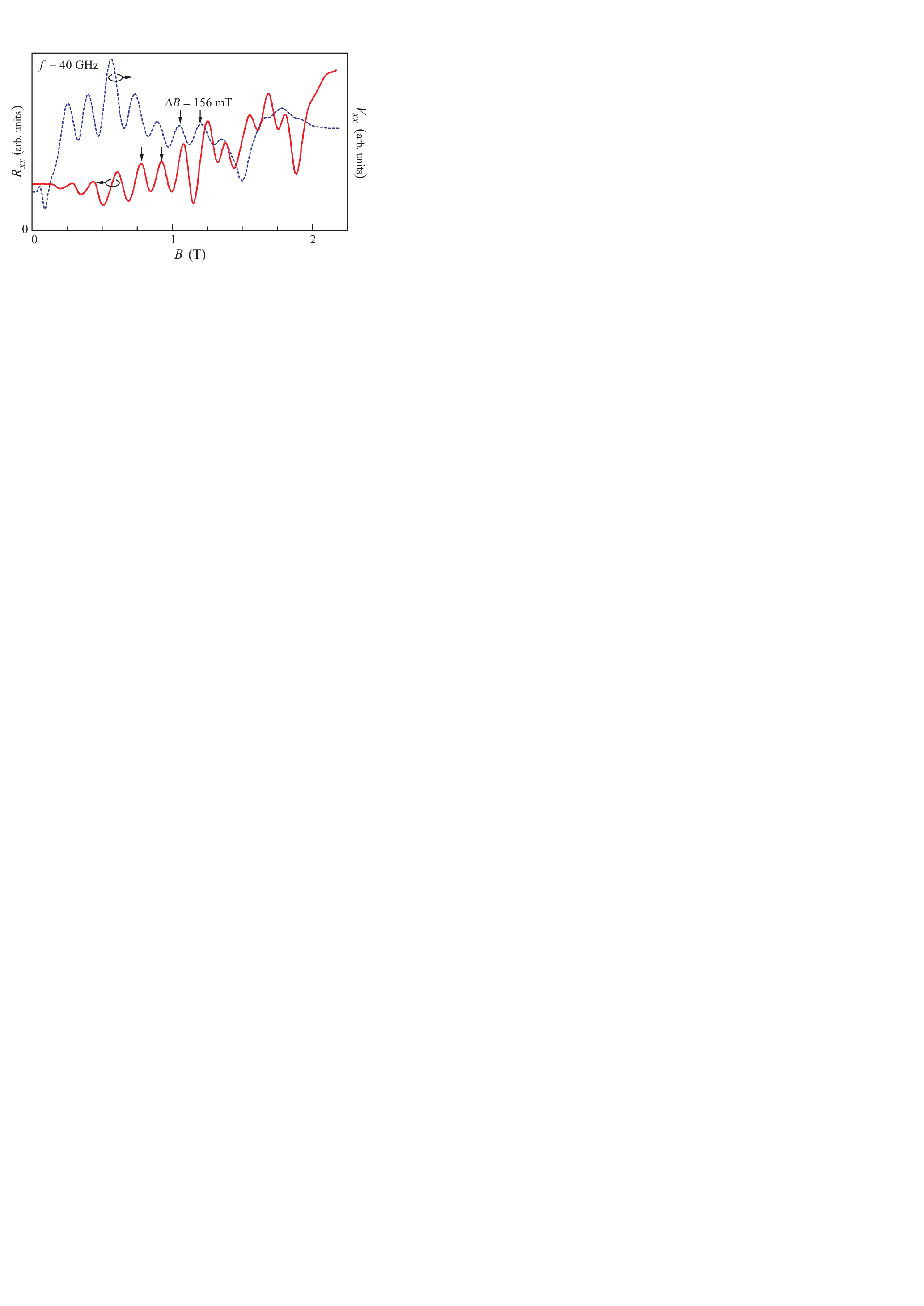}
}
\caption{$B$-periodic magnetoresistance (solid line) and photovoltage (dashed)
oscillations at $f=40\,{\rm GHz}$ measured in a Hall bar with potential probes
separated by $0.5\,{\rm mm}$. From \citet{kukushkin:2004}.}
\label{Bper}
\end{figure}

\citet{kukushkin:2004} discovered a different kind of photoresistance and
photovoltage magnetooscillations in the region $\w<\wc$, see \rfig{Bper}. These $B$-periodic
oscillations were attributed to the interference of edge magnetoplasmons (EMP)
emitted from different potential probes along the Hall-bar edge. The condition
for constructive interference of the EMP injected from two contacts separated by
the distance $L$ reads $q L=\w L/v=2\pi N$, where the EMP velocity $v$ is
proportional to the Hall conductivity $\sigma_{xy}\propto n_s/B$. This
interpretation explains the period $\Delta B\propto n_s/\w L$ found in the
experiment. \citet{kukushkin:2004} established the following properties
of the $B$-periodic oscillations: (i) the amplitude of the resistivity
oscillations shows a threshold behavior as a function of $\cal{P}$, (ii)
the photovoltage scales linearly with $\cal P$ in the limit of small
$\cal P$ and saturates with increasing $\cal P$, (iii) the saturation power for the
photovoltage is close to the threshold power for the resistivity oscillations,
(iv) the resistivity and photovoltage oscillations are phase shifted by $\pi/4$
with respect to each other, and (v) the threshold $\cal{P}$ for the resistivity
oscillations is an order of magnitude lower for the microwave
electric field perpendicular to the Hall-bar edge (which supports the EMP
scenario). The phenomenon showed only a weak dependence on $T$ in the broad
interval $T\sim 1-10\,{\rm K}$. The photovoltage reduced only by one order of
magnitude at $T$ as high as $70\,{\rm K}$. Systematic theory of the $B$-periodic
oscillations remains to be developed.

%%%%%%%%%%%%%%%%%%%%%%%%%%%%%%%%%%%%%%%%%%%%%%%%%%%%%%%%%%%%%
%%%%%%%%%%%%%%%%%%%%%%%%%%%%%%%%%%%%%%%%%%%%%%%%%%%%%%%%%%%%

\citet{kukushkin:2005} also reported a successful operation of the microwave
spectrometer based on the $B$-periodic photovoltaic oscillations. The
$B$-tunable selective detection was demonstrated in the frequency range
$20\,{\rm GHz}<f<150\,{\rm GHz}$ and for temperatures up to $T\sim 80\,{\rm K}$
(with a foreseen possibility of extension to the THz frequency range).
Implementation of plasmon resonances in such a device allows one to overcome the
temperature limitation $T<\w$ of the conventional selective detectors based on
electronic transitions. The major advantage of magnetoplasmons is that they can
be exploited at arbitrary $\w\ttr$ provided $\wc\ttr\gg 1$, while the operation
of the conventional plasmonic $B=0$ detectors in field-effect transistors is
limited to the region $\w\ttr>1$.

In a recent experiment \citet{stone:2007}, where MIRO, ZRS, and $B$-periodic
oscillations were observed simultaneously, the period $\Delta B$ of the
$B$-periodic oscillations was reported to scale as $\Delta B\propto n_s/\w$ and
be independent of the distance $L$ between the contacts. This result is in
contradiction to that by \citet{kukushkin:2004}: the controversy has remained
unresolved. Several other experiments, in particular, by
\citet{dorozhkin:2007b,yuan:2006}, reported no interference of the
magnetoplasmon effects and MIRO with each other. This agrees with the
calculation by \citet{volkov:2007}, according to which the magnetoplasmons can
strongly affect MIRO only in the limit of well-separated LLs---which is out
of the range
explored in the above experiments.

\section{Radiation-induced zero-resistance states (ZRS)}
\label{s4}

\subsection{ZRS: Experimental discovery and basic properties}
\label{s4.1}

\begin{figure}[ht]
\includegraphics[width=\columnwidth]{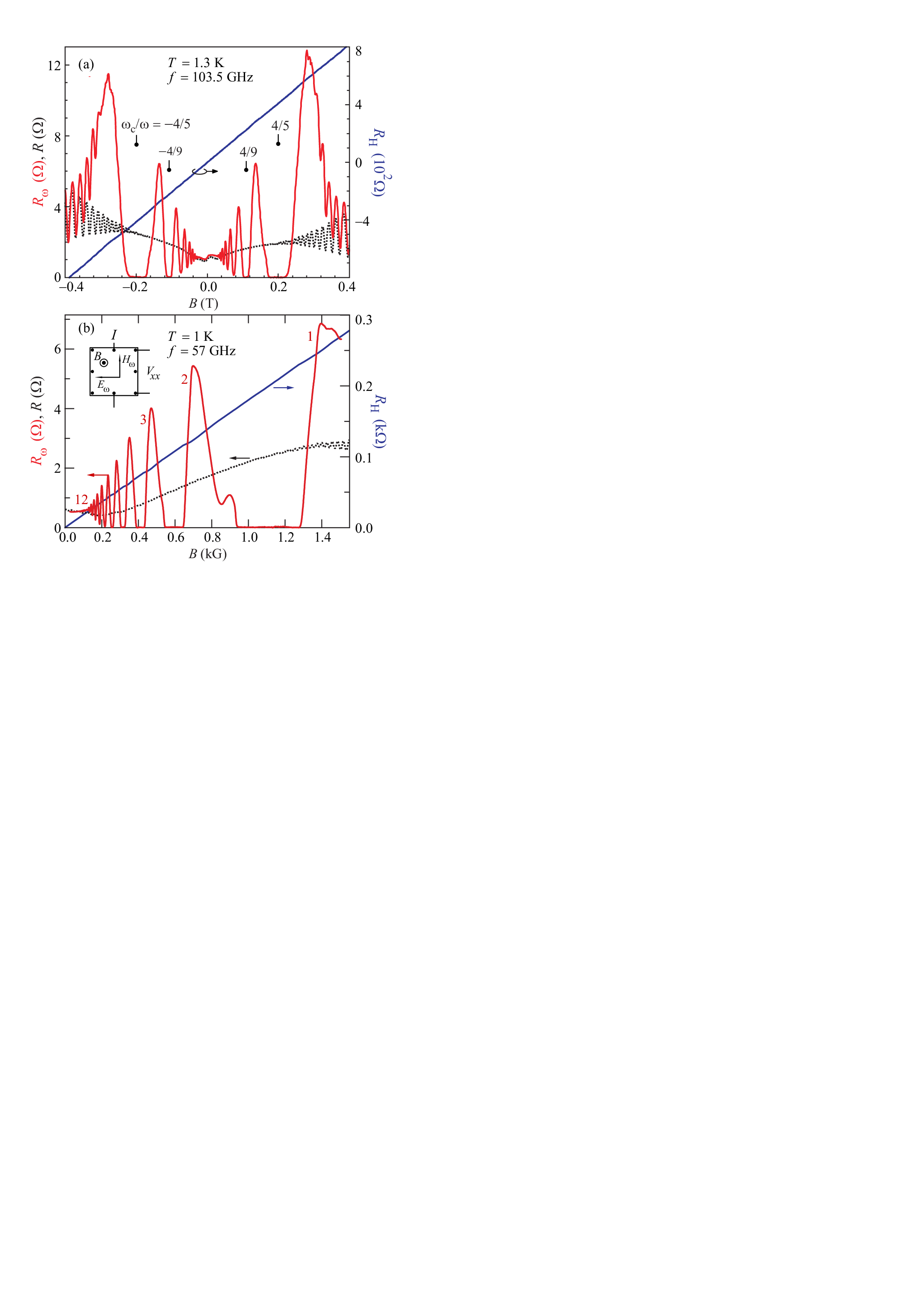}
\caption{Longitudinal [$R_\omega(B)$, left axis] and Hall [$R_H(B)$, right axis] magnetoresistance under microwave irradiation.
Longitudinal magnetoresistance $R(B)$ without irradiation is also shown.
Parameters: (a) microwave frequency $f=103.5$ GHz, temperature $T=1.3$~K, electron density $n_e\simeq 3\times 10^{11}$ cm$^{-2}$, and mobility  $\mu\simeq 1.5 \times 10^{7}$ cm$^{2}$/Vs;
(b) $f=57$ GHz, $T \simeq 1.0$ K, $n_e\simeq 3.5\times 10^{11}$ cm$^{-2}$, and $\mu\simeq 2.5 \times 10^{7}$ cm$^{2}$/Vs.
Adapted from (a) \citet{mani:2002} and (b) \citet{zudov:2003}.}
\label{fig.zrs}
\end{figure}

Zero resistance is a rare occurrence in condensed matter physics, usually signaling a novel state of matter,
such as superconductivity and QH effects \citep{klitzing:1980,tsui:1982}, see \rsec{s4.2.2} for further examples.
Experiments by \textcite{mani:2002} and by \textcite{zudov:2003} on microwave-irradiated
very-high mobility ($\mu \gtrsim 10^7$ cm$^2$/V$\cdot$s) 2DEG revealed that the lower order minima of MIRO (\rsec{s3.1})
can extend all the way to zero forming the zero-resistance states (ZRS), see \rfig{fig.zrs}.
With appropriate microwave intensity, temperature, and sample quality,
ZRS can span magnetic field ranges corresponding to several tens in filling factors.
However, unlike the QH effect, vanishing of diagonal resistance in microwave-irradiated 2DEG is not accompanied by Hall quantization.
%%%%%%%%%%%%%%%%%%%%%%%%%%%%%%%%%%%%%%%%%%%%%%%%%
%%%%%%%%%%%%%%%%%%%%%%%%%%%%%%%%%%%%%%%%%%%%%%%%%

\begin{figure}[ht]
\includegraphics[width=\columnwidth]{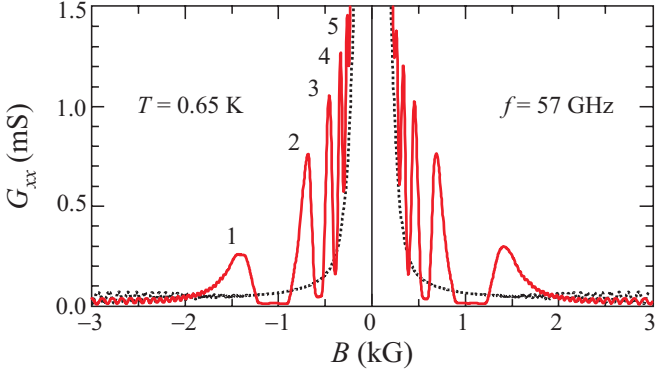}
\caption{Magnetoconductance $G_{xx}(B)$ under microwave irradiation of frequency $f=57$ GHz.
These data were obtained at $T=0.65$ K on a Corbino-shaped 2DEG with inner (outer) diameter 0.5 (3.0) mm,
density $n_e\simeq 3.55\times 10^{11}$ cm$^{-2}$, and mobility  $\mu\simeq 1.28 \times 10^{7}$ cm$^{2}$/Vs.
Adapted from \citet{yang:2003}.
}
\label{fig.zcs}
\end{figure}

Over the past decade ZRS were observed over a wide frequency range, from as low as 9 GHz \citep{willett:2004} to as high as 254 GHz \citep{smet:2005}.
Experiments in ultra-high mobility ($\mu \gtrsim 2 \cdot 10^7$ cm$^2$/V$\cdot$s) 2DEG also revealed ZRS stemming from the minima of fractional MIRO
(\rsec{VI.A.1}), namely near $\eac=3/2,1/2$ and $2/3$ \cite{zudov:2006a}, as well as a ZRS
presumably appearing due to a frequency mixing under bichromatic microwave irradiation \cite{zudov:2006a}.
Recently, ZRS were also observed to emerge from the MISO {\em maxima} in double QWs \cite{wiedmann:2010b}, see \rfig{ZRS_DQW}.
Shortly after the discovery of ZRS, experiments by \textcite{yang:2003} in Corbino-disk shaped 2DEG
demonstrated the existence of corresponding {\em zero-conductance states} (ZCS) (\rfig{fig.zcs}).
Later studies by \textcite{bykov:2010d} reported observation of ZCS in capacitively-coupled 2DEG.
As discussed in \rsec{s7.3.1}, ZCS were also realized in a non-degenerate 2D electron system on liquid $^3$He surface \cite{konstantinov:2010}.

%%%%%%%%%%%%%%%%%%%%%%%%%%%%%%%%%%%%%%%%%%%%%%%%%
%%%%%%%%%%%%%%%%%%%%%%%%%%%%%%%%%%%%%%%%%%%%%%%%%

\begin{figure}[ht]
\includegraphics[width=\columnwidth]{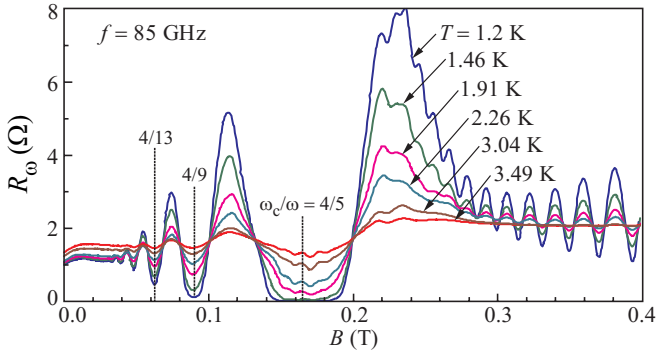}%fig23
\caption{Temperature evolution of ZRS and MIRO obtained under $f=85$~GHz illumination on the same sample as in \rfig{fig.zrs}a.
Adapted from \citet{mani:2002}.
}
\label{maniT}
\end{figure}

The temperature dependence of the ZRS has been examined by several groups \citep{mani:2002,zudov:2003,willett:2004}.
All studies found that ZRS disappear with increasing temperature (see \rfig{maniT}), transforming into MIRO minima
where the resistance approximately followed the Arrhenius law $\rho_\w\propto \exp(-\Delta/T)$.
The ``energy gaps'' $\Delta\propto B$ extracted in this way exceed relevant $T\sim\w$ by an order of magnitude.

%%%%%%%%%%%%%%%%%%%%%%%%%%%%%%%%%%%%%%%%%%%%%%%%%
%%%%%%%%%%%%%%%%%%%%%%%%%%%%%%%%%%%%%%%%%%%%%%%%%

\begin{figure}[ht]
\includegraphics[width=0.85\columnwidth]{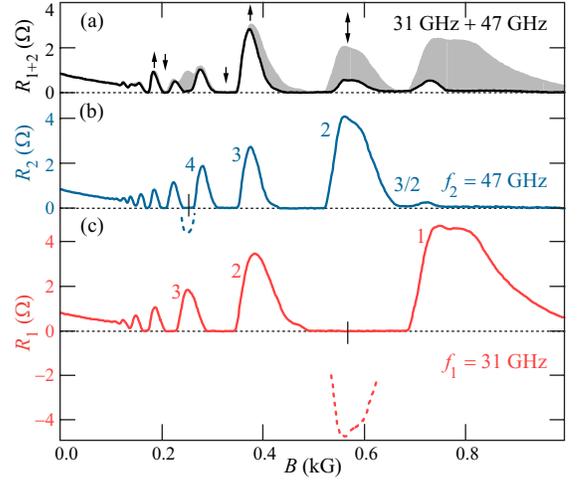}%fig24
\caption{Magnetoresistances (a) $R_1(B)$ at $f_1=31$ GHz, (b) $R_2(B)$ at $f_2=47$ GHz, and (c) $R_{1+2}$ at both $f_1$ and $f_2$.
Maximum-maximum, minimum-minimum and maximum-minimum overlaps are marked by $\uparrow$, $\downarrow$, and $\updownarrow$, respectively.
The upper boundary of the shaded area represents the average of monochromatic resistances $R_{1}$ and $R_{2}$.
% RR
Dotted lines in (b) and (c) represent reconstructed negative resistance.
These data were obtained on a 2DEG ($\sim$ 5 mm $\times$ 5 mm) with density $n_e\simeq 3.6\times 10^{11}$ cm$^{-2}$
and mobility  $\mu\simeq 2.0 \times 10^{7}$ cm$^{2}$/Vs. Adapted from \citet{zudov:2006b}.
}
\label{fig.anc}
\end{figure}

As discussed in \rsec{s4.2}, a homogeneous state with negative resistivity
is electrically unstable and is expected to break into domains with local current density
at which the nonlinear resistance is zero \citep{andreev:2003,auerbach:2005, finkler:2009}.
To date direct confirmation of absolute negative resistivity and the domain model has been limited.
First experimental support of the domain picture was provided by
\textcite{willett:2004} who observed large voltages between an internal and an external contact to 2DEG in the ZRS regime, see \rsec{s4.3}.
Experiments by \textcite{zudov:2006b} employing radiation of two distinct frequencies found that, away from ZRS,
the bichromatic photoresistance can be approximated by a simple superposition of monochromatic photoresistances, e.g.
$\delta \rho_{\omega_1\omega_2} \simeq \alpha_1\delta \rho_{\omega_1}+\alpha_2\delta \rho_{\omega_2}$.
Theoretically the bichromatic photoresponse was addressed by \textcite{lei:2006a,lei:2006c}.
Assuming that such a superposition also holds in the regime where one of the frequencies gives rise to ZRS,
e.g. underlying microscopic resistance is negative $\rho+\delta \rho_{\omega_1}<0$, but $\rho+\delta \rho_{\omega_1\omega_2}>0$,
one can reconstruct negative resistance which is otherwise masked by instability (see \rfig{fig.anc}).
In other words, the experiment uses one of the frequencies to probe the absolute negative resistance corresponding
to a ZRS induced by another frequency.

\begin{figure}[ht]
\includegraphics[width=0.85\columnwidth]{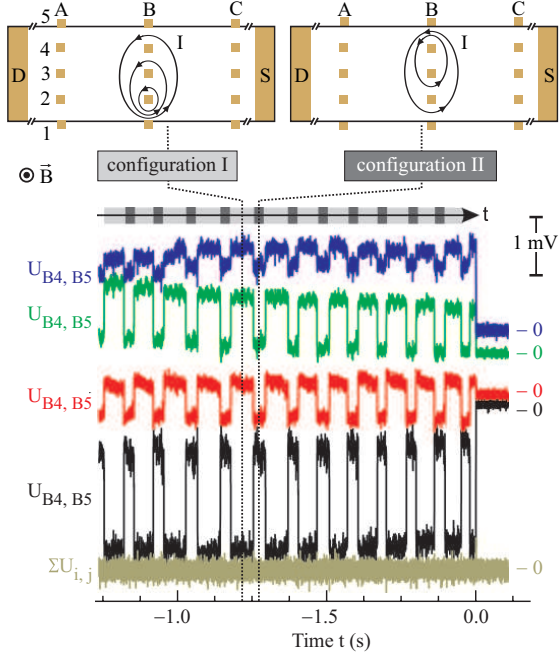}%fig25
\caption{Photovoltages (for $B=−95$~mT, $T=0.5$~K, and $f=48.1$~GHz) across different pairs of adjacent contacts shown in the inset (middle column).
The voltages at $t>0$ (no radiation) correspond to the zero reference level.  Closed loops in the insets indicate the Hall current
flow between the contacts in two different configurations. From \citet{dorozhkin:2011}.
}
\label{fig.telegraph}
\end{figure}

\textcite{dorozhkin:2011} performed time-resolved measurements of Hall voltages between internal probes
and observed random telegraph signals in the ZRS regime. These signals were interpreted in terms of spontaneous switching
between two nearly degenerate configurations of spontaneously formed domains (\rfig{fig.telegraph}).
%%%%%%%%%%%%%%%%%%%%%%%%%%%%%%%%%%%%%%%%%%%%%%%%%
%%%%%%%%%%%%%%%%%%%%%%%%%%%%%%%%%%%%%%%%%%%%%%%%%
%%%%%%%%%%%%%%%%%%%%%%%%%%%%%%%%%%%%%%%%%%%%%%%%%
%%%%%%%%%%%%%%%%%%%%%%%%%%%%%%%%%%%%%%%%%%%%%%%%%

\subsection{Instabilities of a nonequilibrium 2D electron gas and
the emergence of domains in a strong magnetic field: Phenomenological approach}
\label{s4.2}

As discussed in Sec.~\ref{s3}, for a sufficiently strong radiation
power (and other parameters in the appropriate range), the linear
resistivity of the microwave-illuminated 2DEG becomes negative. Such a
state is unstable, which results in formation of spontaneous
current and voltage breaking rotational symmetry
\cite{andreev:2003}; the corresponding theory is presented in Sec.~\ref{s4.2.1}.
In Sec.~\ref{s4.2.2} we review some previously obtained results
on related systems.

\subsubsection{ZRS effective theory: Spontaneous symmetry breaking}
\label{s4.2.1}

\begin{figure}[ht]
\includegraphics[width=0.9\columnwidth]{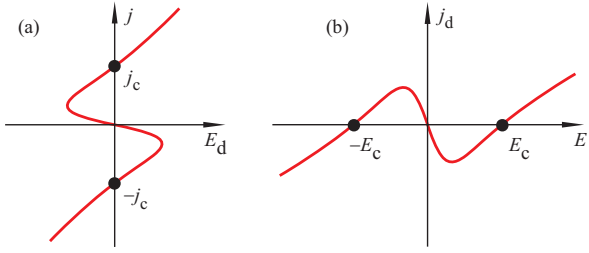}
\caption{S-shaped [(a), Hall bar setup, \req{e4.2.1}] and
N-shaped [(b), Corbino setup, \req{e4.2.2}] current-voltage characteristics.}
\label{f4.2.1}
\end{figure}

The starting point of the theory of \textcite{andreev:2003} is a
model with a constant Hall resistivity $\rho_H$ and a nonlinear
current dependence of the dissipative component  ${\bf E}_d$ of the
electric field:
\be
\label{e4.2.1}
{\bf E}=  \rho_H  {\bf j} \times \hat{\bf z}
+ {{\bf j}\over |{\bf j}|}  E_d(|{\bf j}|).
\ee
It is assumed that the function $E_d(j)$ has a negative derivative at
$j=0$ and a zero at $j = j_c$. The theory is phenomenological, i.e. it
assumes that there is a certain microscopic mechanism that provides such
current-voltage characteristics.  A distinct feature of the
considered strong magnetic field regime is a duality between current
and voltage. Specifically, Eq.~(\ref{e4.2.1}) is appropriate in the
Hall-bar geometry, when only one (say, $j_x$) component of the current
is present; the two terms in  Eq.~(\ref{e4.2.1}) then give the Hall
($E_y$) and the longitudinal ($E_x$) field respectively. In contrast, for the
Corbino-disc geometry one field component is zero (say, $E_x=0$) and
one is interested in ${\bf j}(E_y)$. Since the Hall component is the
dominant one in the conductivity tensor, the dependence of the
field dependence of the  dissipative  component  $j_d(E)$  is obtained
from $E_d(j)$ in Eq.~(\ref{e4.2.1}) by exchanging current and voltage
\cite{bergeret:2003},
\be
\label{e4.2.2}
{\bf j}= \sigma_H \hat{\bf z} \times {\bf E} +  {{\bf E}\over |{\bf
E}|}  j_d(|{\bf E}|),
\ee
with $\sigma_H \simeq (\rho_H)^{-1}$ and $j_d(E) = \sigma_H
E_d(\sigma_H E)$. In other words, the dissipative current-voltage
characteristics  are S-shaped in the first case and N-shaped in the
second case, see Fig.~\ref{f4.2.1}.

%%%%%%%%%%%%%%%%%%%%%%%%%%%%%%%%%%%%%%%%%%%%%%%%%%%%%%%%%%%%%
%%%%%%%%%%%%%%%%%%%%%%%%%%%%%%%%%%%%%%%%%%%%%%%%%%%%%%%%%%%%%

Equation (\ref{e4.2.1}) is supplemented by the continuity equation,
\be
\label{e4.2.3}
\partial n_e/\partial t + \nabla \cdot {\bf j} =0.
\ee
Further,
\be
\label{e4.2.4}
{\bf E} = - \boldsymbol{\nabla} \phi({\bf r}),
\ee
where $\phi({\bf r})$ is
the electrostatic potential; its variations are related to those of
density  according to
\be
\label{e4.2.5}
\delta\phi({\bf r}) = \int d^2 r' U({\bf r},{\bf r'}) \delta n_e ({\bf
r'})\,,
\ee
where $U({\bf r}, {\bf r'})$ is the Coulomb interaction (that may
be screened in the presence of an external gate). To explore the
stability of the system, one considers a small deviation in density $\delta
n_e({\bf r},t)$. Using Eqs.~(\ref{e4.2.2})-(\ref{e4.2.5}) and
linearizing in $\delta n_e$, one gets
\be
\label{e4.2.7}
\partial \delta n_e /\partial t = \boldsymbol{\nabla} \hat{\sigma}_d
\boldsymbol{\nabla} \hat{U} \delta n_e,
\ee
where $\hat{\sigma}_d$ is the dissipative differential conductivity
$(\sigma_d)_{\alpha\beta} = \partial (j_d)_\alpha/\partial E_\beta$
satisfying $(\sigma_d)_{\alpha\beta}=(\sigma_d)_{\beta\alpha}$. Note
that the Hall component of the conductivity dropped out of
Eq.~(\ref{e4.2.7}) because of its antisymmetric character. The
stability requires that the real part of all eigenvalues of the operator on the
r.h.s. of Eq.~(\ref{e4.2.7}) is non-positive.
Transforming it into the momentum space
and using $U(q)>0$, the stability condition reduces to the requirement
that eigenvalues of  $\hat{\sigma}_d$ are positive, which yields
\bea
&& dj_d / dE \ge 0 ; \label{e4.2.8} \\
&& j_d/ E \ge 0 . \label{e4.2.9}
\eea
In other words, the stability requires that both absolute and
differential conductivity are non-negative.
As is clear from the above derivation, Eqs.~(\ref{e4.2.8}) and
(\ref{e4.2.9}) represent the stability
to longitudinal and transverse fluctuations of
the electric field with respect to its direction in the state under
consideration.

\begin{figure}[ht]
\includegraphics[width=\columnwidth]{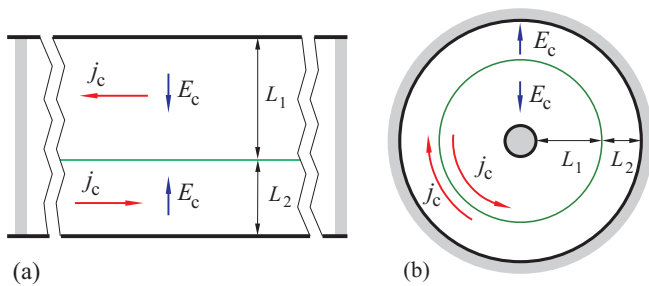}
      \caption{Simplest domain structures for (a) Hall-bar  and (b) Corbino-disc  geometries.
The current density $j_c$ and the electric field $E_c=\rho_H j_c$ in the domains
are fixed by the condition (a)  $E_d=0$  or (b) $j_d=0$, see \rfig{f4.2.1}.
Corbino-disc [Hall-bar] system adapts to an applied bias $V=E_c(L_1-L_2)$
[applied current $I=j_c(L_1-L_2)$] remaining in ZCS [ZRS].
Adapted from \citet{andreev:2003,vavilov:2004}.}
      \label{f4.2.2}
\end{figure}

Therefore, when the linear-response conductivity is negative
the system is unstable near $j=E=0$: fluctuations will grow
until the stability point with current $j_c$ and electric field $E_c =
\rho_H j_c$ is reached, see \rfig{f4.2.1}. This implies that the system will break into
domains with spontaneous currents and fields of these magnitudes. The
simplest possible domain structure consists of two domains as shown in
Fig.~\ref{f4.2.2} for the case of Hall-bar and Corbino-disc setups (see
Sec.~\ref{s4.4} for a discussion of more complicated domain
structures). The values of
spontaneous current and field correspond to dissipative current $j_d=0$
(i.e. zero absolute dissipative conductivity) in the Corbino disc
setup and to dissipative field $E_d=0$ (and thus zero absolute
dissipative resistivity) in the Hall bar geometry. When a finite
current (for the Hall bar) or voltage (Corbino disc) is applied, the
domain wall shifts to accommodate it. The system remains in the ZRS (or
ZCS in the Corbino geometry) until the applied
current (field) exceeds $j_c$ (respectively, $E_c$).

%%%%%%%%%%%%%%%%%%%%%%%%%%%%%%%%%%%%%%%%%%%%%%%%%%%%%%%%%%%%%
%%%%%%%%%%%%%%%%%%%%%%%%%%%%%%%%%%%%%%%%%%%%%%%%%%%%%%%%%%%%%

A further step forward in the study of the ZRS was done by
\textcite{volkov:2004,bergeret:2003} who included on the r.h.s. of
Eq.~(\ref{e4.2.2}) the diffusive term $-e\hat{D}\boldsymbol{\nabla}n_e$
and assumed that, contrary to the conductivity, the diffusion tensor
is not affected by microwave radiation. [This assumption was later
justified by \textcite{dmitriev:2009a}.] This introduces in the problem
the nonequilibrium screening length that governs the width of domain
walls, see \rsec{s3.3.2}. \textcite{volkov:2004} assumed however a 3D
relation between the potential and the density [i.e. they assume
$V(q)\sim 1/q^2$ momentum dependence of the Coulomb interaction
rather than $V(q)\sim 1/q$ appropriate for 2D geometry,
see \cite{dorozhkin:2011a}] and, as a consequence,
obtain the 3D screening length $\lambda_{3D} = (\epsilon D/4\pi|\sigma^{(3D)}_d|)^{1/2}$.
The resulting equations are used to find the spatial profile of the domain
wall. The central result of \textcite{volkov:2004} is the residual
resistivity of a finite sample in the ZRS phase (with a single domain
wall assumed), which is found to be negative and exponentially small:
$\rho_{\rm res} \propto - \exp\{-L/\lambda_{\rm 3D}\}$. Here $L$ is the
sample size in the direction transverse to the domain
wall.

\textcite{dorozhkin:2011a} established the electrical stability
condition in a finite sample with realistic 2D Coulomb interaction.
Instead of conditions (\ref{e4.2.8}) and (\ref{e4.2.9})
(which in the linear regime read $\sigma_d>0$), one gets $\sigma_d>-\epsilon D/2L$,
or $\lambda^{-1}>\pi/L$ in terms of the 2D nonequilibrium screening length (\ref{lambda}).
In the diffusion-stabilized regime of negative
conductivity, $-\epsilon D/2L<\sigma_d<0$, the theory of
\textcite{dorozhkin:2011a} predicts the emergence of two regions with opposite
directions of electric field (this requires broken inversion symmetry, for instance,
due to asymmetric contact configuration).
The amplitude of these fields increases when the system approaches the
instability threshold. This effect is a precursor of the domain structure
in the regime $\sigma_d<-\epsilon D/2L$.

\subsubsection{Earlier results on related problems}
\label{s4.2.2}

The instability of a semiconductor with dc negative conductivity was
first pointed out
by \textcite{zakharov:1960}. Subsequently many performed
detailed analyses of such instabilities
in various systems with negative differential or absolute
conductivity. The corresponding literature is extensive, and we
restrict ourselves to quoting several early
articles \cite{ridley:1963,bonch-bruevich:1965,volkov:1967,elesin:1967},
a review by \textcite{volkov:1969}, and the books by
\textcite{bonch-bruevich:1975,pozhela:1981,schoell:2001}.
The interested reader can find there an overview of various mechanisms of
emergence of negative conductivity (characterized by N-shape or
S-shape current-voltage-characteristics), derivation of stability conditions,
and an  analysis of domain formation as a result of instabilities.
In this context, particularly well known is the Gunn effect \cite{gunn:1963}: an N-type
current-voltage-characteristic leads to an instability  resulting in
the formation of moving domains and in microwave generation.
The corresponding semiconductor device (Gunn diode) has found various
application  in high-frequency electronics.

The absolute negative conductivity was observed by
\textcite{banis:1972} [see also Sec.~VI.3 in
\textcite{pozhela:1981}]. This was achieved by exposing GaAs samples
exhibiting a negative differential conductance to a microwave
radiation. The explanation given by \textcite{banis:1972,pozhela:1981}
assumes that the microwave field can be treated adiabatically and
simply leads to an (oscillating in time) shift of the operation point of the
device. After the time averaging of the corresponding differential
conductivity this leads to a negative dc conductivity. It seems,
however, that this explanation (that would be perfectly correct for a
low-frequency field) does not fully catch the physics of the
experiment. Indeed, the frequency of the microwave field (of order 10
GHz) does not seem to be small compared to inverse characteristic
times in the device, so that the adiabaticity assumption is not
met. Furthermore, the authors report that, in the absence of an
external circuit, a large  spontaneous dc voltage developed on the
sample (whose polarity could be stabilized by a small symmetry
breaking perturbation). The amplitude of the voltage was determined by
the zero of the current-voltage characteristics, $I(V)=0$. This
spontaneous formation of dc field domains resulting from the
microwave-induced absolute
negative conductivity indicates certain analogy between the experiment of
\textcite{banis:1972} and the systems constituting the subject
of this review.

It was experimentally discovered \cite{liao:1980,basun:1983} that
intense laser illumination generates strong electric fields (detected
via the splitting of luminescence lines) in ruby crystals. A
phenomenological  theory of this phenomenon was developed by
\textcite{dyakonov:1984a,dyakonov:1984b}. It is based on an assumption
that the absolute conductivity in ruby becomes negative for a
sufficiently strong illumination power. The resulting instability
leads to formation of electric field domains, thus explaining the
experimental observations. There is a clear analogy
between the theory of ZRS as resulting from regions of
negative resistivity in MIRO and this theory of ruby domains. There
are, however, also essential differences. First, is the presence
of a strong magnetic field  in the MIRO problem. Second, contrary to an
isotropic 2DEG, ruby is anisotropic: the electric field and the
current are directed along the $C_3$ axis. This modifies the analysis
of the stability (only longitudinal fluctuations need to be
considered) and renders the part of the current-voltage-characteristics with
$I/V<0$ but $dI/dV>0$ stable. To our knowledge, no microscopic theory
of negative conductivity in ruby has been developed.

The impact of a strong transverse magnetic field on this class of
phenomena was also appreciated approximately half a century ago.
\textcite{kazarinov:1963} showed that
a quantizing transverse magnetic field may lead under
nonequilibrium conditions to strong nonlinearities in
current-voltage-characteristics, in particular, to a negative differential
resistance. The fact that in the case of large Hall angle
($\omega_c\ttr\gg 1$) current-voltage-charactertics have a dual shape in
the Hall bar
($j_y=0$) and Corbino disk ($E_y=0$) geometries (in particular, an
S-shaped characteristic of the Hall-bar device gives rise to an
N-shaped characteristic for the Corbino disc)  was pointed out by
\textcite{bass:1965,bogomolov:1967}. \textcite{kogan:1968} studied an
instability in a system with negative differential resistance
 at intermediate Hall angles ($\omega_c\ttr\sim 1$).
\textcite{elesin:1969,gladun:1970} and particularly
\textcite{ryzhii:1970,ryzhii:1986}
discussed the photoconductivity of a 2D gas in a
quantizing magnetic field and strong electric field (see \rsec{s3.2})
and concluded that it is possible to reach an absolute negative conductivity in
this class of systems.

Another broad class of semiconductor systems in which similar
phenomena have been intensively studied are single and multiple
QW heterostructures.
% RR
In particular, much work was devoted to
resonant transmission via double-barrier structures, where
a negative differential resistance was found
\cite{tsu:1973,sollner:1983}. More recently, it was predicted that
under laser illumination the system may show
absolute negative resistance \cite{dakhnovskii:1995}.
Further, the absolute negative conductivity was theoretically predicted
\cite{pavlovich:1976,ignatov:1995} and experimentally
observed  \cite{keay:1995} in semiconductor superlattices under
THz radiation; another theoretical work considers a superlattice in a
magnetic field \cite{cannon:2000}. The theoretical predictions include
generation of spontaneous voltages and currents resulting from
instabilities. A negative absolute conductivity
was also found in a related model \cite{hartmann:1997} of a particle
in a periodic lattice driven by an ac field and coupled to a
dissipative bath.

Absolute negative conductance under strongly non-equilibrium conditions
emerges also in other types of systems. In particular, \textcite{aronov:1975}
predicted this effect to happen in a Josephson junction of two
superconductors with different gaps one of which is subjected to light
illumination. This prediction was experimentally confirmed by
\textcite{gershenzon:1986,gershenzon:1988}. In view of the spatially local
character of the junction, the negative conductance does not lead in
this case to an instability (contrary to extended systems that tend to
break into domains in such situations). Finally, an absolutely negative
mobility was obtained in a purely classical model of a Brownian
particle subjected to non-equilibrium noise \cite{eichhorn:2002}.

It is clear from the above discussion that many of the aspects of the
ZRS problem have appeared earlier in related problems. On the other hand,
 to our knowledge, the combination of the key features of
the ZRS problem, namely, (i) a strong magnetic field (large Hall angle),
(ii) instability induced by an absolute negative
dissipative conductivity, and (iii) 2D isotropy of the problem, has
not appeared in any other context.

\subsection{Microscopic theory; determination of currents and fields in domains}
\label{s4.3}

We combine now the phenomenological ZRS theory of Sec.~\ref{s4.2.1}
with microscopic calculations of photoresistivity.
As discussed in Sec.~\ref{s3}, the linear resistivity
in the presence of microwaves may become
% RR
negative around its
minima. Specifically, consider the inelastic mechanism that is dominant
for sufficiently low temperatures and assume first the regime of
overlapping LLs. According to Sec.~\ref{s3.2.2}, the linear
resistivity is negative when the dimensionless
microwave power satisfies ${\cal P} > {\cal P}^*> 0$,
with the threshold value given by \cite{dmitriev:2004,dmitriev:2005}
\be
\label{e4.3.1}
{\cal P}^*=\frac{\ttr}{\tin}
\left(4\delta^2\frac{\pi\omega}{\omega_c}\sin
\frac{2\pi\omega}{\omega_c}
-\sin^2\frac{\pi\omega}{\omega_c}\right)^{-1},
\ee
see Fig.~\ref{MIROov}. The spontaneous field in the ZRS domains is found to be
\be
\label{e4.3.2}
E_c = E_0\sqrt{\frac{\cal P}{{\cal P}^*}-1}, \qquad E_0=\frac{\wc}{\pi e R_c}\sqrt{\frac{\ttr}{2\tin}}\,.
\ee
For the marked MIRO minima in Fig.~\ref{MIROov},
the corresponding
values $E^*/E_0$ are shown by arrows in the inset.

In the regime of separated LLs with width
$2\Gamma=2(2\omega_c/\pi\tau_{{\rm q}})^{1/2}$ and assuming a
small-angle impurity scattering,  the threshold power is
\be
\label{e4.3.3}
{\cal  P}^*\sim\Gamma^2\ttr/\omega\omega_c\tin,
\ee
and the spontaneous field is of the order of
\be
\label{e4.3.4}
{E}_c\sim \left({\ttr\over\tq}\right)^{1/2}\,{\omega_c\over e R_c}\,.
\ee

Within the displacement mechanism (relevant for higher temperatures) and
for a small-angle impurity scattering the corresponding results were
obtained by \textcite{vavilov:2004}. In this case the characteristic value
of the spontaneous field is given by
Eq.~(\ref{e4.3.4}) in limits
of both overlapping and separated LLs.

We point out that the form $(\ttr/\tq)^{1/2}$
of the factor in front of $\omega_c/eR_c$ in
Eq.~(\ref{e4.3.4})
depends on the character of disorder. For a model of mixed disorder
(\ref{mixed-disorder}) and for overlapping LLs
this factor is reduced and becomes of order unity when the weight of
the short-range component becomes sufficiently large, see
\cite{khodas:2008,dmitriev:2009b} and Sec.~\ref{s3.2}.
A related result was obtained by \textcite{auerbach:2007}, who studied
the displacement mechanism for mixed disorder (with short-range
component determining the transport rate)
in the regime of separated LLs
and found ${\cal E}_{\rm dc}^*\sim \Gamma / eR_c$ (if one sets the
correlation length of short-range disorder to be $\sim k_F^{-1}$).
To our knowledge,
the inelastic mechanism has not been studied for the case of separated
LLs and mixed disorder.

These predictions can be confronted with the experiment.
Focusing on the inelastic mechanism and on the
regime of overlapping LLs and assuming characteristic values
of parameters  $\omega/2\pi \simeq 50$~GHz, $T\sim 1\:{\rm K}$, and
$\tau_{\rm in}^{-1}\sim 10\:{\rm mK}$,
$\tq^{-1}\sim 0.3\:{\rm K}$,  $\ttr^{-1}=10$~mK, and
$v_F=2 \cdot 10^7$~cm/s, one gets  \cite{dmitriev:2004,dmitriev:2005}
the threshold microwave intensity required
for the emergence of the ZRS ${P}^* \lesssim 1\:{\rm
mW/cm^2}$, in conformity with the experiments.  Further, for these
parameter values and for ${\cal
P}-{\cal P}^*\simeq {\cal P}^*$ (i.e. for the
microwave power exceeding its threshold by roughly a factor of 2)
the estimated dc electric field in the domains,
Eq.~(\ref{e4.3.2}), is found to be $E_c\sim 1\:{\rm
V/cm}$. In the experiment of
\textcite{willett:2004} the voltage drop between an internal
and an external contact (separated by 200~${\rm \mu m}$) generated by
the radiation in the absence of the drive current was of the order of
5~mV for $\omega_c/2\pi\simeq 20$~GHz. Assuming the simplest domain
geometry (i.e no additional domain walls between the contacts), this
yields $E_c\sim 0.25\:{\rm V/cm}$. On the other hand,
the above theoretical estimate yields $E_c\sim
0.15\:{\rm V/cm}$ [taking into account the $\omega_c^2$ dependence of
$E_c$ following from Eq.~(\ref{e4.3.2})], so that the
experimental value is somewhat larger. In a more
recent experiment \cite{dorozhkin:2011} spontaneous fields of the order
of $E_c\sim 0.15\:{\rm V/cm}$ were reported for
$\omega_c/2\pi\simeq 50$~GHz, which is in this case several times
smaller than the theoretical estimate. In general, the agreement between
theory and experiment appears to be very reasonable if one
takes into account some deviations of parameters (relaxation
rates, ratio of the microwave power to its threshold value) from those used
in the theoretical estimate.

\subsection{Effective theories of the phase transition into ZRS and of ZRS
dynamics}
\label{s4.4}

The transition into ZRS belongs to the class of dynamical phase
transitions \cite{hohenberg:1977}. These phenomena are in general
governed by non-linear differential equations (of hydrodynamic type)
with stochastic (noise) terms and encompass spontaneous pattern formation
in a great variety of systems driven away from equilibrium \cite{cross:1993}.
Contrary to conventional (thermodynamic) phase transitions, such
phenomena in general are not characterized by a free energy
functional. This invalidates, in particular, the Mermin-Wagner
theorem, opening the way to spontaneous breaking of a continuous
symmetry in 2D systems, which is of direct relevance to the ZRS problem.

\textcite{auerbach:2005,finkler:2006} performed an analysis of
possible domain patterns in the ZRS phase. In spirit of
Sec.~\ref{s4.2}, they started with a
model with a constant Hall conductivity $\sigma_H$ and a non-linear
% RR
field dependence of the dissipative current $j_d$,
Eqs.~(\ref{e4.2.2})-(\ref{e4.2.5}).   Their crucial observation
 is that for this problem one
can define a Lyapunov functional
\be
\label{e4.4.4}
G[\phi]  = \int d^2 r g({\bf E}({\bf r})), \ \ g({\bf E}) =
\int_0^{\bf E} d{\bf E'} {\bf j}_d({\bf E'}),
\ee
with the key property $dG/dt \le 0$. This implies that minima of the Lyapunov
functional are stable steady states. To explore the ZRS phase
(negative linear dissipative conductivity),
\textcite{auerbach:2005,finkler:2006} expand $g({\bf E})$ near the
zero $E_c$ of the current-voltage characteristics $j_d(E)$:
\be
\label{e4.4.5}
g({\bf E}) = g(E_c) + {\sigma_c\over 2}(E-E_c)^2 + \kappa
|\boldsymbol{\nabla} \cdot {\bf E}|^2.
\ee
The last term here [penalizing large field gradients and determining the
width of domain walls, $l_{\rm dw}\sim (\kappa/
\sigma_c)^{1/2}$] is added on phenomenological grounds.
On the microscopic level, a finite width of the domain wall emerges
when one takes into account the diffusive contribution
$-e\hat{D}\boldsymbol{\nabla}n_e$ to the current $j$.
However, the local form of
the corresponding term in Eq.~(\ref{e4.4.5}) correspond to the
relation between $E$ and $n_e$ characteristic for 3D rather than for 2D
systems, see the discussion at the end of Sec.~\ref{s4.2.1}. It
remains to be seen, to what extent this may affect the results.

\begin{figure}[ht]
\includegraphics[width=\columnwidth]{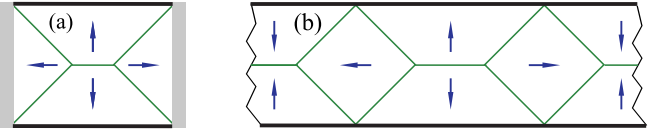}
      \caption{(a) Proposed domain structure near metallic contacts (shaded grey regions) in a rectangular sample;
(b) possible structure with an additional pair of domains. Arrows show the direction of the electric field $\bE_c$,
see also \rfig{f4.2.2}. Adapted from \citet{finkler:2006,finkler:2009}. }
      \label{f4.2.3}
\end{figure}

Minimization
of the Lyapunov functional yields the simple domain structures in the
Corbino and Hall bar geometries, see Figs.~\ref{f4.2.2} and \ref{f4.2.3}a.
Introducing further a weak and smooth disorder field ${\bf
E}_{\rm dis}({\bf r})$, with $|{\bf E}_{\rm dis}|\ll E_c$,
\textcite{auerbach:2005,finkler:2006} generalized
Eq.~(\ref{e4.4.5}) [denoted below as $g_0({\bf E})$] to
\be
\label{e4.4.6}
g({\bf E},{\bf E}_{\rm dis}) = g_0({\bf E}) - \sigma_1(E) {\bf E} {\bf
E}_{\rm dis} +
O(E_{\rm dis}^2).
\ee
The term proportional to  ${\bf E}_{\rm dis}$ induces a correction to the
stability condition, with the result that the current density at a
domain wall is
${\bf j} = - \sigma_1(E_c){\bf E}_{\rm dis} + O(E_{\rm dis}^2)$.  For a
short correlation length of the disorder a comparison of the domain
wall and bulk contribution to the Lyapunov functional in the spirit of
\textcite{imry:1975} shows \cite{auerbach:2005} that no new
domains are formed. On the other hand, disorder with a sufficiently
large correlation length $\xi_{\rm dis}$ favors the breakdown of the system in
multiple domains with a size set by
$\xi_{\rm dis}$. Further, \textcite{auerbach:2005,finkler:2006} used this
approach to estimate the effect of the potential $\phi_{\rm dis}({\bf r})$
(regular or random) that induces multiple domain walls on
the ZRS conductivity. They obtained $\sigma = C \sigma_1 (\langle |{\bf
E}_{\rm dis}|\rangle)/E_c$, with a numerical prefactor $C \sim 1$ depending on the
specific form (or statistics) of the potential.

%%%%%%%%%%%%%%%%%%%%%%%%%%%%%%%%%%%%%%%%%%%%%%%%%%%%%%%%%%%%%
%%%%%%%%%%%%%%%%%%%%%%%%%%%%%%%%%%%%%%%%%%%%%%%%%%%%%%%%%%%%%

\textcite{finkler:2009} found that even in a homogeneous system creating
additional domains may be favorable. They
pointed out that the Lyapunov ``energy'' $\epsilon_{\rm dw}$
of the domain wall depends on the angle $\phi$
between the wall and the field in the domains. If
$\epsilon_{\rm dw}(\pi/4) / \epsilon_{\rm dw}(\pi/2) \le 1/2\sqrt{2}$,
additional domains as shown  in Fig.~\ref{f4.2.3}b  will
emerge. The dependence $\epsilon_{\rm dw}(\phi)$ is
determined by the precise form of the Lyapunov function $g(E)$.
Remarkably, for the simplest choice of this function
\textcite{finkler:2009} obtain for the ratio $\epsilon_{\rm dw}(\pi/4) /
\epsilon_{\rm dw}(\pi/2)$ exactly the value $1/2\sqrt{2}$, which
implies degeneracy between the states with and without additional
domains with $\pi/4$ domain walls.  They conjecture that a more
realistic Lyapunov function might yield a smaller value of this ratio,
thus favoring additional domains. \textcite{finkler:2009} consider
further the effect of spatial variation of the Hall conductivity
$\sigma_H$ and show that in this situation non-stationary (periodic)
solutions may emerge. The simplest example is the Corbino geometry with
Hall conductivity changing linearly along the $y$ axis (radial
coordinate), $d\sigma_H/dy = \alpha$. Then $\boldsymbol{\nabla}
\cdot {\bf j}_H = \alpha E_x$.
Combining this with $j_d=0$ (characteristic for a
homogeneous system away from domain walls), we get $\partial
n/\partial t = - \boldsymbol{\nabla} \cdot {\bf j} = - \alpha E_x
({\bf r})$. In
the simplest domain geometry (two domains with electric field in the
radial direction, Fig.~\ref{f4.2.2}) $E_x=0$ and we do not get any time
dependence. However, in the presence of additional $\pi/4$ domains,
Fig.~\ref{f4.2.3}b, the field there is along the $x$ axis, which implies
that such domains should move. \textcite{finkler:2009} also performed a
numerical modeling of the problem in the torus geometry and with a
spatially varying Hall conductivity and indeed found non-stationary,
time-periodic solutions, in a certain range of the parameter $\alpha$
controlling a typical gradient of  $\sigma_H$.

What is the character of the transition into the ZRS phase? For
equilibrium phase transitions the answer to such a question is
usually based on the Landau (mean-field) theory of phase transitions
complemented by a renormalization group (RG) analysis. A counterpart
of this approach applicable to dynamical critical phenomena (including
those in strongly non-equilibrium systems) is known as the dynamical
renormalization group \cite{forster:1977,hohenberg:1977}. A powerful
framework for its technical implementation is the Martin-Siggia-Rose
formalism
\cite{martin:1973,janssen:1976,de-dominicis:1976,de-dominicis:1978}
closely related to the Keldysh formalism.
This approach allows one to cast the evolution governed by a stochastic
non-linear equation into a Lagrangian form convenient for the
implementation of the RG procedure.

\textcite{alicea:2005} made a first step in application of these ideas
to the problem of ZRS transition. They started with formulating a
general equation describing the long-scale, long-time dynamics of a
microwave-driven system:
\bea
\partial_t{\bf j} &+& \omega_0^{-1} \partial_t^2{\bf j} \nonumber \\
&=& - \mu \boldsymbol{\nabla} \phi + \omega_c \hat{\bf z}\times {\bf
j} - r {\bf j} - u |{\bf j}|^2{\bf j} \nonumber \\
&& + \eta_1 \nabla^2  {\bf j} + \eta_2 \boldsymbol{\nabla}
(\boldsymbol{\nabla} \cdot {\bf j})
- \eta_3 \nabla^4  {\bf j} - \eta_4 \boldsymbol{\nabla}^3
(\boldsymbol{\nabla} \cdot {\bf j}) \nonumber \\
&&-\nu_1({\bf j}\cdot \boldsymbol{\nabla}) {\bf j}
-\nu_2 \boldsymbol{\nabla} ({\bf j}^2) - \nu_3 (\boldsymbol{\nabla}
\cdot {\bf j}) {\bf j} \nonumber \\
&&+ \gamma_1 \phi\boldsymbol{\nabla} \phi + \gamma_2 \phi
{\bf j} + \gamma_3 \phi \hat{\bf z}\times {\bf j} + \boldsymbol{\zeta}
+ \ldots .
\label{e4.4.7}
\eea
Here the $\mu$ and $\omega_c$ terms represent the electric-field and
Lorentz forces, the $r$ and $u$ terms describe the non-linear
resistivity,  the term with $\partial_t^2$ (on the left-hand side)
originates from the frequency dispersion of the resistivity, the
$\eta_i$ terms characterize its momentum dispersion, the $\nu_i$ terms
are convective non-linear contributions, and the $\gamma_i$ terms account
for the density dependence of transport coefficients. Finally,
$\boldsymbol{\zeta}$ is the Langevin noise source with a correlation
function
\be
\label{e4.4.8}
\langle \zeta_\alpha({\bf r}, t) \zeta_\beta({\bf r'},
t')\rangle = 2 g \delta_{\alpha\beta} \delta({\bf r}-{\bf
r'})\delta(t-t').
\ee
Equation (\ref{e4.4.7}) is supplemented by the continuity equation
(\ref{e4.2.3}) and
the relation (\ref{e4.2.5}) between the potential and the
density. \textcite{alicea:2005}  postulate Eq. (\ref{e4.4.7}) on
symmetry grounds: it includes leading terms of the expansion in
gradients, time derivative, and amplitudes of the current and the
potential. It should be pointed out that
 the assumption \cite{alicea:2005} of local relation between ${\bf
j}$ and $\phi$  appears to be an oversimplification, since there are
contributions to $\partial_t{\bf j}({\bf r}, t)$ that depend in a local way on
the density $n_e({\bf r})$ and since Eq.~(\ref{e4.2.5}) linking the
density to the potential is non-local. It remains to be seen how
sensitive is the result to this assumption.

Equation (\ref{e4.4.7}) describes a transition from a conventional
resistive state at $r>0$ to the ZRS at $r<0$.
Within the transition picture of Sec.~\ref{s4.2}, the transition is
continuous (second order), and the spontaneous current in the
symmetry-broken phase is $j = \sqrt{|r|/u}$, i.e. $j \sim |r|^\beta$
with the critical index $\beta = 1/2$. Clearly, this is just the
Landau mean-field description of the transition, and one needs to find out
how fluctuations affect these results. This is what
\textcite{alicea:2005} do; their main findings are as follows:

(i) The symmetry-broken state is stable with respect to
current-density fluctuations (there is no infrared divergence contrary
to usual thermodynamic transitions where the Mermin-Wagner theorem
implies a logarithmic divergence in 2D leading to destruction of the
long-range order).

(ii) For the model with short-range interaction, $U({\bf r}-{\bf r'})
= C^{-1} \delta({\bf r} - {\bf r'})$ (that would correspond to a
system with a screening gate) and with a symmetry
with respect to $\phi \to \phi + {\rm const}$ the Gaussian fixed point
of the mean-field theory is stable, with non-linear terms being
marginally irrelevant. As a result, the mean-field exponents hold,
implying the scaling $j \sim |r|^\beta$ with $\beta=1/2$ on the ZRS
side of the transition, as well as   $j \sim |E|^\delta$ with
$\delta=1/3$ and the overdamped ``diffusion mode'' $\sim 1/(-i\omega +
Dq^4)$ at criticality, with logarithmic corrections to scaling.
The invariance with respect to $\phi \to \phi + {\rm const}$ forbids
in Eq.~(\ref{e4.4.7}) those terms that explicitly depend on the value
of $\phi({\bf r})$ (rather than on its gradients). In the physical
system such invariance does approximately hold; however, it gets
violated when variations of density [linked to those of $\phi$ via
Eq.~(\ref{e4.2.5})] become comparable to the total density of the
electron gas.

(iii) The long-range Coulomb interaction  drives the system
away from the Gaussian fixed point. \textcite{alicea:2005} analyze the
RG equations in the model with $1/k^\epsilon$ interaction (in momentum
space) that has the upper critical dimension $d=2+\epsilon$ and find
no stable fixed points in 2D. This leads them to the conclusion that
the transition becomes first order in the presence of long-range
interaction. Presumably, this should also hold in the physical case
of $1/k$ interaction, i.e. $\epsilon=1$.

(iv) The terms not included in the $\phi \to \phi + {\rm const}$ model
drive the transition first order as well (whether with short-range or
with long-range interaction). These results are obtained by the
analysis of the RG flow $d_{\rm UC}-\epsilon$ dimensions, where the
upper critical dimension $d_{\rm UC}$ is found to be $d_{\rm UC}=4$
and $d_{\rm UC}=7$ for short- and long-range interactions,
respectively. In both cases no stable fixed points are found, implying
that the transition gets first order. This conclusion is then
extrapolated to the physical dimension $d=2$.

(v) \textcite{alicea:2005} also point out a similarity between the ZRS
problem and a phase transition in the 2D bird-flocking model. The latter
is essentially a dynamical generalization of the 2D $XY$-model: birds
move with a constant velocity and at each time step every of them
picks up a new direction governed by average velocity of surrounding
birds, with some level of noise
\cite{vicsek:1995,toner:1995,toner:1998,gregoire:2004}. For a
sufficiently weak noise the
rotational symmetry gets spontaneously broken: the bird flock acquires
a macroscopic collective velocity. While \textcite{vicsek:1995} found
a second-order phase transition, a more recent work
\cite{gregoire:2004} obtains a first-order transition.
\cite{alicea:2005} find that the ZRS and the bird-flocking model are
almost equivalent, up to the Lorentz-force term (present in the former
but not in the latter), so that the first-order transitions in both of
them would be mutually consistent.

\section{Hall field- and phonon-induced resistance oscillations (HIRO and PIRO)}
\label{s5}

\subsection{HIRO: Experimental discovery and basic properties}
\label{s5.1}

\begin{figure}[ht]
\includegraphics[width=\columnwidth]{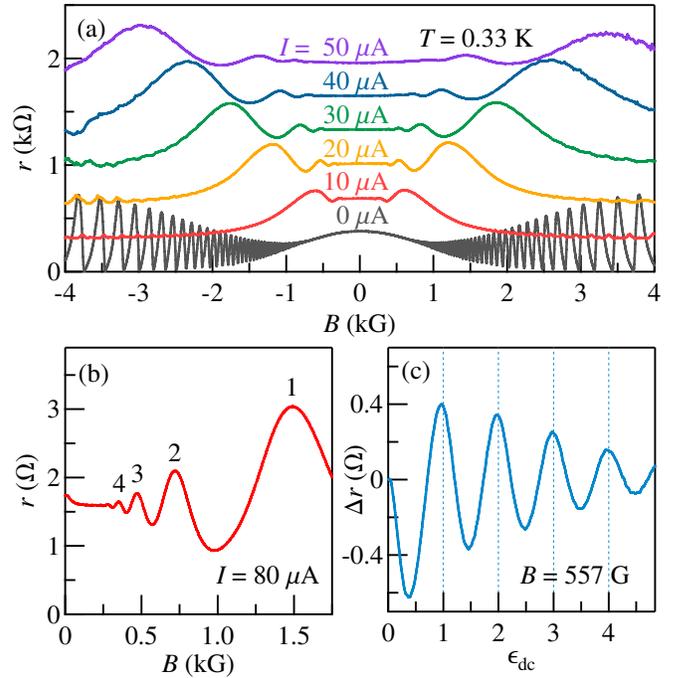}
\caption{(a) Differential magnetoresistance $r(B)$ measured at different dc current $I$, as marked,
in a Hall bar sample of width $w=50$ $\mu m$ with $\mu \simeq 3.0 \times 10^6$ cm$^2$/V s
and $n_e \simeq 2.0 \times 10^{11}$ cm$^{-2}$. Adapted from \citet{yang:2002}.
(b) Differential magnetoresistivity $r$ showing four HIRO maxima marked by
integers measured by sweeping $B$ at fixed $I=80\,\mu{\rm A}$. (c) Correction to
the differential resistivity $\delta r$ versus $\edc$ obtained by sweeping $I$
at fixed $B=557\,{\rm G}$. The data were obtained at $T\simeq 1.5$ K in a
$100\,\mu{\rm m}$ wide Hall bar sample with $n_e\simeq 3.7\times 10^{11}\,{\rm
cm^{-2}}$ and $\mu\simeq 1.2 \times 10^7\,{\rm cm^2/V\,s}$. Adapted from
\citet{zhang:2007a}.}
\label{fig.hiro}
\end{figure}

A decade ago \textcite{yang:2002} discovered prominent $1/B$-oscillations of the
differential resistivity $r \equiv \rho + I(d\rho/dI)$ in a 2DEG with moderate
mobility $\mu\sim 10^{6}\,{\rm cm^{2}/V\,s}$ subject to a constant direct
current $I$, see \rfig{fig.hiro}a. These oscillations, termed the Hall field-induced resistance
oscillations (HIRO), can alternatively be observed at fixed $B$ and varying $I$.
The latter approach allows one to observe more oscillations since the
amplitude of the DOS modulation (fixed by the value $\wc\tq$) does not change
during the measurement. Examples of both realizations of HIRO in a high mobility
($\mu\sim 10^{7}\,{\rm cm^{2}/V\,s}$) 2DEG are shown in \rfig{fig.hiro}b and c.
%%%%%%%%%%%%%%%%%%%%%%%%%%%%%%%%%%%%%%%%%%%%%%%%%
%%%%%%%%%%%%%%%%%%%%%%%%%%%%%%%%%%%%%%%%%%%%%%%%%
\textcite{yang:2002} proposed that HIRO stem from the commensurability between
the cyclotron diameter $2\rc$ and the spatial separation $\Delta y=\wc/eE$
between the Hall field-tilted LLs. Here $E=\rho_{H}j$ is the Hall electric
field, $j=I/w$ is the current density, and $w$ is the Hall bar width. This
yields oscillations with $\edc={2\rc}/{\Delta y}=2e E\rc/\wc$ with a
characteristic dependence $\edc\propto n_e^{-1/2}$ on the electron density,
confirmed by \textcite{yang:2002}. In addition to single-layer 2D electron
systems, HIRO were also observed in double QWs
\cite{bykov:2008a,mamani:2009b,wiedmann:2011c} and in 2D hole systems based on
carbon-doped GaAs/AlGaAs QWs \cite{dai:2009}. HIRO were reported to
retain their dc character under ac excitation with frequencies up to $100\,{\rm
kHz}$ \cite{bykov:2005c}.
Very recently HIRO were also observed in a Corbino ring-shaped 2DEG \cite{bykov:2012}.

At $2\pi\edc\gtrsim 1$, the oscillatory part $\Delta r$ of HIRO in
Fig.~\ref{fig.hiro} is well described \cite{zhang:2007a} by
\be
\Delta r/\rho\simeq a \delta^2  \cos (2\pi\edc)~.
\label{eq.hiro}
\ee
In agreement with other experiments \cite{hatke:2009c,hatke:2010a,hatke:2011a},
the HIRO maxima and minima in \rfig{fig.hiro} occur at integer and half-integer
values of $\edc$, respectively. The amplitude $a$, obtained by
\citet{vavilov:2007} for $\edc\gg 1$ in the form
\be
a=16\ttr/\pi\tpi~,
\label{eq.hiro.a}
\ee
is proportional to the backscattering rate $1/\tpi$, see \req{VAG>>} below. HIRO
can thus be understood in terms of resonant inter-LL transitions resulting from
backscattering off impurities, i.e., HIRO require a sufficient amount of
short-range scatterers for their observation. Dingle-plot analysis using
\reqs{eq.hiro}{eq.hiro.a} allows one to extract both the quantum lifetime $\tq$
and the backscattering time $\tpi\simeq\tsh$. For a 2DEG with $\mu\sim
10^{7}\,{\rm cm^{2}/V\,s}$, such analysis at low $T$ produced $\tq\simeq
20\,{\rm ps}$ and $\tpi\simeq 5\ttr$ \cite{hatke:2009c}.

\citet{hatke:2009c} demonstrated that, similar to MIRO, the decay of HIRO at
high $T$ is primarily due to the LL broadening induced by electron-electron
interactions, which gives $\delta^2=e^{-2\pi/\wc\tq^*}$ in \req{eq.hiro} with
$1/\tq^*(T)-1/\tq\propto T^2$, see \rsecs{s3.1}{damping}. At low $T$, the
experimentally observed $T$ dependence of $1/\tq^*(T)$ deviates from the $T^2$
law, which can be related to the heating of 2DEG by the direct current. The heating
may also explain the decay of HIRO with $\edc$ at constant $B$, see
\rfig{fig.hiro}b. According to \req{eq.hiro}, the decay likely results from a
suppression of $\delta$ as $I$ increases.

\textcite{hatke:2011a} found that the HIRO amplitude is strongly suppressed
[similar to MIRO \cite{yang:2006} (see \rsec{s3.3.2prim})] by the
in-plane magnetic field $B_\parallel \sim 1\,{\rm T}$, while the HIRO period
depends only on the perpendicular component. The experiment demonstrates that
the suppression should be attributed to a $B_\parallel$-induced enhancement of
the quantum scattering rate $1/\tq^*-1/\tq^*|_{B_\parallel=0}\propto
B_\parallel^2$. In the regime of strongly developed SdH oscillations,
$B_\parallel$ starts to manifest itself also via the Zeeman splitting, see
Sec.~\ref{s7.2.4}.

\subsection{Theory of nonlinear dc transport}
\label{s5.2}

Similar to the mechanisms of MIRO (Sec.~\ref{s3.2}), two most important
contributions to the nonlinear electric current in the absence of microwaves
come from the direct effect of the dc field on impurity scattering (displacement
mechanism) and from nonequilibrium changes of the isotropic part of the
distribution function (inelastic mechanism). For smooth disorder,
these effects were systematically studied by \citet{vavilov:2004}
(displacement mechanism) and by \citet{dmitriev:2005} (inelastic). A comprehensive
analysis for a generic model of disorder [Eq.~(\ref{tau_n})] for high $T\gg\wc$
and overlapping LLs was performed in \cite{vavilov:2007}.

Next we provide an overview of the main results of these works using the formulation of
% RR
Sec.~\ref{s3.2.1} (which, in turn, follows from the general quantum kinetic
% RR
approach outlined in Sec.~\ref{s3.2.3}). In the absence of the microwave field,
Eqs.~\eqref{jx}-\eqref{f} reduce to
\be\label{jxDC}
j_d=2\nu_0 e\int\!d\ve\left\langle\Delta
X_{\varphi\varphi^\prime}\tau^{-1}_{\varphi-\varphi^\prime}
\tilde{\nu}_{\ve}\tilde{\nu}_{\ve^\prime}(f_\ve-f_{\ve^\prime})\right\rangle_{
\varphi\varphi^\prime}~,
\ee
where $\ve^\prime=\ve+eE\Delta X_{\varphi\varphi^\prime}$ and the angle brackets
denote averaging over $\varphi$ and $\varphi^\prime$. The isotropic part $f$ of
the distribution function obeys
\be\label{fDC}
\left(f_{\ve}-f^{T}_{\ve}\right)/\tin=\left\langle\tau^{-1}_{
\varphi-\varphi^\prime}
\tilde{\nu}_{\ve^\prime}(f_\ve-f_{\ve^\prime})\right\rangle_{
\varphi\varphi^\prime}~.
\ee

We restrict further analysis to $2\pi^2 T/\wc\gg 1$ (thermally suppressed SdH
oscillations) and $\wc\tq\ll 1$ (overlapping LLs). In this limit,
Eqs.~\eqref{jxDC} and \eqref{fDC} yield the main result of \cite{vavilov:2007}
\be\label{VAGgen}
\frac{j_d}{\sigma_{\rm D} E}=1-2\delta^2
\ttr\gamma^{\prime\prime}(\zeta)+2\delta^2\ttr\frac{\gamma^\prime(\zeta)}{\zeta}
{\cal F}(\tin,\zeta)~,
\ee
where $\gamma(\zeta)$, for $\tau^{-1}_{\varphi-\varphi^\prime}$ parametrized by
Eq.~(\ref{tau_n}), reads
\be
\gamma(\zeta)=\sum\nolimits_n\tau_n^{-1} J_n^2(\zeta),\quad
\zeta\equiv\pi\edc=\frac{2\pi eER_c}{\wc}~,
\label{zeta}
\ee
with $J_n(\zeta)$ being the Bessel functions. The second (displacement) term in
Eq.~(\ref{VAGgen}) comes from Eq.~\eqref{jxDC} when substituting  the
equilibrium distribution $f^{T}_{\ve}$ for $f_{\ve}$. The last (inelastic) term
is proportional to the amplitude
\be\label{F}
{\cal F}(\tin,\zeta)=\frac{-2\zeta\gamma^\prime(\zeta)}{\tau_{\rm
in}^{-1}+\tau_0^{-1}-\gamma(\zeta)}
\ee
of the oscillations of $f_\ve$ in \req{fDC},
\be\label{fF}
f_{\ve}=f^{T}_{\ve}+\delta \sin\left(\frac{2\pi\ve}{\wc}\right){\cal
F}(\tin,\zeta)\frac{\wc}{2\pi}\partial_\ve f^{T}_{\ve}~.
\ee
In the large-$\zeta$ limit, Eq.~(\ref{VAGgen}) is significantly simplified:
\be\label{VAG>>}
j_d\to\sigma_{\rm D} E+2\delta^2 \sigma_{\rm D} E\frac{4\ttr}{\pi\tau_\pi \zeta}
\sin 2\zeta,
\ee
where the backscattering rate $\tau^{-1}_\pi=\sum_n \tau^{-1}_n \exp(in\pi)$
(the actual condition on $\zeta$ depends on the type of disorder, see below).
The nonlinear current is then dominated by the displacement term and shows
oscillations proportional to $\sin 2\zeta$, which yields the $\cos 2\zeta$
dependence of the differential resistivity observed in the experiment, cf.\
\req{eq.hiro}.

To explain the HIRO phase, let us look at Eq.~\eqref{jxDC} for $T\gg\wc\zeta\sim
eER_c$ and neglect the oscillations of $f_\ve$ by assuming
$f_{\ve}=f^{T}_{\ve}$. Then, Eq.~(\ref{jxDC}) is expressible as an energy
averaging over the period $\wc$:
\be\label{jxDC1}
\frac{j_d}{2\nu_0 e^2 E}=\Big\langle\frac{\left(\Delta
X_{\varphi\varphi^\prime}\right)^2}{\tau_{\varphi-\varphi^\prime}}
\left\langle\tilde{\nu}_{\ve}\tilde{\nu}_{\ve+eE\Delta
X_{\varphi\varphi^\prime}}\right\rangle_\ve\Big\rangle_{\varphi\varphi^\prime}
~.
\ee
In overlapping LLs,
$\langle\tilde{\nu}_{\ve}\tilde{\nu}_{\ve+\w}
\rangle_\ve=1+2\delta^2\cos(2\pi\w/\wc)$. The angular integrations in
Eq.~\eqref{jxDC1} within the stationary-phase approximation in the limit $E\to
\infty$ yield \req{VAG>>}. The main contribution to HIRO comes from the
backscattering processes with $\varphi\simeq\pm\pi/2$ and
$|\varphi-\varphi^\prime|\simeq \pi$. These correspond to the maximum possible
shift $\Delta X_{\varphi\varphi^\prime}\simeq 2R_c$ along the electric field.
The average $\langle\tilde{\nu}_{\ve}\tilde{\nu}_{\ve+2eER_c}\rangle_\ve$ is
maximized at integer $\edc= 2eER_c/\wc$, which results in HIRO maxima in the
differential resistivity at integer $\edc$, cf.\ \req{eq.hiro}.

The amplitude of the resistivity oscillations in Eq.~\eqref{VAG>>} does not
depend on $E$ and $T$, while experiments show considerable suppression for
strong current and/or high $T$. As discussed in \rsec{s5.1}, this suppression can be
explained as a result of additional broadening of LLs by electron-electron
interactions, combined with the effect of heating of the 2DEG by the dc field
which is controlled by electron-phonon interactions.

To analyze the obtained results, \textcite{vavilov:2007} introduced the
mixed-disorder model \eqref{mixed-disorder} (used in Sec.~\ref{s3.2} to describe
MIRO). The smooth (characterized by $\tsm^{-1}$) and sharp ($\tsh^{-1}$)
components of disorder yield two separate contributions to
\be\label{g_mixed}
\gamma(\zeta)=\frac{J_0^2(\zeta)}{\tsh}+\frac{1}{\tsm\sqrt{1+\chi\zeta^2}}~.
\ee
The contribution of the sharp component changes on a scale of $\zeta\sim 1$ and
shows strong oscillations for $\zeta\agt 1$. The second term describes the
nonlinear effects, studied by \citet{vavilov:2004}, that result from small-angle
scattering by the long-range component ($\chi^{1/2}\ll 1$ gives a typical value
of the scattering angle, see Sec.~\ref{s3.2}). These become relevant in a much
stronger dc field,  $\zeta\sim\chi^{-1/2}\gg 1$, and do not contribute
to HIRO.

At $\zeta\ll 1$, Eqs.~\eqref{VAGgen} and \eqref{F} reduce to
\bea\label{VAG<}
&&\frac{j_d}{\sigma_{\rm D} E}=1+2\delta^2-{3\ttr\over
4\tau_\star}\delta^2\zeta^2-2\delta^2{\cal F}(\tin,\zeta)~,\\
\label{F<}
&&{\cal F}(\tin,\zeta)=\frac{2\tin\zeta^2/\ttr}{1+\tin\zeta^2/2\ttr}~,
\eea
where $1/\tau_\star\simeq 3/\tsh +12\chi^2/\tsm$ and $1/\ttr\simeq
1/\tsh+\chi/{\tsm}$.
Note that the nonlinear terms $\propto \zeta^2$ in Eqs.~\eqref{VAG<} and \eqref{F<}
can be equivalently obtained by taking the limit
of $\omega\to 0$ in the MIRO terms in Eqs.~(\ref{sigmadis}), (\ref{sigmain}),
and ({\ref{RROvLLs}). At order $E^3$ in $j_d$, the ratio of the inelastic and
displacement contributions in the case of smooth disorder is of the order of
$\tin\tau_\star/\ttr^2\sim\tin/\tq\gg 1$ [similar to MIRO,
Eq.~\eqref{BtoA}]. That is, the inelastic term, obtained for smooth disorder by
\citet{dmitriev:2005}, dominates and Eqs.~\eqref{VAG<} and \eqref{F<} are valid
for all $\zeta\ll\chi^{-1/2}$. The suppression of $j_d/E$ in a relatively weak
dc field at $\zeta\sim (\ttr/\tin)^{1/2}$ is the strongest effect in the
nonlinear dc response for the case of smooth disorder.

It is important to note the dual role that the electric field plays in
Eqs.~\eqref{F} and \eqref{F<}. On the one hand, it creates the oscillations
\eqref{fF} in the energy distribution of electrons; on the other, it also opens
an additional channel of inelastic relaxation, thus controlling the magnitude of
the oscillations. Indeed, in the presence of the dc field, an electron can
change its kinetic energy via elastic collisions with impurities. At $\zeta\gg
(\ttr/\tin)^{1/2}$, the resulting ``spectral diffusion" becomes more efficient
than the inelastic relaxation due to electron-electron collisions. This leads to
the saturation of the oscillations in Eqs.~\eqref{F<} and \eqref{fF} and makes
the nonlinear response independent of $T$ for $\zeta\gg (\ttr/\tin)^{1/2}$. Because
of the spectral diffusion, ${\cal F}(\tin,\zeta)$ remains smaller than or of the
order of unity for arbitrary $\zeta$. Note that the oscillatory term in
Eq.~\eqref{fF} contains additionally two small factors $\delta$ and
$\wc\partial_\ve f^{T}_\ve\sim-\wc/T$ and, therefore, remains small at any
$\zeta$, which makes the expansion \eqref{fF} legitimate. Unlike
Eq.~\eqref{VAG<}, which contains the ``pure" displacement and inelastic
contributions, the last term in Eq.~\eqref{VAGgen} is, strictly speaking, a
result of the interplay of the inelastic and displacement mechanisms in the
strongly nonlinear response.

Let us now consider the case $\chi\tsh\ll\tsm\ll\tsh$ and $\zeta\ll\chi^{-1/2}$,
when the short-range component of disorder determines the transport scattering
rate, i.e., $\tsh\simeq\ttr$, while the nonlinear corrections generated by the
smooth component are negligible. Equations \eqref{VAGgen} and \eqref{F} then
reduce to
\be\label{VAG><}
\frac{j_d}{\sigma_{\rm D} E}\simeq
1-2\delta^2[J_0^2(\zeta)]^{\prime\prime}-\frac{16\,\delta^2
J_0^2(\zeta)J_1^2(\zeta)}{\ttr/\tin+1-J_0^2(\zeta)}~.
\ee
In this limit, $\tsm\simeq\tq$ enters $j_d$ only via the Dingle factor
$\delta=\exp(-\pi/\wc\tq)$. The above expression captures both the high-field
limit \eqref{VAG>>} for HIRO and its essential modifications at intermediate
($\zeta\sim 1$) and small ($\zeta\sim\sqrt{\ttr/\tin}$) electric fields. Unless
$\tin\ll\ttr$ (which may occur at elevated $T$), the inelastic and displacement
terms in Eq.~(\ref{VAG><}) are equally important at $\zeta\sim 1$. Provided
$\tsh\sim\ttr$, the smooth component only slightly modifies Eq.~\eqref{VAG><} in
the relevant range of $\zeta\ll\chi^{-1/2}$, see Eq.~(3.8) and Figs.~2,3 in
\cite{vavilov:2007}.

\subsection{Nonlinear resistivity: Inelastic effects}
\label{s5.3}
In accordance with the theoretical predictions, the most pronounced inelastic
effects in the nonlinear resistivity were observed in 2DEGs with high density
$n_e\sim 10^{12}\,{\rm cm^{-2}}$ and moderate mobility $\mu\sim 10^6\,{\rm
cm^2/V\,s}$
\cite{bykov:2005c,zhang:2007b,bykov:2007,kalmanovitz:2008a,zhang:2009,
vitkalov:2009,mamani:2009b}. Indeed, according to Eqs.~\eqref{VAG<} and
\eqref{F<}, the differential resistivity at order $\zeta^2$ reads
\be\label{zeta2}
r=\rho_{\rm D}(1+2\delta^2) -12\delta^2\rho_{\rm
D}\left(\frac{3\ttr}{16\tau_\star}+\frac{\tin}{\ttr}\right)\zeta^2~.
\ee
The maximum value of the factor $3\ttr/16\tau_\star$
% RR
in the displacement
term is 9/16 (which corresponds to $\ttr=\tsh$, see \rfig{tau_star}a),
while the factor $\tin/\ttr$ in the inelastic contribution can be estimated as
$\ve_F/T^2\ttr\simeq 130(1\,{\rm K}/T)^2$ (for $\mu=10^6\,{\rm cm^2/V\,s}$ and
$n_e=10^{12}\,{\rm cm^{-2}}$). This gives $16\tau_\star\tin/3\ttr^2\gtrsim ( 15\, {\rm K}/T)^2$
for the relative magnitude meaning the inelastic contribution
dominates up to high $T\gtrsim 15\,{\rm K}$.
This justifies the analysis of the low-field nonlinear resistivity
performed by \citet{zhang:2007b,zhang:2009,vitkalov:2009,mamani:2009b} solely in
terms of the inelastic mechanism [and also justifies retaining the
higher-order terms in \req{F<}]. All these experiments reproduced the theoretical
predictions in the range of applicability of the theory\footnote{The analytical
results by \cite{dmitriev:2005,mamani:2009b} are applicable for $2\pi^2T/\wc\gg
1$ and overlapping LLs. In the regime of SdH oscillations ($2\pi^2T/\wc\alt1$),
which was also studied experimentally by
\citet{zhang:2007b,bykov:2007,kalmanovitz:2008a}, such a description is not
parametrically justified and additional theoretical analysis is required, see \cite{dmitriev:2011a}.} and
reported values of $\tin\propto T^{-2}$ close to those calculated by
\citet{dmitriev:2005}. In particular, \citet{mamani:2009b} generalized the theory
\cite{dmitriev:2005} to the two-subband case and found that the MISO peaks in
the nonlinear response are inverted at $\zeta=(2\ttr/3\tin)^{1/2}$, which was
used to accurately determine $\tin$. The effect is similar to the interplay of
MISO and MIRO discussed in \rsec{s3.3.3}. In the nonlinear dc response,
interaction of MISO and MIRO was observed by \citet{wiedmann:2011c} and
explained therein in terms of the inelastic mechanism. The data
\cite{wiedmann:2011c} indicate also a two-subband counterpart of the nonlinear
mixing of HIRO and MIRO described in \rsec{sec.miro.dc.exp}.

In the ultra-high mobility 2DEG with $n_e\simeq 3.95 \times 10^{11}\,{\rm
cm^{-2}}$ and $\mu\simeq 8.9 \times 10^{6}\,{\rm cm^2/V\,s}$ used by
\cite{hatke:2012b}, the estimated ratio $\tin/\tau\simeq\ve_F/T^2\ttr\simeq
(2\,{\rm K}/T)^2$ is much smaller than in the samples discussed above and, as a
result, both contributions in \req{zeta2} are relevant. In this case, the Dingle
analysis of the low-field differential resistivity at fixed $T=1.5\,{\rm K}$
yielded a value of 2.25 for the expression in the
brackets in the nonlinear term of \req{zeta2}. Using \req{t*},
for $\ttr=20\tq$ and $\tsh\simeq 5\ttr$ \cite{hatke:2009c} one obtains
$3\ttr/16\tau_\star\simeq 0.23$ which divides almost equally between the sharp
and smooth components of disorder, see \rfig{tau_star}a. It follows that the
inelastic term $\tin/\tau\simeq 2$ at $T=1.5\,{\rm K}$, which agrees well with the
estimate above. This example shows that measurements of the nonlinear
magnetoresistivity in high LLs in both limits $\zeta\gg 1$ (HIRO) and $\zeta\ll
1$ [\req{zeta2}] provide a method to determine various scattering rates---in
particular, $T$-dependent $\tq$ and $\tin$, as well as $\tst$ and
$\tau_\pi$---which bring valuable information about disorder and interactions in
2D electron systems.

\subsection{Zero-differential resistance states (ZdRS)}
\label{s5.4}

Experiments by \citet{bykov:2007,zhang:2008} revealed that the differential
resistance can drop all the way to zero (Fig.~\ref{fig.zdrs}), leading to the
formation of the zero-differential resistance states (ZdRS).
\citet{bykov:2007,kalmanovitz:2008a} (in samples with $n_e\simeq 8\times
10^{11}\,{\rm cm^{-2}}$ and $\mu\simeq 8\times 10^5\,{\rm cm^2/V\,s}$) observed
ZdRS emerging from the maxima of SdH oscillations. The transition to ZdRS was
accompanied by a reproducible negative spike (\rfig{fig.zdrs}a) and by temporal
fluctuations at higher dc bias.

%%%%%%%%%%%%%%%%%%%%%%%%%%%%%%%%%%%%%%%%%%%%%%%%%
\begin{figure}[ht]
\includegraphics[width=\columnwidth]{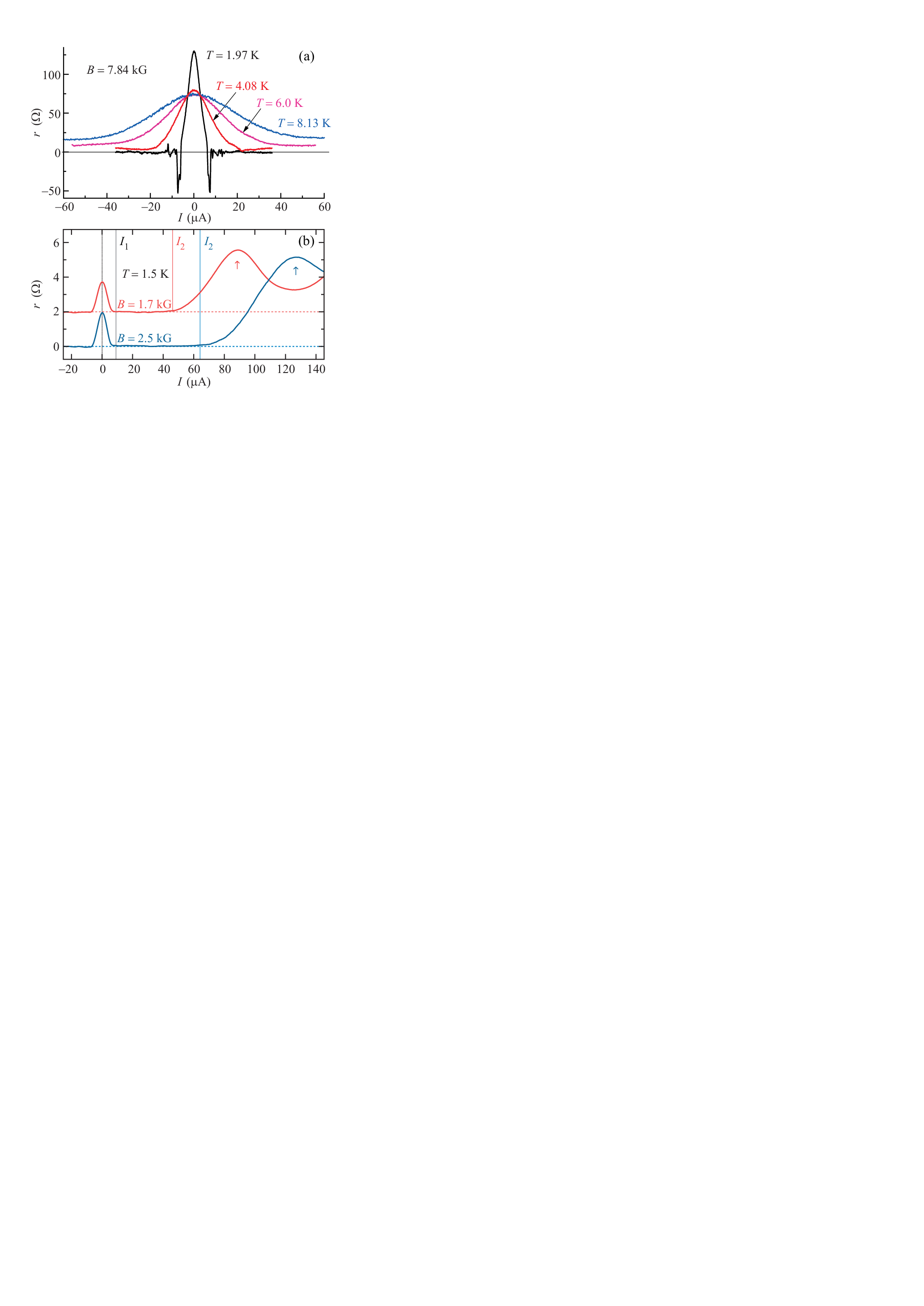}
\caption{
(a) Differential resistivity $r$ vs current $I$ for several values of $T$ at
fixed $B=7.64\,{\rm kG}$, measured in a $50\,\mu{\rm m}$ wide Hall bar sample.
Adapted from \citet{bykov:2007}. (b) Similar dependence at $B=1.7$ and
$B=2.5\,{\rm kG}$, measured at $T\simeq 1.5\,{\rm K}$ in a $100\,\mu{\rm m}$
wide Hall bar sample. The first HIRO maximum is marked by $\uparrow$. The curve
for $B=1.7\,{\rm kG}$ is shifted by $2\,\Omega$ upward.
Adapted from \citet{hatke:2010a}.}
\label{fig.zdrs}
\end{figure}
%%%%%%%%%%%%%%%%%%%%%%%%%%%%%%%%%%%%%%%%%%%%%%%%%

By contrast, \textcite{zhang:2008} (in a sample with $n_e\simeq 4.8\times
10^{11}\,{\rm cm^{-2}}$ and $\mu\simeq 4.4\times 10^6\,{\rm cm^2/V\,s}$) and
\citet{hatke:2010a} (in samples with $n_e\simeq 3.8\times10^{11}\,{\rm cm^{-2}}$
and $\mu\simeq 1.0\times 10^7\,{\rm cm^2/V\,s}$) demonstrated the possibility of
ZdRS evolving from the principal minimum of HIRO in the regime of suppressed SdH
oscillations. Neither overshoot to negative values nor temporal fluctuations
were detected (\rfig{fig.zdrs}b). \citet{hatke:2010a} reported ZdRS over a
continuous range of currents and in magnetic fields extending well below the
onset of SdH oscillations. Similar to ZRS, ZdRS were found to disappear with
increasing temperature ($T\gtrsim 2\,{\rm K}$) and with increasing overlap
between LLs ($B\lesssim 1\,{\rm kG}$). The minimum current $I_1$
(Fig.~\ref{fig.zdrs}b) required to support ZdRS was found to be roughly
$B$-independent. The maximum current $I_2$ was found to increase roughly
linearly with $B$, tracing the fundamental HIRO peak (cf.\ $\uparrow$ in
\rfig{fig.zdrs}b). ZdRS were also observed to develop from the maxima of MIRO
\cite{zhang:2007c} and MISO \cite{bykov:2010c,gusev:2011,wiedmann:2011c}.

On the theoretical side, ZdRS originate from the parts of (local) $I$-$V$
characteristics where the differential resistivity is negative, which violates
the stability condition (\ref{e4.2.8}). Similar to the case of ZRS (\rsec{s4}),
this results in an instability leading to the formation of current and field
domains. Unlike ZRS, however, the absolute resistivity remains positive, which
leads to an essential difference between the Corbino (N-shaped $I$-$V$
characteristics) and Hall bar (S-shaped $I$-$V$ characteristics) geometries. While
in the Hall bar geometry a stationary domain configuration with the current
flowing along the domain walls is expected \cite{bykov:2007}, in the Corbino
geometry nonstationary (moving) domains are predicted \cite{vavilov:2004}
similar to the Gunn effect \cite{gunn:1963}.

\subsection{PIRO: Experimental discovery and basic properties}
\label{s5.5}

Resonant interaction of 2D electrons with longitudinal optical (LO) phonons was
predicted \citep{gurevich:1961} and confirmed in magnetotransport measurements
in GaAs/AlGaAs heterostructures \citep{tsui:1980} long time ago. This
interaction manifests itself as an enhancement of the longitudinal resistivity
whenever the LO-phonon frequency $\olo \simeq n \wc,~n=1,2,3,\ldots$. In GaAs,
$\olo \sim 10^{13}\,{\rm s}^{-1}$ and observation of the LO-phonon-induced
oscillations requires high $T\gtrsim 10^2$ K and strong $B \gtrsim 10^2\,{\rm
kG}$ \cite{tsui:1980}.

%%%%%%%%%%%%%%%%%%%%%%%%%%%%%%%%%%%%%%%%%%%%%%%%%
\begin{figure}[ht]
\includegraphics[width=\columnwidth]{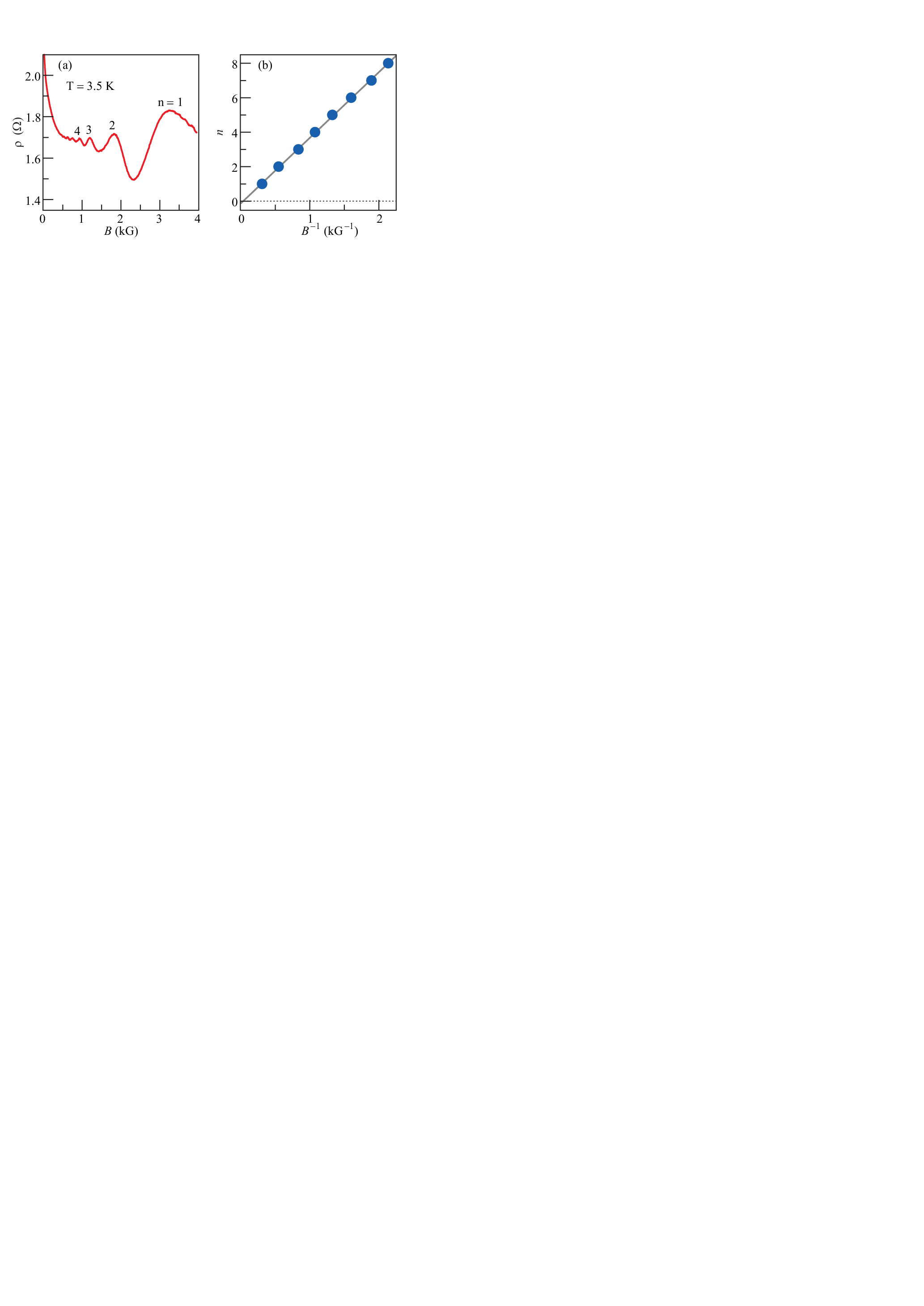}
\caption{(a) Magnetoresistivity $\rho$ as a function of $B$, showing eight PIRO
maxima, four of which are marked by integers. (b) PIRO order $n$ versus
$B^{-1}$. The linear fit yields $s\simeq 3.4\,{\rm km/s}$. The data were
obtained at $T = 3.5\,{\rm K}$ in a $100\,\mu{\rm m}$ wide Hall bar sample with
$n_e\simeq 3.8\times 10^{11}\,{\rm cm^{-2}}$ and $\mu\simeq 1.2 \times
10^7\,{\rm cm^2/V\,s}$. Adapted from \citet{hatke:2009b}.}
\label{fig.piro}
\end{figure}
%%%%%%%%%%%%%%%%%%%%%%%%%%%%%%%%%%%%%%%%%%%%%%%%%

A decade ago, another class of phonon-induced oscillations, termed the
phonon-induced resistance oscillations (PIRO), was discovered in the
linear-response resistivity of a 2DEG with mobility $\mu\sim 10^6\,{\rm
cm^2/V\,s}$ \citep{zudov:2001b}. PIRO emerged at much lower $T \sim 1-10\,{\rm
K}$ and much lower $B \sim 1-10\,{\rm kG}$ and are understood in terms of
resonant interaction of 2D electrons in high LLs with acoustic phonons which
carry momentum $2k_F$ and have a characteristic frequency $\os = 2 k_F s$. Here
$s$ is the speed of sound and the out-of-plane component of the phonon momentum
is neglected. Such interaction causes a correction to the resistivity $\Delta
\rho^{\rm ph}$ which oscillates with the ratio
\be
\eph = {\os}/{\wc} \propto n_e^{1/2}~.
\label{eph}
\ee
The $n_e^{1/2}$ dependence was confirmed experimentally \cite{zudov:2001b}.

Electron backscattering by an acoustic phonon is most effective when $\os=n
\wc$, $n=1,2\ldots$, which maximizes the (thermally averaged) product of the
initial and final densities of states. In the simplest model of 2D isotropic
phonons, the oscillatory part of the resistivity for $\eph \gtrsim 1$ and
$T\gg\wc,\os$ reads \cite{dmitriev:2010}
\be
\frac {\Delta \rho^{\rm ph}}{\rho_{\rm D}} \simeq \frac {2g^2 T \ttr}
{\pi\sqrt{\eph}}  \delta^2 \cos (2\pi\eph - \pi/4)
\label{eq.piro}
\ee
[see also \req{jphlinear}]. The phase $-\pi/4$ was confirmed experimentally by
\citet{hatke:2011up} in a variety of high mobility samples. Many other
experiments
\cite{zudov:2001b,yang:2002b,zhang:2008,hatke:2009b,zudov:2009,bykov:2005b,
bykov:2009a} reported PIRO maxima at integer $\eph$. As discussed in
\rsec{s5.7}, the phase of PIRO (unlike that for MIRO and HIRO) is not expected
to be universal and is sensitive to the anisotropy of relevant phonon modes,
crystallographic orientation of the sample, width of the QW, etc.
\cite{raichev:2009,dmitriev:2010}.

In contrast to other low-$B$ magnetoresistance oscillations (SdH oscillations,
MIRO, MISO, and HIRO), which are observed at low $T$, PIRO are best
resolved at $T \sim \os$ and get strongly suppressed at both low and high $T$
\cite{zudov:2001b,yang:2002b,hatke:2009b,bykov:2009a}. At low $T$, both emission
and absorption of $2k_F$-phonons is exponentially suppressed by the thermal
factors for relevant electron and phonon states, see \rsec{s5.7}.
On the other hand---similar to MIRO, MISO, and HIRO---PIRO are insensitive to
the temperature smearing of the Fermi surface which suppresses SdH oscillations
at $2\pi^2T/\wc\gg 1$. A detailed study of the $T$ dependence of PIRO revealed
that higher-$B$ (lower-$\eph$) oscillations are best developed at progressively
higher $T$: the temperature at which a given oscillation reaches its maximum
amplitude was found to scale with $\sqrt{B}$\cite{hatke:2009b,bykov:2009a}.
This suggests that, at high $T$,
the suppression of PIRO (similar to MIRO and HIRO, see \rsecs{s3.1}{damping} and
\ref{s5.1}, but unlike the SdH oscillations) is due to interaction-induced
broadening of LLs.

Similar to HIRO and MIRO, PIRO are most pronounced in high-quality 2DEGs. Early
experiments in moderate mobility samples ($\mu \sim 10^6\,{\rm cm^{2}/V\,s}$)
required temperatures in the range about $5-20\,{\rm K}$ and revealed only a few
oscillations \cite{zudov:2001b,bykov:2005b}. Recent experiments in high
mobility samples ($\mu \sim 10^7\,{\rm cm^2/V\,s}$) showed up to eight
oscillations (\rfig{fig.piro}) which remained visible down to $T \simeq 2\,{\rm
K}$ \cite{hatke:2009b}.

Related phonon-induced oscillations in the magnetothermopower were observed by
\textcite{zhang:2004}. Interplay of PIRO and MISO was investigated by
\textcite{bykov:2010a} in a single GaAs QW with two populated
subbands.

\subsection{PIRO in a strong Hall field}
\label{s5.6}
%%%%%%%%%%%%%%%%%%%%%%%%%%%%%%%%%%%%%%%%%%%%%%%%%
\begin{figure}[ht]
\includegraphics[width=\columnwidth]{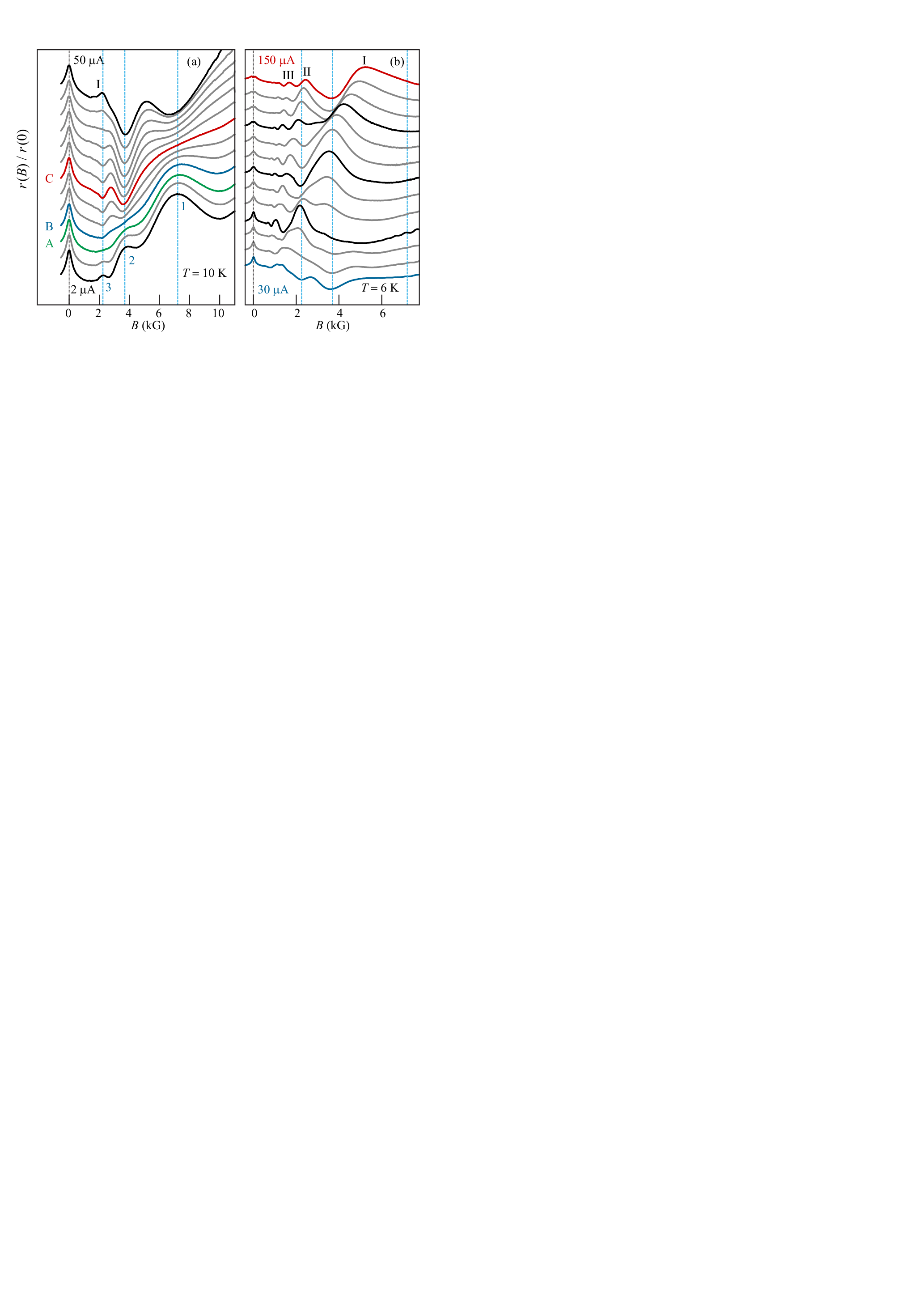}
\caption{Differential magnetoresistivity $r$ as a function of $B$, normalized to
the resistivity at $B=0$, for different $I$ (a) from 2 to $50\,\mu{\rm
A}$ in steps of $4\,\mu{\rm A}$ at $T = 10\,{\rm K}$ and (b) from 30 to $150\,\mu{\rm A}$
in steps of $10\,\mu{\rm A}$ at $T = 6\,{\rm K}$. The
traces are offset for clarity. Integers  I, II, and III at the top traces
mark the order of the HIRO peaks. The vertical lines correspond to integer
$\eph=1,2$ and 3. The curves labeled by (A), (B), and (C) in (a) are discussed in
the caption to Fig.~\ref{phiro}. The data were obtained in a $50\,\mu{\rm m}$
wide Hall bar sample with $n_e\simeq 4.8\times 10^{11}\,{\rm cm^{-2}}$ and
$\mu\simeq 4.4 \times 10^6\,{\rm cm^2/V\,s}$. Adapted from \citet{zhang:2008}.}
\label{fig.piro.dc}
\end{figure}
%%%%%%%%%%%%%%%%%%%%%%%%%%%%%%%%%%%%%%%%%%%%%%%%%

Recent studies \cite{zhang:2008} of PIRO in a strong Hall field established that
the dc field leads to the evolution of the PIRO maxima into minima and back, see
\rfig{fig.piro.dc}. It was also found that the strong dc field enables detection
of PIRO at low $T$ where the linear-response PIRO are exponentially suppressed.
One additional experimental finding was the observation of a pronounced resistance
maximum at $\edc = \eph$, where the Hall velocity equals the speed of sound. All
these results are reproduced within the theoretical model of
\textcite{dmitriev:2010}, see \rsec{s5.7}.

\subsection{Microscopic theory of PIRO}
\label{s5.7}
In the case of MIRO (Sec.~\ref{s3.2}) or nonlinear dc transport
(Sec.~\ref{s5.2}), the inelastic effects that lead to oscillations in the energy
distribution of electrons play an essential and often dominant role. By
contrast, PIRO are observed at elevated $T$, where a fast inelastic relaxation
makes effects of this type much less important. Therefore, one can consider only
effects that are similar to the displacement contribution to HIRO and MIRO. In
terms of migration of the guiding centers of cyclotron orbits
[Eqs.~\eqref{jx}-\eqref{M}], the phonon-assisted dissipative current $j_d^{(p)}$
has the form of Eq.~\eqref{jx} with
\bea \nonumber
&&W_{x_1\to x_2}=\left\langle\ \!\int\!d\ve_1\!\int\!d\ve_2\,{\cal M}_{\ve_1
\ve_2}\delta\left(x_1-x_2+\Delta X_{\varphi_1\varphi_2}\right)\right.
\\
&&\left.\times\sum\nolimits_\pm\Gamma_{\varphi_1\varphi_2}^{(\rm sp)}\left({\cal
N}_{\w_{12}}
+\delta_{1,\pm 1}\right)\delta\left(\Delta\ve^{\rm
tot}_{12}\mp\w_{12}\right)\right\rangle_{\varphi_1\varphi_2}~,\nonumber
\\ \label{Wph}
\eea
where $\Gamma_{\varphi_1\varphi_2}^{(\rm sp)}$ denotes the probability of
spontaneous emission of a phonon with frequency $\w_{12}$ and ${\cal N}_\omega$
is the Planck distribution function. In the factor ${\cal M}_{\ve \ve^\prime}$
[Eq.~(\ref{M})], one can substitute, in accordance with the above, $f^T_\ve$ for
$f_\ve$. Similar to Eq.~\eqref{W}, the delta function containing $\Delta\ve^{\rm
tot}_{12}$ ensures energy conservation.

We start with a simplified single-mode model which assumes interaction with 2D
isotropic acoustic phonons via a deformation potential. We also assume
that the out-of-plane component of the phonon momentum is negligible, i.e., the
width $b$ of the QW to which the 2DEG is confined is large, $b\gg
k_F^{-1}$. In this case,
\be\label{g}
\Gamma_{\varphi_1\varphi_2}^{(\rm sp)}=g^2 \w_{12}/2, \qquad g^2=m {\cal D}^2/\rho b
s^2~,
\ee
where phonon frequency for quasielastic scattering is given by $\w_{12}=2 k_F
s\sin|(\varphi_1-\varphi_2)/2|$, $\rho $ is the mass density, and ${\cal D}$ is the
deformation-potential constant. In overlapping LLs and at high $T\gg\wc,\w_\pi$,
Eqs.~\eqref{jx}, \eqref{Wph}, and \eqref{g} yield \cite{dmitriev:2010}
\be\label{jphlinear}
j_d^{(p)}=\sigma_{\rm D} E \ttr g^2
T(1+2\delta^2[J_0(2\pi\eph)-J_2(2\pi\eph)])~.
\ee
The linear-in-$T$ dependence of $j_d^{(p)}$ is due to the fact that at
$T\gg\w_\pi$ the occupation number for relevant phonon modes is large, so that
$\Gamma_{\varphi_1\varphi_2}^{(\rm sp)}{\cal N}_{\w_{12}}\simeq g^2 T/2$, while
the contribution of spontaneous emission is negligible. The current shows
oscillations with $\eph=\w_\pi/\wc$, controlled by commensurability between the
cyclotron energy and the maximum possible---in the process of
scattering---phonon energy. That is, in accord with the original interpretation
\citep{zudov:2001b}, PIRO originate from resonant inter-LL transitions caused by
the backscattering of electrons by acoustic phonons. The position of
higher-order maxima of PIRO in Eq.~\eqref{jphlinear} at $\eph\gg 1$ is given by
\req{eq.piro}. In narrow QWs ($k_F b\sim 1$) or in wide QWs
for $\eph\gg(k_F b)^2$, one should take into account the 3D character of acoustic phonons.
Still in the isotropic approximation, the $\cos(2\pi\eph-\pi/4)$ behavior of PIRO
[\req{eq.piro}] changes then to $\cos(2\pi\eph)$, while the amplitude
of PIRO reduces by a factor of $\pi\eph^{1/2}/2b k_F$ \cite{raichev:2009}.

While the isotropic single-branch model captures the essential physics of PIRO,
in real structures the electron-phonon interaction is more
complicated. In bulk GaAs, there are three anisotropic phonon branches and two
mechanisms (via the deformation and piezoelectric potentials) of electron-phonon
interaction. A comprehensive study of PIRO, using the general form of
interaction with bulk acoustic phonons, was performed by \citet{raichev:2009}.
For higher harmonics of PIRO, the effects of anisotropy were treated there
analytically, producing three distinct contributions with different phases and
periods \cite{raichev:2009}. Additional contributions can arise from interaction
with interface phonon modes \cite{zudov:2001b}.

The analysis by \citet{raichev:2009} shows that PIRO in the ultra-high mobility
and moderate electron density sample used by \citet{hatke:2009b} are dominated
by the bulk transverse acoustic mode propagating along the 2D plane and
polarized perpendicular to the plane. The sound velocity of this mode
$s_{TO}=3.40\,{\rm km/s}$ is indeed very close to $s=3.44\,{\rm km/s}$ extracted
from the period of PIRO by \cite{hatke:2009b}. A subleading contribution of the
longitudinal mode with a noticeably higher velocity produces an extra peak
observed by \citet{hatke:2009b} at sufficiently high $B$, and also explains
the beating pattern at lower $B$, caused by the interference of these two
contributions to PIRO [see Fig.~1 in \cite{raichev:2009}]. By contrast, in the
moderate mobility and high electron density sample studied by
\cite{bykov:2005b}, PIRO are dominated by the higher-energy longitudinal mode.
In another recent experiment \cite{zhang:2008}, which studied the effects of a
strong dc field on PIRO (discussed below), the longitudinal mode also prevailed,
which explains the good agreement between experiment \cite{zhang:2008} and
theory \cite{dmitriev:2010} based on the single-mode approximation
\eqref{g}.

The Hall field tilts LLs and changes the commensurability condition for
electron-phonon scattering. This results in oscillations with
$\epsilon_\pm\equiv\edc\pm\eph$, where $\edc$ is the parameter that controls
HIRO (Sec.~\ref{s5.2}). The change of the oscillations period by the dc
field can be viewed as a Doppler shift of the phonon modes in the frame moving
with the Hall velocity $v_H=cE/B$ across the field. In the moving frame, the
electric field is absent while the phonon dispersion becomes anisotropic, $s\to
s-v_H\cos[(\varphi_1+\varphi_2)/2]$ (this formulation is particularly useful in
the case of a complicated dispersion relation of phonons). For $1\ll\edc<\eph$
and $1\ll\eph-\edc$, Eqs.~\eqref{jx}, \eqref{Wph}, and \eqref{g} yield the following
for the oscillating part of $j_d^{\rm (p)}$:
\be\label{jph<}
\frac{j_d^{\rm (p, osc)}}{\sigma_{\rm D} E}= \frac{4\delta^2}{\pi^2} g^2 T\tau
\left(\frac{\sin 2\pi\epsilon_+}{\sqrt{\edc\epsilon_+}}
+\frac{\cos 2\pi\epsilon_-}{\sqrt{\edc|\epsilon_-|}}\right)~.
\ee
The evolution of the differential resistivity $r=\partial E/\partial j_d$ with
varying $\edc$ in this regime was studied experimentally by
\textcite{zhang:2008}. The results of calculation according to Eqs.~\eqref{Wph}
and \eqref{g}, illustrated in Fig.~\ref{phiro}, quantitatively reproduce the
experimental data in Fig.~\ref{fig.piro.dc} without fitting parameters. Note
that the theoretical plot does not include HIRO which become relevant at higher
dc bias, see \rfig{fig.piro.dc}b, but are negligible in the region of interest,
$B>2\,{\rm kG}$ and $I<40\,\mu{\rm A}$. In qualitative agreement with the
experimental results of \textcite{zhang:2008}, similar oscillations with
$\eph-\edc$ were also obtained numerically, using the balance-equation approach,
by \textcite{lei:2008}.

%%%%%%%%%%%%%%%%%%%%%%%%%%%%%%%%%%%%%%%%%%%
\begin{figure}
\includegraphics[width=\columnwidth]{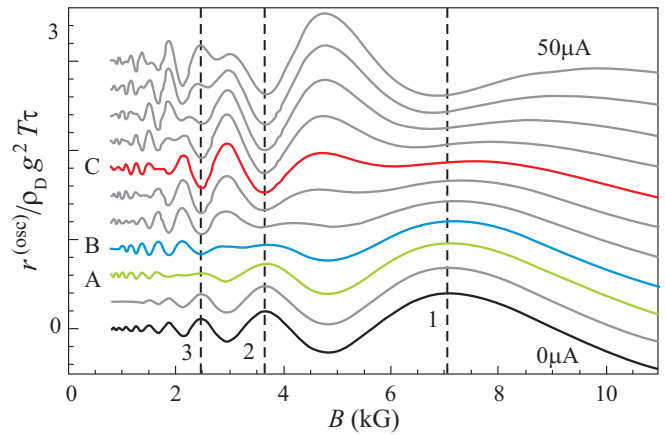}
\caption{Calculated oscillatory part $r^{(\rm osc)}$ of the differential
resistivity $\partial E/\partial j_d$ versus $B$, in units of $\rho_{\rm
D}g^2T\tau$, for several values of current $I$ from 0 to
$50\,\mu{\rm A}$ in steps of $5\,\mu{\rm A}$. The traces are offset for clarity.
The vertical lines correspond to integer $\eph$. The traces for (A) $10\,\mu{\rm
A}$, (B) $15\,\mu{\rm A}$, and (C) $30\,\mu{\rm A}$ correspond approximately to
the values of $I$ at which the third, second, and first PIRO peaks,
respectively, disappear before evolving into minima as $I$ increases. The curves
are very similar to those measured in the experiment: cf.\ the curves in
\rfig{fig.piro.dc} labeled as A ($10\,\mu{\rm A}$), B ($14\,\mu{\rm A}$), and C
($28\,\mu{\rm A}$). From \citet{dmitriev:2010}.}
\label{phiro}
\end{figure}
%%%%%%%%%%%%%%%%%%%%%%%%%%%%%%%%%%%%%%%%%%%%

In a stronger dc field, $\edc>\eph$, corresponding to a supersonic Hall velocity
$v_H>s$, the theory of \citet{dmitriev:2010} predicts that the phase of the
oscillations changes in the second term of Eq.~\eqref{jph<} from $\cos
2\pi\epsilon_-$ to $\sin 2\pi\epsilon_-$ for $\edc>\eph$. Note that at $v_H>s$,
the effective sound velocity $s-v_H\cos[(\varphi_1+\varphi_2)/2]$ in the moving
frame becomes negative for certain transitions. Emission of a phonon with energy
$\w$ in such transitions increases the electron kinetic energy by $\Delta>0$ at
the expense of the electrostatic energy which is decreased by $\Delta+\w$.

The physical meaning of the ``sound barrier" is clear from the low-temperature
behavior of the phonon-assisted transport. At $T\ll\w_\pi$, stimulated emission
and absorption of phonons are exponentially suppressed by the Planck factor
${\cal N}_{\w_\pi}\simeq\exp(-\w_\pi/T)$ in Eq.~\eqref{Wph}. On the other hand,
spontaneous emission in the scattering processes reducing the kinetic energy is
also strongly suppressed by the factor ${\cal M}_{\ve_1 \ve_2}$.
However, in the spontaneous emission processes that increase the kinetic energy,
the constraint imposed by the factor ${\cal M}_{\ve_1 \ve_2}$ is not effective.
As a result, at $\edc>\eph$ the phonon-assisted transport due to spontaneous
phonon emission survives even at $T=0$. Specifically, for $1\ll\epsilon_-\ll\edc$
the phonon-induced oscillations at $T\ll\w_\pi$ are described by
\be\label{jphT0}
\frac{j_d^{\rm (p, osc)}}{\sigma_{\rm D} E\ttr}=
\frac{2\delta^2\eph\epsilon_-^{1/2}}{\pi^2\edc^{3/2}} g^2 \wc
\sin 2\pi\epsilon_-.
\ee
The evolution of the oscillations with temperature is illustrated in
Fig.~\ref{phiroT} (where also the smooth part of the phonon-assisted current at
$T=0$ is shown).

Nonlinear mixing of PIRO and MIRO was investigated by \textcite{raichev:2010b}.
This study predicted oscillations with $\omega_\pi\pm\omega$ and discussed the
feasibility of their experimental observation. Further, the analysis by
\citet{raichev:2010b} showed that heating of the 2DEG leads to additional
$\omega_\pi/\wc$-oscillations with a phase shift with respect to the equilibrium
PIRO. Separately, nonlinear mixing of PIRO and MISO in two-subband systems
was shown by \citet{bykov:2010a,raichev:2010a} to lead to oscillations at
frequencies determined by $\w_\pi$ and the subband splitting energy. A further
example of frequency mixing for the case of MIRO and HIRO is discussed in
\rsec{VI.B.2}.

%%%%%%%%%%%%%%%%%%%%%%%%%%%%%%%%%%%%%%%%%%%
\begin{figure}
\includegraphics[width=\columnwidth]{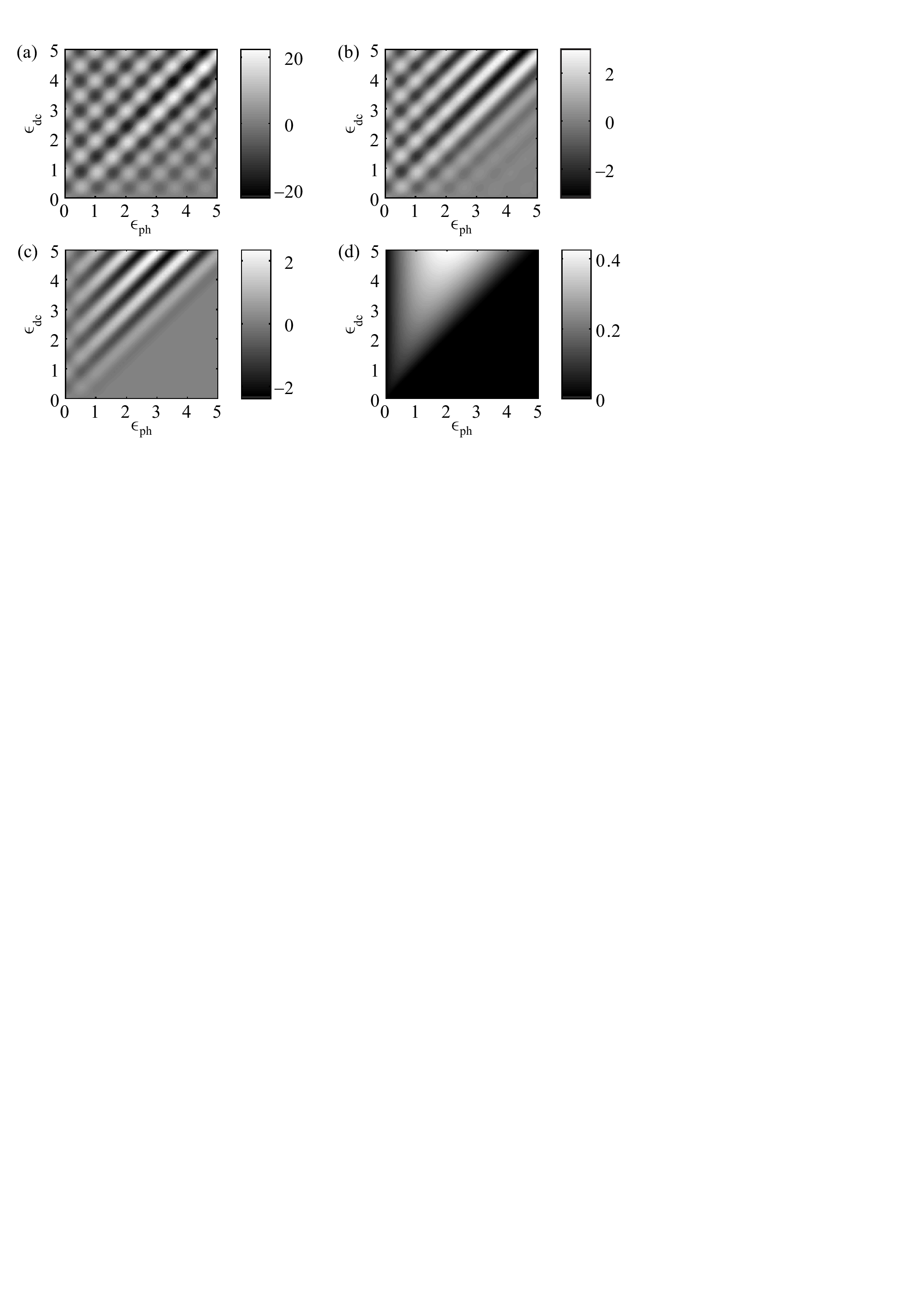}
\caption{(a)-(c) Oscillatory part of the phonon-assisted current $j_d^{\rm (p,
osc)}$ in units of $2\delta^2 g^2\wc\ttr\sigma_{\rm D} E$, calculated for (a)
$T=5\wc$, (b) $T=0.7 \wc$, and (c) $T=0.25 \wc$. (d) Smooth part of
the phonon-assisted current at $T=0$. From \citet{dmitriev:2010}.}
\label{phiroT}
\end{figure}
%%%%%%%%%%%%%%%%%%%%%%%%%%%%%%%%%%%%%%%%%%%%

%\input{s6.tex}

\section{Transport in strong ac and dc fields}
\label{s6}

In the preceding sections, we discussed two types of nonequilibrium
magnetooscillations, MIRO (\rsec{s3}) and HIRO (\rsec{s5}). The nature of MIRO
has so far been considered at the level of single-photon processes. We now turn
to mode-mixing phenomena; specifically, to multiphoton processes in MIRO and to
nonlinear mixing of MIRO and HIRO.

\subsection{Fractional MIRO}
\label{VI.A}

\subsubsection{Experiments at high microwave power levels}
\label{VI.A.1}

In addition to MIRO, whose extrema in the vicinity of integer
$\epsilon_{\rm ac}$ values are described by Eq.~(\ref{eq.miro.max}) and the oscillatory structure is shown in
Fig.~\ref{fig.miro}, similar oscillatory features have been observed near
$\epsilon_{\rm ac}$ given by certain rational fractions, with maxima ${(+)}$ and
minima ${(-)}$ of the photoresistance at
\be
\epsilon_{\rm ac}^\pm =\frac n m \mp \varphi_{\rm ac}^\pm~,
\label{VI.1}
\ee
where $n$ and $m$ are integers (with noninteger $n/m$) and $\varphi_{\rm
ac}^\pm>0$ is, in typical experiments, considerably smaller than 1/4
(Fig.~\ref{fig.fzrs}). The oscillatory feature at $\epsilon_{\rm ac}=1/2$ was
observed by
\citet{dorozhkin:2003,willett:2004,dorozhkin:2005,dorozhkin:2007,wiedmann:2009a}
[in retrospective, the first signature of the half-integer MIRO can be
recognized already in early experiments on moderate-mobility samples
\cite{zudov:2001a}, see
a weak oscillation at $B\simeq 2\,{\rm kG}$ in the $f=45\,{\rm GHz}$ trace
there]. Similar fractional MIRO (fMIRO) for other fractions in the $n=1$ series
$\epsilon_{\rm ac} = 1/m$ were reported for $m=3,4$
\cite{dorozhkin:2007,wiedmann:2009a} and $m=5,6,7,8$ \cite{wiedmann:2009a}.
Fractions that have been observed in the fMIRO series with $n=2$ (apart from the
even denominator fractions coinciding with the members of the $n=1$ series) are
represented by $m=3$ \cite{zudov:2004,zudov:2006b,wiedmann:2009a,dorozhkin:2007}
and $m=5,7$ \cite{wiedmann:2009a}. Note that the fMIRO series with $n=1,2$ belong to
the high-$B$ side of the CR and thus do not overlap with the integer MIRO which
occur at $\epsilon_{\rm ac}>1$. A third series of fMIRO among those observed is,
in contrast to the variable-denominator series above, characterized by fixed
$m=2$ and variable $n$. The reported oscillatory features in the $\epsilon_{\rm
ac}=n/2$ series that fall in the intervals between integer $\epsilon_{\rm ac}$
are (apart from $\epsilon_{\rm ac}=1/2$ mentioned above) those with $n=3$ and 5
\cite{zudov:2003,zudov:2004,zudov:2006b}. The half-integer fMIRO with $n\geq 3$
lie on the low-$B$ side of the CR and therefore require for their observation
(other experimental conditions being equal) higher mobilities compared to the fMIRO
series with $n=1$.

Similar to MIRO, the most prominent minima from all three fMIRO series, namely
those in the vicinity of $\epsilon_{\rm ac}=1/2,2/3,3/2$, were shown to evolve
into ZRS with increasing microwave power in ultra-high mobility samples
\cite{zudov:2006b}. As seen in Fig.~\ref{fig.fzrs}a, the formation of these
fractional ZRS was accompanied by the diminishing and narrowing of the
neighboring photoresistance peaks, as well as by the overall suppression of the
photoresistance at $\epsilon_{\rm ac}<1/2$. A similar strong suppression of the
resistance by radiation on the high-$B$ side of the CR was examined by
\citet{dorozhkin:2005}.

%%%%%%%%%%%%%%%%%%%%%%%%%%%%%%%%%%%%%%%%%%%%%%%%%
\begin{figure}[t]
\includegraphics[width=\columnwidth]{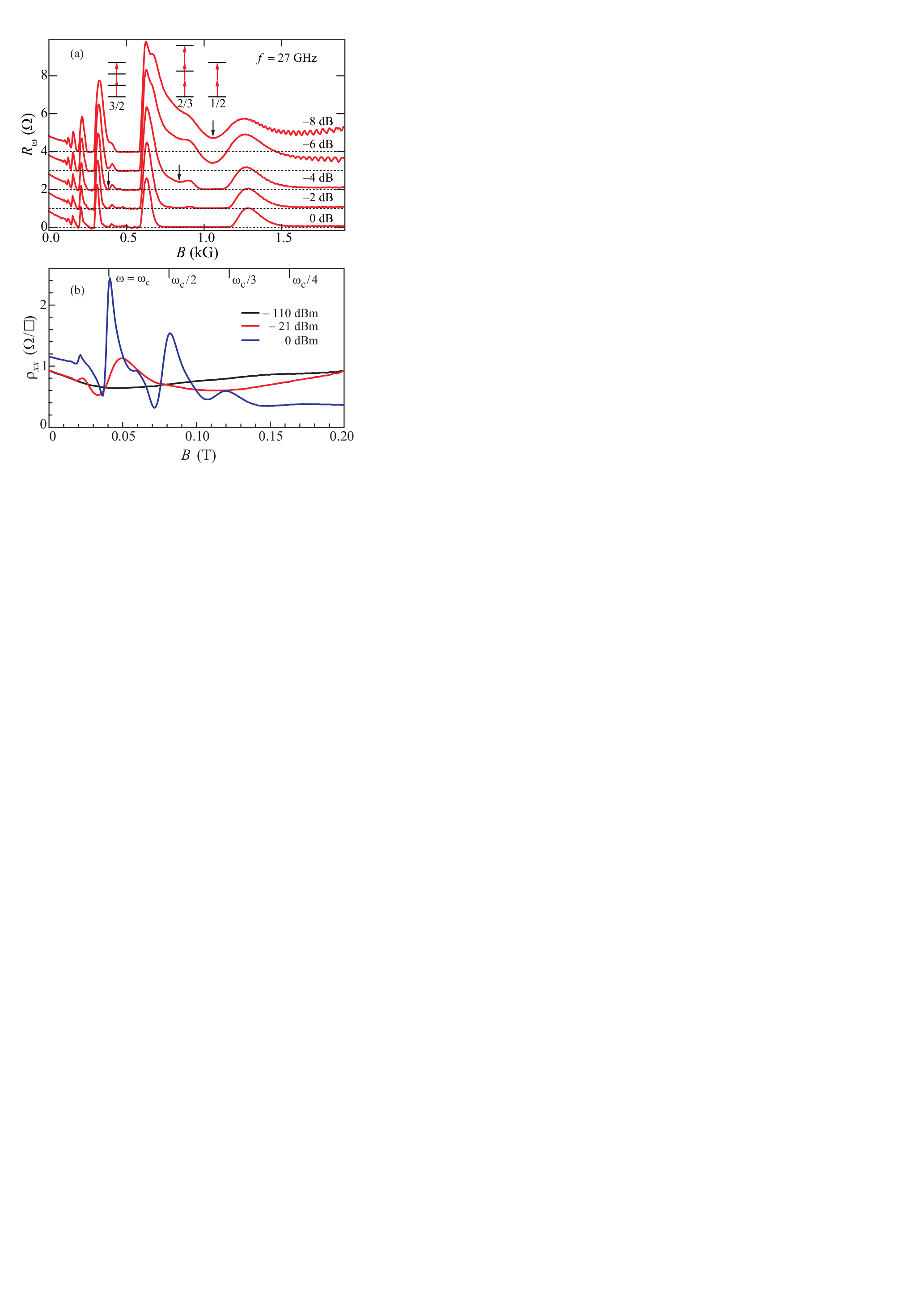}
\caption{(a) Magnetoresistance under microwave illumination at $f=27\,{\rm GHz}$
for different radiation intensities. The traces, offset vertically for clarity,
are labeled according to the attenuation levels. The downward arrows mark the
minima near $\epsilon_{\rm ac}= 3/2$, 2/3, and 1/2,
which develop into fractional ZRS at higher radiation intensities. The inset
illustrates multiphoton processes responsible for these features. The data were
obtained at $T \simeq 1.0\,{\rm K}$ in a sample with $n_e\simeq 3.6\times
10^{11}\,{\rm cm^{-2}}$ and $\mu\simeq 2\times 10^7\,{\rm cm^2/V\,s}$. From
\citet{zudov:2006a}. (b) Similar data for $f=17\,{\rm GHz}$ at $T \simeq
1.4\,{\rm K}$ in a sample with $n_e\simeq 2.7\times 10^{11}\,{\rm cm^{-2}}$ and
$\mu\simeq 1.7\times 10^7\,{\rm cm^2/V\,s}$. From \citet{dorozhkin:2007}.
}
\label{fig.fzrs}
\end{figure}
%%%%%%%%%%%%%%%%%%%%%%%%%%%%%%%%%%%%%%%%%%%%%%%%%

\subsubsection{Multiphoton absorption via virtual states}
\label{VI.A.2}

The theoretical framework developed to describe fMIRO relates them to
multiphoton processes in which $m$ photons are absorbed in transitions between
electron states separated in energy by $n\omega_c$
\cite{dmitriev:2007b,lei:2006b,pechenezhskii:2007,lei:2003,torres:2005}, as
illustrated in the inset to Fig.~\ref{fig.fzrs}a. One can distinguish two
(generally, interfering with each other) scattering channels for multiphoton
transitions: sequential absorption of single photons via {\it real} (resonant)
intermediate electron states and multiphoton absorption via {\it virtual}
intermediate states. For overlapping LLs, both types of multiphoton transitions
yield in the limit of a strong ac field oscillatory features in the vicinity of
$\epsilon_{\rm ac}=n/m$---with an amplitude proportional at $T\gg m\omega$ to
$\delta^{2m}$---which, for given $\delta$, are damped by disorder much more
strongly than the integer MIRO \cite{pechenezhskii:2007}. For separated LLs,
however, single-photon real transitions are forbidden if $\omega$ is not
sufficiently large to bridge the gap between the LLs (see Fig.~\ref{windows}),
so that microwave absorption occurs then by means of multiphoton transitions via
virtual states. This is, in particular, the case for two-photon transitions
between neighboring LLs at $\omega=\omega_c/2\gg 1/\tau_{\rm q}$.

In the limit of well-separated LLs, the mechanism of multiphoton absorption via
virtual intermediate states is much similar to that of multiphoton ionization in
atomic or semiconductor systems \cite{keldysh:1965}. In particular, a
perturbative expansion in the radiation power is justified if $\omega$ is large
compared to the transition rate between LLs the slope of which oscillates in
phase with the ac field. The parameter that governs the validity of perturbation
theory is $(\tau_{\rm q}/\tau){\cal P}\ll 1$, where $\cal P$ is given by
Eq.~(72) \cite{dmitriev:2007}. Here and in the rest of Sec.~\ref{VI.A}, we
overview the approach developed by \citet{dmitriev:2007b} and
\citet{pechenezhskii:2007} to describe fMIRO for the case of smooth disorder. We
expect that the picture that results from this approach remains qualitatively
correct for more general models of disorder. Similar to the integer MIRO, fMIRO
induced by the inelastic mechanism \cite{dmitriev:2007b} are much stronger than
those induced by the displacement mechanism\footnote{\label{lei_footnote} The method of ``force- and
energy-balance equations" \cite{lei:2006b,lei:2003,lei:2010} tacitly assumes an instant
thermal equilibration in the frame moving with the drift velocity, i.e., the
inelastic mechanism of photoconductivity is lost within this formalism ``by
construction". Similarly, the argument by \citet{torres:2005} that the
``inelastic processes can be safely ignored" if they are slower than the elastic
ones is in contradiction with the fact that without them the stationary regime
of photoconductivity cannot be established: in effect, \citet{torres:2005} also
implicitly assume an infinitely fast equilibration of the same kind.}
\cite{lei:2006b,lei:2003,torres:2005,dmitriev:2007b}, provided $\tau_{\rm
in}/\tau_{\rm q}\gg 1$. Under this condition, the half-integer fMIRO near
$\epsilon_{\rm ac}=n/2$ with $n=1,3,5,\ldots$ are described at order ${\cal
P}^2$ for well-separated LLs \cite{dmitriev:2007b} by
\be
{\rho_{\rm ph}\over \rho_{\rm D}}={3\tau_{\rm in}\tau_{\rm q}\over
32\tau^2}\,{\cal P}^2{\cal R}_2(2\omega)(1+\vartheta)~,
\label{VI.2}
\ee
where the function ${\cal R}_2(\omega)$, odd in the detuning $2\omega-n\omega_c$,
is defined in Eq.~(85), and $\vartheta=2|{\cal E}_+{\cal E}_-|^2/{\cal
P}^2$ depends on the polarization of microwaves; specifically,
$\vartheta=[(\omega^2-\omega_c^2)/(\omega^2+\omega_c^2)]^2/2$ and $\vartheta=0$
for the cases of linear and circular polarization, respectively. For $n=1$,
Eq.~(\ref{VI.2}) describes the two-photon CR in the photoresistivity.

The two-photon contribution of the displacement mechanism is smaller (in the
parameter $\tau_{\rm q}/\tau_{\rm in}\ll 1$) but possesses, in the case of
linear polarization, anisotropy with respect to the mutual orientation of the dc
and ac fields. At order ${\cal P}^2$, the depth of the modulation, as the
orientation is changed, between the principal values of the photoresistivity
tensor induced by the displacement mechanism is given by the factor $1\pm
\sqrt{2\vartheta}/(1+\vartheta)$ \citep[for the full analytical expression,
see][]{dmitriev:2007b}. The orientation dependence thus provides a measure of
the role of the displacement mechanism in fMIRO.

\subsubsection{Multiphoton absorption via sidebands}
\label{VI.A.3}

The multiphoton absorption via virtual states [Eq.~(\ref{VI.2})] is the main
mechanism of fMIRO in the limit of well-separated LLs. A subleading (in the
parameter $1/\omega_c\tau_{\rm q}\ll 1$) contribution to fMIRO comes from
single-photon transitions via radiation-induced ``sidebands" formed in the gaps
of the DOS between the LLs \cite{dmitriev:2007b}. The sideband mechanism can be
viewed as resulting from disorder-induced transitions between the quasienergy
levels \cite{zeldovich:1967} that characterize the stationary states of a
homogeneous electron system under the action of a periodic-in-time perturbation.
A systematic approach to studying the spectral properties of a 2DEG driven by
the ac field in the presence of disorder is formalized in terms of a
time-dependent self-energy of the Wigner-transformed retarded Green's function
\cite{vavilov:2004,dmitriev:2007,dmitriev:2007b}. The notion of a time-dependent
DOS was also introduced to describe sidebands in a closely related problem of
the dynamical Franz-Keldysh effect \cite{yacoby:1968} by
\citet{jauho:1996,johnsen:1998}. In the fMIRO problem, the nonequilibrium DOS of
each LL acquires, to first order in $1/\omega_c\tau_{\rm q}$, two static
satellite peaks
\be
\nu^{\rm sb}_\pm(\epsilon)=(\pi {\cal
P}/8\omega_c\tau)\nu^{(N)}(\epsilon\pm\omega)~,
\label{VI.3}
\ee
where $\nu^{(N)}(\epsilon)$ is given by Eq.~(27), centered for
$\omega=\omega_c/2$ in the middle between LLs, and also satellites centered at
the same energies and oscillating in time with a period $2\pi/\omega$
\cite{dmitriev:2007b}. Note that the DOS peaks (\ref{VI.3}) can be probed by
resonant spectroscopy in the same way as if they were present at equilibrium.
For the half-integer fMIRO, single-photon transitions to and from the static
sidebands yield a contribution which, at order ${\cal P}^2$, has the same shape
as in Eq.~(\ref{VI.2}) and is of order $\omega_c\tau_{\rm q}$ times smaller
\cite{dmitriev:2007b}. On the other hand, the time-dependent sidebands give a
contribution which is smaller than that in Eq.~(\ref{VI.2}) by only a factor of
order $(\omega_c\tau_{\rm q})^{1/2}$ and has a distinctly different shape; in
particular, it is even in the detuning from the fMIRO resonance \citep[for details,
see][]{dmitriev:2007b}.

\subsubsection{Multiphoton absorption via real states}
\label{VI.A.4}

Multiphoton absorption via virtual states or sidebands provides the mechanisms
of fMIRO in the case of well-separated LLs. If, however, gaps between LLs are
opened (i.e., $\omega_c\tau_{\rm q}>\pi/2$, Sec.~II.C.1), but the width of the
gaps $\Delta_g<\omega_c/2$, then for any given $\omega$ in the interval
$\Delta_g<\omega<\omega_c-\Delta_g$ real single-photon transitions are possible
{\it both} between neighboring LLs and within the same LL. For $\omega$ from the
above interval, an electron can thus be transferred between neighboring LLs via
two sequential single-photon transitions. More generally, for
$(n-1)\omega_c+\Delta_g<\omega<n\omega_c-\Delta_g$, single-photon real
transitions are possible with a change of the LL index by both $n-1$
and $n$, where $n=1,2,3,\ldots$. For given $n$, therefore, there exists a finite
range of $\omega$ and $\omega_c$ within which $m\neq n$ photons can bridge the
gap between $n$ LLs via real intermediate states, which leads to the
emergence of an additional mechanism of fMIRO
\cite{dorozhkin:2007,pechenezhskii:2007}. fMIRO induced by this mechanism of
sequential single-photon absorption are specific to the crossover between the
regimes of overlapping and well-separated LLs and disappear altogether for
$2\Delta_g>\omega_c$. At the crossover, however, they are the main fMIRO channel
at order ${\cal P}^2$ if $\tau_{\rm in}/\tau_{\rm q}\gg 1$, with an amplitude
proportional to $(\tau_{\rm in}{\cal P}/\tau)^2$, similar to the second-order
expansion in ${\cal P}$ of the integer MIRO contribution in Eq.~(97).

\subsection{Nonlinear mixing of MIRO and HIRO}
\label{sec.miro.dc.exp}

The physics of high LLs out of equilibrium is further enriched by the
possibility of simultaneously applying both ac and strong dc fields to make MIRO
and HIRO superimpose onto each other. A nonlinear response of the 2DEG to the
fields provides a way of mixing the ``frequencies" $\epsilon_{\rm ac}$ and
$\epsilon_{\rm dc}$ of the two types of magnetooscillations.

\subsubsection{Experiments on MIRO in a strong dc field}
\label{VI.B.1}

The nonlinear mixing of MIRO and HIRO was demonstrated in experiments on
ultra-high mobility structures by \citet{zhang:2007c} and \citet{hatke:2008a}.
Figure~\ref{fig.miro.dc}a shows the effect of a direct current $I$ varied under continuous
microwave irradiation on the differential resistivity $r_\omega$: the MIRO
maxima are seen to evolve into minima, and vice versa, with increasing $I$.
One of the most remarkable features in Fig.~\ref{fig.miro.dc}a is the
crossing points (marked by arrows) of the curves measured at different currents:
the resistivity shows little change at $\epsilon_{\rm ac}=n$ and $n+1/2$, where
$n$ is an integer.

%%%%%%%%%%%%%%%%%%%%%%%%%%%%%%%%%%%%%%%%%%%%%%%%%
\begin{figure}[b]
\includegraphics[width=\columnwidth]{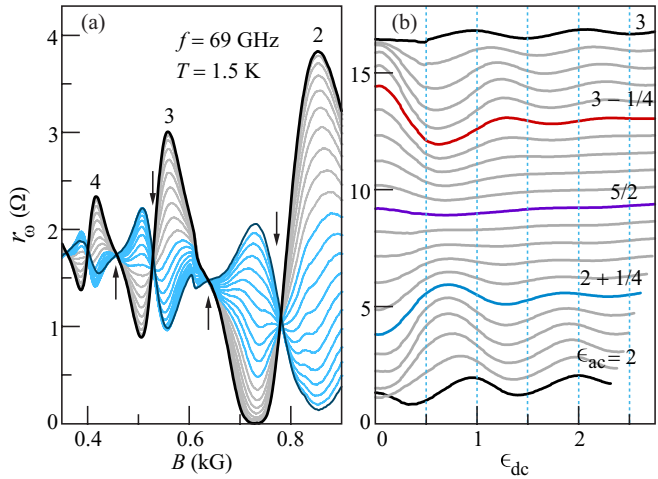}
\caption{(a) Differential resistivity $r_\omega$ versus magnetic field at a
radiation frequency of $69\,{\rm GHz}$ for different levels of dc excitation,
from $I = 0$ (thick black curve) to $20\,\mu$A in steps of $1\,\mu$A.
Integers show the order of the MIRO peaks. The arrows mark the zero-response nodes
that remain largely immune to dc excitation. (b) Part of the data in
(a), represented as the dependence of $r_\omega$ on
$\epsilon_{\rm dc}$ at different fixed $\epsilon_{\rm ac}$ ranging from 2 to 3
in steps of 0.05. The traces are offset in steps of $0.75\,\Omega$. The data
were obtained at $T = 1.5\,{\rm K}$ in a $100\,\mu{\rm m}$ wide Hall bar sample
with $n_e\simeq 3.7\times 10^{11}\,{\rm cm^{-2}}$ and $\mu\simeq 1.2 \times
10^7\,{\rm cm^2/V\,s}$. Adapted from \citet{hatke:2008a}.
}
\label{fig.miro.dc}
\end{figure}
%%%%%%%%%%%%%%%%%%%%%%%%%%%%%%%%%%%%%%%%%%%%%%%%%

More insight into this phenomenon is gained if one plots $r_\omega$ as a
function of $\epsilon_{\rm dc}$ for different fixed $\epsilon_{\rm ac}$, as in
Fig.~\ref{fig.miro.dc}b. The amplitude of the oscillations of $r_\omega$ in
Fig.~\ref{fig.miro.dc}b is seen to be much larger around $\epsilon_{\rm ac}=n\pm
1/4$ (near which $\epsilon_{\rm ac}$ the extrema of MIRO are located) than
around $\epsilon_{\rm ac}= n, n + 1/2$ (near which $\epsilon_{\rm ac}$ MIRO
exhibit zero response). Figure~\ref{fig.miro.dc}b also reveals a substantial
difference between integer and half-integer $\epsilon_{\rm ac}$: while at
$\epsilon_{\rm ac}=2,3$ the oscillations are suppressed but still clearly
visible, they virtually disappear at $\epsilon_{\rm ac}=5/2,7/2$. The phase of
the oscillations of $r_\omega$ as a function of $\epsilon_{\rm dc}$ also shows a
strong dependence on $\epsilon_{\rm ac}$. Specifically, at integer
$\epsilon_{\rm ac}=2,3$, $r_\omega$ behaves very similar to HIRO, with maxima
occurring at $\epsilon_{\rm dc} \simeq n$, Eq.~(\ref{eq.hiro}). In contrast, the
oscillations of $r_\omega$ at the MIRO maximum for $\epsilon_{\rm ac}\simeq 3-1/4$
and the MIRO minimum for $\epsilon_{\rm ac}\simeq 3+1/4$ are phase shifted by about
a quarter cycle in opposite directions compared to HIRO.

As discussed in \rsec{s5.4},
a minimum of HIRO may evolve into a ZdRS \cite{hatke:2010a}. The nonlinear
mixing of MIRO and HIRO gives rise to an interplay between ZRS associated with
MIRO and ZdRS. Namely, as $\epsilon_{\rm dc}$ increases, ZRS that are present at
zero dc field transform into maxima of $r_\omega$, as shown in
Fig.~\ref{fig.miro.dc}a for the case of ZRS near $\epsilon_{\rm ac}=2+1/4$. In
turn, a ``strong maximum" of MIRO may be converted by the strong dc field into a
ZdRS. The ZdRS in a microwave-driven 2DEG, originating from the maximum of MIRO
near $\epsilon_{\rm ac}=2-1/4$, was observed by \citet{zhang:2007c,zhang:2008b}.
In fact, this observation appears to be an example of a more general (not
relying on the presence of microwaves) phenomenon which relates ZdRS to the {\it
maxima} of magnetooscillations at zero dc field. The vanishing of $r_\omega$ of
a similar nature---with ZdRS evolving from the maxima of oscillations---has also
been observed in the absence of radiation: in particular, the emergence of ZdRS
from the maxima of strongly developed SdH oscillations was demonstrated by
\citet{bykov:2007}. The ZdRS was also found to emerge from the maxima of MISO
\cite{bykov:2010c,gusev:2011,wiedmann:2011c}.

\subsubsection{Frequency mixing for magnetooscillations}
\label{VI.B.2}

Analysis of the experimental data in Fig.~\ref{fig.miro.dc}b led
\citet{hatke:2008a} to conclude that the local maxima of $r_\omega$ in the
$\epsilon_{\rm ac}$--$\epsilon_{\rm dc}$ plane occur at the points given by
\be
\epsilon_{\rm ac}+\epsilon_{\rm dc} \simeq n~,\quad
\epsilon_{\rm ac}-\epsilon_{\rm dc} \simeq m-1/2~,
\label{VI.4}
\ee
where $n$ and $m$ are integers [the first condition was established in earlier
work by \citet{zhang:2007c}]. \citet{hatke:2008a} also argued that
Eqs.~(\ref{VI.4}) point to the key role played by backscattering off disorder in
shaping the oscillations in Fig.~\ref{fig.miro.dc}b (similar to HIRO in the
limit of $\epsilon_{\rm dc}\gg 1$). Within a qualitative picture proposed by
\citet{hatke:2008a}, $r_\omega$ has a maximum when disorder-induced scattering,
accompanied by shifts of the cyclotron orbit in real space by $2R_c$, is
maximized for shifts in the direction of the dc field (first condition) and
minimized for shifts in the opposite direction (second condition).

As the calculation by \citet{khodas:2008} shows, the above picture is
qualitatively accurate (to first order in $\cal P$ and for $\epsilon_{\rm
ac}>\epsilon_{\rm dc}$) in describing the mechanism of ``frequency mixing" of
MIRO and HIRO; however, it misses an important microwave-induced correction to
the amplitude of HIRO which arises at the same order in $\cal P$. Specifically,
to order ${\cal O}({\cal P})$ and at $\epsilon_{\rm dc}\gg 1$, the oscillatory
correction to the resistivity $\delta r_\omega=r_\omega-\rho_{\rm D}$ for
overlapping LLs reads \cite{khodas:2008}
\bea
&&\delta r_\omega/\rho_{\rm D}=(16\delta^2/\pi\epsilon_{\rm
dc})(\tau/\tau_\pi)\left(C_0+C_++C_-\right)~,\nonumber\\
&&C_0=(1-2{\cal P})\,\epsilon_{\rm dc}\cos 2\pi\epsilon_{\rm dc}~,\nonumber\\
&&C_\pm={\cal P}\,(\epsilon_{\rm dc}\pm\epsilon_{\rm ac})\cos 2\pi(\epsilon_{\rm
dc}\pm\epsilon_{\rm ac})~.
\label{VI.5}
\eea
The frequency-mixing terms $C_\pm$ are seen to be exactly canceled by the
linear-in-$\cal P$ correction to the amplitude of HIRO in $C_0$ at integer
$\epsilon_{\rm ac}$ (in which case the oscillatory correction $\delta r_\omega$
reduces to the plain HIRO).

Moreover, Eqs.~(\ref{VI.4}) can be obtained from Eqs.~(\ref{VI.5}) only if the
modified HIRO, described by the term $C_0$, are neglected. In general, the
location of the extrema of $r_\omega$ given by Eqs.~(\ref{VI.5}), and the whole
picture of the oscillations for that matter, depend in an essential way on $\cal
P$. In particular, the most characteristic feature in
Figs.~\ref{fig.miro.dc}---the vanishing of the oscillations at half-integer
$\epsilon_{\rm ac}$---can be obtained, within the theoretical framework on which
Eqs.~(\ref{VI.5}) are based, at special values of $\cal P$ only. Specifically,
Eqs.~(\ref{VI.5}) yield for half-integer $\epsilon_{\rm ac}$:
\be
\delta r_\omega/\rho_{\rm D}=(16\delta^2/\pi)(\tau/\tau_\pi)(1-4{\cal P})\cos
2\pi\epsilon_{\rm dc}~.
\label{VI.6}
\ee
The amplitude of the oscillations in Eq.~(\ref{VI.6}) is seen to be reduced by
microwaves. In fact, extending the calculation of $\delta r_\omega$ for
$\epsilon_{\rm dc}\gg 1$ to arbitrary $\cal P$ \cite{khodas:2008,khodas:2010}
gives for half-integer $\epsilon_{\rm ac}$ Eq.~(\ref{VI.6}) with the factor
$1-4{\cal P}$ being substituted by the Bessel function $J_0(4\sqrt{\cal P})$
[see Eq.~(\ref{VI.7}) below]. The nodes of the oscillations of $r_\omega$ as a
function of $\epsilon_{\rm dc}$ at half-integer $\epsilon_{\rm ac}$ are thus
only reproduced at zeros\footnote{A reversal of the sign of the oscillations
with $\epsilon_{\rm dc}$ when $\cal P$ reaches a certain threshold, reported by
\citet{zudov:2009}, may provide a valuable calibration point for the microwave
intensity seen by the 2DEG, which is difficult to obtain from other
experiments.} of $J_0(4\sqrt{\cal P})$. Given the accuracy of the reported nodes
at half-integer $\epsilon_{\rm ac}$, it appears that the data in
Fig.~\ref{fig.miro.dc} were likely obtained in the very close vicinity of a zero
of $J_0(4\sqrt{\cal P})$.

\subsubsection{Fractional HIRO}
\label{VI.B.3}

The nonlinear response of a 2DEG to the dc field was also investigated
experimentally in the regime of fractional MIRO, in the vicinity of the most
prominent fraction $\epsilon_{\rm ac}=1/2$, by \citet{hatke:2008b}. Remarkably,
the experiment revealed the emergence of fractional HIRO at $\epsilon_{\rm
ac}=1/2$, with ``frequency-doubling" for the oscillations of $r_\omega$ as a
function of $\epsilon_{\rm dc}$. As seen in Fig.~\ref{fig.fmiro.dc}, the overall
behavior of $r_\omega$ at $\epsilon_{\rm ac}=1/2$ in the interval
$0<\epsilon_{\rm dc}<1$ closely replicates that at the CR in the interval
$0<\epsilon_{\rm dc}<2$---with the period of the oscillations in $\epsilon_{\rm
dc}$ at $\epsilon_{\rm ac}=1/2$ being half that at $\epsilon_{\rm ac}=1$ (which
means that the period of the oscillations of $r_\omega$ as a function of the
direct current is the same).

%%%%%%%%%%%%%%%%%%%%%%%%%%%%%%%%%%%%%%%%%%%%%%%%%
\begin{figure}[t]
\includegraphics[width=\columnwidth]{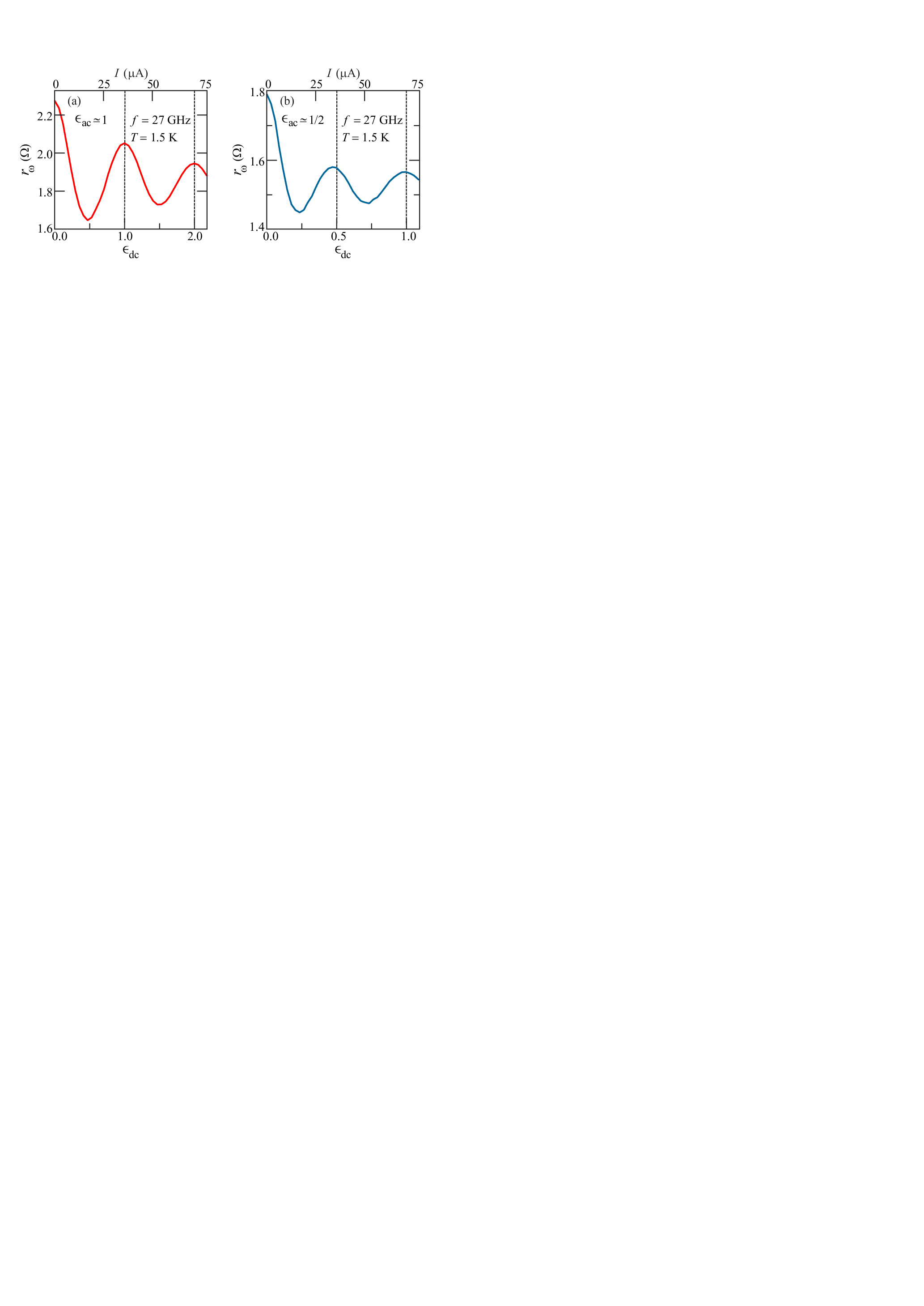}
\caption{Differential resistivity $r_\omega$ versus $\epsilon_{\rm dc}$ (bottom
axis) and the direct current $I$ (top axis) at (a) $\epsilon_{\rm ac}\simeq 1/2$ and
(b) $\epsilon_{\rm ac}\simeq 1$. The data were obtained at $f=27\,{\rm GHz}$ and $T =
1.5\,{\rm K}$ in a $100\,\mu{\rm m}$ wide Hall bar sample with $n_e\simeq
3.8\times 10^{11}\,{\rm cm^{-2}}$ and $\mu\simeq 1.3 \times 10^7\,{\rm
cm^2/V\,s}$. From \citet{hatke:2008b}.
}
\label{fig.fmiro.dc}
\end{figure}
%%%%%%%%%%%%%%%%%%%%%%%%%%%%%%%%%%%%%%%%%%%%%%%%%

A description of the fractional HIRO in a microwave-excited 2DEG was suggested
by \citet{lei:2009} in terms of a postulated ansatz which assumes an infinitely
fast thermal equilibration in the frame moving with the drift velocity (see
footnote \ref{lei_footnote}). \citet{lei:2009} argued that the maxima of $r_\omega$ should obey
the condition  $\epsilon_{\rm dc}+n\epsilon_{\rm ac}=m$ with integer $n$ and
$m$, and the numerically obtained solution to the ansatz equation yielded a
picture similar to the one in Fig.~\ref{fig.fmiro.dc}; however, the dependence
of the strength and shape of the oscillations on the major parameters, such as
$\cal P$ or the degree of overlap of LLs, remained unclarified.

On the other hand, the observed frequency-doubling for HIRO may be a signal of
the inelastic mechanism of nonequilibrium magnetooscillations.
Indeed, in overlapping LLs, a controlled calculation by \citet{khodas:2008} at
order $\delta^2$ and at first order in $\cal P$ in the collision integral showed
that, while the displacement mechanism is dominant in the limit of
$\epsilon_{\rm dc}\gg 1$, the inelastic mechanism for $\tau_{\rm in}/\tau\agt 1$
is equally important at $\epsilon_{\rm dc}\sim 1$ (see also Sec.~V.C) and
produces strong subharmonics in the dependence of $r_\omega$ on $\epsilon_{\rm
dc}$---in contrast to the displacement mechanism at this order in $\delta$.
Importantly, the strength of the subharmonics in the presence of radiation
depends on $\cal P$ and oscillates with $\epsilon_{\rm ac}$ \cite{khodas:2008}.
In particular, for overlapping LLs, the subharmonic with a period in
$\epsilon_{\rm dc}$ equal to 1/2 is as strong at $\epsilon_{\rm ac}=1/2$,
$\epsilon_{\rm dc}\sim 1$, and ${\cal P}\sim 1$ as the harmonic with period 1.
In the case of separated LLs, the half-integer MIRO and half-integer HIRO are
also likely to strengthen each other in analogy with the theory of two-photon
absorption (Sec.~VI.A). Further theoretical work is necessary to study the exact
form of the interplay of the two types of fractional oscillations.

\subsubsection{AC field-periodic oscillations}
\label{sec.miro.dc.hp.exp}

%%%%%%%%%%%%%%%%%%%%%%%%%%%%%%%%%%%%%%%%%%%%%%%%%
\begin{figure}[t]
\includegraphics[width=\columnwidth]{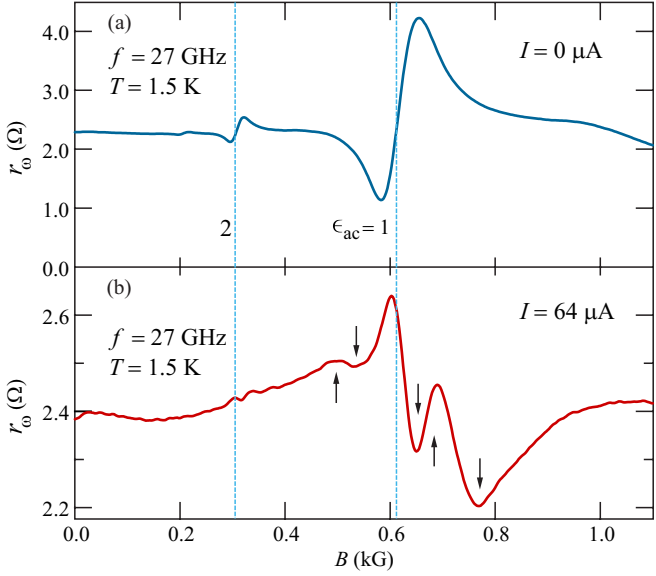}
\caption{Differential resistivity $r_\omega (B)$ (a) for zero direct current $I$
and (b) for $I=64\,\mu{\rm A}$. The vertical lines mark the position of the CR
and its second harmonic. Parameters of the sample, frequency $f=27\,{\rm GHz}$,
and temperature $T=1.5\,{\rm K}$ are the same as in Fig.~\ref{fig.fmiro.dc}.
Adapted from \citet{khodas:2010}.
}
\label{fig.miro.dc.hp.1}
\end{figure}
%%%%%%%%%%%%%%%%%%%%%%%%%%%%%%%%%%%%%%%%%%%%%%%%%

In Sec.~\ref{VI.B.3} the focus was on the regime of a moderate microwave power
$({\cal P}\alt 1)$. The nonlinear mixing of HIRO and MIRO in the case of
high-power radiation $({\cal P}\gg 1$) results in the emergence of oscillations
of $r_\omega$ as a function of $\cal P$ (in addition to the oscillations in
$\epsilon_{\rm dc}$ and $\epsilon_{\rm ac}$). These were observed
\cite{khodas:2010} as a series of oscillations grouped around the CR and
decaying away from it, with much less pronounced but similar behavior also
detected around $\epsilon_{\rm ac}=2$. In contrast to the integer MIRO, which
have one maximum and one minimum per unit interval of $\epsilon_{\rm ac}$
(Fig.~\ref{fig.miro.dc.hp.1}a), or the fractional MIRO with their oscillatory
features around rational values of $\epsilon_{\rm ac}$, these oscillations
exhibit multiple maxima and minima around the CR (and its second harmonic) with
a period which depends on $\cal P$, as shown in Fig.~\ref{fig.miro.dc.hp.1}b.

A theoretical description by \citet{khodas:2008,khodas:2010} of the reported oscillations in $\cal P$
 gives for the oscillatory correction $\delta
r_\omega$ at order $\delta^2$ for $\epsilon_{\rm dc}\gg \max\{1,\sqrt{{\cal
P}}\}$:
\bea
&&\hspace{-7mm}{\delta r_\omega\over\rho_{\rm
D}}={16\delta^2\over\pi\epsilon_{\rm
dc}}{\tau\over\tau_\pi}\left[\,\epsilon_{\rm dc}\cos (2\pi\epsilon_{\rm
dc})J_0\!\left(4{\sqrt{\cal P}}\sin\pi\epsilon_{\rm ac}\right)
\right.\nonumber\\
&&\hspace{-7mm}-\left.2\epsilon_{\rm ac}\sin (2\pi\epsilon_{\rm dc})\cos
(\pi\epsilon_{\rm ac})\sqrt{{\cal P}}J_1\!\left(4{\sqrt{\cal
P}}\sin\pi\epsilon_{\rm ac}\right)\,\right]~,\nonumber\\
\label{VI.7}
\eea
which generalizes Eqs.~(\ref{VI.5}) to the case of large $\cal P$, with the
Bessel functions $J_{0,1}$ oscillating with a period which depends on $\cal P$
and $\epsilon_{\rm ac}$. Oscillations in $\cal P$ of this form are a
characteristic signature of multiphoton transitions between stationary states of
a periodically (in time) driven electron system. Namely, the absorption
rate for $N$-photon transitions between these states is proportional\footnote{A
similar structure arises in a variety of closely related problems, e.g., in studying
the multiphoton CR in three dimensions \cite{seely:1974}
or the dynamical Franz-Keldysh effect \cite{johnsen:1998}.} to
$J_N^2[2\sqrt{{\cal P}}\sin(\theta/2)]$, where $\theta$ is the angle between the
initial and final directions of the velocity (in the quasiclassical formulation)
\cite{khodas:2010,lei:2009}. In the limit $\epsilon_{\rm dc}\gg 1$, where the
main contribution to HIRO comes from $\theta\simeq \pi$, summing over $N$
\cite{khodas:2010} reproduces Eq.~(\ref{VI.7}). The characteristic number of
photons $N$ in the transitions that yield Eq.~(\ref{VI.7}) for ${\cal P}\gg 1$
is\footnote{More specifically, for $1/{\cal P}^{1/2}\ll |\bar{\epsilon}_{\rm ac}|$, where $\bar{\epsilon}_{\rm ac}=\eac-n$ and $n$ is the nearest integer, the sum over $N$, which is of the type $\sum_N\!J_N^2(2\sqrt{\cal P})(\edc+N\eac)\cos[2\pi(\edc+N\eac)]$ and contains the factor oscillating in $N$ with a period $1/|\bar{\epsilon}_{\rm ac}|$, is determined by a close vicinity of the sharp peak in $J_N^2(2\sqrt{{\cal P}})$ at $N=2\sqrt{\cal P}$ whose width is of order ${\cal P}^{1/6}$. The peak falls off exponentially on the side of larger $N$ and as $(2\sqrt{\cal P}-N)^{-1/2}$ on the other side. For $1/{\cal P}^{1/2}\ll |\bar{\epsilon}_{\rm ac}|\ll 1/{\cal P}^{1/6}$, the main contribution to $\delta r_\omega$ comes from a narrow interval of positive $2\sqrt{\cal P}-N$ whose width is of order $1/|\bar{\epsilon}_{\rm ac}|$. } $2\sqrt{{\cal P}}$. This conclusion is further corroborated by the observation
that $r_\omega$ from Eq.~(\ref{VI.7}) in the limit ${\cal P}\gg 1$ oscillates,
as a function of $\epsilon_{\rm ac}$ in the vicinity of integer $\epsilon_{\rm
ac}$, with a period $1/2\sqrt{{\cal P}}$. The experimental data
\cite{khodas:2010}, shown in Fig.~\ref{miro_dc_hp_fig_2}a, demonstrate that the
oscillations of $r_\omega$ with $\epsilon_{\rm ac}$ become more rapid with
increasing microwave power, in close agreement with the result of
Eq.~(\ref{VI.7}), shown in Fig.~\ref{miro_dc_hp_fig_2}b.

%%%%%%%%%%%%%%%%%%%%%%%%%%%%%%%%%%%%%%%%%%%%%%%%%
\begin{figure}[t]
\includegraphics[width=\columnwidth]{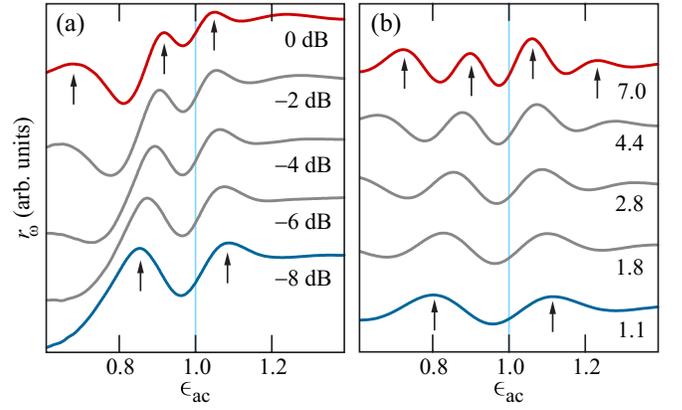}
\vspace{-0.1in}
\caption{(a) Measured and (b) calculated differential resistivity $r_\omega$
versus $\epsilon_{\rm ac}$ for $I=54\,\mu{\rm A}$ and different microwave
intensities. The traces are labeled according to (a) the attenuation levels and
(b) the value of ${\cal P}$. Parameters of the sample, frequency $f=27\,{\rm
GHz}$, and temperature $T=1.5$~K are the same as in
Figs.~\ref{fig.fmiro.dc} and \ref{fig.miro.dc.hp.1}. From \citet{khodas:2010}.
}
\label{miro_dc_hp_fig_2}
\end{figure}
%%%%%%%%%%%%%%%%%%%%%%%%%%%%%%%%%%%%%%%%%%%%%%%%%

%\input{s7.tex}

\section{Outlook}
\label{s7}

\subsection{Perspectives}
\label{s7.2}

Here we briefly outline several particularly promising directions
of ongoing and future research in the field of nonequilibrium magnetotransport
phenomena in semiconductor structures.

\subsubsection{Domain structure of ZRS and ZdRS}
\label{s7.2.1}

The present status of research in this area has been reviewed in Sec.~\ref{s4}.
The physics of the broken-symmetry states with spontaneously formed domains
remains one of the central research directions in the field. Particularly
important open questions include the character of the transition into ZRS, the
corresponding critical behavior (in the case when the transition is continuous)
beyond the mean-field picture, the residual resistance of a finite-size system,
% RR
the origin of the experimentally observed
Arrhenius-like temperature dependence of the ZRS resistivity,
the effect of thermal and nonequilibrium noise on ZRS, as well as dynamics of
the domain structure.

\subsubsection{Giant photoresponse at the second harmonic of the cyclotron
resonance }
\label{s7.2.1prim}
Recent experiments on ultra-high mobility structures reported a narrow spike at
$\w\simeq 2\wc$ superposed on the $n=2$ maximum of MIRO
\cite{dai:2010,hatke:2011b}. The height of the spike was several times (up to an
order of magnitude) larger than the height of other MIRO maxima.
\citet{dai:2010} also observed an enhancement of higher-order even MIRO maxima
($n=4,6,8$), see \rfig{fig:dai}. It is important to note that these effects were
observed in samples which also show a very strong negative MR (also in the
absence of microwaves) attributed to the quasiclassical memory effects, see
\rsec{s2.2.1}, which indicates a possible connection between these phenomena. As
a matter of fact, the memory effects that are responsible for the QCMR may also
produce spikes in the photoconductivity at the CR harmonics, see
\rsecs{s2.2.1}{s2.2.3}, and \ref{s3.2.5}. However, it is unclear why only even
harmonics (and, most prominently, the one at $n=2$) are enhanced. Further
theoretical work is thus needed.

%%%%%%%%%%%%%%%%%%%%%%%%%%%%%%%%%%%%%%%%%%%%%%%%%%%%%%%%%%%%%
\begin{figure}[ht]
\includegraphics[width=\columnwidth]{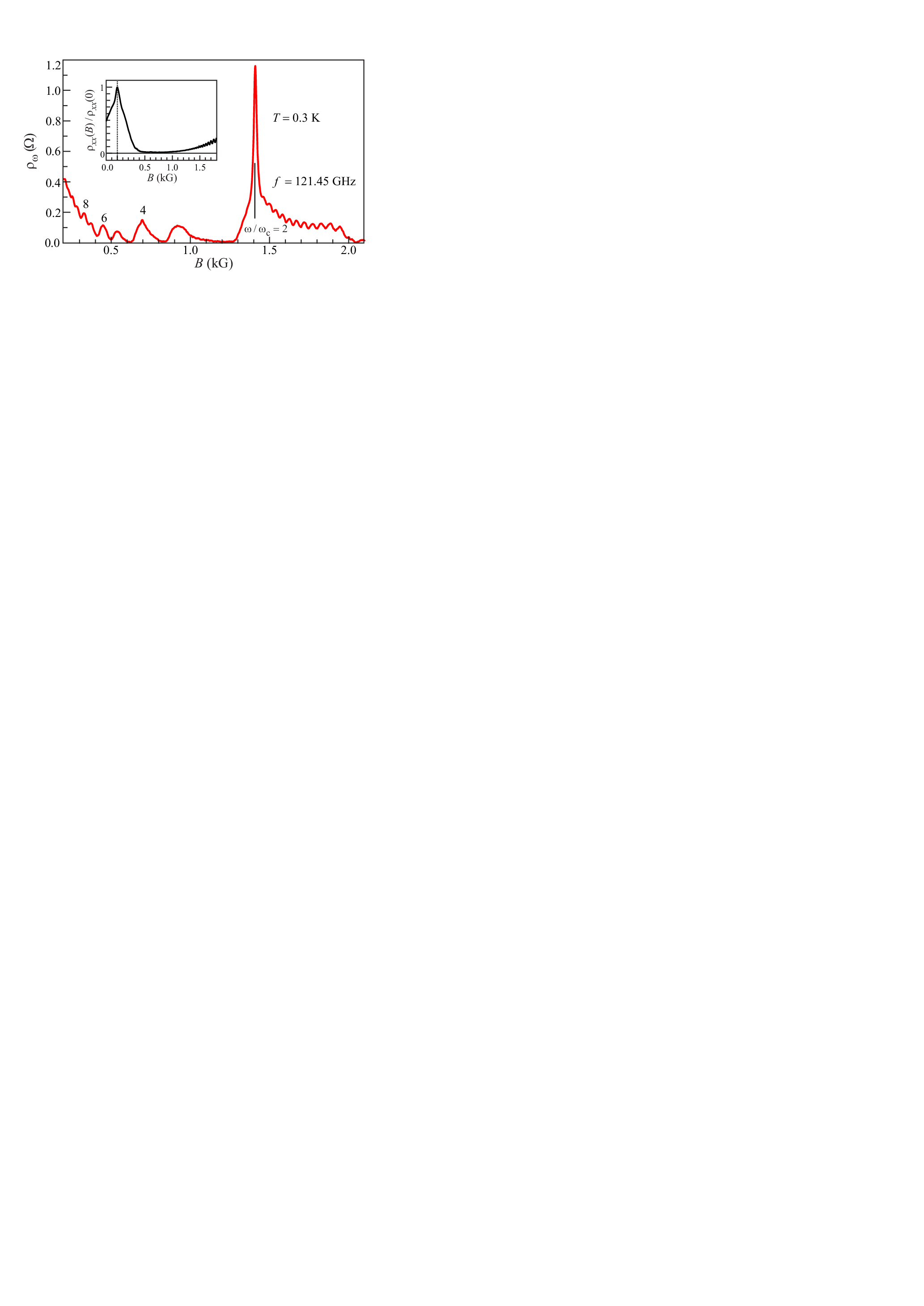}
\caption{Magnetoresistivity measured at $f=121.45\,{\rm GHz}$ and $T=0.3\,{\rm
K}$ in a Hall bar sample with $n_e\simeq 2.9\times 10^{11}\,{\rm cm^{-2}}$ and
$\mu\simeq 3.0\times 10^7\,{\rm cm^2/V\,s}$. The dark resistivity, normalized to
its value at $B=0$, is shown in the inset. Adapted from \citet{dai:2010}.}
\label{fig:dai}
\end{figure}
%%%%%%%%%%%%%%%%%%%%%%%%%%%%%%%%%%%%%%%%%%%%%%%%%%%%%%%%%%%%

\subsubsection{Nonequilibrium phenomena in separated Landau levels}
\label{s7.2.2}

Nonequilibrium phenomena in the regime of fully separated LLs deserve further
study. On the theoretical side, the theory for mixed disorder, which is the type
of disorder characteristic of most of the experimentally relevant samples (see
Secs.~\ref{s3} and \ref{s5}), needs to be extended into this regime. On the
experimental side, this corresponds to studying the nonequilibrium phenomena in
stronger magnetic fields. It is further interesting to see how the phenomena
discussed in this review evolve when entering the QH regime, see also
Sec.~\ref{s7.2.5}.

\subsubsection{Nonequilibrium transport at low temperature}
\label{s7.2.3}

Most of the experiments and theories presented in this review focussed on
the temperature regime where SdH oscillations are suppressed. The interplay of
the nonequilibrium physics and SdH oscillations at lower $T$ is also of
interest. A remarkable phenomenon of this kind that was studied both
experimentally
\cite{zhang:2007b,kalmanovitz:2008a,zhang:2009,studenikin:2010,zudov:2011,bykov:2007} and
theoretically \cite{dmitriev:2011a} is the dc field-induced inversion of the
phase of SdH oscillations. In particular, \citet{bykov:2007} reported the
formation of ZdRS that arise, as the strength of the dc field increases, from
the maxima of SdH oscillations, see Sec.~\ref{s5.4}. A similar effect on the
SdH oscillations by the microwave field was observed by \textcite{mani:2004d}.

\subsubsection{Spin-related phenomena}
\label{s7.2.4}

In the regime of strongly developed SdH oscillations, the spin degree of freedom
starts to play an important role. \textcite{romero:2008} explored the effect of
a parallel magnetic field $B_\parallel$ on ZdRS in this regime (Sec.~\ref{s4.4})
and found that the ZdRS are destroyed for sufficiently strong $B_\parallel$. The
result was explained \cite{romero:2008} within the kinetic equation formalism
described in Secs.~\ref{s3},\ref{s5} in terms of a Zeeman splitting-induced
modification of the oscillatory component of the DOS. \textcite{studenikin:2010}
further extended the study of the effect of Zeeman splitting on nonlinear
magnetotransport to the regime of strong magnetic fields normal to the 2DEG.
Still, spin-related phenomena in nonequilibrium magnetotransport remain largely
unexplored. This research direction overlaps with the investigations of the
regime of separated LLs (Sec.~\ref{s7.2.2}) and of the QH regime
(Sec.~\ref{s7.2.5}).

\subsubsection{Polarization dependence of MIRO}
\label{s7.2.4prim}

As discussed in \rsec{polarization}, the relative orientation of the dc field
and the linearly polarized microwave field affects considerably the MIRO
amplitude within the framework of the displacement mechanism, while the
inelastic contribution is insensitive to the direction of the linear
polarization.
\textcite{mani:2002,mani:2004f,wiedmann:2011c} reported equal MIRO signal
strength for the microwave field polarized along and perpendicular to the
current direction, which is consistent with the dominance of the inelastic
contribution. On the other hand, a recent experimental study \cite{mani:2011}
reported a pronounced difference of the MIRO amplitude for two directions of the
linear polarization, indicating the relevance of the displacement contribution.
Systematic experimental study of the dependence of this effect on the
temperature and microwave power would permit a quantitative comparison with
theoretical predictions (see \rsec{in_vs_dis}).

\textcite{smet:2005} used an optical setup in order to study MIRO induced by
circularly polarized microwaves. According to the theories presented in
Sec.~\ref{s3}, the amplitude of MIRO depends on the direction of circular
polarization through the factor $1/[(\omega\pm \omega_c)^2 + 1/\ttr^2]$, where
the minus (plus) sign corresponds to active (inactive) polarization.
Surprisingly, \textcite{smet:2005} reported very similar results for two
directions of circular polarization, with a difference being much smaller than
expected from this factor, see Fig.~\ref{smet05}.
% RR
To understand a possible reason for the failure to
see a large effect after changing the direction of circular polarization of the
incoming radiation, it is important to take into account that the electric field
in \reqs{P}{sigmadis}, and (\ref{sigmain}) is the {\it total} electric field
acting on electrons in the sample.
% RR
%%%%%%%%%%%%%%%%%%%%%%%%%%%%%%%%%%%%%%%%%%%%%%%%%%%%%%%%%%%%%
\begin{figure}[ht]
\includegraphics[width=\columnwidth]{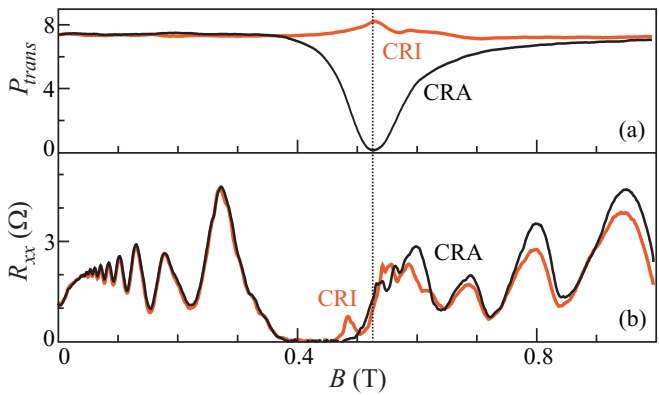}
\caption{(a) Transmitted power (arb.\ units) and (b) magnetoresistivity
measured at $f=200\,{\rm GHz}$ and $T\simeq 1.7\,{\rm K}$
in a van der Pauw sample with $n_e\simeq 2.6\times 10^{11}\,{\rm cm^{-2}}$ and
$\mu\simeq 1.8\times 10^7\,{\rm cm^2/V\,s}$
for both active (CRA)
and inactive (CRI) senses of circular polarization. Adapted from \citet{smet:2005}.}
\label{smet05}
\end{figure}
%%%%%%%%%%%%%%%%%%%%%%%%%%%%%%%%%%%%%%%%%%%%%%%%%%%%%%%%%%%%

The total field is different from the source field because of
screening by the 2DEG, see \rsecs{s2.3.2}{screening}. This leads to a strong
broadening of the CR: the above factors are modified to $1/[(\omega\pm
\omega_c)^2 + \gamma^2]$, where $\gamma = 2\pi e^2 n_e/m c n_r$ [\req{II.51}]
for $\gamma \gg 1/\ttr$. For a typical concentration $n_e=3\times 10^{11}\:{\rm
cm}^{-2}$ and the effective refractive index $n_r=(\sqrt{12.4}+1)/2\simeq 2.3$
for the case of a GaAs-based structure one finds $\gamma/2\pi\simeq 17\,{\rm
GHz}$, which is comparable to characteristic microwave frequencies. While this
explains why no sharp resonance features is experimentally observed near
$\omega=\omega_c$, this is not sufficient to explain why the difference in the
amplitude of the MIRO harmonics with $n=2,3,4$ for two directions of the
circular polarization in \textcite{smet:2005} was within a few percent
(Fig.~\ref{smet05}).
% RR
Indeed,
since $\gamma$ is several times smaller than the frequencies $\omega/2\pi=100
\div 250\:{\rm GHz}$ used in this experiment, one would expect a much larger
difference (a factor of about 9 for $n=2$ and of about 4 for $n=3$, etc.). This
discrepancy can be most likely attributed to the fact that the metallic
electrodes and other elements of the sample setup may strongly affect the
polarization of the electric field. In particular, \textcite{mikhailov:2006}
showed, for a simple geometry, that metallic contacts may strongly reduce the
degree of circular polarization of microwaves [the actual geometry of the
experiment \cite{smet:2005} is much more complex, and a quantitative evaluation
of the polarization of the actual electric field would require a numerical
modeling of the whole setup].

% RR
In fact, close inspection of the transmission data shown in Fig.~\ref{smet05}a
[see also Fig.~1 of \textcite{smet:2005}]
provides additional support to this explanation. While the
data do show, as expected, a CR dip around $\omega_c=\omega$ for the active
polarization, the $- 1/(\omega -\omega_c)^2$ tail of the Lorentzian for the
active polarization is not observed when $|\omega_c|$ becomes smaller than
$\omega$ by more than $\gamma$. Instead, the data for both polarizations of the
incoming wave become practically identical, with the transmission amplitude for
the active polarization being even slightly higher. This suggests that,
in this range of magnetic fields, the polarization of the incoming radiation was
essentially immaterial for the actual polarization of the electric field. Since
this is exactly the range of magnetic fields where MIRO are measured, this
observation is consistent with the fact that the MIRO signal was almost
identical for both directions of the circular polarization of the incoming wave.

It appears that the presence of nearby metallic contacts and, more generally,
the effects of a finite sample geometry may strongly affect the actual
polarization of microwaves, thus seriously complicating the study of the
polarization dependence of MIRO. Devising an experimental setup that would be
free from these drawbacks remains a challenge for future work.

\subsubsection{Nonequilibrium magnetotransport in the integer and fractional
quantum Hall regime}
\label{s7.2.5}

It is natural to ask whether the phenomena described in this review have their
counterparts in the QH regime, both on a QH plateau and near
a QH transition. In this respect, it is worth recalling another
nonequilibrium phenomenon---the breakdown of the QH effect at a
high current density---that has been extensively studied, see
\textcite{nachtwei:1999} for a review. While this is an essentially different
phenomenon, certain aspects bear similarity with the MIRO and HIRO and, especially,
ZRS and ZdRS physics (the nonequilibrium phase transition between states with zero
and finite resistivity in a transverse magnetic field; the inhomogeneous spatial
distribution of the electric field and current; the importance of relaxation
processes; etc.).
% RR
Recent experiments \cite{stellmach:2007,nachtwei:2008} addressed
the photoresponse of QH structures on both sides of the breakdown bias.
It remains to be seen whether there is a deeper connection
between the two classes of phenomena.

It would be further highly interesting to observe phenomena analogous to those
reviewed here near half-filling of the lowest LL, where transport is
associated with composite fermions (CFs) \cite{heinonen:1998,jain:2007}. A possibility
to observe MIRO of CFs was discussed by \textcite{park:2004}.
There is, however, a serious difficulty related to the fact that
CFs are much more strongly scattered by disorder than electrons. The
dominant type of disorder for CFs is an effective random magnetic
field produced by the random scalar potential through the flux attachment
\cite{halperin:1993}. The damping of quantum magnetooscillations induced by the
random magnetic field is much stronger than the one ($\propto
e^{-\pi/\omega_c\tq}$) characteristic of the regime of low magnetic fields
\cite{aronov:1995,mirlin:1996,mirlin:1998b,evers:1999}. This appears to be an
essential obstacle for the observation of pronounced MIRO or HIRO in this
regime. On the other hand, while the DOS oscillations around $\nu=1/2$ are very
strongly damped, one does observe conductivity oscillations which are of a
different origin: they are induced by quantum localization of CFs,
leading eventually to the fractional QH effect
\cite{mirlin:1998b,evers:1999}. An intriguing question, deserving theoretical
and experimental investigation, is whether these oscillations would be
sufficient to induce MIRO or HIRO under nonequilibrium conditions.

\subsection{Nonequilibrium magnetotransport phenomena in other 2D systems}
\label{s7.3}

\subsubsection{Electrons on liquid helium}
\label{s7.3.1}

Electrons on a surface of liquid helium represent a remarkable 2D electronic
system with parameters strongly different from those of 2DEGs in semiconductor
structures, see \textcite{grimes:1978,andrei:1997,monarkha:2004} for reviews. First, because
of a very low density, correlations play an essential role: in the
zero-temperature limit this system would be a Wigner crystal. Second, the
electron gas on helium becomes nondegenerate already for very low temperatures.
Finally, relaxation times at Kelvin temperatures are several orders of magnitude
longer than in semiconductor 2DEGs. Quasiclassical magnetotransport in this
system has been studied both experimentally and theoretically during the last
two decades \cite{dykman:1993,dykman:1997,lea:1998,dykman:2003,monarkha:2004}. The focus of a
more recent activity was on the photoresponse of the electron fluid. A
characteristic feature of the photoresponse is resonant absorption at a
frequency of order $100\,{\rm GHz}$
\cite{konstantinov:2007,konstantinov:2008a,konstantinov:2008b,konstantinov:2009a,konstantinov:2012b},
corresponding to the transition from the lowest to the first excited subband
(i.e., the first excited state for motion in the direction transverse to the
surface). It is worth noting that the conductivity of electrons on the
helium surface is measured in a contactless way, by capacitively coupling the 2D
electron system to electrodes and measuring the complex conductance (admittance)
of the circuit.

In recent experiments \cite{konstantinov:2009b,konstantinov:2010}, the
resonant intersubband excitation measurements were performed in a transverse
magnetic field. Remarkably, the system showed $1/B$-periodic magnetoconductance
oscillations and zero-conductance states (the experiments were performed in the
Corbino-disk geometry), similar to those found in the semiconductor 2DEG
(\rsecs{s3}{s4}). The similarity includes the same periodicity and phase, as
well as a strong temperature dependence. Further, \textcite{konstantinov:2011}
reported spontaneous currents (in the absence of driving voltage) in the regime
of zero-conductance states. An explanation of these oscillations put forward by
\textcite{konstantinov:2009b,konstantinov:2010,monarkha:2011,monarkha:2011b} is analogous to
the MIRO theory (Sec.\ref{s3}) and combines features of both the inelastic and
displacement mechanisms. Specifically, it was proposed that the resonant
microwave absorption leads to a strongly nonequilibrium occupation of the first
excited subband, which is followed by a quasielastic transition from the lowest
LL of the first excited subband to one of higher LLs of the lowest subband.

The above observations appear truly remarkable as they show that the
nonequilibrium phenomena in high LLs, discussed in the review, are also relevant
to strongly interacting and nondegenerate 2D electron systems. On the other
hand, a systematic theory of the nonequilibrium magnetotransport phenomena in
the electron liquids on helium remains to be developed. In particular, one has
to clarify the role of resonant absorption to the excited subband, the relative
weight of the contributions of different mechanisms, relevant relaxation
processes, etc.

\subsubsection{Graphene}
\label{s7.3.2}

The experimental breakthrough in fabrication of graphene \cite{novoselov:2004}
has brought a new 2D system with fascinating properties into condensed-matter
physics \cite{geim:2007}. The hallmark of graphene is the massless Dirac-fermion
character of charge carriers. It manifests itself, in particular, in the
unconventional QH effect in both graphene
\cite{novoselov:2005,zhang:2005} and graphene bilayers
\cite{novoselov:2006,mccann:2006}. The Dirac spectrum implies also a linearly
vanishing DOS at the neutrality point, i.e., the effective mass vanishes as
$\sqrt{|n_e|}$ as a function of the electron concentration or, equivalently, of
the gate voltage. As a result, the LLs are not equidistant, with the distance
between them (cyclotron frequency) being strongly enhanced near the Dirac point.
One of the manifestations of this is the observation of the QH effect
up to room temperature \cite{novoselov:2007}. For a recent review of the
electronic properties of graphene see \cite{castro_neto:2009}. One of the most
active directions in the graphene research is devoted to the optoelectronic
properties of graphene, including a variety of prospective applications
(photodetectors, light emitters, solar cells, lasers, etc)
\cite{mueller:2010,bonaccorso:2010}.

Spectroscopy studies of graphene in a transverse magnetic field confirmed the
Dirac-fermion LL spectrum \cite{sadowski:2006}. \citet{neugebauer:2009}
presented a spectroscopic manifestation of the exceptionally high purity of
graphene samples: the LL quantization was observed at $B$ as low as  $1\,{\rm
mT}$ and $T$ up to $50\,{\rm K}$. For a review of LL spectroscopy of graphene
see \citet{orlita:2010}.

The properties of graphene make it a very interesting and promising system for
studying nonequilibrium physics in quantizing magnetic fields. Several
differences compared to the semiconductor systems may be foreseen. First,
according to the above, one can expect that in high-quality graphene samples
with the chemical potential located near the Dirac point the quantum phenomena
will become pronounced at relatively high $T$ and relatively low $B$. Second,
the nonequidistant character of LLs may lead to a modification of the functional
dependence of magnetooscillations on $T$ and the strength of disorder. In fact,
a first experiment in this direction appeared recently: \textcite{tan:2011}
studied SdH oscillations in graphene under dc bias and observed a phase
inversion analogous to that found in semiconductor structures, see
Sec.~\ref{s7.2.3}

\section{Acknowledgments}
\label{s8}

We are grateful to Y. Adamov, I.L.~Aleiner, H.-S.~Chiang, S.I.~Dorozhkin, R.R.~Du, A.L.~Efros, F.~Evers, O.M.~Fedorych,
R.~Gellmann, I.V.~Gornyi, J.A.~Gupta, A.T.~Hatke, M.~Khodas, T.A.~Knuuttila, F.~von~Oppen, I.V.~Pechenezhskii,
L.N.~Pfeiffer, I.V.~Ponomarev, M.~Potemski, J.L.~Reno, J.A.~Simmons, S.A.~Studenikin, M.G.~Vavilov, Z.R.~Wasilevski,
K.W.~West, J.~Wilke, P.~W\"olfle, C.L.~Yang, Z.Q.~Yuan, and W.~Zhang for fruitful collaborations on the topics included
in this review. We also thank many of our colleagues, in particular, A.~Auerbach, A.A.~Bykov, K.~von~Klitzing, I.V.~Kukushkin,
R.G.~Mani, O.E.~Raichev, M.E.~Raikh, V.I.~Ryzhii, B.I.~Shklovskii, J.H.~Smet, R.A.~Suris, S.A.~Vitkalov, V.A.~Volkov,
and R.L.~Willett, for illuminating discussions over the past years.
This work was supported by DFG, DFG-CFN, DFG-RFBR, DFG-SPP "Halbleiter Spintronik", DOE Grant No. DE-SC002567, and NSF Grant No. DMR-0548014.

{ \it Note added in proof.}---
Several important developments occurred after the submission of this review. References to some of them are added to the main text, others are listed below.

{\begin{itemize}

\item In a series of papers \cite{dmitriev:2012,dietrich:2012a,dietrich:2012b,dietrich:2012c},
 the transport properties of
a 2DEG in a GaAs quantum well surrounded by AlAs/GaAs superlattices were
investigated. A peculiar feature of these structures is a substantial
concentration of mobile (but highly resistive) $X$ electrons in the AlAs layers. These layers
serve as additional screening gates and induce a spatially inhomogeneous electron density profile
across the Hall bar when a direct current is driven through the 2DEG. The obtained results include the
suppression of the transport and quantum scattering rates (particularly strong
for the latter) due to the screening by $X$ electrons \cite{dmitriev:2012}, as
well as the observation of a current-induced modulation of SdH oscillations
\cite{dietrich:2012b} and MISO \cite{dietrich:2012c} in the nonlinear regime
resulting from the electron density variation induced by the $X$-electron layer.
Unlike HIRO, the period of these oscillations is independent of $B$.

\item SdH oscillations in ultrahigh mobility quantum wells in tilted magnetic
fields were measured by \onlinecite{hatke:2012c} at very high tilt angles.
The data were explained in terms of the dependence of the
effective mass on the in-plane magnetic-field component $B_\parallel$ due to a finite thickness of the quantum wells. Recall
that the effect of $B_\parallel$ on the scattering rates has also been observed, see Sec.~\ref{s3.3.2prim} and \ref{s5.1}.

\item  A strong photoresistance peak that
originates from the magnetoplasmon resonance (MPR) superimposed on ZRS was
observed by \onlinecite{hatke:2012d}.
The experiment suggests that the contributions of MPR and MIRO to the resistivity
sum up and can be independently controlled by tuning the magnetic
field and microwave frequency. This makes it possible to extract separately the
MPR and MIRO contributions. Similar to the bichromatic MIRO experiments (see Sec.~\ref{s4.1}),
the extrapolation of the extracted individual contributions
to the region where the observed resistance is zero yields a negative resistivity.

\item For electrons on liquid helium (see Sec.~\ref{s7.3.1}), the theory of
microwave-induced magnetoconductance oscillations was further
developed \cite{monarkha:2012} to include strong Coulomb interaction between
electrons. The theory demonstrates the dramatic effect that slow thermal fluctuations
of the electric field in the electron liquid have on inter-LL scattering.
The theoretical results agree well with the experimental findings by \onlinecite{konstantinov:2009b,konstantinov:2010}.
\end{itemize}}


\begin{thebibliography}{380}
\expandafter\ifx\csname natexlab\endcsname\relax\def\natexlab#1{#1}\fi
\expandafter\ifx\csname bibnamefont\endcsname\relax
  \def\bibnamefont#1{#1}\fi
\expandafter\ifx\csname bibfnamefont\endcsname\relax
  \def\bibfnamefont#1{#1}\fi
\expandafter\ifx\csname citenamefont\endcsname\relax
  \def\citenamefont#1{#1}\fi
\expandafter\ifx\csname url\endcsname\relax
  \def\url#1{\texttt{#1}}\fi
\expandafter\ifx\csname urlprefix\endcsname\relax\def\urlprefix{URL }\fi
\providecommand{\bibinfo}[2]{#2}
\providecommand{\eprint}[2][]{\url{#2}}

\bibitem[{\citenamefont{Abrahams} \emph{et~al.}(2001)\citenamefont{Abrahams,
  Kravchenko, and Sarachik}}]{abrahams:2001}
\bibinfo{author}{\bibnamefont{Abrahams}, \bibfnamefont{E.}},
  \bibinfo{author}{\bibfnamefont{S.~V.} \bibnamefont{Kravchenko}}, and
  \bibinfo{author}{\bibfnamefont{M.~P.} \bibnamefont{Sarachik}},
  \bibinfo{year}{2001}, \bibinfo{journal}{Rev. Mod. Phys.}
  \textbf{\bibinfo{volume}{73}}, \bibinfo{pages}{251}.

\bibitem[{\citenamefont{Abstreiter}
  \emph{et~al.}(1976)\citenamefont{Abstreiter, Kotthaus, Koch, and
  Dorda}}]{abstreiter:1976}
\bibinfo{author}{\bibnamefont{Abstreiter}, \bibfnamefont{G.}},
  \bibinfo{author}{\bibfnamefont{J.~P.} \bibnamefont{Kotthaus}},
  \bibinfo{author}{\bibfnamefont{J.~F.} \bibnamefont{Koch}}, and
  \bibinfo{author}{\bibfnamefont{G.}~\bibnamefont{Dorda}},
  \bibinfo{year}{1976}, \bibinfo{journal}{Phys. Rev. B}
  \textbf{\bibinfo{volume}{14}}, \bibinfo{pages}{2480}.

\bibitem[{\citenamefont{Adamov} \emph{et~al.}(2006)\citenamefont{Adamov,
  Gornyi, and Mirlin}}]{adamov:2006}
\bibinfo{author}{\bibnamefont{Adamov}, \bibfnamefont{Y.}},
  \bibinfo{author}{\bibfnamefont{I.~V.} \bibnamefont{Gornyi}}, and
  \bibinfo{author}{\bibfnamefont{A.~D.} \bibnamefont{Mirlin}},
  \bibinfo{year}{2006}, \bibinfo{journal}{Phys. Rev. B}
  \textbf{\bibinfo{volume}{73}}, \bibinfo{pages}{045426}.

\bibitem[{\citenamefont{Alferov}(2001)}]{alferov:2001}
\bibinfo{author}{\bibnamefont{Alferov}, \bibfnamefont{Z.~I.}},
  \bibinfo{year}{2001}, \bibinfo{journal}{Rev. Mod. Phys.}
  \textbf{\bibinfo{volume}{73}}, \bibinfo{pages}{767}.

\bibitem[{\citenamefont{Alicea} \emph{et~al.}(2005)\citenamefont{Alicea,
  Balents, Fisher, Paramekanti, and Radzihovsky}}]{alicea:2005}
\bibinfo{author}{\bibnamefont{Alicea}, \bibfnamefont{J.}},
  \bibinfo{author}{\bibfnamefont{L.}~\bibnamefont{Balents}},
  \bibinfo{author}{\bibfnamefont{M.~P.~A.} \bibnamefont{Fisher}},
  \bibinfo{author}{\bibfnamefont{A.}~\bibnamefont{Paramekanti}}, and
  \bibinfo{author}{\bibfnamefont{L.}~\bibnamefont{Radzihovsky}},
  \bibinfo{year}{2005}, \bibinfo{journal}{Phys. Rev. B}
  \textbf{\bibinfo{volume}{71}}, \bibinfo{pages}{235322}.

\bibitem[{\citenamefont{Altshuler and Aronov}(1985)}]{altshuler:1985}
\bibinfo{author}{\bibnamefont{Altshuler}, \bibfnamefont{B.~L.}}, and
  \bibinfo{author}{\bibfnamefont{A.~G.} \bibnamefont{Aronov}},
  \bibinfo{year}{1985}, in \emph{\bibinfo{booktitle}{Electron-Electron
  Interactions in Disordered Systems}}, edited by
  \bibinfo{editor}{\bibfnamefont{A.~L.} \bibnamefont{Efros}} and
  \bibinfo{editor}{\bibfnamefont{M.}~\bibnamefont{Pollak}}
  (\bibinfo{publisher}{North-Holland, Amsterdam}).

\bibitem[{\citenamefont{Anderson and Brinkman}(2003)}]{anderson:2003}
\bibinfo{author}{\bibnamefont{Anderson}, \bibfnamefont{P.~W.}}, and
  \bibinfo{author}{\bibfnamefont{W.~F.} \bibnamefont{Brinkman}},
  \bibinfo{year}{2003}, \bibinfo{journal}{arXiv:cond-mat/0302129} .

\bibitem[{\citenamefont{Ando}(1974{\natexlab{a}})}]{ando:1974a}
\bibinfo{author}{\bibnamefont{Ando}, \bibfnamefont{T.}},
  \bibinfo{year}{1974}{\natexlab{a}}, \bibinfo{journal}{J. Phys. Soc. Jpn.}
  \textbf{\bibinfo{volume}{36}}, \bibinfo{pages}{1521}.

\bibitem[{\citenamefont{Ando}(1974{\natexlab{b}})}]{ando:1974b}
\bibinfo{author}{\bibnamefont{Ando}, \bibfnamefont{T.}},
  \bibinfo{year}{1974}{\natexlab{b}}, \bibinfo{journal}{J. Phys. Soc. Jpn.}
  \textbf{\bibinfo{volume}{37}}, \bibinfo{pages}{1233}.

\bibitem[{\citenamefont{Ando}(1975)}]{ando:1975a}
\bibinfo{author}{\bibnamefont{Ando}, \bibfnamefont{T.}}, \bibinfo{year}{1975},
  \bibinfo{journal}{J. Phys. Soc. Jpn.} \textbf{\bibinfo{volume}{38}},
  \bibinfo{pages}{989}.

\bibitem[{\citenamefont{Ando}(1982)}]{ando:1982}
\bibinfo{author}{\bibnamefont{Ando}, \bibfnamefont{T.}}, \bibinfo{year}{1982},
  \bibinfo{journal}{J. Phys. Soc. Jpn.} \textbf{\bibinfo{volume}{51}},
  \bibinfo{pages}{3900}.

\bibitem[{\citenamefont{Ando} \emph{et~al.}(1982)\citenamefont{Ando, Fowler,
  and Stern}}]{ando:1982a}
\bibinfo{author}{\bibnamefont{Ando}, \bibfnamefont{T.}},
  \bibinfo{author}{\bibfnamefont{A.~B.} \bibnamefont{Fowler}}, and
  \bibinfo{author}{\bibfnamefont{F.}~\bibnamefont{Stern}},
  \bibinfo{year}{1982}, \bibinfo{journal}{Rev. Mod. Phys.}
  \textbf{\bibinfo{volume}{54}}, \bibinfo{pages}{437}.

\bibitem[{\citenamefont{Ando} \emph{et~al.}(1975)\citenamefont{Ando, Matsumoto,
  and Uemura}}]{ando:1975}
\bibinfo{author}{\bibnamefont{Ando}, \bibfnamefont{T.}},
  \bibinfo{author}{\bibfnamefont{Y.}~\bibnamefont{Matsumoto}}, and
  \bibinfo{author}{\bibfnamefont{Y.}~\bibnamefont{Uemura}},
  \bibinfo{year}{1975}, \bibinfo{journal}{J. Phys. Soc. Jpn.}
  \textbf{\bibinfo{volume}{39}}, \bibinfo{pages}{279}.

\bibitem[{\citenamefont{Ando and Uemura}(1974)}]{ando:1974}
\bibinfo{author}{\bibnamefont{Ando}, \bibfnamefont{T.}}, and
  \bibinfo{author}{\bibfnamefont{Y.}~\bibnamefont{Uemura}},
  \bibinfo{year}{1974}, \bibinfo{journal}{J. Phys. Soc. Jpn.}
  \textbf{\bibinfo{volume}{36}}, \bibinfo{pages}{959}.

\bibitem[{\citenamefont{Andreev} \emph{et~al.}(2003)\citenamefont{Andreev,
  Aleiner, and Millis}}]{andreev:2003}
\bibinfo{author}{\bibnamefont{Andreev}, \bibfnamefont{A.~V.}},
  \bibinfo{author}{\bibfnamefont{I.~L.} \bibnamefont{Aleiner}}, and
  \bibinfo{author}{\bibfnamefont{A.~J.} \bibnamefont{Millis}},
  \bibinfo{year}{2003}, \bibinfo{journal}{Phys. Rev. Lett.}
  \textbf{\bibinfo{volume}{91}}, \bibinfo{pages}{056803}.

\bibitem[{\citenamefont{Andreev} \emph{et~al.}(2011)\citenamefont{Andreev,
  Muravev, Kukushkin, Schmult, and Dietsche}}]{andreev:2011}
\bibinfo{author}{\bibnamefont{Andreev}, \bibfnamefont{I.~V.}},
  \bibinfo{author}{\bibfnamefont{V.~M.} \bibnamefont{Muravev}},
  \bibinfo{author}{\bibfnamefont{I.~V.} \bibnamefont{Kukushkin}},
  \bibinfo{author}{\bibfnamefont{S.}~\bibnamefont{Schmult}}, and
  \bibinfo{author}{\bibfnamefont{W.}~\bibnamefont{Dietsche}},
  \bibinfo{year}{2011}, \bibinfo{journal}{Phys. Rev. B}
  \textbf{\bibinfo{volume}{83}}, \bibinfo{pages}{121308(R)}.

\bibitem[{\citenamefont{Andrei}(1997)}]{andrei:1997}
\bibinfo{editor}{\bibnamefont{Andrei}, \bibfnamefont{E.~Y.}} (ed.),
  \bibinfo{year}{1997}, \emph{\bibinfo{title}{Two-dimensional electron systems
  on helium and other cryogenic substrates}} (\bibinfo{publisher}{Springer
  Verlag}).

\bibitem[{\citenamefont{Aronov} \emph{et~al.}(1995)\citenamefont{Aronov,
  Altshuler, Mirlin, and W\"olfle}}]{aronov:1995}
\bibinfo{author}{\bibnamefont{Aronov}, \bibfnamefont{A.~G.}},
  \bibinfo{author}{\bibfnamefont{E.}~\bibnamefont{Altshuler}},
  \bibinfo{author}{\bibfnamefont{A.~D.} \bibnamefont{Mirlin}}, and
  \bibinfo{author}{\bibfnamefont{P.}~\bibnamefont{W\"olfle}},
  \bibinfo{year}{1995}, \bibinfo{journal}{Europhys. Lett.}
  \textbf{\bibinfo{volume}{29}}, \bibinfo{pages}{239}.

\bibitem[{\citenamefont{Aronov and Spivak}(1975)}]{aronov:1975}
\bibinfo{author}{\bibnamefont{Aronov}, \bibfnamefont{A.~G.}}, and
  \bibinfo{author}{\bibfnamefont{B.~Z.} \bibnamefont{Spivak}},
  \bibinfo{year}{1975}, \bibinfo{journal}{JETP Lett.}
  \textbf{\bibinfo{volume}{22}}, \bibinfo{pages}{101}.

\bibitem[{\citenamefont{Auerbach} \emph{et~al.}(2005)\citenamefont{Auerbach,
  Finkler, Halperin, and Yacoby}}]{auerbach:2005}
\bibinfo{author}{\bibnamefont{Auerbach}, \bibfnamefont{A.}},
  \bibinfo{author}{\bibfnamefont{I.}~\bibnamefont{Finkler}},
  \bibinfo{author}{\bibfnamefont{B.~I.} \bibnamefont{Halperin}}, and
  \bibinfo{author}{\bibfnamefont{A.}~\bibnamefont{Yacoby}},
  \bibinfo{year}{2005}, \bibinfo{journal}{Phys. Rev. Lett.}
  \textbf{\bibinfo{volume}{94}}, \bibinfo{pages}{196801}.

\bibitem[{\citenamefont{Auerbach and Pai}(2007)}]{auerbach:2007}
\bibinfo{author}{\bibnamefont{Auerbach}, \bibfnamefont{A.}}, and
  \bibinfo{author}{\bibfnamefont{G.~V.} \bibnamefont{Pai}},
  \bibinfo{year}{2007}, \bibinfo{journal}{Phys. Rev. B}
  \textbf{\bibinfo{volume}{76}}, \bibinfo{pages}{205318}.

\bibitem[{\citenamefont{Averkiev} \emph{et~al.}(2001)\citenamefont{Averkiev,
  Golub, Tarasenko, and Willander}}]{averkiev:2001}
\bibinfo{author}{\bibnamefont{Averkiev}, \bibfnamefont{N.~S.}},
  \bibinfo{author}{\bibfnamefont{L.~E.} \bibnamefont{Golub}},
  \bibinfo{author}{\bibfnamefont{S.~A.} \bibnamefont{Tarasenko}}, and
  \bibinfo{author}{\bibfnamefont{M.}~\bibnamefont{Willander}},
  \bibinfo{year}{2001}, \bibinfo{journal}{J. Phys.: Condens. Matter}
  \textbf{\bibinfo{volume}{13}}, \bibinfo{pages}{2517}.

\bibitem[{\citenamefont{Banis} \emph{et~al.}(1972)\citenamefont{Banis,
  Parshelyunas, and Pozhela}}]{banis:1972}
\bibinfo{author}{\bibnamefont{Banis}, \bibfnamefont{T.~Y.}},
  \bibinfo{author}{\bibfnamefont{I.~V.} \bibnamefont{Parshelyunas}}, and
  \bibinfo{author}{\bibfnamefont{Y.~K.} \bibnamefont{Pozhela}},
  \bibinfo{year}{1972}, \bibinfo{journal}{Sov. Phys. Semicond.}
  \textbf{\bibinfo{volume}{5}}, \bibinfo{pages}{1727}.

\bibitem[{\citenamefont{Baskin} \emph{et~al.}(1978)\citenamefont{Baskin,
  Magarill, and Entin}}]{baskin:1978}
\bibinfo{author}{\bibnamefont{Baskin}, \bibfnamefont{E.~M.}},
  \bibinfo{author}{\bibfnamefont{L.~N.} \bibnamefont{Magarill}}, and
  \bibinfo{author}{\bibfnamefont{M.~V.} \bibnamefont{Entin}},
  \bibinfo{year}{1978}, \bibinfo{journal}{Sov. Phys. JETP}
  \textbf{\bibinfo{volume}{48}}, \bibinfo{pages}{365}.

\bibitem[{\citenamefont{Bass}(1965)}]{bass:1965}
\bibinfo{author}{\bibnamefont{Bass}, \bibfnamefont{F.~G.}},
  \bibinfo{year}{1965}, \bibinfo{journal}{Sov. Phys. JETP}
  \textbf{\bibinfo{volume}{21}}, \bibinfo{pages}{181}.

\bibitem[{\citenamefont{Basun} \emph{et~al.}(1983)\citenamefont{Basun,
  Kaplyanskii, and Feofilov}}]{basun:1983}
\bibinfo{author}{\bibnamefont{Basun}, \bibfnamefont{S.~A.}},
  \bibinfo{author}{\bibfnamefont{A.~A.} \bibnamefont{Kaplyanskii}}, and
  \bibinfo{author}{\bibfnamefont{S.~P.} \bibnamefont{Feofilov}},
  \bibinfo{year}{1983}, \bibinfo{journal}{JETP Lett.}
  \textbf{\bibinfo{volume}{37}}, \bibinfo{pages}{586}.

\bibitem[{\citenamefont{Beenakker}(1989)}]{beenakker:1989}
\bibinfo{author}{\bibnamefont{Beenakker}, \bibfnamefont{C.~W.~J.}},
  \bibinfo{year}{1989}, \bibinfo{journal}{Phys. Rev. Lett.}
  \textbf{\bibinfo{volume}{62}}, \bibinfo{pages}{2020}.

\bibitem[{\citenamefont{Beenakker and van Houten}(1991)}]{beenakker:1991}
\bibinfo{author}{\bibnamefont{Beenakker}, \bibfnamefont{C.~W.~J.}}, and
  \bibinfo{author}{\bibfnamefont{H.}~\bibnamefont{van Houten}},
  \bibinfo{year}{1991} (\bibinfo{publisher}{North-Holland, Amsterdam}),
  volume~\bibinfo{volume}{44} of \emph{\bibinfo{series}{Solid State Physics}},
  p.~\bibinfo{pages}{1}.

\bibitem[{\citenamefont{Benedict}(1987)}]{benedict:1987}
\bibinfo{author}{\bibnamefont{Benedict}, \bibfnamefont{K.~A.}},
  \bibinfo{year}{1987}, \bibinfo{journal}{Nucl. Phys. B}
  \textbf{\bibinfo{volume}{280}}, \bibinfo{pages}{549}.

\bibitem[{\citenamefont{Benedict and Chalker}(1986)}]{benedict:1986}
\bibinfo{author}{\bibnamefont{Benedict}, \bibfnamefont{K.~A.}}, and
  \bibinfo{author}{\bibfnamefont{J.~T.} \bibnamefont{Chalker}},
  \bibinfo{year}{1986}, \bibinfo{journal}{J. Phys. C}
  \textbf{\bibinfo{volume}{19}}, \bibinfo{pages}{3587}.

\bibitem[{\citenamefont{Bergeret} \emph{et~al.}(2003)\citenamefont{Bergeret,
  Huckestein, and Volkov}}]{bergeret:2003}
\bibinfo{author}{\bibnamefont{Bergeret}, \bibfnamefont{F.~S.}},
  \bibinfo{author}{\bibfnamefont{B.}~\bibnamefont{Huckestein}}, and
  \bibinfo{author}{\bibfnamefont{A.~F.} \bibnamefont{Volkov}},
  \bibinfo{year}{2003}, \bibinfo{journal}{Phys. Rev. B}
  \textbf{\bibinfo{volume}{67}}, \bibinfo{pages}{241303(R)}.

\bibitem[{\citenamefont{Bobylev} \emph{et~al.}(1995)\citenamefont{Bobylev,
  Maa{\o}, Hansen, and Hauge}}]{bobylev:1995}
\bibinfo{author}{\bibnamefont{Bobylev}, \bibfnamefont{A.~V.}},
  \bibinfo{author}{\bibfnamefont{F.~A.} \bibnamefont{Maa{\o}}},
  \bibinfo{author}{\bibfnamefont{A.}~\bibnamefont{Hansen}}, and
  \bibinfo{author}{\bibfnamefont{E.~H.} \bibnamefont{Hauge}},
  \bibinfo{year}{1995}, \bibinfo{journal}{Phys. Rev. Lett.}
  \textbf{\bibinfo{volume}{75}}, \bibinfo{pages}{197}.

\bibitem[{\citenamefont{Bockhorn} \emph{et~al.}(2011)\citenamefont{Bockhorn,
  Barthold, Schuh, Wegscheider, and Haug}}]{bockhorn:2011}
\bibinfo{author}{\bibnamefont{Bockhorn}, \bibfnamefont{L.}},
  \bibinfo{author}{\bibfnamefont{P.}~\bibnamefont{Barthold}},
  \bibinfo{author}{\bibfnamefont{D.}~\bibnamefont{Schuh}},
  \bibinfo{author}{\bibfnamefont{W.}~\bibnamefont{Wegscheider}}, and
  \bibinfo{author}{\bibfnamefont{R.~J.} \bibnamefont{Haug}},
  \bibinfo{year}{2011}, \bibinfo{journal}{Phys. Rev. B}
  \textbf{\bibinfo{volume}{83}}, \bibinfo{pages}{113301}.

\bibitem[{\citenamefont{Bogomolov} \emph{et~al.}(1967)\citenamefont{Bogomolov,
  Shulman, Aronov, and Pikus}}]{bogomolov:1967}
\bibinfo{author}{\bibnamefont{Bogomolov}, \bibfnamefont{V.~N.}},
  \bibinfo{author}{\bibfnamefont{S.~G.} \bibnamefont{Shulman}},
  \bibinfo{author}{\bibfnamefont{A.~G.} \bibnamefont{Aronov}}, and
  \bibinfo{author}{\bibfnamefont{G.~E.} \bibnamefont{Pikus}},
  \bibinfo{year}{1967}, \bibinfo{journal}{JETP Lett.}
  \textbf{\bibinfo{volume}{5}}, \bibinfo{pages}{169}.

\bibitem[{\citenamefont{Bonaccorso}
  \emph{et~al.}(2010)\citenamefont{Bonaccorso, Sun, Hasan, and
  Ferrari}}]{bonaccorso:2010}
\bibinfo{author}{\bibnamefont{Bonaccorso}, \bibfnamefont{F.}},
  \bibinfo{author}{\bibfnamefont{Z.}~\bibnamefont{Sun}},
  \bibinfo{author}{\bibfnamefont{T.}~\bibnamefont{Hasan}}, and
  \bibinfo{author}{\bibfnamefont{A.~C.} \bibnamefont{Ferrari}},
  \bibinfo{year}{2010}, \bibinfo{journal}{Nature Phot.}
  \textbf{\bibinfo{volume}{4}}, \bibinfo{pages}{611}.

\bibitem[{\citenamefont{Bonch-Bruevich and Kogan}(1965)}]{bonch-bruevich:1965}
\bibinfo{author}{\bibnamefont{Bonch-Bruevich}, \bibfnamefont{V.~L.}}, and
  \bibinfo{author}{\bibfnamefont{S.~M.} \bibnamefont{Kogan}},
  \bibinfo{year}{1965}, \bibinfo{journal}{Sov. Phys. Solid State}
  \textbf{\bibinfo{volume}{7}}, \bibinfo{pages}{15}.

\bibitem[{\citenamefont{Bonch-Bruevich}
  \emph{et~al.}(1975)\citenamefont{Bonch-Bruevich, Zvyagin, and
  Mironov}}]{bonch-bruevich:1975}
\bibinfo{author}{\bibnamefont{Bonch-Bruevich}, \bibfnamefont{V.~L.}},
  \bibinfo{author}{\bibfnamefont{I.~P.} \bibnamefont{Zvyagin}}, and
  \bibinfo{author}{\bibfnamefont{A.~G.} \bibnamefont{Mironov}},
  \bibinfo{year}{1975}, \emph{\bibinfo{title}{Domain electrical instabilities
  in semiconductors}} (\bibinfo{publisher}{Consultants Bureau, New York}).

\bibitem[{\citenamefont{Br{\'e}zin}
  \emph{et~al.}(1984)\citenamefont{Br{\'e}zin, Gross, and
  Itzykson}}]{brezin:1984}
\bibinfo{author}{\bibnamefont{Br{\'e}zin}, \bibfnamefont{E.}},
  \bibinfo{author}{\bibfnamefont{D.~J.} \bibnamefont{Gross}}, and
  \bibinfo{author}{\bibfnamefont{C.}~\bibnamefont{Itzykson}},
  \bibinfo{year}{1984}, \bibinfo{journal}{Nucl. Phys. B}
  \textbf{\bibinfo{volume}{235}}, \bibinfo{pages}{24}.

\bibitem[{\citenamefont{Buks}
  \emph{et~al.}(1994{\natexlab{a}})\citenamefont{Buks, Heiblum, Levinson, and
  Shtrikman}}]{buks:1994a}
\bibinfo{author}{\bibnamefont{Buks}, \bibfnamefont{E.}},
  \bibinfo{author}{\bibfnamefont{M.}~\bibnamefont{Heiblum}},
  \bibinfo{author}{\bibfnamefont{Y.}~\bibnamefont{Levinson}}, and
  \bibinfo{author}{\bibfnamefont{H.}~\bibnamefont{Shtrikman}},
  \bibinfo{year}{1994}{\natexlab{a}}, \bibinfo{journal}{Semicond. Sci.
  Technol.} \textbf{\bibinfo{volume}{9}}, \bibinfo{pages}{2031}.

\bibitem[{\citenamefont{Buks}
  \emph{et~al.}(1994{\natexlab{b}})\citenamefont{Buks, Heiblum, and
  Shtrikman}}]{buks:1994}
\bibinfo{author}{\bibnamefont{Buks}, \bibfnamefont{E.}},
  \bibinfo{author}{\bibfnamefont{M.}~\bibnamefont{Heiblum}}, and
  \bibinfo{author}{\bibfnamefont{H.}~\bibnamefont{Shtrikman}},
  \bibinfo{year}{1994}{\natexlab{b}}, \bibinfo{journal}{Phys. Rev. B}
  \textbf{\bibinfo{volume}{49}}, \bibinfo{pages}{14790(R)}.

\bibitem[{\citenamefont{Bykov}(2008{\natexlab{a}})}]{bykov:2008d}
\bibinfo{author}{\bibnamefont{Bykov}, \bibfnamefont{A.~A.}},
  \bibinfo{year}{2008}{\natexlab{a}}, \bibinfo{journal}{JETP Lett.}
  \textbf{\bibinfo{volume}{87}}, \bibinfo{pages}{233}.

\bibitem[{\citenamefont{Bykov}(2008{\natexlab{b}})}]{bykov:2008b}
\bibinfo{author}{\bibnamefont{Bykov}, \bibfnamefont{A.~A.}},
  \bibinfo{year}{2008}{\natexlab{b}}, \bibinfo{journal}{JETP Lett.}
  \textbf{\bibinfo{volume}{88}}, \bibinfo{pages}{64}.

\bibitem[{\citenamefont{Bykov}(2008{\natexlab{c}})}]{bykov:2008a}
\bibinfo{author}{\bibnamefont{Bykov}, \bibfnamefont{A.~A.}},
  \bibinfo{year}{2008}{\natexlab{c}}, \bibinfo{journal}{JETP Lett.}
  \textbf{\bibinfo{volume}{88}}, \bibinfo{pages}{394}.

\bibitem[{\citenamefont{Bykov}(2010)}]{bykov:2010b}
\bibinfo{author}{\bibnamefont{Bykov}, \bibfnamefont{A.~A.}},
  \bibinfo{year}{2010}, \bibinfo{journal}{JETP Lett.}
  \textbf{\bibinfo{volume}{91}}, \bibinfo{pages}{361}.

\bibitem[{\citenamefont{Bykov} \emph{et~al.}(2012)\citenamefont{Bykov,
  Dmitriev, Marchishin, Byrnes, and Vitkalov}}]{bykov:2012}
\bibinfo{author}{\bibnamefont{Bykov}, \bibfnamefont{A.~A.}},
  \bibinfo{author}{\bibfnamefont{D.~V.} \bibnamefont{Dmitriev}},
  \bibinfo{author}{\bibfnamefont{I.}~\bibnamefont{Marchishin}},
  \bibinfo{author}{\bibfnamefont{S.}~\bibnamefont{Byrnes}}, and
  \bibinfo{author}{\bibfnamefont{S.~A.} \bibnamefont{Vitkalov}},
  \bibinfo{year}{2012}, \bibinfo{journal}{Appl. Phys. Lett.}
  \textbf{\bibinfo{volume}{100}}, \bibinfo{pages}{251602}.

\bibitem[{\citenamefont{Bykov and Goran}(2009)}]{bykov:2009a}
\bibinfo{author}{\bibnamefont{Bykov}, \bibfnamefont{A.~A.}}, and
  \bibinfo{author}{\bibfnamefont{A.~V.} \bibnamefont{Goran}},
  \bibinfo{year}{2009}, \bibinfo{journal}{JETP Lett.}
  \textbf{\bibinfo{volume}{90}}, \bibinfo{pages}{578}.

\bibitem[{\citenamefont{Bykov}
  \emph{et~al.}(2010{\natexlab{a}})\citenamefont{Bykov, Goran, and
  Vitkalov}}]{bykov:2010a}
\bibinfo{author}{\bibnamefont{Bykov}, \bibfnamefont{A.~A.}},
  \bibinfo{author}{\bibfnamefont{A.~V.} \bibnamefont{Goran}}, and
  \bibinfo{author}{\bibfnamefont{S.~A.} \bibnamefont{Vitkalov}},
  \bibinfo{year}{2010}{\natexlab{a}}, \bibinfo{journal}{Phys. Rev. B}
  \textbf{\bibinfo{volume}{81}}, \bibinfo{pages}{155322}.

\bibitem[{\citenamefont{Bykov} \emph{et~al.}(2008)\citenamefont{Bykov, Islamov,
  Goran, and Toropov}}]{bykov:2008e}
\bibinfo{author}{\bibnamefont{Bykov}, \bibfnamefont{A.~A.}},
  \bibinfo{author}{\bibfnamefont{D.~R.} \bibnamefont{Islamov}},
  \bibinfo{author}{\bibfnamefont{A.~V.} \bibnamefont{Goran}}, and
  \bibinfo{author}{\bibfnamefont{A.~I.} \bibnamefont{Toropov}},
  \bibinfo{year}{2008}, \bibinfo{journal}{JETP Lett.}
  \textbf{\bibinfo{volume}{87}}, \bibinfo{pages}{477}.

\bibitem[{\citenamefont{Bykov}
  \emph{et~al.}(2005{\natexlab{a}})\citenamefont{Bykov, Kalagin, and
  Bakarov}}]{bykov:2005b}
\bibinfo{author}{\bibnamefont{Bykov}, \bibfnamefont{A.~A.}},
  \bibinfo{author}{\bibfnamefont{A.~K.} \bibnamefont{Kalagin}}, and
  \bibinfo{author}{\bibfnamefont{A.~K.} \bibnamefont{Bakarov}},
  \bibinfo{year}{2005}{\natexlab{a}}, \bibinfo{journal}{JETP Lett.}
  \textbf{\bibinfo{volume}{81}}, \bibinfo{pages}{523}.

\bibitem[{\citenamefont{Bykov}
  \emph{et~al.}(2010{\natexlab{b}})\citenamefont{Bykov, Marchishin, Goran, and
  Dmitriev}}]{bykov:2010d}
\bibinfo{author}{\bibnamefont{Bykov}, \bibfnamefont{A.~A.}},
  \bibinfo{author}{\bibfnamefont{I.~V.} \bibnamefont{Marchishin}},
  \bibinfo{author}{\bibfnamefont{A.~V.} \bibnamefont{Goran}}, and
  \bibinfo{author}{\bibfnamefont{D.~V.} \bibnamefont{Dmitriev}},
  \bibinfo{year}{2010}{\natexlab{b}}, \bibinfo{journal}{Appl. Phys. Lett.}
  \textbf{\bibinfo{volume}{97}}, \bibinfo{pages}{082107}.

\bibitem[{\citenamefont{Bykov}
  \emph{et~al.}(2010{\natexlab{c}})\citenamefont{Bykov, Mozulev, and
  Kalagin}}]{bykov:2010e}
\bibinfo{author}{\bibnamefont{Bykov}, \bibfnamefont{A.~A.}},
  \bibinfo{author}{\bibfnamefont{E.~G.} \bibnamefont{Mozulev}}, and
  \bibinfo{author}{\bibfnamefont{A.~K.} \bibnamefont{Kalagin}},
  \bibinfo{year}{2010}{\natexlab{c}}, \bibinfo{journal}{JETP Lett.}
  \textbf{\bibinfo{volume}{92}}, \bibinfo{pages}{379}.

\bibitem[{\citenamefont{Bykov}
  \emph{et~al.}(2010{\natexlab{d}})\citenamefont{Bykov, Mozulev, and
  Vitkalov}}]{bykov:2010c}
\bibinfo{author}{\bibnamefont{Bykov}, \bibfnamefont{A.~A.}},
  \bibinfo{author}{\bibfnamefont{E.~G.} \bibnamefont{Mozulev}}, and
  \bibinfo{author}{\bibfnamefont{S.~A.} \bibnamefont{Vitkalov}},
  \bibinfo{year}{2010}{\natexlab{d}}, \bibinfo{journal}{JETP Lett.}
  \textbf{\bibinfo{volume}{92}}, \bibinfo{pages}{475}.

\bibitem[{\citenamefont{Bykov}
  \emph{et~al.}(2005{\natexlab{b}})\citenamefont{Bykov, Zhang, Vitkalov,
  Kalagin, and Bakarov}}]{bykov:2005c}
\bibinfo{author}{\bibnamefont{Bykov}, \bibfnamefont{A.~A.}},
  \bibinfo{author}{\bibfnamefont{J.-Q.} \bibnamefont{Zhang}},
  \bibinfo{author}{\bibfnamefont{S.}~\bibnamefont{Vitkalov}},
  \bibinfo{author}{\bibfnamefont{A.~K.} \bibnamefont{Kalagin}}, and
  \bibinfo{author}{\bibfnamefont{A.~K.} \bibnamefont{Bakarov}},
  \bibinfo{year}{2005}{\natexlab{b}}, \bibinfo{journal}{Phys. Rev. B}
  \textbf{\bibinfo{volume}{72}}, \bibinfo{pages}{245307}.

\bibitem[{\citenamefont{Bykov} \emph{et~al.}(2007)\citenamefont{Bykov, Zhang,
  Vitkalov, Kalagin, and Bakarov}}]{bykov:2007}
\bibinfo{author}{\bibnamefont{Bykov}, \bibfnamefont{A.~A.}},
  \bibinfo{author}{\bibfnamefont{J.-Q.} \bibnamefont{Zhang}},
  \bibinfo{author}{\bibfnamefont{S.}~\bibnamefont{Vitkalov}},
  \bibinfo{author}{\bibfnamefont{A.~K.} \bibnamefont{Kalagin}}, and
  \bibinfo{author}{\bibfnamefont{A.~K.} \bibnamefont{Bakarov}},
  \bibinfo{year}{2007}, \bibinfo{journal}{Phys. Rev. Lett.}
  \textbf{\bibinfo{volume}{99}}, \bibinfo{pages}{116801}.

\bibitem[{\citenamefont{Cannon} \emph{et~al.}(2000)\citenamefont{Cannon,
  Kusmartsev, Alekseev, and Campbell}}]{cannon:2000}
\bibinfo{author}{\bibnamefont{Cannon}, \bibfnamefont{E.~H.}},
  \bibinfo{author}{\bibfnamefont{F.~V.} \bibnamefont{Kusmartsev}},
  \bibinfo{author}{\bibfnamefont{K.~N.} \bibnamefont{Alekseev}}, and
  \bibinfo{author}{\bibfnamefont{D.~K.} \bibnamefont{Campbell}},
  \bibinfo{year}{2000}, \bibinfo{journal}{Phys. Rev. Lett.}
  \textbf{\bibinfo{volume}{85}}, \bibinfo{pages}{1302}.

\bibitem[{\citenamefont{Carra} \emph{et~al.}(1989)\citenamefont{Carra, Chalker,
  and Benedict}}]{carra:1989}
\bibinfo{author}{\bibnamefont{Carra}, \bibfnamefont{P.}},
  \bibinfo{author}{\bibfnamefont{J.~T.} \bibnamefont{Chalker}}, and
  \bibinfo{author}{\bibfnamefont{K.~A.} \bibnamefont{Benedict}},
  \bibinfo{year}{1989}, \bibinfo{journal}{Ann. Phys. (N.Y.)}
  \textbf{\bibinfo{volume}{194}}, \bibinfo{pages}{1}.

\bibitem[{\citenamefont{{Castro Neto}} \emph{et~al.}(2009)\citenamefont{{Castro
  Neto}, Guinea, Peres, Novoselov, and Geim}}]{castro_neto:2009}
\bibinfo{author}{\bibnamefont{{Castro Neto}}, \bibfnamefont{A.~H.}},
  \bibinfo{author}{\bibfnamefont{F.}~\bibnamefont{Guinea}},
  \bibinfo{author}{\bibfnamefont{N.~M.~R.} \bibnamefont{Peres}},
  \bibinfo{author}{\bibfnamefont{K.~S.} \bibnamefont{Novoselov}}, and
  \bibinfo{author}{\bibfnamefont{A.~K.} \bibnamefont{Geim}},
  \bibinfo{year}{2009}, \bibinfo{journal}{Rev. Mod. Phys.}
  \textbf{\bibinfo{volume}{81}}, \bibinfo{pages}{109}.

\bibitem[{\citenamefont{Chaplik}(1971)}]{chaplik:1971}
\bibinfo{author}{\bibnamefont{Chaplik}, \bibfnamefont{A.~V.}},
  \bibinfo{year}{1971}, \bibinfo{journal}{Sov. Phys. JETP}
  \textbf{\bibinfo{volume}{33}}, \bibinfo{pages}{997}.

\bibitem[{\citenamefont{Cheianov} \emph{et~al.}(2004)\citenamefont{Cheianov,
  Dmitriev, and Kachorovskii}}]{cheianov:2004}
\bibinfo{author}{\bibnamefont{Cheianov}, \bibfnamefont{V.~V.}},
  \bibinfo{author}{\bibfnamefont{A.~P.} \bibnamefont{Dmitriev}}, and
  \bibinfo{author}{\bibfnamefont{V.~Y.} \bibnamefont{Kachorovskii}},
  \bibinfo{year}{2004}, \bibinfo{journal}{Phys. Rev. B}
  \textbf{\bibinfo{volume}{70}}, \bibinfo{pages}{245307}.

\bibitem[{\citenamefont{Chepelianskii and
  Shepelyansky}(2009)}]{chepelianskii:2009}
\bibinfo{author}{\bibnamefont{Chepelianskii}, \bibfnamefont{A.~D.}}, and
  \bibinfo{author}{\bibfnamefont{D.~L.} \bibnamefont{Shepelyansky}},
  \bibinfo{year}{2009}, \bibinfo{journal}{Phys. Rev. B}
  \textbf{\bibinfo{volume}{80}}, \bibinfo{pages}{241308(R)}.

\bibitem[{\citenamefont{Chiu} \emph{et~al.}(1976)\citenamefont{Chiu, Lee, and
  Quinn}}]{chiu:1976}
\bibinfo{author}{\bibnamefont{Chiu}, \bibfnamefont{K.~W.}},
  \bibinfo{author}{\bibfnamefont{T.~K.} \bibnamefont{Lee}}, and
  \bibinfo{author}{\bibfnamefont{J.~J.} \bibnamefont{Quinn}},
  \bibinfo{year}{1976}, \bibinfo{journal}{Surf. Sci.}
  \textbf{\bibinfo{volume}{58}}, \bibinfo{pages}{182}.

\bibitem[{\citenamefont{Choi} \emph{et~al.}(1986)\citenamefont{Choi, Tsui, and
  Palmateer}}]{choi:1986}
\bibinfo{author}{\bibnamefont{Choi}, \bibfnamefont{K.~K.}},
  \bibinfo{author}{\bibfnamefont{D.~C.} \bibnamefont{Tsui}}, and
  \bibinfo{author}{\bibfnamefont{S.~C.} \bibnamefont{Palmateer}},
  \bibinfo{year}{1986}, \bibinfo{journal}{Phys. Rev. B}
  \textbf{\bibinfo{volume}{33}}, \bibinfo{pages}{8216}.

\bibitem[{\citenamefont{Coleridge}(1990)}]{coleridge:1990}
\bibinfo{author}{\bibnamefont{Coleridge}, \bibfnamefont{P.~T.}},
  \bibinfo{year}{1990}, \bibinfo{journal}{Semicond. Sci. Technol.}
  \textbf{\bibinfo{volume}{5}}, \bibinfo{pages}{961}.

\bibitem[{\citenamefont{Coleridge}(1991)}]{coleridge:1991}
\bibinfo{author}{\bibnamefont{Coleridge}, \bibfnamefont{P.~T.}},
  \bibinfo{year}{1991}, \bibinfo{journal}{Phys. Rev. B}
  \textbf{\bibinfo{volume}{44}}, \bibinfo{pages}{3793}.

\bibitem[{\citenamefont{Coleridge}(1997)}]{coleridge:1997}
\bibinfo{author}{\bibnamefont{Coleridge}, \bibfnamefont{P.~T.}},
  \bibinfo{year}{1997}, \bibinfo{journal}{Semicond. Sci. Technol.}
  \textbf{\bibinfo{volume}{12}}, \bibinfo{pages}{22}.

\bibitem[{\citenamefont{Coleridge} \emph{et~al.}(1989)\citenamefont{Coleridge,
  Stoner, and Fletcher}}]{coleridge:1989}
\bibinfo{author}{\bibnamefont{Coleridge}, \bibfnamefont{P.~T.}},
  \bibinfo{author}{\bibfnamefont{R.}~\bibnamefont{Stoner}}, and
  \bibinfo{author}{\bibfnamefont{R.}~\bibnamefont{Fletcher}},
  \bibinfo{year}{1989}, \bibinfo{journal}{Phys. Rev. B}
  \textbf{\bibinfo{volume}{39}}, \bibinfo{pages}{1120}.

\bibitem[{\citenamefont{Cooper} \emph{et~al.}(2001)\citenamefont{Cooper, Lilly,
  Eisenstein, Jungwirth, Pfeiffer, and West}}]{cooper:2001}
\bibinfo{author}{\bibnamefont{Cooper}, \bibfnamefont{K.~B.}},
  \bibinfo{author}{\bibfnamefont{M.~P.} \bibnamefont{Lilly}},
  \bibinfo{author}{\bibfnamefont{J.~P.} \bibnamefont{Eisenstein}},
  \bibinfo{author}{\bibfnamefont{T.}~\bibnamefont{Jungwirth}},
  \bibinfo{author}{\bibfnamefont{L.~N.} \bibnamefont{Pfeiffer}}, and
  \bibinfo{author}{\bibfnamefont{K.~W.} \bibnamefont{West}},
  \bibinfo{year}{2001}, \bibinfo{journal}{Solid State Commun.}
  \textbf{\bibinfo{volume}{119}}, \bibinfo{pages}{89}.

\bibitem[{\citenamefont{Cross and Hohenberg}(1993)}]{cross:1993}
\bibinfo{author}{\bibnamefont{Cross}, \bibfnamefont{M.~C.}}, and
  \bibinfo{author}{\bibfnamefont{P.~C.} \bibnamefont{Hohenberg}},
  \bibinfo{year}{1993}, \bibinfo{journal}{Rev. Mod. Phys.}
  \textbf{\bibinfo{volume}{65}}, \bibinfo{pages}{851}.

\bibitem[{\citenamefont{Dai} \emph{et~al.}(2010)\citenamefont{Dai, Du,
  Pfeiffer, and West}}]{dai:2010}
\bibinfo{author}{\bibnamefont{Dai}, \bibfnamefont{Y.}},
  \bibinfo{author}{\bibfnamefont{R.~R.} \bibnamefont{Du}},
  \bibinfo{author}{\bibfnamefont{L.~N.} \bibnamefont{Pfeiffer}}, and
  \bibinfo{author}{\bibfnamefont{K.~W.} \bibnamefont{West}},
  \bibinfo{year}{2010}, \bibinfo{journal}{Phys. Rev. Lett.}
  \textbf{\bibinfo{volume}{105}}, \bibinfo{pages}{246802}.

\bibitem[{\citenamefont{Dai} \emph{et~al.}(2009)\citenamefont{Dai, Yuan, Yang,
  Du, Manfra, Pfeiffer, and West}}]{dai:2009}
\bibinfo{author}{\bibnamefont{Dai}, \bibfnamefont{Y.}},
  \bibinfo{author}{\bibfnamefont{Z.~Q.} \bibnamefont{Yuan}},
  \bibinfo{author}{\bibfnamefont{C.~L.} \bibnamefont{Yang}},
  \bibinfo{author}{\bibfnamefont{R.~R.} \bibnamefont{Du}},
  \bibinfo{author}{\bibfnamefont{M.~J.} \bibnamefont{Manfra}},
  \bibinfo{author}{\bibfnamefont{L.~N.} \bibnamefont{Pfeiffer}}, and
  \bibinfo{author}{\bibfnamefont{K.~W.} \bibnamefont{West}},
  \bibinfo{year}{2009}, \bibinfo{journal}{Phys. Rev. B}
  \textbf{\bibinfo{volume}{80}}, \bibinfo{pages}{041310(R)}.

\bibitem[{\citenamefont{Dakhnovskii and Metiu}(1995)}]{dakhnovskii:1995}
\bibinfo{author}{\bibnamefont{Dakhnovskii}, \bibfnamefont{Y.}}, and
  \bibinfo{author}{\bibfnamefont{H.}~\bibnamefont{Metiu}},
  \bibinfo{year}{1995}, \bibinfo{journal}{Phys. Rev. B}
  \textbf{\bibinfo{volume}{51}}, \bibinfo{pages}{4193}.

\bibitem[{\citenamefont{De~Dominicis}(1976)}]{de-dominicis:1976}
\bibinfo{author}{\bibnamefont{De~Dominicis}, \bibfnamefont{C.}},
  \bibinfo{year}{1976}, \bibinfo{journal}{J. Phys. Colloques}
  \textbf{\bibinfo{volume}{37-C1}}, \bibinfo{pages}{247}.

\bibitem[{\citenamefont{De~Dominicis and Peliti}(1978)}]{de-dominicis:1978}
\bibinfo{author}{\bibnamefont{De~Dominicis}, \bibfnamefont{C.}}, and
  \bibinfo{author}{\bibfnamefont{L.}~\bibnamefont{Peliti}},
  \bibinfo{year}{1978}, \bibinfo{journal}{Phys. Rev. B}
  \textbf{\bibinfo{volume}{18}}, \bibinfo{pages}{353}.

\bibitem[{\citenamefont{Dietel}(2006)}]{dietel:2006}
\bibinfo{author}{\bibnamefont{Dietel}, \bibfnamefont{J.}},
  \bibinfo{year}{2006}, \bibinfo{journal}{Phys. Rev. B}
  \textbf{\bibinfo{volume}{73}}, \bibinfo{pages}{125350}.

\bibitem[{\citenamefont{Dietel} \emph{et~al.}(2005)\citenamefont{Dietel,
  Glazman, Hekking, and von Oppen}}]{dietel:2005}
\bibinfo{author}{\bibnamefont{Dietel}, \bibfnamefont{J.}},
  \bibinfo{author}{\bibfnamefont{L.~I.} \bibnamefont{Glazman}},
  \bibinfo{author}{\bibfnamefont{F.~W.~J.} \bibnamefont{Hekking}}, and
  \bibinfo{author}{\bibfnamefont{F.}~\bibnamefont{von Oppen}},
  \bibinfo{year}{2005}, \bibinfo{journal}{Phys. Rev. B}
  \textbf{\bibinfo{volume}{71}}, \bibinfo{pages}{045329}.

\bibitem[{\citenamefont{Dietrich}
  \emph{et~al.}(2012{\natexlab{a}})\citenamefont{Dietrich, Byrnes, Vitkalov,
  Dmitriev, and Bykov}}]{dietrich:2012b}
\bibinfo{author}{\bibnamefont{Dietrich}, \bibfnamefont{S.}},
  \bibinfo{author}{\bibfnamefont{S.}~\bibnamefont{Byrnes}},
  \bibinfo{author}{\bibfnamefont{S.}~\bibnamefont{Vitkalov}},
  \bibinfo{author}{\bibfnamefont{D.~V.} \bibnamefont{Dmitriev}}, and
  \bibinfo{author}{\bibfnamefont{A.~A.} \bibnamefont{Bykov}},
  \bibinfo{year}{2012}{\natexlab{a}}, \bibinfo{journal}{Phys. Rev. B}
  \textbf{\bibinfo{volume}{85}}, \bibinfo{pages}{155307}.

\bibitem[{\citenamefont{Dietrich}
  \emph{et~al.}(2012{\natexlab{b}})\citenamefont{Dietrich, Byrnes, Vitkalov,
  Goran, and Bykov}}]{dietrich:2012c}
\bibinfo{author}{\bibnamefont{Dietrich}, \bibfnamefont{S.}},
  \bibinfo{author}{\bibfnamefont{S.}~\bibnamefont{Byrnes}},
  \bibinfo{author}{\bibfnamefont{S.}~\bibnamefont{Vitkalov}},
  \bibinfo{author}{\bibfnamefont{A.~V.} \bibnamefont{Goran}}, and
  \bibinfo{author}{\bibfnamefont{A.~A.} \bibnamefont{Bykov}},
  \bibinfo{year}{2012}{\natexlab{b}}, \bibinfo{journal}{Phys. Rev. B}
  \textbf{\bibinfo{volume}{86}}, \bibinfo{pages}{075471}.

\bibitem[{\citenamefont{Dietrich}
  \emph{et~al.}(2012{\natexlab{c}})\citenamefont{Dietrich, Vitkalov, Dmitriev,
  and Bykov}}]{dietrich:2012a}
\bibinfo{author}{\bibnamefont{Dietrich}, \bibfnamefont{S.}},
  \bibinfo{author}{\bibfnamefont{S.}~\bibnamefont{Vitkalov}},
  \bibinfo{author}{\bibfnamefont{D.~V.} \bibnamefont{Dmitriev}}, and
  \bibinfo{author}{\bibfnamefont{A.~A.} \bibnamefont{Bykov}},
  \bibinfo{year}{2012}{\natexlab{c}}, \bibinfo{journal}{Phys. Rev. B}
  \textbf{\bibinfo{volume}{85}}, \bibinfo{pages}{115312}.

\bibitem[{\citenamefont{Dingle} \emph{et~al.}(1978)\citenamefont{Dingle,
  St{\"o}rmer, Gossard, and Wiegmann}}]{dingle:1978}
\bibinfo{author}{\bibnamefont{Dingle}, \bibfnamefont{R.}},
  \bibinfo{author}{\bibfnamefont{H.~L.} \bibnamefont{St{\"o}rmer}},
  \bibinfo{author}{\bibfnamefont{A.~C.} \bibnamefont{Gossard}}, and
  \bibinfo{author}{\bibfnamefont{W.}~\bibnamefont{Wiegmann}},
  \bibinfo{year}{1978}, \bibinfo{journal}{Appl. Phys. Lett.}
  \textbf{\bibinfo{volume}{33}}, \bibinfo{pages}{665}.

\bibitem[{\citenamefont{Dmitriev} \emph{et~al.}(2012)\citenamefont{Dmitriev,
  Strygin, Bykov, Dietrich, and Vitkalov}}]{dmitriev:2012}
\bibinfo{author}{\bibnamefont{Dmitriev}, \bibfnamefont{D.~V.}},
  \bibinfo{author}{\bibfnamefont{I.~S.} \bibnamefont{Strygin}},
  \bibinfo{author}{\bibfnamefont{A.~A.} \bibnamefont{Bykov}},
  \bibinfo{author}{\bibfnamefont{S.}~\bibnamefont{Dietrich}}, and
  \bibinfo{author}{\bibfnamefont{S.~A.} \bibnamefont{Vitkalov}},
  \bibinfo{year}{2012}, \bibinfo{journal}{JETP Letters}
  \textbf{\bibinfo{volume}{95}}, \bibinfo{pages}{420}.

\bibitem[{\citenamefont{Dmitriev}(2011)}]{dmitriev:2011a}
\bibinfo{author}{\bibnamefont{Dmitriev}, \bibfnamefont{I.~A.}},
  \bibinfo{year}{2011}, \bibinfo{journal}{J. Phys.: Conf. Ser.}
  \textbf{\bibinfo{volume}{334}}, \bibinfo{pages}{012015}.

\bibitem[{\citenamefont{Dmitriev}
  \emph{et~al.}(2009{\natexlab{a}})\citenamefont{Dmitriev, Dorozhkin, and
  Mirlin}}]{dmitriev:2009a}
\bibinfo{author}{\bibnamefont{Dmitriev}, \bibfnamefont{I.~A.}},
  \bibinfo{author}{\bibfnamefont{S.~I.} \bibnamefont{Dorozhkin}}, and
  \bibinfo{author}{\bibfnamefont{A.~D.} \bibnamefont{Mirlin}},
  \bibinfo{year}{2009}{\natexlab{a}}, \bibinfo{journal}{Phys. Rev. B}
  \textbf{\bibinfo{volume}{80}}, \bibinfo{pages}{125418}.

\bibitem[{\citenamefont{Dmitriev} \emph{et~al.}(2008)\citenamefont{Dmitriev,
  Evers, Gornyi, Mirlin, Polyakov, and W\"olfle}}]{dmitriev:2008a}
\bibinfo{author}{\bibnamefont{Dmitriev}, \bibfnamefont{I.~A.}},
  \bibinfo{author}{\bibfnamefont{F.}~\bibnamefont{Evers}},
  \bibinfo{author}{\bibfnamefont{I.~V.} \bibnamefont{Gornyi}},
  \bibinfo{author}{\bibfnamefont{A.~D.} \bibnamefont{Mirlin}},
  \bibinfo{author}{\bibfnamefont{D.~G.} \bibnamefont{Polyakov}}, and
  \bibinfo{author}{\bibfnamefont{P.}~\bibnamefont{W\"olfle}},
  \bibinfo{year}{2008}, \bibinfo{journal}{Phys. Status Solidi (b)}
  \textbf{\bibinfo{volume}{245}}, \bibinfo{pages}{239}.

\bibitem[{\citenamefont{Dmitriev} \emph{et~al.}(2010)\citenamefont{Dmitriev,
  Gellmann, and Vavilov}}]{dmitriev:2010}
\bibinfo{author}{\bibnamefont{Dmitriev}, \bibfnamefont{I.~A.}},
  \bibinfo{author}{\bibfnamefont{R.}~\bibnamefont{Gellmann}}, and
  \bibinfo{author}{\bibfnamefont{M.~G.} \bibnamefont{Vavilov}},
  \bibinfo{year}{2010}, \bibinfo{journal}{Phys. Rev. B}
  \textbf{\bibinfo{volume}{82}}, \bibinfo{pages}{201311(R)}.

\bibitem[{\citenamefont{Dmitriev}
  \emph{et~al.}(2009{\natexlab{b}})\citenamefont{Dmitriev, Khodas, Mirlin,
  Polyakov, and Vavilov}}]{dmitriev:2009b}
\bibinfo{author}{\bibnamefont{Dmitriev}, \bibfnamefont{I.~A.}},
  \bibinfo{author}{\bibfnamefont{M.}~\bibnamefont{Khodas}},
  \bibinfo{author}{\bibfnamefont{A.~D.} \bibnamefont{Mirlin}},
  \bibinfo{author}{\bibfnamefont{D.~G.} \bibnamefont{Polyakov}}, and
  \bibinfo{author}{\bibfnamefont{M.~G.} \bibnamefont{Vavilov}},
  \bibinfo{year}{2009}{\natexlab{b}}, \bibinfo{journal}{Phys. Rev. B}
  \textbf{\bibinfo{volume}{80}}, \bibinfo{pages}{165327}.

\bibitem[{\citenamefont{Dmitriev} \emph{et~al.}(2003)\citenamefont{Dmitriev,
  Mirlin, and Polyakov}}]{dmitriev:2003}
\bibinfo{author}{\bibnamefont{Dmitriev}, \bibfnamefont{I.~A.}},
  \bibinfo{author}{\bibfnamefont{A.~D.} \bibnamefont{Mirlin}}, and
  \bibinfo{author}{\bibfnamefont{D.~G.} \bibnamefont{Polyakov}},
  \bibinfo{year}{2003}, \bibinfo{journal}{Phys. Rev. Lett.}
  \textbf{\bibinfo{volume}{91}}, \bibinfo{pages}{226802}.

\bibitem[{\citenamefont{Dmitriev} \emph{et~al.}(2004)\citenamefont{Dmitriev,
  Mirlin, and Polyakov}}]{dmitriev:2004}
\bibinfo{author}{\bibnamefont{Dmitriev}, \bibfnamefont{I.~A.}},
  \bibinfo{author}{\bibfnamefont{A.~D.} \bibnamefont{Mirlin}}, and
  \bibinfo{author}{\bibfnamefont{D.~G.} \bibnamefont{Polyakov}},
  \bibinfo{year}{2004}, \bibinfo{journal}{Phys. Rev. B}
  \textbf{\bibinfo{volume}{70}}, \bibinfo{pages}{165305}.

\bibitem[{\citenamefont{Dmitriev}
  \emph{et~al.}(2007{\natexlab{a}})\citenamefont{Dmitriev, Mirlin, and
  Polyakov}}]{dmitriev:2007}
\bibinfo{author}{\bibnamefont{Dmitriev}, \bibfnamefont{I.~A.}},
  \bibinfo{author}{\bibfnamefont{A.~D.} \bibnamefont{Mirlin}}, and
  \bibinfo{author}{\bibfnamefont{D.~G.} \bibnamefont{Polyakov}},
  \bibinfo{year}{2007}{\natexlab{a}}, \bibinfo{journal}{Phys. Rev. B}
  \textbf{\bibinfo{volume}{75}}, \bibinfo{pages}{245320}.

\bibitem[{\citenamefont{Dmitriev}
  \emph{et~al.}(2007{\natexlab{b}})\citenamefont{Dmitriev, Mirlin, and
  Polyakov}}]{dmitriev:2007b}
\bibinfo{author}{\bibnamefont{Dmitriev}, \bibfnamefont{I.~A.}},
  \bibinfo{author}{\bibfnamefont{A.~D.} \bibnamefont{Mirlin}}, and
  \bibinfo{author}{\bibfnamefont{D.~G.} \bibnamefont{Polyakov}},
  \bibinfo{year}{2007}{\natexlab{b}}, \bibinfo{journal}{Phys. Rev. Lett.}
  \textbf{\bibinfo{volume}{99}}, \bibinfo{pages}{206805}.

\bibitem[{\citenamefont{Dmitriev} \emph{et~al.}(2005)\citenamefont{Dmitriev,
  Vavilov, Aleiner, Mirlin, and Polyakov}}]{dmitriev:2005}
\bibinfo{author}{\bibnamefont{Dmitriev}, \bibfnamefont{I.~A.}},
  \bibinfo{author}{\bibfnamefont{M.~G.} \bibnamefont{Vavilov}},
  \bibinfo{author}{\bibfnamefont{I.~L.} \bibnamefont{Aleiner}},
  \bibinfo{author}{\bibfnamefont{A.~D.} \bibnamefont{Mirlin}}, and
  \bibinfo{author}{\bibfnamefont{D.~G.} \bibnamefont{Polyakov}},
  \bibinfo{year}{2005}, \bibinfo{journal}{Phys. Rev. B}
  \textbf{\bibinfo{volume}{71}}, \bibinfo{pages}{115316}.

\bibitem[{\citenamefont{Dorozhkin}(2003)}]{dorozhkin:2003}
\bibinfo{author}{\bibnamefont{Dorozhkin}, \bibfnamefont{S.~I.}},
  \bibinfo{year}{2003}, \bibinfo{journal}{JETP Lett.}
  \textbf{\bibinfo{volume}{77}}, \bibinfo{pages}{577}.

\bibitem[{\citenamefont{Dorozhkin}
  \emph{et~al.}(2007{\natexlab{a}})\citenamefont{Dorozhkin, Bykov,
  Pechenezhskii, and Bakarov}}]{dorozhkin:2007b}
\bibinfo{author}{\bibnamefont{Dorozhkin}, \bibfnamefont{S.~I.}},
  \bibinfo{author}{\bibfnamefont{A.~A.} \bibnamefont{Bykov}},
  \bibinfo{author}{\bibfnamefont{I.~V.} \bibnamefont{Pechenezhskii}}, and
  \bibinfo{author}{\bibfnamefont{A.~K.} \bibnamefont{Bakarov}},
  \bibinfo{year}{2007}{\natexlab{a}}, \bibinfo{journal}{JETP Lett.}
  \textbf{\bibinfo{volume}{85}}, \bibinfo{pages}{576}.

\bibitem[{\citenamefont{Dorozhkin}
  \emph{et~al.}(2011{\natexlab{a}})\citenamefont{Dorozhkin, Dmitriev, and
  Mirlin}}]{dorozhkin:2011a}
\bibinfo{author}{\bibnamefont{Dorozhkin}, \bibfnamefont{S.~I.}},
  \bibinfo{author}{\bibfnamefont{I.~A.} \bibnamefont{Dmitriev}}, and
  \bibinfo{author}{\bibfnamefont{A.~D.} \bibnamefont{Mirlin}},
  \bibinfo{year}{2011}{\natexlab{a}}, \bibinfo{journal}{Phys. Rev. B}
  \textbf{\bibinfo{volume}{84}}, \bibinfo{pages}{125448}.

\bibitem[{\citenamefont{Dorozhkin} \emph{et~al.}(2009)\citenamefont{Dorozhkin,
  Pechenezhskiy, Pfeiffer, West, Umansky, von Klitzing, and
  Smet}}]{dorozhkin:2009}
\bibinfo{author}{\bibnamefont{Dorozhkin}, \bibfnamefont{S.~I.}},
  \bibinfo{author}{\bibfnamefont{I.~V.} \bibnamefont{Pechenezhskiy}},
  \bibinfo{author}{\bibfnamefont{L.~N.} \bibnamefont{Pfeiffer}},
  \bibinfo{author}{\bibfnamefont{K.~W.} \bibnamefont{West}},
  \bibinfo{author}{\bibfnamefont{V.}~\bibnamefont{Umansky}},
  \bibinfo{author}{\bibfnamefont{K.}~\bibnamefont{von Klitzing}}, and
  \bibinfo{author}{\bibfnamefont{J.~H.} \bibnamefont{Smet}},
  \bibinfo{year}{2009}, \bibinfo{journal}{Phys. Rev. Lett.}
  \textbf{\bibinfo{volume}{102}}, \bibinfo{pages}{036602}.

\bibitem[{\citenamefont{Dorozhkin}
  \emph{et~al.}(2011{\natexlab{b}})\citenamefont{Dorozhkin, Pfeiffer, West, von
  Klitzing, and Smet}}]{dorozhkin:2011}
\bibinfo{author}{\bibnamefont{Dorozhkin}, \bibfnamefont{S.~I.}},
  \bibinfo{author}{\bibfnamefont{L.}~\bibnamefont{Pfeiffer}},
  \bibinfo{author}{\bibfnamefont{K.}~\bibnamefont{West}},
  \bibinfo{author}{\bibfnamefont{K.}~\bibnamefont{von Klitzing}}, and
  \bibinfo{author}{\bibfnamefont{J.~H.} \bibnamefont{Smet}},
  \bibinfo{year}{2011}{\natexlab{b}}, \bibinfo{journal}{Nature Phys.}
  \textbf{\bibinfo{volume}{7}}, \bibinfo{pages}{336}.

\bibitem[{\citenamefont{Dorozhkin}
  \emph{et~al.}(2007{\natexlab{b}})\citenamefont{Dorozhkin, Smet, von Klitzing,
  Pfeiffer, and West}}]{dorozhkin:2007}
\bibinfo{author}{\bibnamefont{Dorozhkin}, \bibfnamefont{S.~I.}},
  \bibinfo{author}{\bibfnamefont{J.~H.} \bibnamefont{Smet}},
  \bibinfo{author}{\bibfnamefont{K.}~\bibnamefont{von Klitzing}},
  \bibinfo{author}{\bibfnamefont{L.~N.} \bibnamefont{Pfeiffer}}, and
  \bibinfo{author}{\bibfnamefont{K.~W.} \bibnamefont{West}},
  \bibinfo{year}{2007}{\natexlab{b}}, \bibinfo{journal}{JETP Lett.}
  \textbf{\bibinfo{volume}{86}}, \bibinfo{pages}{543}.

\bibitem[{\citenamefont{Dorozhkin} \emph{et~al.}(2005)\citenamefont{Dorozhkin,
  Smet, Umansky, and von Klitzing}}]{dorozhkin:2005}
\bibinfo{author}{\bibnamefont{Dorozhkin}, \bibfnamefont{S.~I.}},
  \bibinfo{author}{\bibfnamefont{J.~H.} \bibnamefont{Smet}},
  \bibinfo{author}{\bibfnamefont{V.}~\bibnamefont{Umansky}}, and
  \bibinfo{author}{\bibfnamefont{K.}~\bibnamefont{von Klitzing}},
  \bibinfo{year}{2005}, \bibinfo{journal}{Phys. Rev. B}
  \textbf{\bibinfo{volume}{71}}, \bibinfo{pages}{201306(R)}.

\bibitem[{\citenamefont{Dreizin and Dykhne}(1973)}]{dreizin:1973}
\bibinfo{author}{\bibnamefont{Dreizin}, \bibfnamefont{Y.~A.}}, and
  \bibinfo{author}{\bibfnamefont{A.~M.} \bibnamefont{Dykhne}},
  \bibinfo{year}{1973}, \bibinfo{journal}{Sov. Phys. JETP}
  \textbf{\bibinfo{volume}{36}}, \bibinfo{pages}{127}.

\bibitem[{\citenamefont{Du} \emph{et~al.}(2004)\citenamefont{Du, Zudov, Yang,
  Yuan, Pfeiffer, and West}}]{du:2004a}
\bibinfo{author}{\bibnamefont{Du}, \bibfnamefont{R.~R.}},
  \bibinfo{author}{\bibfnamefont{M.~A.} \bibnamefont{Zudov}},
  \bibinfo{author}{\bibfnamefont{C.~L.} \bibnamefont{Yang}},
  \bibinfo{author}{\bibfnamefont{Z.~Q.} \bibnamefont{Yuan}},
  \bibinfo{author}{\bibfnamefont{L.~N.} \bibnamefont{Pfeiffer}}, and
  \bibinfo{author}{\bibfnamefont{K.~W.} \bibnamefont{West}},
  \bibinfo{year}{2004}, \bibinfo{journal}{Int. J. Mod. Phys. B}
  \textbf{\bibinfo{volume}{18}}, \bibinfo{pages}{3465}.

\bibitem[{\citenamefont{Durst}(2006)}]{durst:2006}
\bibinfo{author}{\bibnamefont{Durst}, \bibfnamefont{A.~C.}},
  \bibinfo{year}{2006}, \bibinfo{journal}{Nature (London)}
  \textbf{\bibinfo{volume}{442}}, \bibinfo{pages}{752}.

\bibitem[{\citenamefont{Durst and Girvin}(2004)}]{durst:2004}
\bibinfo{author}{\bibnamefont{Durst}, \bibfnamefont{A.~C.}}, and
  \bibinfo{author}{\bibfnamefont{S.~M.} \bibnamefont{Girvin}},
  \bibinfo{year}{2004}, \bibinfo{journal}{Science}
  \textbf{\bibinfo{volume}{304}}, \bibinfo{pages}{1752}.

\bibitem[{\citenamefont{Durst} \emph{et~al.}(2003)\citenamefont{Durst, Sachdev,
  Read, and Girvin}}]{durst:2003}
\bibinfo{author}{\bibnamefont{Durst}, \bibfnamefont{A.~C.}},
  \bibinfo{author}{\bibfnamefont{S.}~\bibnamefont{Sachdev}},
  \bibinfo{author}{\bibfnamefont{N.}~\bibnamefont{Read}}, and
  \bibinfo{author}{\bibfnamefont{S.~M.} \bibnamefont{Girvin}},
  \bibinfo{year}{2003}, \bibinfo{journal}{Phys. Rev. Lett.}
  \textbf{\bibinfo{volume}{91}}, \bibinfo{pages}{086803}.

\bibitem[{\citenamefont{Dyakonov}(1984)}]{dyakonov:1984a}
\bibinfo{author}{\bibnamefont{Dyakonov}, \bibfnamefont{M.~I.}},
  \bibinfo{year}{1984}, \bibinfo{journal}{JETP Lett.}
  \textbf{\bibinfo{volume}{39}}, \bibinfo{pages}{185}.

\bibitem[{\citenamefont{Dyakonov and Furman}(1984)}]{dyakonov:1984b}
\bibinfo{author}{\bibnamefont{Dyakonov}, \bibfnamefont{M.~I.}}, and
  \bibinfo{author}{\bibfnamefont{A.~S.} \bibnamefont{Furman}},
  \bibinfo{year}{1984}, \bibinfo{journal}{Sov. Phys. JETP}
  \textbf{\bibinfo{volume}{60}}, \bibinfo{pages}{1191}.

\bibitem[{\citenamefont{Dykman} \emph{et~al.}(1997)\citenamefont{Dykman,
  Fang-Yen, and Lea}}]{dykman:1997}
\bibinfo{author}{\bibnamefont{Dykman}, \bibfnamefont{M.~I.}},
  \bibinfo{author}{\bibfnamefont{C.}~\bibnamefont{Fang-Yen}}, and
  \bibinfo{author}{\bibfnamefont{M.~J.} \bibnamefont{Lea}},
  \bibinfo{year}{1997}, \bibinfo{journal}{Phys. Rev. B}
  \textbf{\bibinfo{volume}{55}}, \bibinfo{pages}{16249}.

\bibitem[{\citenamefont{Dykman} \emph{et~al.}(1993)\citenamefont{Dykman, Lea,
  Fozooni, and Frost}}]{dykman:1993}
\bibinfo{author}{\bibnamefont{Dykman}, \bibfnamefont{M.~I.}},
  \bibinfo{author}{\bibfnamefont{M.~J.} \bibnamefont{Lea}},
  \bibinfo{author}{\bibfnamefont{P.}~\bibnamefont{Fozooni}}, and
  \bibinfo{author}{\bibfnamefont{J.}~\bibnamefont{Frost}},
  \bibinfo{year}{1993}, \bibinfo{journal}{Phys. Rev. Lett.}
  \textbf{\bibinfo{volume}{70}}, \bibinfo{pages}{3975}.

\bibitem[{\citenamefont{Dykman and Pryadko}(2003)}]{dykman:2003}
\bibinfo{author}{\bibnamefont{Dykman}, \bibfnamefont{M.~I.}}, and
  \bibinfo{author}{\bibfnamefont{L.~P.} \bibnamefont{Pryadko}},
  \bibinfo{year}{2003}, \bibinfo{journal}{Phys. Rev. B}
  \textbf{\bibinfo{volume}{67}}, \bibinfo{pages}{235104}.

\bibitem[{\citenamefont{Efetov and Marikhin}(1989)}]{efetov:1989}
\bibinfo{author}{\bibnamefont{Efetov}, \bibfnamefont{K.~B.}}, and
  \bibinfo{author}{\bibfnamefont{V.~G.} \bibnamefont{Marikhin}},
  \bibinfo{year}{1989}, \bibinfo{journal}{Phys. Rev. B}
  \textbf{\bibinfo{volume}{40}}, \bibinfo{pages}{12126}.

\bibitem[{\citenamefont{Efros} \emph{et~al.}(1990)\citenamefont{Efros, Pikus,
  and Samsonidze}}]{efros:1990}
\bibinfo{author}{\bibnamefont{Efros}, \bibfnamefont{A.~L.}},
  \bibinfo{author}{\bibfnamefont{F.~G.} \bibnamefont{Pikus}}, and
  \bibinfo{author}{\bibfnamefont{G.~G.} \bibnamefont{Samsonidze}},
  \bibinfo{year}{1990}, \bibinfo{journal}{Phys. Rev. B}
  \textbf{\bibinfo{volume}{41}}, \bibinfo{pages}{8295}.

\bibitem[{\citenamefont{Eichhorn} \emph{et~al.}(2002)\citenamefont{Eichhorn,
  Reimann, and H\"anggi}}]{eichhorn:2002}
\bibinfo{author}{\bibnamefont{Eichhorn}, \bibfnamefont{R.}},
  \bibinfo{author}{\bibfnamefont{P.}~\bibnamefont{Reimann}}, and
  \bibinfo{author}{\bibfnamefont{P.}~\bibnamefont{H\"anggi}},
  \bibinfo{year}{2002}, \bibinfo{journal}{Phys. Rev. E}
  \textbf{\bibinfo{volume}{66}}, \bibinfo{pages}{066132}.

\bibitem[{\citenamefont{Elesin}(1969)}]{elesin:1969}
\bibinfo{author}{\bibnamefont{Elesin}, \bibfnamefont{V.~F.}},
  \bibinfo{year}{1969}, \bibinfo{journal}{Sov. Phys. JETP}
  \textbf{\bibinfo{volume}{28}}, \bibinfo{pages}{410}.

\bibitem[{\citenamefont{Elesin and Manykin}(1967)}]{elesin:1967}
\bibinfo{author}{\bibnamefont{Elesin}, \bibfnamefont{V.~F.}}, and
  \bibinfo{author}{\bibfnamefont{E.~A.} \bibnamefont{Manykin}},
  \bibinfo{year}{1967}, \bibinfo{journal}{Sov. Phys. Solid State}
  \textbf{\bibinfo{volume}{8}}, \bibinfo{pages}{2891}.

\bibitem[{\citenamefont{Evers} \emph{et~al.}(1999)\citenamefont{Evers, Mirlin,
  Polyakov, and W{\"o}lfle}}]{evers:1999}
\bibinfo{author}{\bibnamefont{Evers}, \bibfnamefont{F.}},
  \bibinfo{author}{\bibfnamefont{A.~D.} \bibnamefont{Mirlin}},
  \bibinfo{author}{\bibfnamefont{D.~G.} \bibnamefont{Polyakov}}, and
  \bibinfo{author}{\bibfnamefont{P.}~\bibnamefont{W{\"o}lfle}},
  \bibinfo{year}{1999}, \bibinfo{journal}{Phys. Rev. B}
  \textbf{\bibinfo{volume}{60}}, \bibinfo{pages}{8951}.

\bibitem[{\citenamefont{Fal'ko and Khmel'nitskii}(1989)}]{falko:1989}
\bibinfo{author}{\bibnamefont{Fal'ko}, \bibfnamefont{V.~I.}}, and
  \bibinfo{author}{\bibfnamefont{D.~E.} \bibnamefont{Khmel'nitskii}},
  \bibinfo{year}{1989}, \bibinfo{journal}{Sov. Phys. JETP}
  \textbf{\bibinfo{volume}{68}}, \bibinfo{pages}{1150}.

\bibitem[{\citenamefont{Fedorych} \emph{et~al.}(2010)\citenamefont{Fedorych,
  Potemski, Studenikin, Gupta, Wasilewski, and Dmitriev}}]{fedorych:2010}
\bibinfo{author}{\bibnamefont{Fedorych}, \bibfnamefont{O.~M.}},
  \bibinfo{author}{\bibfnamefont{M.}~\bibnamefont{Potemski}},
  \bibinfo{author}{\bibfnamefont{S.~A.} \bibnamefont{Studenikin}},
  \bibinfo{author}{\bibfnamefont{J.~A.} \bibnamefont{Gupta}},
  \bibinfo{author}{\bibfnamefont{Z.~R.} \bibnamefont{Wasilewski}}, and
  \bibinfo{author}{\bibfnamefont{I.~A.} \bibnamefont{Dmitriev}},
  \bibinfo{year}{2010}, \bibinfo{journal}{Phys. Rev. B}
  \textbf{\bibinfo{volume}{81}}, \bibinfo{pages}{201302(R)}.

\bibitem[{\citenamefont{Ferry and Goodnick}(1997)}]{ferry:1997}
\bibinfo{author}{\bibnamefont{Ferry}, \bibfnamefont{D.}}, and
  \bibinfo{author}{\bibfnamefont{S.}~\bibnamefont{Goodnick}},
  \bibinfo{year}{1997}, \emph{\bibinfo{title}{Transport in Nanostructures}}
  (\bibinfo{publisher}{Cambridge University, Cambridge}).

\bibitem[{\citenamefont{Finkler} \emph{et~al.}(2006)\citenamefont{Finkler,
  Halperin, Auerbach, and Yacoby}}]{finkler:2006}
\bibinfo{author}{\bibnamefont{Finkler}, \bibfnamefont{I.}},
  \bibinfo{author}{\bibfnamefont{B.~I.} \bibnamefont{Halperin}},
  \bibinfo{author}{\bibfnamefont{A.}~\bibnamefont{Auerbach}}, and
  \bibinfo{author}{\bibfnamefont{A.}~\bibnamefont{Yacoby}},
  \bibinfo{year}{2006}, \bibinfo{journal}{J. Stat. Phys.}
  \textbf{\bibinfo{volume}{125}}, \bibinfo{pages}{1093}.

\bibitem[{\citenamefont{Finkler and Halperin}(2009)}]{finkler:2009}
\bibinfo{author}{\bibnamefont{Finkler}, \bibfnamefont{I.~G.}}, and
  \bibinfo{author}{\bibfnamefont{B.~I.} \bibnamefont{Halperin}},
  \bibinfo{year}{2009}, \bibinfo{journal}{Phys. Rev. B}
  \textbf{\bibinfo{volume}{79}}, \bibinfo{pages}{085315}.

\bibitem[{\citenamefont{Fitzgerald}(2003)}]{fitzgerald:2003}
\bibinfo{author}{\bibnamefont{Fitzgerald}, \bibfnamefont{R.}},
  \bibinfo{year}{2003}, \bibinfo{journal}{Physics Today}
  \textbf{\bibinfo{volume}{56}}, \bibinfo{pages}{24}.

\bibitem[{\citenamefont{Fogler} \emph{et~al.}(1997)\citenamefont{Fogler, Dobin,
  Perel, and Shklovskii}}]{fogler:1997}
\bibinfo{author}{\bibnamefont{Fogler}, \bibfnamefont{M.~M.}},
  \bibinfo{author}{\bibfnamefont{A.~Y.} \bibnamefont{Dobin}},
  \bibinfo{author}{\bibfnamefont{V.~I.} \bibnamefont{Perel}}, and
  \bibinfo{author}{\bibfnamefont{B.~I.} \bibnamefont{Shklovskii}},
  \bibinfo{year}{1997}, \bibinfo{journal}{Phys. Rev. B}
  \textbf{\bibinfo{volume}{56}}, \bibinfo{pages}{6823}.

\bibitem[{\citenamefont{Fogler and Shklovskii}(1998)}]{fogler:1998}
\bibinfo{author}{\bibnamefont{Fogler}, \bibfnamefont{M.~M.}}, and
  \bibinfo{author}{\bibfnamefont{B.~I.} \bibnamefont{Shklovskii}},
  \bibinfo{year}{1998}, \bibinfo{journal}{Phys. Rev. Lett.}
  \textbf{\bibinfo{volume}{80}}, \bibinfo{pages}{4749}.

\bibitem[{\citenamefont{Forster} \emph{et~al.}(1977)\citenamefont{Forster,
  Nelson, and Stephen}}]{forster:1977}
\bibinfo{author}{\bibnamefont{Forster}, \bibfnamefont{D.}},
  \bibinfo{author}{\bibfnamefont{D.~R.} \bibnamefont{Nelson}}, and
  \bibinfo{author}{\bibfnamefont{M.~J.} \bibnamefont{Stephen}},
  \bibinfo{year}{1977}, \bibinfo{journal}{Phys. Rev. A}
  \textbf{\bibinfo{volume}{16}}, \bibinfo{pages}{732}.

\bibitem[{\citenamefont{Friedland} \emph{et~al.}(1996)\citenamefont{Friedland,
  Hey, Kostial, Klann, and Ploog}}]{friedland:1996}
\bibinfo{author}{\bibnamefont{Friedland}, \bibfnamefont{K.-J.}},
  \bibinfo{author}{\bibfnamefont{R.}~\bibnamefont{Hey}},
  \bibinfo{author}{\bibfnamefont{H.}~\bibnamefont{Kostial}},
  \bibinfo{author}{\bibfnamefont{R.}~\bibnamefont{Klann}}, and
  \bibinfo{author}{\bibfnamefont{K.}~\bibnamefont{Ploog}},
  \bibinfo{year}{1996}, \bibinfo{journal}{Phys. Rev. Lett.}
  \textbf{\bibinfo{volume}{77}}, \bibinfo{pages}{4616}.

\bibitem[{\citenamefont{Galaktionov}
  \emph{et~al.}(2006)\citenamefont{Galaktionov, Savchenko, and
  Ritchie}}]{galaktionov:2006}
\bibinfo{author}{\bibnamefont{Galaktionov}, \bibfnamefont{E.~A.}},
  \bibinfo{author}{\bibfnamefont{A.~K.} \bibnamefont{Savchenko}}, and
  \bibinfo{author}{\bibfnamefont{D.~A.} \bibnamefont{Ritchie}},
  \bibinfo{year}{2006}, \bibinfo{journal}{Phys. Status Solidi (c)}
  \textbf{\bibinfo{volume}{3}}, \bibinfo{pages}{304}.

\bibitem[{\citenamefont{Geim and Novoselov}(2007)}]{geim:2007}
\bibinfo{author}{\bibnamefont{Geim}, \bibfnamefont{A.~K.}}, and
  \bibinfo{author}{\bibfnamefont{K.~S.} \bibnamefont{Novoselov}},
  \bibinfo{year}{2007}, \bibinfo{journal}{Nature Mat.}
  \textbf{\bibinfo{volume}{6}}, \bibinfo{pages}{183}.

\bibitem[{\citenamefont{Gerhardts}(1975)}]{gerhardts:1975}
\bibinfo{author}{\bibnamefont{Gerhardts}, \bibfnamefont{R.~R.}},
  \bibinfo{year}{1975}, \bibinfo{journal}{Z. Phys. B}
  \textbf{\bibinfo{volume}{21}}, \bibinfo{pages}{285}.

\bibitem[{\citenamefont{Gerhardts} \emph{et~al.}(1989)\citenamefont{Gerhardts,
  Weiss, and von Klitzing}}]{gerhards:1989}
\bibinfo{author}{\bibnamefont{Gerhardts}, \bibfnamefont{R.~R.}},
  \bibinfo{author}{\bibfnamefont{D.}~\bibnamefont{Weiss}}, and
  \bibinfo{author}{\bibfnamefont{K.}~\bibnamefont{von Klitzing}},
  \bibinfo{year}{1989}, \bibinfo{journal}{Phys. Rev. Lett.}
  \textbf{\bibinfo{volume}{62}}, \bibinfo{pages}{1173}.

\bibitem[{\citenamefont{Gershenzon and Falei}(1986)}]{gershenzon:1986}
\bibinfo{author}{\bibnamefont{Gershenzon}, \bibfnamefont{M.~E.}}, and
  \bibinfo{author}{\bibfnamefont{M.~I.} \bibnamefont{Falei}},
  \bibinfo{year}{1986}, \bibinfo{journal}{JETP Lett.}
  \textbf{\bibinfo{volume}{44}}, \bibinfo{pages}{682}.

\bibitem[{\citenamefont{Gershenzon and Falei}(1988)}]{gershenzon:1988}
\bibinfo{author}{\bibnamefont{Gershenzon}, \bibfnamefont{M.~E.}}, and
  \bibinfo{author}{\bibfnamefont{M.~I.} \bibnamefont{Falei}},
  \bibinfo{year}{1988}, \bibinfo{journal}{Sov. Phys. JETP}
  \textbf{\bibinfo{volume}{67}}, \bibinfo{pages}{389}.

\bibitem[{\citenamefont{Girvin} \emph{et~al.}(1982)\citenamefont{Girvin,
  Jonson, and Lee}}]{girvin:1982}
\bibinfo{author}{\bibnamefont{Girvin}, \bibfnamefont{S.~M.}},
  \bibinfo{author}{\bibfnamefont{M.}~\bibnamefont{Jonson}}, and
  \bibinfo{author}{\bibfnamefont{P.~A.} \bibnamefont{Lee}},
  \bibinfo{year}{1982}, \bibinfo{journal}{Phys. Rev. B}
  \textbf{\bibinfo{volume}{26}}, \bibinfo{pages}{1651}.

\bibitem[{\citenamefont{Gladun and Ryzhii}(1970)}]{gladun:1970}
\bibinfo{author}{\bibnamefont{Gladun}, \bibfnamefont{A.~D.}}, and
  \bibinfo{author}{\bibfnamefont{V.~I.} \bibnamefont{Ryzhii}},
  \bibinfo{year}{1970}, \bibinfo{journal}{Sov. Phys. JETP}
  \textbf{\bibinfo{volume}{30}}, \bibinfo{pages}{534}.

\bibitem[{\citenamefont{Gold}(1989)}]{gold:1989}
\bibinfo{author}{\bibnamefont{Gold}, \bibfnamefont{A.}}, \bibinfo{year}{1989},
  \bibinfo{journal}{Appl. Phys. Lett.} \textbf{\bibinfo{volume}{54}},
  \bibinfo{pages}{2100}.

\bibitem[{\citenamefont{Goran} \emph{et~al.}(2009)\citenamefont{Goran, Bykov,
  Toropov, and Vitkalov}}]{goran:2009}
\bibinfo{author}{\bibnamefont{Goran}, \bibfnamefont{A.~V.}},
  \bibinfo{author}{\bibfnamefont{A.~A.} \bibnamefont{Bykov}},
  \bibinfo{author}{\bibfnamefont{A.~I.} \bibnamefont{Toropov}}, and
  \bibinfo{author}{\bibfnamefont{S.~A.} \bibnamefont{Vitkalov}},
  \bibinfo{year}{2009}, \bibinfo{journal}{Phys. Rev. B}
  \textbf{\bibinfo{volume}{80}}, \bibinfo{pages}{193305}.

\bibitem[{\citenamefont{Gornyi and Mirlin}(2003)}]{gornyi:2003}
\bibinfo{author}{\bibnamefont{Gornyi}, \bibfnamefont{I.~V.}}, and
  \bibinfo{author}{\bibfnamefont{A.~D.} \bibnamefont{Mirlin}},
  \bibinfo{year}{2003}, \bibinfo{journal}{Phys. Rev. Lett.}
  \textbf{\bibinfo{volume}{90}}, \bibinfo{pages}{076801}.

\bibitem[{\citenamefont{Gornyi and Mirlin}(2004)}]{gornyi:2004}
\bibinfo{author}{\bibnamefont{Gornyi}, \bibfnamefont{I.~V.}}, and
  \bibinfo{author}{\bibfnamefont{A.~D.} \bibnamefont{Mirlin}},
  \bibinfo{year}{2004}, \bibinfo{journal}{Phys. Rev. B}
  \textbf{\bibinfo{volume}{69}}, \bibinfo{pages}{045313}.

\bibitem[{\citenamefont{Gr\'egoire and Chat\'e}(2004)}]{gregoire:2004}
\bibinfo{author}{\bibnamefont{Gr\'egoire}, \bibfnamefont{G.}}, and
  \bibinfo{author}{\bibfnamefont{H.}~\bibnamefont{Chat\'e}},
  \bibinfo{year}{2004}, \bibinfo{journal}{Phys. Rev. Lett.}
  \textbf{\bibinfo{volume}{92}}, \bibinfo{pages}{025702}.

\bibitem[{\citenamefont{Grimes}(1978)}]{grimes:1978}
\bibinfo{author}{\bibnamefont{Grimes}, \bibfnamefont{C.~C.}},
  \bibinfo{year}{1978}, \bibinfo{journal}{Surface Science}
  \textbf{\bibinfo{volume}{73}}, \bibinfo{pages}{379}.

\bibitem[{\citenamefont{Gunn}(1963)}]{gunn:1963}
\bibinfo{author}{\bibnamefont{Gunn}, \bibfnamefont{J.~B.}},
  \bibinfo{year}{1963}, \bibinfo{journal}{Solid State Commun.}
  \textbf{\bibinfo{volume}{1}}, \bibinfo{pages}{88}.

\bibitem[{\citenamefont{Gurevich and Firsov}(1961)}]{gurevich:1961}
\bibinfo{author}{\bibnamefont{Gurevich}, \bibfnamefont{V.~L.}}, and
  \bibinfo{author}{\bibfnamefont{Y.~A.} \bibnamefont{Firsov}},
  \bibinfo{year}{1961}, \bibinfo{journal}{Sov. Phys. JETP}
  \textbf{\bibinfo{volume}{13}}, \bibinfo{pages}{137}.

\bibitem[{\citenamefont{Gusev} \emph{et~al.}(2011)\citenamefont{Gusev,
  Wiedmann, Raichev, Bakarov, and Portal}}]{gusev:2011}
\bibinfo{author}{\bibnamefont{Gusev}, \bibfnamefont{G.~M.}},
  \bibinfo{author}{\bibfnamefont{S.}~\bibnamefont{Wiedmann}},
  \bibinfo{author}{\bibfnamefont{O.~E.} \bibnamefont{Raichev}},
  \bibinfo{author}{\bibfnamefont{A.~K.} \bibnamefont{Bakarov}}, and
  \bibinfo{author}{\bibfnamefont{J.~C.} \bibnamefont{Portal}},
  \bibinfo{year}{2011}, \bibinfo{journal}{Phys. Rev. B}
  \textbf{\bibinfo{volume}{83}}, \bibinfo{pages}{041306(R)}.

\bibitem[{\citenamefont{Halperin} \emph{et~al.}(1993)\citenamefont{Halperin,
  Lee, and Read}}]{halperin:1993}
\bibinfo{author}{\bibnamefont{Halperin}, \bibfnamefont{B.~I.}},
  \bibinfo{author}{\bibfnamefont{P.~A.} \bibnamefont{Lee}}, and
  \bibinfo{author}{\bibfnamefont{N.}~\bibnamefont{Read}}, \bibinfo{year}{1993},
  \bibinfo{journal}{Phys. Rev. B} \textbf{\bibinfo{volume}{47}},
  \bibinfo{pages}{7312}.

\bibitem[{\citenamefont{Hartmann} \emph{et~al.}(1997)\citenamefont{Hartmann,
  Grifoni, and H\"anggi}}]{hartmann:1997}
\bibinfo{author}{\bibnamefont{Hartmann}, \bibfnamefont{L.}},
  \bibinfo{author}{\bibfnamefont{M.}~\bibnamefont{Grifoni}}, and
  \bibinfo{author}{\bibfnamefont{P.}~\bibnamefont{H\"anggi}},
  \bibinfo{year}{1997}, \bibinfo{journal}{Europhys. Lett.}
  \textbf{\bibinfo{volume}{38}}, \bibinfo{pages}{497}.

\bibitem[{\citenamefont{Hatke}
  \emph{et~al.}(2008{\natexlab{a}})\citenamefont{Hatke, Chiang, Zudov,
  Pfeiffer, and West}}]{hatke:2008b}
\bibinfo{author}{\bibnamefont{Hatke}, \bibfnamefont{A.~T.}},
  \bibinfo{author}{\bibfnamefont{H.-S.} \bibnamefont{Chiang}},
  \bibinfo{author}{\bibfnamefont{M.~A.} \bibnamefont{Zudov}},
  \bibinfo{author}{\bibfnamefont{L.~N.} \bibnamefont{Pfeiffer}}, and
  \bibinfo{author}{\bibfnamefont{K.~W.} \bibnamefont{West}},
  \bibinfo{year}{2008}{\natexlab{a}}, \bibinfo{journal}{Phys. Rev. Lett.}
  \textbf{\bibinfo{volume}{101}}, \bibinfo{pages}{246811}.

\bibitem[{\citenamefont{Hatke}
  \emph{et~al.}(2008{\natexlab{b}})\citenamefont{Hatke, Chiang, Zudov,
  Pfeiffer, and West}}]{hatke:2008a}
\bibinfo{author}{\bibnamefont{Hatke}, \bibfnamefont{A.~T.}},
  \bibinfo{author}{\bibfnamefont{H.-S.} \bibnamefont{Chiang}},
  \bibinfo{author}{\bibfnamefont{M.~A.} \bibnamefont{Zudov}},
  \bibinfo{author}{\bibfnamefont{L.~N.} \bibnamefont{Pfeiffer}}, and
  \bibinfo{author}{\bibfnamefont{K.~W.} \bibnamefont{West}},
  \bibinfo{year}{2008}{\natexlab{b}}, \bibinfo{journal}{Phys. Rev. B}
  \textbf{\bibinfo{volume}{77}}, \bibinfo{pages}{201304(R)}.

\bibitem[{\citenamefont{Hatke} \emph{et~al.}(2010)\citenamefont{Hatke, Chiang,
  Zudov, Pfeiffer, and West}}]{hatke:2010a}
\bibinfo{author}{\bibnamefont{Hatke}, \bibfnamefont{A.~T.}},
  \bibinfo{author}{\bibfnamefont{H.-S.} \bibnamefont{Chiang}},
  \bibinfo{author}{\bibfnamefont{M.~A.} \bibnamefont{Zudov}},
  \bibinfo{author}{\bibfnamefont{L.~N.} \bibnamefont{Pfeiffer}}, and
  \bibinfo{author}{\bibfnamefont{K.~W.} \bibnamefont{West}},
  \bibinfo{year}{2010}, \bibinfo{journal}{Phys. Rev. B}
  \textbf{\bibinfo{volume}{82}}, \bibinfo{pages}{041304(R)}.

\bibitem[{\citenamefont{Hatke}
  \emph{et~al.}(2011{\natexlab{a}})\citenamefont{Hatke, Khodas, Zudov,
  Pfeiffer, and West}}]{hatke:2011up1}
\bibinfo{author}{\bibnamefont{Hatke}, \bibfnamefont{A.~T.}},
  \bibinfo{author}{\bibfnamefont{M.}~\bibnamefont{Khodas}},
  \bibinfo{author}{\bibfnamefont{M.~A.} \bibnamefont{Zudov}},
  \bibinfo{author}{\bibfnamefont{L.~N.} \bibnamefont{Pfeiffer}}, and
  \bibinfo{author}{\bibfnamefont{K.~W.} \bibnamefont{West}},
  \bibinfo{year}{2011}{\natexlab{a}}, \bibinfo{journal}{Phys. Rev. B}
  \textbf{\bibinfo{volume}{84}}, \bibinfo{pages}{241302(R)}.

\bibitem[{\citenamefont{Hatke}
  \emph{et~al.}(2009{\natexlab{a}})\citenamefont{Hatke, Zudov, Pfeiffer, and
  West}}]{hatke:2009b}
\bibinfo{author}{\bibnamefont{Hatke}, \bibfnamefont{A.~T.}},
  \bibinfo{author}{\bibfnamefont{M.~A.} \bibnamefont{Zudov}},
  \bibinfo{author}{\bibfnamefont{L.~N.} \bibnamefont{Pfeiffer}}, and
  \bibinfo{author}{\bibfnamefont{K.~W.} \bibnamefont{West}},
  \bibinfo{year}{2009}{\natexlab{a}}, \bibinfo{journal}{Phys. Rev. Lett.}
  \textbf{\bibinfo{volume}{102}}, \bibinfo{pages}{086808}.

\bibitem[{\citenamefont{Hatke}
  \emph{et~al.}(2009{\natexlab{b}})\citenamefont{Hatke, Zudov, Pfeiffer, and
  West}}]{hatke:2009c}
\bibinfo{author}{\bibnamefont{Hatke}, \bibfnamefont{A.~T.}},
  \bibinfo{author}{\bibfnamefont{M.~A.} \bibnamefont{Zudov}},
  \bibinfo{author}{\bibfnamefont{L.~N.} \bibnamefont{Pfeiffer}}, and
  \bibinfo{author}{\bibfnamefont{K.~W.} \bibnamefont{West}},
  \bibinfo{year}{2009}{\natexlab{b}}, \bibinfo{journal}{Phys. Rev. B}
  \textbf{\bibinfo{volume}{79}}, \bibinfo{pages}{161308(R)}.

\bibitem[{\citenamefont{Hatke}
  \emph{et~al.}(2009{\natexlab{c}})\citenamefont{Hatke, Zudov, Pfeiffer, and
  West}}]{hatke:2009a}
\bibinfo{author}{\bibnamefont{Hatke}, \bibfnamefont{A.~T.}},
  \bibinfo{author}{\bibfnamefont{M.~A.} \bibnamefont{Zudov}},
  \bibinfo{author}{\bibfnamefont{L.~N.} \bibnamefont{Pfeiffer}}, and
  \bibinfo{author}{\bibfnamefont{K.~W.} \bibnamefont{West}},
  \bibinfo{year}{2009}{\natexlab{c}}, \bibinfo{journal}{Phys. Rev. Lett.}
  \textbf{\bibinfo{volume}{102}}, \bibinfo{pages}{066804}.

\bibitem[{\citenamefont{Hatke}
  \emph{et~al.}(2011{\natexlab{b}})\citenamefont{Hatke, Zudov, Pfeiffer, and
  West}}]{hatke:2011b}
\bibinfo{author}{\bibnamefont{Hatke}, \bibfnamefont{A.~T.}},
  \bibinfo{author}{\bibfnamefont{M.~A.} \bibnamefont{Zudov}},
  \bibinfo{author}{\bibfnamefont{L.~N.} \bibnamefont{Pfeiffer}}, and
  \bibinfo{author}{\bibfnamefont{K.~W.} \bibnamefont{West}},
  \bibinfo{year}{2011}{\natexlab{b}}, \bibinfo{journal}{Phys. Rev. B}
  \textbf{\bibinfo{volume}{83}}, \bibinfo{pages}{121301(R)}.

\bibitem[{\citenamefont{Hatke}
  \emph{et~al.}(2011{\natexlab{c}})\citenamefont{Hatke, Zudov, Pfeiffer, and
  West}}]{hatke:2011d}
\bibinfo{author}{\bibnamefont{Hatke}, \bibfnamefont{A.~T.}},
  \bibinfo{author}{\bibfnamefont{M.~A.} \bibnamefont{Zudov}},
  \bibinfo{author}{\bibfnamefont{L.~N.} \bibnamefont{Pfeiffer}}, and
  \bibinfo{author}{\bibfnamefont{K.~W.} \bibnamefont{West}},
  \bibinfo{year}{2011}{\natexlab{c}}, \bibinfo{journal}{Phys. Rev. B}
  \textbf{\bibinfo{volume}{84}}, \bibinfo{pages}{241304(R)}.

\bibitem[{\citenamefont{Hatke}
  \emph{et~al.}(2011{\natexlab{d}})\citenamefont{Hatke, Zudov, Pfeiffer, and
  West}}]{hatke:2011c}
\bibinfo{author}{\bibnamefont{Hatke}, \bibfnamefont{A.~T.}},
  \bibinfo{author}{\bibfnamefont{M.~A.} \bibnamefont{Zudov}},
  \bibinfo{author}{\bibfnamefont{L.~N.} \bibnamefont{Pfeiffer}}, and
  \bibinfo{author}{\bibfnamefont{K.~W.} \bibnamefont{West}},
  \bibinfo{year}{2011}{\natexlab{d}}, \bibinfo{journal}{Phys. Rev. B}
  \textbf{\bibinfo{volume}{83}}, \bibinfo{pages}{201301(R)}.

\bibitem[{\citenamefont{Hatke}
  \emph{et~al.}(2011{\natexlab{e}})\citenamefont{Hatke, Zudov, Pfeiffer, and
  West}}]{hatke:2011up}
\bibinfo{author}{\bibnamefont{Hatke}, \bibfnamefont{A.~T.}},
  \bibinfo{author}{\bibfnamefont{M.~A.} \bibnamefont{Zudov}},
  \bibinfo{author}{\bibfnamefont{L.~N.} \bibnamefont{Pfeiffer}}, and
  \bibinfo{author}{\bibfnamefont{K.~W.} \bibnamefont{West}},
  \bibinfo{year}{2011}{\natexlab{e}}, \bibinfo{journal}{Phys. Rev. B}
  \textbf{\bibinfo{volume}{84}}, \bibinfo{pages}{121301(R)}.

\bibitem[{\citenamefont{Hatke}
  \emph{et~al.}(2011{\natexlab{f}})\citenamefont{Hatke, Zudov, Pfeiffer, and
  West}}]{hatke:2011a}
\bibinfo{author}{\bibnamefont{Hatke}, \bibfnamefont{A.~T.}},
  \bibinfo{author}{\bibfnamefont{M.~A.} \bibnamefont{Zudov}},
  \bibinfo{author}{\bibfnamefont{L.~N.} \bibnamefont{Pfeiffer}}, and
  \bibinfo{author}{\bibfnamefont{K.~W.} \bibnamefont{West}},
  \bibinfo{year}{2011}{\natexlab{f}}, \bibinfo{journal}{Phys. Rev. B}
  \textbf{\bibinfo{volume}{83}}, \bibinfo{pages}{081301(R)}.

\bibitem[{\citenamefont{Hatke}
  \emph{et~al.}(2012{\natexlab{a}})\citenamefont{Hatke, Zudov, Pfeiffer, and
  West}}]{hatke:2012b}
\bibinfo{author}{\bibnamefont{Hatke}, \bibfnamefont{A.~T.}},
  \bibinfo{author}{\bibfnamefont{M.~A.} \bibnamefont{Zudov}},
  \bibinfo{author}{\bibfnamefont{L.~N.} \bibnamefont{Pfeiffer}}, and
  \bibinfo{author}{\bibfnamefont{K.~W.} \bibnamefont{West}},
  \bibinfo{year}{2012}{\natexlab{a}}, \bibinfo{journal}{Phys. Rev. B}
  \textbf{\bibinfo{volume}{86}}, \bibinfo{pages}{081307(R)}.

\bibitem[{\citenamefont{Hatke}
  \emph{et~al.}(2012{\natexlab{b}})\citenamefont{Hatke, Zudov, Pfeiffer, and
  West}}]{hatke:2012c}
\bibinfo{author}{\bibnamefont{Hatke}, \bibfnamefont{A.~T.}},
  \bibinfo{author}{\bibfnamefont{M.~A.} \bibnamefont{Zudov}},
  \bibinfo{author}{\bibfnamefont{L.~N.} \bibnamefont{Pfeiffer}}, and
  \bibinfo{author}{\bibfnamefont{K.~W.} \bibnamefont{West}},
  \bibinfo{year}{2012}{\natexlab{b}}, \bibinfo{journal}{Phys. Rev. B}
  \textbf{\bibinfo{volume}{85}}, \bibinfo{pages}{241305}.

\bibitem[{\citenamefont{Hatke}
  \emph{et~al.}(2012{\natexlab{c}})\citenamefont{Hatke, Zudov, Reno, Pfeiffer,
  and West}}]{hatke:2012a}
\bibinfo{author}{\bibnamefont{Hatke}, \bibfnamefont{A.~T.}},
  \bibinfo{author}{\bibfnamefont{M.~A.} \bibnamefont{Zudov}},
  \bibinfo{author}{\bibfnamefont{J.~L.} \bibnamefont{Reno}},
  \bibinfo{author}{\bibfnamefont{L.~N.} \bibnamefont{Pfeiffer}}, and
  \bibinfo{author}{\bibfnamefont{K.~W.} \bibnamefont{West}},
  \bibinfo{year}{2012}{\natexlab{c}}, \bibinfo{journal}{Phys. Rev. B}
  \textbf{\bibinfo{volume}{85}}, \bibinfo{pages}{081304(R)}.

\bibitem[{\citenamefont{Hatke}
  \emph{et~al.}(2012{\natexlab{d}})\citenamefont{Hatke, Zudov, Watson, and
  Manfra}}]{hatke:2012d}
\bibinfo{author}{\bibnamefont{Hatke}, \bibfnamefont{A.~T.}},
  \bibinfo{author}{\bibfnamefont{M.~A.} \bibnamefont{Zudov}},
  \bibinfo{author}{\bibfnamefont{J.~D.} \bibnamefont{Watson}}, and
  \bibinfo{author}{\bibfnamefont{M.~J.} \bibnamefont{Manfra}},
  \bibinfo{year}{2012}{\natexlab{d}}, \bibinfo{journal}{Phys. Rev. B}
  \textbf{\bibinfo{volume}{85}}, \bibinfo{pages}{121306}.

\bibitem[{\citenamefont{Heiblum} \emph{et~al.}(1984)\citenamefont{Heiblum,
  Mendez, and Stern}}]{heiblum:1984}
\bibinfo{author}{\bibnamefont{Heiblum}, \bibfnamefont{M.}},
  \bibinfo{author}{\bibfnamefont{E.~E.} \bibnamefont{Mendez}}, and
  \bibinfo{author}{\bibfnamefont{F.}~\bibnamefont{Stern}},
  \bibinfo{year}{1984}, \bibinfo{journal}{Appl. Phys. Lett.}
  \textbf{\bibinfo{volume}{44}}, \bibinfo{pages}{1064}.

\bibitem[{\citenamefont{Heinonen}(1998)}]{heinonen:1998}
\bibinfo{editor}{\bibnamefont{Heinonen}, \bibfnamefont{O.}} (ed.),
  \bibinfo{year}{1998}, \emph{\bibinfo{title}{Composite Fermions: A Unified
  View of the Quantum Hall Regime}} (\bibinfo{publisher}{World Scientific,
  Singapore}).

\bibitem[{\citenamefont{Herring}(1960)}]{herring:1960}
\bibinfo{author}{\bibnamefont{Herring}, \bibfnamefont{C.}},
  \bibinfo{year}{1960}, \bibinfo{journal}{J. Appl. Phys.}
  \textbf{\bibinfo{volume}{31}}, \bibinfo{pages}{1939}.

\bibitem[{\citenamefont{Hohenberg and Halperin}(1977)}]{hohenberg:1977}
\bibinfo{author}{\bibnamefont{Hohenberg}, \bibfnamefont{P.~C.}}, and
  \bibinfo{author}{\bibfnamefont{B.~I.} \bibnamefont{Halperin}},
  \bibinfo{year}{1977}, \bibinfo{journal}{Rev. Mod. Phys.}
  \textbf{\bibinfo{volume}{49}}, \bibinfo{pages}{435}.

\bibitem[{\citenamefont{Houghton} \emph{et~al.}(1982)\citenamefont{Houghton,
  Senna, and Ying}}]{houghton:1982}
\bibinfo{author}{\bibnamefont{Houghton}, \bibfnamefont{A.}},
  \bibinfo{author}{\bibfnamefont{J.~R.} \bibnamefont{Senna}}, and
  \bibinfo{author}{\bibfnamefont{S.~C.} \bibnamefont{Ying}},
  \bibinfo{year}{1982}, \bibinfo{journal}{Phys. Rev. B}
  \textbf{\bibinfo{volume}{25}}, \bibinfo{pages}{2196}.

\bibitem[{\citenamefont{Ignatov} \emph{et~al.}(1995)\citenamefont{Ignatov,
  Schomburg, Grenzer, Renk, and Dodin}}]{ignatov:1995}
\bibinfo{author}{\bibnamefont{Ignatov}, \bibfnamefont{A.~A.}},
  \bibinfo{author}{\bibfnamefont{E.}~\bibnamefont{Schomburg}},
  \bibinfo{author}{\bibfnamefont{J.}~\bibnamefont{Grenzer}},
  \bibinfo{author}{\bibfnamefont{K.~F.} \bibnamefont{Renk}}, and
  \bibinfo{author}{\bibfnamefont{E.~P.} \bibnamefont{Dodin}},
  \bibinfo{year}{1995}, \bibinfo{journal}{Z. Phys. B}
  \textbf{\bibinfo{volume}{98}}, \bibinfo{pages}{187}.

\bibitem[{\citenamefont{Ilani} \emph{et~al.}(2000)\citenamefont{Ilani, Yacoby,
  Mahalu, and Shtrikman}}]{ilani:2000}
\bibinfo{author}{\bibnamefont{Ilani}, \bibfnamefont{S.}},
  \bibinfo{author}{\bibfnamefont{A.}~\bibnamefont{Yacoby}},
  \bibinfo{author}{\bibfnamefont{D.}~\bibnamefont{Mahalu}}, and
  \bibinfo{author}{\bibfnamefont{H.}~\bibnamefont{Shtrikman}},
  \bibinfo{year}{2000}, \bibinfo{journal}{Phys. Rev. Lett.}
  \textbf{\bibinfo{volume}{84}}, \bibinfo{pages}{3133}.

\bibitem[{\citenamefont{Ilani} \emph{et~al.}(2001)\citenamefont{Ilani, Yacoby,
  Mahalu, and Shtrikman}}]{ilani:2001}
\bibinfo{author}{\bibnamefont{Ilani}, \bibfnamefont{S.}},
  \bibinfo{author}{\bibfnamefont{A.}~\bibnamefont{Yacoby}},
  \bibinfo{author}{\bibfnamefont{D.}~\bibnamefont{Mahalu}}, and
  \bibinfo{author}{\bibfnamefont{H.}~\bibnamefont{Shtrikman}},
  \bibinfo{year}{2001}, \bibinfo{journal}{Science}
  \textbf{\bibinfo{volume}{292}}, \bibinfo{pages}{1354}.

\bibitem[{\citenamefont{Imry and Ma}(1975)}]{imry:1975}
\bibinfo{author}{\bibnamefont{Imry}, \bibfnamefont{Y.}}, and
  \bibinfo{author}{\bibfnamefont{S.-K.} \bibnamefont{Ma}},
  \bibinfo{year}{1975}, \bibinfo{journal}{Phys. Rev. Lett.}
  \textbf{\bibinfo{volume}{35}}, \bibinfo{pages}{1399}.

\bibitem[{\citenamefont{Isichenko}(1992)}]{isichenko:1992}
\bibinfo{author}{\bibnamefont{Isichenko}, \bibfnamefont{M.~B.}},
  \bibinfo{year}{1992}, \bibinfo{journal}{Rev. Mod. Phys.}
  \textbf{\bibinfo{volume}{64}}, \bibinfo{pages}{961}.

\bibitem[{\citenamefont{Isihara and Smr\v{c}ka}(1986)}]{isihara:1986}
\bibinfo{author}{\bibnamefont{Isihara}, \bibfnamefont{A.}}, and
  \bibinfo{author}{\bibfnamefont{L.}~\bibnamefont{Smr\v{c}ka}},
  \bibinfo{year}{1986}, \bibinfo{journal}{J. Phys. C}
  \textbf{\bibinfo{volume}{19}}, \bibinfo{pages}{6777}.

\bibitem[{\citenamefont{Jain}(2007)}]{jain:2007}
\bibinfo{author}{\bibnamefont{Jain}, \bibfnamefont{J.~K.}},
  \bibinfo{year}{2007}, \emph{\bibinfo{title}{Composite Fermions}}
  (\bibinfo{publisher}{Cambridge University, Cambridge}).

\bibitem[{\citenamefont{Janssen}(1976)}]{janssen:1976}
\bibinfo{author}{\bibnamefont{Janssen}, \bibfnamefont{H.-K.}},
  \bibinfo{year}{1976}, \bibinfo{journal}{Z. Physik B}
  \textbf{\bibinfo{volume}{23}}, \bibinfo{pages}{377}.

\bibitem[{\citenamefont{Jauho and Johnsen}(1996)}]{jauho:1996}
\bibinfo{author}{\bibnamefont{Jauho}, \bibfnamefont{A.~P.}}, and
  \bibinfo{author}{\bibfnamefont{K.}~\bibnamefont{Johnsen}},
  \bibinfo{year}{1996}, \bibinfo{journal}{Phys. Rev. Lett.}
  \textbf{\bibinfo{volume}{76}}, \bibinfo{pages}{4576}.

\bibitem[{\citenamefont{Joas} \emph{et~al.}(2005)\citenamefont{Joas, Dietel,
  and von Oppen}}]{joas:2005}
\bibinfo{author}{\bibnamefont{Joas}, \bibfnamefont{C.}},
  \bibinfo{author}{\bibfnamefont{J.}~\bibnamefont{Dietel}}, and
  \bibinfo{author}{\bibfnamefont{F.}~\bibnamefont{von Oppen}},
  \bibinfo{year}{2005}, \bibinfo{journal}{Phys. Rev. B}
  \textbf{\bibinfo{volume}{72}}, \bibinfo{pages}{165323}.

\bibitem[{\citenamefont{Joas} \emph{et~al.}(2004)\citenamefont{Joas, Raikh, and
  von Oppen}}]{joas:2004}
\bibinfo{author}{\bibnamefont{Joas}, \bibfnamefont{C.}},
  \bibinfo{author}{\bibfnamefont{M.~E.} \bibnamefont{Raikh}}, and
  \bibinfo{author}{\bibfnamefont{F.}~\bibnamefont{von Oppen}},
  \bibinfo{year}{2004}, \bibinfo{journal}{Phys. Rev. B}
  \textbf{\bibinfo{volume}{70}}, \bibinfo{pages}{235302}.

\bibitem[{\citenamefont{Johnsen and Jauho}(1998)}]{johnsen:1998}
\bibinfo{author}{\bibnamefont{Johnsen}, \bibfnamefont{K.}}, and
  \bibinfo{author}{\bibfnamefont{A.-P.} \bibnamefont{Jauho}},
  \bibinfo{year}{1998}, \bibinfo{journal}{Phys. Rev. B}
  \textbf{\bibinfo{volume}{57}}, \bibinfo{pages}{8860}.

\bibitem[{\citenamefont{Kalmanovitz}
  \emph{et~al.}(2008)\citenamefont{Kalmanovitz, Bykov, Vitkalov, and
  Toropov}}]{kalmanovitz:2008a}
\bibinfo{author}{\bibnamefont{Kalmanovitz}, \bibfnamefont{N.~R.}},
  \bibinfo{author}{\bibfnamefont{A.~A.} \bibnamefont{Bykov}},
  \bibinfo{author}{\bibfnamefont{S.~A.} \bibnamefont{Vitkalov}}, and
  \bibinfo{author}{\bibfnamefont{A.~I.} \bibnamefont{Toropov}},
  \bibinfo{year}{2008}, \bibinfo{journal}{Phys. Rev. B}
  \textbf{\bibinfo{volume}{78}}, \bibinfo{pages}{085306}.

\bibitem[{\citenamefont{Kashuba}(2006{\natexlab{a}})}]{kashuba:2006b}
\bibinfo{author}{\bibnamefont{Kashuba}, \bibfnamefont{A.}},
  \bibinfo{year}{2006}{\natexlab{a}}, \bibinfo{journal}{JETP Lett.}
  \textbf{\bibinfo{volume}{83}}, \bibinfo{pages}{293}.

\bibitem[{\citenamefont{Kashuba}(2006{\natexlab{b}})}]{kashuba:2006}
\bibinfo{author}{\bibnamefont{Kashuba}, \bibfnamefont{A.}},
  \bibinfo{year}{2006}{\natexlab{b}}, \bibinfo{journal}{Phys. Rev. B}
  \textbf{\bibinfo{volume}{73}}, \bibinfo{pages}{125340}.

\bibitem[{\citenamefont{Kazarinov and Skobov}(1963)}]{kazarinov:1963}
\bibinfo{author}{\bibnamefont{Kazarinov}, \bibfnamefont{R.~F.}}, and
  \bibinfo{author}{\bibfnamefont{V.~G.} \bibnamefont{Skobov}},
  \bibinfo{year}{1963}, \bibinfo{journal}{Sov. Phys. JETP}
  \textbf{\bibinfo{volume}{17}}, \bibinfo{pages}{921}.

\bibitem[{\citenamefont{Keay} \emph{et~al.}(1995)\citenamefont{Keay, Zeuner,
  S.~J.~Allen, Maranowski, Gossard, Bhattacharaya, and Rodwell}}]{keay:1995}
\bibinfo{author}{\bibnamefont{Keay}, \bibfnamefont{B.~J.}},
  \bibinfo{author}{\bibfnamefont{S.}~\bibnamefont{Zeuner}},
  \bibinfo{author}{\bibfnamefont{J.}~\bibnamefont{S.~J.~Allen}},
  \bibinfo{author}{\bibfnamefont{K.~D.} \bibnamefont{Maranowski}},
  \bibinfo{author}{\bibfnamefont{A.~C.} \bibnamefont{Gossard}},
  \bibinfo{author}{\bibfnamefont{U.}~\bibnamefont{Bhattacharaya}}, and
  \bibinfo{author}{\bibfnamefont{M.~J.~W.} \bibnamefont{Rodwell}},
  \bibinfo{year}{1995}, \bibinfo{journal}{Phys. Rev. Lett.}
  \textbf{\bibinfo{volume}{75}}, \bibinfo{pages}{4102}.

\bibitem[{\citenamefont{Keldysh}(1965)}]{keldysh:1965}
\bibinfo{author}{\bibnamefont{Keldysh}, \bibfnamefont{L.~V.}},
  \bibinfo{year}{1965}, \bibinfo{journal}{Sov. Phys. JETP}
  \textbf{\bibinfo{volume}{20}}, \bibinfo{pages}{1307}.

\bibitem[{\citenamefont{Kennett} \emph{et~al.}(2005)\citenamefont{Kennett,
  Robinson, Cooper, and Fal'ko}}]{kennett:2005}
\bibinfo{author}{\bibnamefont{Kennett}, \bibfnamefont{M.~P.}},
  \bibinfo{author}{\bibfnamefont{J.~P.} \bibnamefont{Robinson}},
  \bibinfo{author}{\bibfnamefont{N.~R.} \bibnamefont{Cooper}}, and
  \bibinfo{author}{\bibfnamefont{V.~I.} \bibnamefont{Fal'ko}},
  \bibinfo{year}{2005}, \bibinfo{journal}{Phys. Rev. B}
  \textbf{\bibinfo{volume}{71}}, \bibinfo{pages}{195420}.

\bibitem[{\citenamefont{Khodas} \emph{et~al.}(2010)\citenamefont{Khodas,
  Chiang, Hatke, Zudov, Vavilov, Pfeiffer, and West}}]{khodas:2010}
\bibinfo{author}{\bibnamefont{Khodas}, \bibfnamefont{M.}},
  \bibinfo{author}{\bibfnamefont{H.-S.} \bibnamefont{Chiang}},
  \bibinfo{author}{\bibfnamefont{A.~T.} \bibnamefont{Hatke}},
  \bibinfo{author}{\bibfnamefont{M.~A.} \bibnamefont{Zudov}},
  \bibinfo{author}{\bibfnamefont{M.~G.} \bibnamefont{Vavilov}},
  \bibinfo{author}{\bibfnamefont{L.~N.} \bibnamefont{Pfeiffer}}, and
  \bibinfo{author}{\bibfnamefont{K.~W.} \bibnamefont{West}},
  \bibinfo{year}{2010}, \bibinfo{journal}{Phys. Rev. Lett.}
  \textbf{\bibinfo{volume}{104}}, \bibinfo{pages}{206801}.

\bibitem[{\citenamefont{Khodas and Vavilov}(2008)}]{khodas:2008}
\bibinfo{author}{\bibnamefont{Khodas}, \bibfnamefont{M.}}, and
  \bibinfo{author}{\bibfnamefont{M.~G.} \bibnamefont{Vavilov}},
  \bibinfo{year}{2008}, \bibinfo{journal}{Phys. Rev. B}
  \textbf{\bibinfo{volume}{78}}, \bibinfo{pages}{245319}.

\bibitem[{\citenamefont{Khveshchenko}(1996)}]{khveshchenko:1996}
\bibinfo{author}{\bibnamefont{Khveshchenko}, \bibfnamefont{D.~V.}},
  \bibinfo{year}{1996}, \bibinfo{journal}{Phys. Rev. Lett.}
  \textbf{\bibinfo{volume}{77}}, \bibinfo{pages}{1817}.

\bibitem[{\citenamefont{von Klitzing}(1986)}]{klitzing:1986}
\bibinfo{author}{\bibnamefont{von Klitzing}, \bibfnamefont{K.}},
  \bibinfo{year}{1986}, \bibinfo{journal}{Rev. Mod. Phys.}
  \textbf{\bibinfo{volume}{58}}, \bibinfo{pages}{519}.

\bibitem[{\citenamefont{von Klitzing} \emph{et~al.}(1980)\citenamefont{von
  Klitzing, Dorda, and Pepper}}]{klitzing:1980}
\bibinfo{author}{\bibnamefont{von Klitzing}, \bibfnamefont{K.}},
  \bibinfo{author}{\bibfnamefont{G.}~\bibnamefont{Dorda}}, and
  \bibinfo{author}{\bibfnamefont{M.}~\bibnamefont{Pepper}},
  \bibinfo{year}{1980}, \bibinfo{journal}{Phys. Rev. Lett.}
  \textbf{\bibinfo{volume}{45}}, \bibinfo{pages}{494}.

\bibitem[{\citenamefont{Kogan}(1968)}]{kogan:1968}
\bibinfo{author}{\bibnamefont{Kogan}, \bibfnamefont{S.~M.}},
  \bibinfo{year}{1968}, \bibinfo{journal}{Sov. Phys. JETP}
  \textbf{\bibinfo{volume}{10}}, \bibinfo{pages}{1213}.

\bibitem[{\citenamefont{Konstantinov}
  \emph{et~al.}(2012{\natexlab{a}})\citenamefont{Konstantinov, Chepelianskii,
  and Kono}}]{konstantinov:2011}
\bibinfo{author}{\bibnamefont{Konstantinov}, \bibfnamefont{D.}},
  \bibinfo{author}{\bibfnamefont{A.}~\bibnamefont{Chepelianskii}}, and
  \bibinfo{author}{\bibfnamefont{K.}~\bibnamefont{Kono}},
  \bibinfo{year}{2012}{\natexlab{a}}, \bibinfo{journal}{J. Phys. Soc. Jpn.}
  \textbf{\bibinfo{volume}{81}}, \bibinfo{pages}{093601}.

\bibitem[{\citenamefont{Konstantinov}
  \emph{et~al.}(2009)\citenamefont{Konstantinov, Dykman, Lea, Monarkha, and
  Kono}}]{konstantinov:2009a}
\bibinfo{author}{\bibnamefont{Konstantinov}, \bibfnamefont{D.}},
  \bibinfo{author}{\bibfnamefont{M.~I.} \bibnamefont{Dykman}},
  \bibinfo{author}{\bibfnamefont{M.~J.} \bibnamefont{Lea}},
  \bibinfo{author}{\bibfnamefont{Y.}~\bibnamefont{Monarkha}}, and
  \bibinfo{author}{\bibfnamefont{K.}~\bibnamefont{Kono}}, \bibinfo{year}{2009},
  \bibinfo{journal}{Phys. Rev. Lett.} \textbf{\bibinfo{volume}{103}},
  \bibinfo{pages}{096801}.

\bibitem[{\citenamefont{Konstantinov}
  \emph{et~al.}(2012{\natexlab{b}})\citenamefont{Konstantinov, Dykman, Lea,
  Monarkha, and Kono}}]{konstantinov:2012b}
\bibinfo{author}{\bibnamefont{Konstantinov}, \bibfnamefont{D.}},
  \bibinfo{author}{\bibfnamefont{M.~I.} \bibnamefont{Dykman}},
  \bibinfo{author}{\bibfnamefont{M.~J.} \bibnamefont{Lea}},
  \bibinfo{author}{\bibfnamefont{Y.~P.} \bibnamefont{Monarkha}}, and
  \bibinfo{author}{\bibfnamefont{K.}~\bibnamefont{Kono}},
  \bibinfo{year}{2012}{\natexlab{b}}, \bibinfo{journal}{Phys. Rev. B}
  \textbf{\bibinfo{volume}{85}}, \bibinfo{pages}{155416}.

\bibitem[{\citenamefont{Konstantinov}
  \emph{et~al.}(2007)\citenamefont{Konstantinov, Isshiki, Monarkha, Akimoto,
  Shirahama, and Kono}}]{konstantinov:2007}
\bibinfo{author}{\bibnamefont{Konstantinov}, \bibfnamefont{D.}},
  \bibinfo{author}{\bibfnamefont{H.}~\bibnamefont{Isshiki}},
  \bibinfo{author}{\bibfnamefont{Y.}~\bibnamefont{Monarkha}},
  \bibinfo{author}{\bibfnamefont{H.}~\bibnamefont{Akimoto}},
  \bibinfo{author}{\bibfnamefont{K.}~\bibnamefont{Shirahama}}, and
  \bibinfo{author}{\bibfnamefont{K.}~\bibnamefont{Kono}}, \bibinfo{year}{2007},
  \bibinfo{journal}{Phys. Rev. Lett.} \textbf{\bibinfo{volume}{98}},
  \bibinfo{pages}{235302}.

\bibitem[{\citenamefont{Konstantinov}
  \emph{et~al.}(2008{\natexlab{a}})\citenamefont{Konstantinov, Isshiki,
  Monarkha, Akimoto, Shirahama, and Kono}}]{konstantinov:2008a}
\bibinfo{author}{\bibnamefont{Konstantinov}, \bibfnamefont{D.}},
  \bibinfo{author}{\bibfnamefont{H.}~\bibnamefont{Isshiki}},
  \bibinfo{author}{\bibfnamefont{Y.}~\bibnamefont{Monarkha}},
  \bibinfo{author}{\bibfnamefont{H.}~\bibnamefont{Akimoto}},
  \bibinfo{author}{\bibfnamefont{K.}~\bibnamefont{Shirahama}}, and
  \bibinfo{author}{\bibfnamefont{K.}~\bibnamefont{Kono}},
  \bibinfo{year}{2008}{\natexlab{a}}, \bibinfo{journal}{J. Phys. Soc. Jpn.}
  \textbf{\bibinfo{volume}{77}}, \bibinfo{pages}{034705}.

\bibitem[{\citenamefont{Konstantinov and Kono}(2009)}]{konstantinov:2009b}
\bibinfo{author}{\bibnamefont{Konstantinov}, \bibfnamefont{D.}}, and
  \bibinfo{author}{\bibfnamefont{K.}~\bibnamefont{Kono}}, \bibinfo{year}{2009},
  \bibinfo{journal}{Phys. Rev. Lett.} \textbf{\bibinfo{volume}{103}},
  \bibinfo{pages}{266808}.

\bibitem[{\citenamefont{Konstantinov and Kono}(2010)}]{konstantinov:2010}
\bibinfo{author}{\bibnamefont{Konstantinov}, \bibfnamefont{D.}}, and
  \bibinfo{author}{\bibfnamefont{K.}~\bibnamefont{Kono}}, \bibinfo{year}{2010},
  \bibinfo{journal}{Phys. Rev. Lett.} \textbf{\bibinfo{volume}{105}},
  \bibinfo{pages}{226801}.

\bibitem[{\citenamefont{Konstantinov}
  \emph{et~al.}(2008{\natexlab{b}})\citenamefont{Konstantinov, Kono, and
  Monarkha}}]{konstantinov:2008b}
\bibinfo{author}{\bibnamefont{Konstantinov}, \bibfnamefont{D.}},
  \bibinfo{author}{\bibfnamefont{K.}~\bibnamefont{Kono}}, and
  \bibinfo{author}{\bibfnamefont{Y.}~\bibnamefont{Monarkha}},
  \bibinfo{year}{2008}{\natexlab{b}}, \bibinfo{journal}{Low Temp. Phys.}
  \textbf{\bibinfo{volume}{34}}, \bibinfo{pages}{377}.

\bibitem[{\citenamefont{Koulakov and Raikh}(2003)}]{koulakov:2003}
\bibinfo{author}{\bibnamefont{Koulakov}, \bibfnamefont{A.~A.}}, and
  \bibinfo{author}{\bibfnamefont{M.~E.} \bibnamefont{Raikh}},
  \bibinfo{year}{2003}, \bibinfo{journal}{Phys. Rev. B}
  \textbf{\bibinfo{volume}{68}}, \bibinfo{pages}{115324}.

\bibitem[{\citenamefont{Kroemer}(2001)}]{kroemer:2001}
\bibinfo{author}{\bibnamefont{Kroemer}, \bibfnamefont{H.}},
  \bibinfo{year}{2001}, \bibinfo{journal}{Rev. Mod. Phys.}
  \textbf{\bibinfo{volume}{73}}, \bibinfo{pages}{783}.

\bibitem[{\citenamefont{Kukushkin} \emph{et~al.}(2004)\citenamefont{Kukushkin,
  Akimov, Smet, Mikhailov, von Klitzing, Aleiner, and Falko}}]{kukushkin:2004}
\bibinfo{author}{\bibnamefont{Kukushkin}, \bibfnamefont{I.~V.}},
  \bibinfo{author}{\bibfnamefont{M.~Y.} \bibnamefont{Akimov}},
  \bibinfo{author}{\bibfnamefont{J.~H.} \bibnamefont{Smet}},
  \bibinfo{author}{\bibfnamefont{S.~A.} \bibnamefont{Mikhailov}},
  \bibinfo{author}{\bibfnamefont{K.}~\bibnamefont{von Klitzing}},
  \bibinfo{author}{\bibfnamefont{I.~L.} \bibnamefont{Aleiner}}, and
  \bibinfo{author}{\bibfnamefont{V.~I.} \bibnamefont{Falko}},
  \bibinfo{year}{2004}, \bibinfo{journal}{Phys. Rev. Lett.}
  \textbf{\bibinfo{volume}{92}}, \bibinfo{pages}{236803}.

\bibitem[{\citenamefont{Kukushkin} \emph{et~al.}(2005)\citenamefont{Kukushkin,
  Mikhailov, Smet, and von Klitzing}}]{kukushkin:2005}
\bibinfo{author}{\bibnamefont{Kukushkin}, \bibfnamefont{I.~V.}},
  \bibinfo{author}{\bibfnamefont{S.~A.} \bibnamefont{Mikhailov}},
  \bibinfo{author}{\bibfnamefont{J.~H.} \bibnamefont{Smet}}, and
  \bibinfo{author}{\bibfnamefont{K.}~\bibnamefont{von Klitzing}},
  \bibinfo{year}{2005}, \bibinfo{journal}{Appl. Phys. Lett.}
  \textbf{\bibinfo{volume}{86}}, \bibinfo{pages}{044101}.

\bibitem[{\citenamefont{Laikhtman and Altshuler}(1994)}]{laikhtman:1994}
\bibinfo{author}{\bibnamefont{Laikhtman}, \bibfnamefont{B.}}, and
  \bibinfo{author}{\bibfnamefont{E.~L.} \bibnamefont{Altshuler}},
  \bibinfo{year}{1994}, \bibinfo{journal}{Ann. Phys. (N.Y.)}
  \textbf{\bibinfo{volume}{232}}, \bibinfo{pages}{332}.

\bibitem[{\citenamefont{Langenbuch}
  \emph{et~al.}(2004)\citenamefont{Langenbuch, Suhrke, and
  R\"ossler}}]{langenbuch:2004}
\bibinfo{author}{\bibnamefont{Langenbuch}, \bibfnamefont{M.}},
  \bibinfo{author}{\bibfnamefont{M.}~\bibnamefont{Suhrke}}, and
  \bibinfo{author}{\bibfnamefont{U.}~\bibnamefont{R\"ossler}},
  \bibinfo{year}{2004}, \bibinfo{journal}{Phys. Rev. B}
  \textbf{\bibinfo{volume}{69}}, \bibinfo{pages}{125303}.

\bibitem[{\citenamefont{Laughlin}(1999)}]{laughlin:1999}
\bibinfo{author}{\bibnamefont{Laughlin}, \bibfnamefont{R.~B.}},
  \bibinfo{year}{1999}, \bibinfo{journal}{Rev. Mod. Phys.}
  \textbf{\bibinfo{volume}{71}}, \bibinfo{pages}{863}.

\bibitem[{\citenamefont{Lea and Dykman}(1998)}]{lea:1998}
\bibinfo{author}{\bibnamefont{Lea}, \bibfnamefont{M.~J.}}, and
  \bibinfo{author}{\bibfnamefont{M.~I.} \bibnamefont{Dykman}},
  \bibinfo{year}{1998}, \bibinfo{journal}{Physica B}
  \textbf{\bibinfo{volume}{249-251}}, \bibinfo{pages}{628}.

\bibitem[{\citenamefont{Leadley} \emph{et~al.}(1992)\citenamefont{Leadley,
  Fletcher, Nicholas, Tao, Foxon, and Harris}}]{leadley:1992}
\bibinfo{author}{\bibnamefont{Leadley}, \bibfnamefont{D.~R.}},
  \bibinfo{author}{\bibfnamefont{R.}~\bibnamefont{Fletcher}},
  \bibinfo{author}{\bibfnamefont{R.~J.} \bibnamefont{Nicholas}},
  \bibinfo{author}{\bibfnamefont{F.}~\bibnamefont{Tao}},
  \bibinfo{author}{\bibfnamefont{C.~T.} \bibnamefont{Foxon}}, and
  \bibinfo{author}{\bibfnamefont{J.~J.} \bibnamefont{Harris}},
  \bibinfo{year}{1992}, \bibinfo{journal}{Phys. Rev. B}
  \textbf{\bibinfo{volume}{46}}, \bibinfo{pages}{12439}.

\bibitem[{\citenamefont{Leadley} \emph{et~al.}(1989)\citenamefont{Leadley,
  Nicholas, Harris, and Foxon}}]{leadely:1989}
\bibinfo{author}{\bibnamefont{Leadley}, \bibfnamefont{D.~R.}},
  \bibinfo{author}{\bibfnamefont{R.~J.} \bibnamefont{Nicholas}},
  \bibinfo{author}{\bibfnamefont{J.~J.} \bibnamefont{Harris}}, and
  \bibinfo{author}{\bibfnamefont{C.~T.} \bibnamefont{Foxon}},
  \bibinfo{year}{1989}, \bibinfo{journal}{Semicond. Sci. Technol.}
  \textbf{\bibinfo{volume}{4}}, \bibinfo{pages}{885}.

\bibitem[{\citenamefont{Lee and Leinaas}(2004)}]{lee:2004}
\bibinfo{author}{\bibnamefont{Lee}, \bibfnamefont{D.-H.}}, and
  \bibinfo{author}{\bibfnamefont{J.~M.} \bibnamefont{Leinaas}},
  \bibinfo{year}{2004}, \bibinfo{journal}{Phys. Rev. B}
  \textbf{\bibinfo{volume}{69}}, \bibinfo{pages}{115336}.

\bibitem[{\citenamefont{Lee} \emph{et~al.}(1983)\citenamefont{Lee, Shur,
  Drummond, and Morko{\c{c}}}}]{lee:1983}
\bibinfo{author}{\bibnamefont{Lee}, \bibfnamefont{K.}},
  \bibinfo{author}{\bibfnamefont{M.~S.} \bibnamefont{Shur}},
  \bibinfo{author}{\bibfnamefont{T.~J.} \bibnamefont{Drummond}}, and
  \bibinfo{author}{\bibfnamefont{H.}~\bibnamefont{Morko{\c{c}}}},
  \bibinfo{year}{1983}, \bibinfo{journal}{J. Appl. Phys.}
  \textbf{\bibinfo{volume}{54}}, \bibinfo{pages}{6432}.

\bibitem[{\citenamefont{Lei}(2006)}]{lei:2006c}
\bibinfo{author}{\bibnamefont{Lei}, \bibfnamefont{X.~L.}},
  \bibinfo{year}{2006}, \bibinfo{journal}{Phys. Rev. B}
  \textbf{\bibinfo{volume}{73}}, \bibinfo{pages}{235322}.

\bibitem[{\citenamefont{Lei}(2008)}]{lei:2008}
\bibinfo{author}{\bibnamefont{Lei}, \bibfnamefont{X.~L.}},
  \bibinfo{year}{2008}, \bibinfo{journal}{Phys. Rev. B}
  \textbf{\bibinfo{volume}{77}}, \bibinfo{pages}{205309}.

\bibitem[{\citenamefont{Lei}(2009)}]{lei:2009}
\bibinfo{author}{\bibnamefont{Lei}, \bibfnamefont{X.~L.}},
  \bibinfo{year}{2009}, \bibinfo{journal}{Phys. Rev. B}
  \textbf{\bibinfo{volume}{79}}, \bibinfo{eid}{115308}.

\bibitem[{\citenamefont{Lei}(2010)}]{lei:2010}
\bibinfo{author}{\bibnamefont{Lei}, \bibfnamefont{X.~L.}},
  \bibinfo{year}{2010}, \bibinfo{journal}{Mater. Sci. Eng., R}
  \textbf{\bibinfo{volume}{70}}, \bibinfo{pages}{126}.

\bibitem[{\citenamefont{Lei and Liu}(2003)}]{lei:2003}
\bibinfo{author}{\bibnamefont{Lei}, \bibfnamefont{X.~L.}}, and
  \bibinfo{author}{\bibfnamefont{S.~Y.} \bibnamefont{Liu}},
  \bibinfo{year}{2003}, \bibinfo{journal}{Phys. Rev. Lett.}
  \textbf{\bibinfo{volume}{91}}, \bibinfo{pages}{226805}.

\bibitem[{\citenamefont{Lei and Liu}(2006{\natexlab{a}})}]{lei:2006a}
\bibinfo{author}{\bibnamefont{Lei}, \bibfnamefont{X.~L.}}, and
  \bibinfo{author}{\bibfnamefont{S.~Y.} \bibnamefont{Liu}},
  \bibinfo{year}{2006}{\natexlab{a}}, \bibinfo{journal}{Appl. Phys. Lett.}
  \textbf{\bibinfo{volume}{89}}, \bibinfo{pages}{182117}.

\bibitem[{\citenamefont{Lei and Liu}(2006{\natexlab{b}})}]{lei:2006b}
\bibinfo{author}{\bibnamefont{Lei}, \bibfnamefont{X.~L.}}, and
  \bibinfo{author}{\bibfnamefont{S.~Y.} \bibnamefont{Liu}},
  \bibinfo{year}{2006}{\natexlab{b}}, \bibinfo{journal}{Appl. Phys. Lett.}
  \textbf{\bibinfo{volume}{88}}, \bibinfo{pages}{212109}.

\bibitem[{\citenamefont{Levinson} \emph{et~al.}(1998)\citenamefont{Levinson,
  Entin-Wohlman, Mirlin, and W\"olfle}}]{levinson:1998}
\bibinfo{author}{\bibnamefont{Levinson}, \bibfnamefont{Y.}},
  \bibinfo{author}{\bibfnamefont{O.}~\bibnamefont{Entin-Wohlman}},
  \bibinfo{author}{\bibfnamefont{A.~D.} \bibnamefont{Mirlin}}, and
  \bibinfo{author}{\bibfnamefont{P.}~\bibnamefont{W\"olfle}},
  \bibinfo{year}{1998}, \bibinfo{journal}{Phys. Rev. B}
  \textbf{\bibinfo{volume}{58}}, \bibinfo{pages}{7113}.

\bibitem[{\citenamefont{Li} \emph{et~al.}(2003)\citenamefont{Li, Proskuryakov,
  Savchenko, Linfield, and Ritchie}}]{li:2003}
\bibinfo{author}{\bibnamefont{Li}, \bibfnamefont{L.}},
  \bibinfo{author}{\bibfnamefont{Y.~Y.} \bibnamefont{Proskuryakov}},
  \bibinfo{author}{\bibfnamefont{A.~K.} \bibnamefont{Savchenko}},
  \bibinfo{author}{\bibfnamefont{E.~H.} \bibnamefont{Linfield}}, and
  \bibinfo{author}{\bibfnamefont{D.~A.} \bibnamefont{Ritchie}},
  \bibinfo{year}{2003}, \bibinfo{journal}{Phys. Rev. Lett.}
  \textbf{\bibinfo{volume}{90}}, \bibinfo{pages}{076802}.

\bibitem[{\citenamefont{Liao} \emph{et~al.}(1980)\citenamefont{Liao, Glass, and
  Humphrey}}]{liao:1980}
\bibinfo{author}{\bibnamefont{Liao}, \bibfnamefont{P.~F.}},
  \bibinfo{author}{\bibfnamefont{A.~M.} \bibnamefont{Glass}}, and
  \bibinfo{author}{\bibfnamefont{L.~M.} \bibnamefont{Humphrey}},
  \bibinfo{year}{1980}, \bibinfo{journal}{Phys. Rev. B}
  \textbf{\bibinfo{volume}{22}}, \bibinfo{pages}{2276}.

\bibitem[{\citenamefont{Lifshitz} \emph{et~al.}(1973)\citenamefont{Lifshitz,
  Azbel, and Kaganov}}]{lifshits:1973}
\bibinfo{author}{\bibnamefont{Lifshitz}, \bibfnamefont{I.~M.}},
  \bibinfo{author}{\bibfnamefont{M.~Y.} \bibnamefont{Azbel}}, and
  \bibinfo{author}{\bibfnamefont{M.~I.} \bibnamefont{Kaganov}},
  \bibinfo{year}{1973}, \emph{\bibinfo{title}{Electron Theory of Metals}}
  (\bibinfo{publisher}{Consultants Bureau, New York}).

\bibitem[{\citenamefont{Lyapilin and Patrakov}(2004)}]{lyapilin:2004}
\bibinfo{author}{\bibnamefont{Lyapilin}, \bibfnamefont{I.~I.}}, and
  \bibinfo{author}{\bibfnamefont{A.~E.} \bibnamefont{Patrakov}},
  \bibinfo{year}{2004}, \bibinfo{journal}{Low Temp. Phys.}
  \textbf{\bibinfo{volume}{30}}, \bibinfo{pages}{834}.

\bibitem[{\citenamefont{Lyapilin and Patrakov}(2006)}]{lyapilin:2006}
\bibinfo{author}{\bibnamefont{Lyapilin}, \bibfnamefont{I.~I.}}, and
  \bibinfo{author}{\bibfnamefont{A.~E.} \bibnamefont{Patrakov}},
  \bibinfo{year}{2006}, \bibinfo{journal}{Phys. Met. Metallogr.}
  \textbf{\bibinfo{volume}{102}}, \bibinfo{pages}{560}.

\bibitem[{\citenamefont{Mamani} \emph{et~al.}(2008)\citenamefont{Mamani, Gusev,
  Lamas, Bakarov, and Raichev}}]{mamani:2008}
\bibinfo{author}{\bibnamefont{Mamani}, \bibfnamefont{N.~C.}},
  \bibinfo{author}{\bibfnamefont{G.~M.} \bibnamefont{Gusev}},
  \bibinfo{author}{\bibfnamefont{T.~E.} \bibnamefont{Lamas}},
  \bibinfo{author}{\bibfnamefont{A.~K.} \bibnamefont{Bakarov}}, and
  \bibinfo{author}{\bibfnamefont{O.~E.} \bibnamefont{Raichev}},
  \bibinfo{year}{2008}, \bibinfo{journal}{Phys. Rev. B}
  \textbf{\bibinfo{volume}{77}}, \bibinfo{pages}{205327}.

\bibitem[{\citenamefont{Mamani}
  \emph{et~al.}(2009{\natexlab{a}})\citenamefont{Mamani, Gusev, Raichev, Lamas,
  and Bakarov}}]{mamani:2009b}
\bibinfo{author}{\bibnamefont{Mamani}, \bibfnamefont{N.~C.}},
  \bibinfo{author}{\bibfnamefont{G.~M.} \bibnamefont{Gusev}},
  \bibinfo{author}{\bibfnamefont{O.~E.} \bibnamefont{Raichev}},
  \bibinfo{author}{\bibfnamefont{T.~E.} \bibnamefont{Lamas}}, and
  \bibinfo{author}{\bibfnamefont{A.~K.} \bibnamefont{Bakarov}},
  \bibinfo{year}{2009}{\natexlab{a}}, \bibinfo{journal}{Phys. Rev. B}
  \textbf{\bibinfo{volume}{80}}, \bibinfo{pages}{075308}.

\bibitem[{\citenamefont{Mamani}
  \emph{et~al.}(2009{\natexlab{b}})\citenamefont{Mamani, Gusev, da~Silva,
  Raichev, Quivy, and Bakarov}}]{mamani:2009a}
\bibinfo{author}{\bibnamefont{Mamani}, \bibfnamefont{N.~C.}},
  \bibinfo{author}{\bibfnamefont{G.~M.} \bibnamefont{Gusev}},
  \bibinfo{author}{\bibfnamefont{E.~C.~F.} \bibnamefont{da~Silva}},
  \bibinfo{author}{\bibfnamefont{O.~E.} \bibnamefont{Raichev}},
  \bibinfo{author}{\bibfnamefont{A.~A.} \bibnamefont{Quivy}}, and
  \bibinfo{author}{\bibfnamefont{A.~K.} \bibnamefont{Bakarov}},
  \bibinfo{year}{2009}{\natexlab{b}}, \bibinfo{journal}{Phys. Rev. B}
  \textbf{\bibinfo{volume}{80}}, \bibinfo{pages}{085304}.

\bibitem[{\citenamefont{Mani}(2004{\natexlab{a}})}]{mani:2004f}
\bibinfo{author}{\bibnamefont{Mani}, \bibfnamefont{R.}},
  \bibinfo{year}{2004}{\natexlab{a}}, \bibinfo{journal}{Physica E}
  \textbf{\bibinfo{volume}{22}}, \bibinfo{pages}{1}.

\bibitem[{\citenamefont{Mani}(2004{\natexlab{b}})}]{mani:2004d}
\bibinfo{author}{\bibnamefont{Mani}, \bibfnamefont{R.~G.}},
  \bibinfo{year}{2004}{\natexlab{b}}, \bibinfo{journal}{Appl. Phys. Lett.}
  \textbf{\bibinfo{volume}{85}}, \bibinfo{pages}{4962}.

\bibitem[{\citenamefont{Mani}(2005)}]{mani:2005}
\bibinfo{author}{\bibnamefont{Mani}, \bibfnamefont{R.~G.}},
  \bibinfo{year}{2005}, \bibinfo{journal}{Phys. Rev. B}
  \textbf{\bibinfo{volume}{72}}, \bibinfo{pages}{075327}.

\bibitem[{\citenamefont{Mani}(2007{\natexlab{a}})}]{mani:2007b}
\bibinfo{author}{\bibnamefont{Mani}, \bibfnamefont{R.~G.}},
  \bibinfo{year}{2007}{\natexlab{a}}, \bibinfo{journal}{Solid State Commun.}
  \textbf{\bibinfo{volume}{144}}, \bibinfo{pages}{409}.

\bibitem[{\citenamefont{Mani}(2007{\natexlab{b}})}]{mani:2007}
\bibinfo{author}{\bibnamefont{Mani}, \bibfnamefont{R.~G.}},
  \bibinfo{year}{2007}{\natexlab{b}}, \bibinfo{journal}{Appl. Phys. Lett.}
  \textbf{\bibinfo{volume}{91}}, \bibinfo{pages}{132103}.

\bibitem[{\citenamefont{Mani}(2008)}]{mani:2008}
\bibinfo{author}{\bibnamefont{Mani}, \bibfnamefont{R.~G.}},
  \bibinfo{year}{2008}, \bibinfo{journal}{Appl. Phys. Lett.}
  \textbf{\bibinfo{volume}{92}}, \bibinfo{pages}{102107}.

\bibitem[{\citenamefont{Mani} \emph{et~al.}(2010)\citenamefont{Mani, Gerl,
  Schmult, Wegscheider, and Umansky}}]{mani:2010}
\bibinfo{author}{\bibnamefont{Mani}, \bibfnamefont{R.~G.}},
  \bibinfo{author}{\bibfnamefont{C.}~\bibnamefont{Gerl}},
  \bibinfo{author}{\bibfnamefont{S.}~\bibnamefont{Schmult}},
  \bibinfo{author}{\bibfnamefont{W.}~\bibnamefont{Wegscheider}}, and
  \bibinfo{author}{\bibfnamefont{V.}~\bibnamefont{Umansky}},
  \bibinfo{year}{2010}, \bibinfo{journal}{Phys. Rev. B}
  \textbf{\bibinfo{volume}{81}}, \bibinfo{pages}{125320}.

\bibitem[{\citenamefont{Mani} \emph{et~al.}(2009)\citenamefont{Mani, Johnson,
  Umansky, Narayanamurti, and Ploog}}]{mani:2009}
\bibinfo{author}{\bibnamefont{Mani}, \bibfnamefont{R.~G.}},
  \bibinfo{author}{\bibfnamefont{W.~B.} \bibnamefont{Johnson}},
  \bibinfo{author}{\bibfnamefont{V.}~\bibnamefont{Umansky}},
  \bibinfo{author}{\bibfnamefont{V.}~\bibnamefont{Narayanamurti}}, and
  \bibinfo{author}{\bibfnamefont{K.}~\bibnamefont{Ploog}},
  \bibinfo{year}{2009}, \bibinfo{journal}{Phys. Rev. B}
  \textbf{\bibinfo{volume}{79}}, \bibinfo{pages}{205320}.

\bibitem[{\citenamefont{Mani}
  \emph{et~al.}(2004{\natexlab{a}})\citenamefont{Mani, Narayanamurti, von
  Klitzing, Smet, Johnson, and Umansky}}]{mani:2004b}
\bibinfo{author}{\bibnamefont{Mani}, \bibfnamefont{R.~G.}},
  \bibinfo{author}{\bibfnamefont{V.}~\bibnamefont{Narayanamurti}},
  \bibinfo{author}{\bibfnamefont{K.}~\bibnamefont{von Klitzing}},
  \bibinfo{author}{\bibfnamefont{J.~H.} \bibnamefont{Smet}},
  \bibinfo{author}{\bibfnamefont{W.~B.} \bibnamefont{Johnson}}, and
  \bibinfo{author}{\bibfnamefont{V.}~\bibnamefont{Umansky}},
  \bibinfo{year}{2004}{\natexlab{a}}, \bibinfo{journal}{Phys. Rev. B}
  \textbf{\bibinfo{volume}{69}}, \bibinfo{pages}{161306(R)}.

\bibitem[{\citenamefont{Mani}
  \emph{et~al.}(2004{\natexlab{b}})\citenamefont{Mani, Narayanamurti, von
  Klitzing, Smet, Johnson, and Umansky}}]{mani:2004a}
\bibinfo{author}{\bibnamefont{Mani}, \bibfnamefont{R.~G.}},
  \bibinfo{author}{\bibfnamefont{V.}~\bibnamefont{Narayanamurti}},
  \bibinfo{author}{\bibfnamefont{K.}~\bibnamefont{von Klitzing}},
  \bibinfo{author}{\bibfnamefont{J.~H.} \bibnamefont{Smet}},
  \bibinfo{author}{\bibfnamefont{W.~B.} \bibnamefont{Johnson}}, and
  \bibinfo{author}{\bibfnamefont{V.}~\bibnamefont{Umansky}},
  \bibinfo{year}{2004}{\natexlab{b}}, \bibinfo{journal}{Phys. Rev. B}
  \textbf{\bibinfo{volume}{70}}, \bibinfo{pages}{155310}.

\bibitem[{\citenamefont{Mani} \emph{et~al.}(2011)\citenamefont{Mani,
  Ramanayaka, and Wegscheider}}]{mani:2011}
\bibinfo{author}{\bibnamefont{Mani}, \bibfnamefont{R.~G.}},
  \bibinfo{author}{\bibfnamefont{A.~N.} \bibnamefont{Ramanayaka}}, and
  \bibinfo{author}{\bibfnamefont{W.}~\bibnamefont{Wegscheider}},
  \bibinfo{year}{2011}, \bibinfo{journal}{Phys. Rev. B}
  \textbf{\bibinfo{volume}{84}}, \bibinfo{pages}{085308}.

\bibitem[{\citenamefont{Mani} \emph{et~al.}(2002)\citenamefont{Mani, Smet, von
  Klitzing, Narayanamurti, Johnson, and Umansky}}]{mani:2002}
\bibinfo{author}{\bibnamefont{Mani}, \bibfnamefont{R.~G.}},
  \bibinfo{author}{\bibfnamefont{J.~H.} \bibnamefont{Smet}},
  \bibinfo{author}{\bibfnamefont{K.}~\bibnamefont{von Klitzing}},
  \bibinfo{author}{\bibfnamefont{V.}~\bibnamefont{Narayanamurti}},
  \bibinfo{author}{\bibfnamefont{W.~B.} \bibnamefont{Johnson}}, and
  \bibinfo{author}{\bibfnamefont{V.}~\bibnamefont{Umansky}},
  \bibinfo{year}{2002}, \bibinfo{journal}{Nature (London)}
  \textbf{\bibinfo{volume}{420}}, \bibinfo{pages}{646}.

\bibitem[{\citenamefont{Mani}
  \emph{et~al.}(2004{\natexlab{c}})\citenamefont{Mani, Smet, von Klitzing,
  Narayanamurti, Johnson, and Umansky}}]{mani:2004e}
\bibinfo{author}{\bibnamefont{Mani}, \bibfnamefont{R.~G.}},
  \bibinfo{author}{\bibfnamefont{J.~H.} \bibnamefont{Smet}},
  \bibinfo{author}{\bibfnamefont{K.}~\bibnamefont{von Klitzing}},
  \bibinfo{author}{\bibfnamefont{V.}~\bibnamefont{Narayanamurti}},
  \bibinfo{author}{\bibfnamefont{W.~B.} \bibnamefont{Johnson}}, and
  \bibinfo{author}{\bibfnamefont{V.}~\bibnamefont{Umansky}},
  \bibinfo{year}{2004}{\natexlab{c}}, \bibinfo{journal}{Phys. Rev. Lett.}
  \textbf{\bibinfo{volume}{92}}, \bibinfo{pages}{146801}.

\bibitem[{\citenamefont{Mani}
  \emph{et~al.}(2004{\natexlab{d}})\citenamefont{Mani, Smet, von Klitzing,
  Narayanamurti, Johnson, and Umansky}}]{mani:2004c}
\bibinfo{author}{\bibnamefont{Mani}, \bibfnamefont{R.~G.}},
  \bibinfo{author}{\bibfnamefont{J.~H.} \bibnamefont{Smet}},
  \bibinfo{author}{\bibfnamefont{K.}~\bibnamefont{von Klitzing}},
  \bibinfo{author}{\bibfnamefont{V.}~\bibnamefont{Narayanamurti}},
  \bibinfo{author}{\bibfnamefont{W.~B.} \bibnamefont{Johnson}}, and
  \bibinfo{author}{\bibfnamefont{V.}~\bibnamefont{Umansky}},
  \bibinfo{year}{2004}{\natexlab{d}}, \bibinfo{journal}{Phys. Rev. B}
  \textbf{\bibinfo{volume}{69}}, \bibinfo{pages}{193304}.

\bibitem[{\citenamefont{Markus} \emph{et~al.}(1994)\citenamefont{Markus,
  Meirav, Shtrikman, and Laikhtman}}]{markus:1994}
\bibinfo{author}{\bibnamefont{Markus}, \bibfnamefont{Y.}},
  \bibinfo{author}{\bibfnamefont{U.}~\bibnamefont{Meirav}},
  \bibinfo{author}{\bibfnamefont{H.}~\bibnamefont{Shtrikman}}, and
  \bibinfo{author}{\bibfnamefont{B.}~\bibnamefont{Laikhtman}},
  \bibinfo{year}{1994}, \bibinfo{journal}{Semicond. Sci. Technol.}
  \textbf{\bibinfo{volume}{9}}, \bibinfo{pages}{1297}.

\bibitem[{\citenamefont{Martin} \emph{et~al.}(2003)\citenamefont{Martin,
  Maslov, and Reizer}}]{martin:2003}
\bibinfo{author}{\bibnamefont{Martin}, \bibfnamefont{G.~W.}},
  \bibinfo{author}{\bibfnamefont{D.~L.} \bibnamefont{Maslov}}, and
  \bibinfo{author}{\bibfnamefont{M.~Y.} \bibnamefont{Reizer}},
  \bibinfo{year}{2003}, \bibinfo{journal}{Phys. Rev. B}
  \textbf{\bibinfo{volume}{68}}, \bibinfo{pages}{241309}.

\bibitem[{\citenamefont{Martin} \emph{et~al.}(1973)\citenamefont{Martin,
  Siggia, and Rose}}]{martin:1973}
\bibinfo{author}{\bibnamefont{Martin}, \bibfnamefont{P.~C.}},
  \bibinfo{author}{\bibfnamefont{E.~D.} \bibnamefont{Siggia}}, and
  \bibinfo{author}{\bibfnamefont{H.~A.} \bibnamefont{Rose}},
  \bibinfo{year}{1973}, \bibinfo{journal}{Phys. Rev. A}
  \textbf{\bibinfo{volume}{8}}, \bibinfo{pages}{423}.

\bibitem[{\citenamefont{McCann and Fal'ko}(2006)}]{mccann:2006}
\bibinfo{author}{\bibnamefont{McCann}, \bibfnamefont{E.}}, and
  \bibinfo{author}{\bibfnamefont{V.~I.} \bibnamefont{Fal'ko}},
  \bibinfo{year}{2006}, \bibinfo{journal}{Phys. Rev. Lett.}
  \textbf{\bibinfo{volume}{96}}, \bibinfo{pages}{086805}.

\bibitem[{\citenamefont{Mikhailov}(2004)}]{mikhailov:2004}
\bibinfo{author}{\bibnamefont{Mikhailov}, \bibfnamefont{S.~A.}},
  \bibinfo{year}{2004}, \bibinfo{journal}{Phys. Rev. B}
  \textbf{\bibinfo{volume}{70}}, \bibinfo{pages}{165311}.

\bibitem[{\citenamefont{Mikhailov}(2011)}]{mikhailov:2011}
\bibinfo{author}{\bibnamefont{Mikhailov}, \bibfnamefont{S.~A.}},
  \bibinfo{year}{2011}, \bibinfo{journal}{Phys. Rev. B}
  \textbf{\bibinfo{volume}{83}}, \bibinfo{pages}{155303}.

\bibitem[{\citenamefont{Mikhailov and Savostianova}(2006)}]{mikhailov:2006}
\bibinfo{author}{\bibnamefont{Mikhailov}, \bibfnamefont{S.~A.}}, and
  \bibinfo{author}{\bibfnamefont{N.~A.} \bibnamefont{Savostianova}},
  \bibinfo{year}{2006}, \bibinfo{journal}{Phys. Rev. B}
  \textbf{\bibinfo{volume}{74}}, \bibinfo{pages}{045325}.

\bibitem[{\citenamefont{Mirlin} \emph{et~al.}(1996)\citenamefont{Mirlin,
  Altshuler, and W\"olfle}}]{mirlin:1996}
\bibinfo{author}{\bibnamefont{Mirlin}, \bibfnamefont{A.~D.}},
  \bibinfo{author}{\bibfnamefont{E.}~\bibnamefont{Altshuler}}, and
  \bibinfo{author}{\bibfnamefont{P.}~\bibnamefont{W\"olfle}},
  \bibinfo{year}{1996}, \bibinfo{journal}{Ann. Phys. (Leipzig)}
  \textbf{\bibinfo{volume}{508}}, \bibinfo{pages}{281}.

\bibitem[{\citenamefont{Mirlin} \emph{et~al.}(2001)\citenamefont{Mirlin,
  Polyakov, Evers, and W\"olfle}}]{mirlin:2001}
\bibinfo{author}{\bibnamefont{Mirlin}, \bibfnamefont{A.~D.}},
  \bibinfo{author}{\bibfnamefont{D.~G.} \bibnamefont{Polyakov}},
  \bibinfo{author}{\bibfnamefont{F.}~\bibnamefont{Evers}}, and
  \bibinfo{author}{\bibfnamefont{P.}~\bibnamefont{W\"olfle}},
  \bibinfo{year}{2001}, \bibinfo{journal}{Phys. Rev. Lett.}
  \textbf{\bibinfo{volume}{87}}, \bibinfo{pages}{126805}.

\bibitem[{\citenamefont{Mirlin} \emph{et~al.}(1998)\citenamefont{Mirlin,
  Polyakov, and W\"olfle}}]{mirlin:1998b}
\bibinfo{author}{\bibnamefont{Mirlin}, \bibfnamefont{A.~D.}},
  \bibinfo{author}{\bibfnamefont{D.~G.} \bibnamefont{Polyakov}}, and
  \bibinfo{author}{\bibfnamefont{P.}~\bibnamefont{W\"olfle}},
  \bibinfo{year}{1998}, \bibinfo{journal}{Phys. Rev. Lett.}
  \textbf{\bibinfo{volume}{80}}, \bibinfo{pages}{2429}.

\bibitem[{\citenamefont{Mirlin} \emph{et~al.}(1999)\citenamefont{Mirlin, Wilke,
  Evers, Polyakov, and W\"olfle}}]{mirlin:1999}
\bibinfo{author}{\bibnamefont{Mirlin}, \bibfnamefont{A.~D.}},
  \bibinfo{author}{\bibfnamefont{J.}~\bibnamefont{Wilke}},
  \bibinfo{author}{\bibfnamefont{F.}~\bibnamefont{Evers}},
  \bibinfo{author}{\bibfnamefont{D.~G.} \bibnamefont{Polyakov}}, and
  \bibinfo{author}{\bibfnamefont{P.}~\bibnamefont{W\"olfle}},
  \bibinfo{year}{1999}, \bibinfo{journal}{Phys. Rev. Lett.}
  \textbf{\bibinfo{volume}{83}}, \bibinfo{pages}{2801}.

\bibitem[{\citenamefont{Mirlin and W\"olfle}(1998)}]{mirlin:1998}
\bibinfo{author}{\bibnamefont{Mirlin}, \bibfnamefont{A.~D.}}, and
  \bibinfo{author}{\bibfnamefont{P.}~\bibnamefont{W\"olfle}},
  \bibinfo{year}{1998}, \bibinfo{journal}{Phys. Rev. B}
  \textbf{\bibinfo{volume}{58}}, \bibinfo{pages}{12986}.

\bibitem[{\citenamefont{Mkhitaryan and Raikh}(2011)}]{mkhitaryan:2011}
\bibinfo{author}{\bibnamefont{Mkhitaryan}, \bibfnamefont{V.~V.}}, and
  \bibinfo{author}{\bibfnamefont{M.~E.} \bibnamefont{Raikh}},
  \bibinfo{year}{2011}, \bibinfo{journal}{Phys. Rev. B}
  \textbf{\bibinfo{volume}{83}}, \bibinfo{pages}{045406}.

\bibitem[{\citenamefont{Monarkha and Kono}(2004)}]{monarkha:2004}
\bibinfo{author}{\bibnamefont{Monarkha}, \bibfnamefont{Y.}}, and
  \bibinfo{author}{\bibfnamefont{K.}~\bibnamefont{Kono}}, \bibinfo{year}{2004},
  \emph{\bibinfo{title}{Two-Dimensional Coulomb Liquids and Solids}}
  (\bibinfo{publisher}{Springer Verlag}).

\bibitem[{\citenamefont{Monarkha}(2011{\natexlab{a}})}]{monarkha:2011}
\bibinfo{author}{\bibnamefont{Monarkha}, \bibfnamefont{Y.~P.}},
  \bibinfo{year}{2011}{\natexlab{a}}, \bibinfo{journal}{Low Temp. Phys.}
  \textbf{\bibinfo{volume}{37}}, \bibinfo{pages}{90}.

\bibitem[{\citenamefont{Monarkha}(2011{\natexlab{b}})}]{monarkha:2011b}
\bibinfo{author}{\bibnamefont{Monarkha}, \bibfnamefont{Y.~P.}},
  \bibinfo{year}{2011}{\natexlab{b}}, \bibinfo{journal}{Low Temp. Phys.}
  \textbf{\bibinfo{volume}{37}}, \bibinfo{pages}{655}.

\bibitem[{\citenamefont{Monarkha}(2012)}]{monarkha:2012}
\bibinfo{author}{\bibnamefont{Monarkha}, \bibfnamefont{Y.~P.}},
  \bibinfo{year}{2012}, \bibinfo{journal}{Low Temp. Phys.}
  \textbf{\bibinfo{volume}{38}}, \bibinfo{pages}{451}.

\bibitem[{\citenamefont{Mueller} \emph{et~al.}(2010)\citenamefont{Mueller, Xia,
  and Avouris}}]{mueller:2010}
\bibinfo{author}{\bibnamefont{Mueller}, \bibfnamefont{T.}},
  \bibinfo{author}{\bibfnamefont{F.}~\bibnamefont{Xia}}, and
  \bibinfo{author}{\bibfnamefont{P.}~\bibnamefont{Avouris}},
  \bibinfo{year}{2010}, \bibinfo{journal}{Nature Phot.}
  \textbf{\bibinfo{volume}{4}}, \bibinfo{pages}{297}.

\bibitem[{\citenamefont{Murzin}(1984)}]{murzin:1984}
\bibinfo{author}{\bibnamefont{Murzin}, \bibfnamefont{S.~S.}},
  \bibinfo{year}{1984}, \bibinfo{journal}{JETP Lett.}
  \textbf{\bibinfo{volume}{39}}, \bibinfo{pages}{695}.

\bibitem[{\citenamefont{Nachtwei}(1999)}]{nachtwei:1999}
\bibinfo{author}{\bibnamefont{Nachtwei}, \bibfnamefont{G.}},
  \bibinfo{year}{1999}, \bibinfo{journal}{Physica E}
  \textbf{\bibinfo{volume}{4}}, \bibinfo{pages}{79}.

\bibitem[{\citenamefont{Nachtwei} \emph{et~al.}(2008)\citenamefont{Nachtwei,
  Gouider, Stellmach, Vasile, Vasilyev, Hein, and Gerhardts}}]{nachtwei:2008}
\bibinfo{author}{\bibnamefont{Nachtwei}, \bibfnamefont{G.}},
  \bibinfo{author}{\bibfnamefont{F.}~\bibnamefont{Gouider}},
  \bibinfo{author}{\bibfnamefont{C.}~\bibnamefont{Stellmach}},
  \bibinfo{author}{\bibfnamefont{G.}~\bibnamefont{Vasile}},
  \bibinfo{author}{\bibfnamefont{Y.~B.} \bibnamefont{Vasilyev}},
  \bibinfo{author}{\bibfnamefont{G.}~\bibnamefont{Hein}}, and
  \bibinfo{author}{\bibfnamefont{R.~R.} \bibnamefont{Gerhardts}},
  \bibinfo{year}{2008}, \bibinfo{journal}{Phys. Rev. B}
  \textbf{\bibinfo{volume}{78}}, \bibinfo{pages}{174305}.

\bibitem[{\citenamefont{Neugebauer}
  \emph{et~al.}(2009)\citenamefont{Neugebauer, Orlita, Faugeras, Barra, and
  Potemski}}]{neugebauer:2009}
\bibinfo{author}{\bibnamefont{Neugebauer}, \bibfnamefont{P.}},
  \bibinfo{author}{\bibfnamefont{M.}~\bibnamefont{Orlita}},
  \bibinfo{author}{\bibfnamefont{C.}~\bibnamefont{Faugeras}},
  \bibinfo{author}{\bibfnamefont{A.-L.} \bibnamefont{Barra}}, and
  \bibinfo{author}{\bibfnamefont{M.}~\bibnamefont{Potemski}},
  \bibinfo{year}{2009}, \bibinfo{journal}{Phys. Rev. Lett.}
  \textbf{\bibinfo{volume}{103}}, \bibinfo{pages}{136403}.

\bibitem[{\citenamefont{Novoselov} \emph{et~al.}(2005)\citenamefont{Novoselov,
  Geim, Morozov, Jiang, Katsnelson, Grigorieva, Dubonos, and
  Firsov}}]{novoselov:2005}
\bibinfo{author}{\bibnamefont{Novoselov}, \bibfnamefont{K.~S.}},
  \bibinfo{author}{\bibfnamefont{A.~K.} \bibnamefont{Geim}},
  \bibinfo{author}{\bibfnamefont{S.~V.} \bibnamefont{Morozov}},
  \bibinfo{author}{\bibfnamefont{D.}~\bibnamefont{Jiang}},
  \bibinfo{author}{\bibfnamefont{M.~I.} \bibnamefont{Katsnelson}},
  \bibinfo{author}{\bibfnamefont{I.~V.} \bibnamefont{Grigorieva}},
  \bibinfo{author}{\bibfnamefont{S.~V.} \bibnamefont{Dubonos}}, and
  \bibinfo{author}{\bibfnamefont{A.~A.} \bibnamefont{Firsov}},
  \bibinfo{year}{2005}, \bibinfo{journal}{Nature (London)}
  \textbf{\bibinfo{volume}{438}}, \bibinfo{pages}{197}.

\bibitem[{\citenamefont{Novoselov} \emph{et~al.}(2004)\citenamefont{Novoselov,
  Geim, Morozov, Jiang, Zhang, Dubonos, Grigorieva, and
  Firsov}}]{novoselov:2004}
\bibinfo{author}{\bibnamefont{Novoselov}, \bibfnamefont{K.~S.}},
  \bibinfo{author}{\bibfnamefont{A.~K.} \bibnamefont{Geim}},
  \bibinfo{author}{\bibfnamefont{S.~V.} \bibnamefont{Morozov}},
  \bibinfo{author}{\bibfnamefont{D.}~\bibnamefont{Jiang}},
  \bibinfo{author}{\bibfnamefont{Y.}~\bibnamefont{Zhang}},
  \bibinfo{author}{\bibfnamefont{S.~V.} \bibnamefont{Dubonos}},
  \bibinfo{author}{\bibfnamefont{I.~V.} \bibnamefont{Grigorieva}}, and
  \bibinfo{author}{\bibfnamefont{A.~A.} \bibnamefont{Firsov}},
  \bibinfo{year}{2004}, \bibinfo{journal}{Science}
  \textbf{\bibinfo{volume}{306}}, \bibinfo{pages}{666}.

\bibitem[{\citenamefont{Novoselov} \emph{et~al.}(2007)\citenamefont{Novoselov,
  Jiang, Zhang, Morozov, Stormer, Zeitler, Maan, Boebinger, Kim, and
  Geim}}]{novoselov:2007}
\bibinfo{author}{\bibnamefont{Novoselov}, \bibfnamefont{K.~S.}},
  \bibinfo{author}{\bibfnamefont{Z.}~\bibnamefont{Jiang}},
  \bibinfo{author}{\bibfnamefont{Y.}~\bibnamefont{Zhang}},
  \bibinfo{author}{\bibfnamefont{S.~V.} \bibnamefont{Morozov}},
  \bibinfo{author}{\bibfnamefont{H.~L.} \bibnamefont{Stormer}},
  \bibinfo{author}{\bibfnamefont{U.}~\bibnamefont{Zeitler}},
  \bibinfo{author}{\bibfnamefont{J.~C.} \bibnamefont{Maan}},
  \bibinfo{author}{\bibfnamefont{G.~S.} \bibnamefont{Boebinger}},
  \bibinfo{author}{\bibfnamefont{P.}~\bibnamefont{Kim}}, and
  \bibinfo{author}{\bibfnamefont{A.~K.} \bibnamefont{Geim}},
  \bibinfo{year}{2007}, \bibinfo{journal}{Science}
  \textbf{\bibinfo{volume}{315}}, \bibinfo{pages}{1379}.

\bibitem[{\citenamefont{Novoselov} \emph{et~al.}(2006)\citenamefont{Novoselov,
  McCann, Morozov, Fal'ko, Katsnelson, Zeitler, Jiang, Schedin, and
  Geim}}]{novoselov:2006}
\bibinfo{author}{\bibnamefont{Novoselov}, \bibfnamefont{K.~S.}},
  \bibinfo{author}{\bibfnamefont{E.}~\bibnamefont{McCann}},
  \bibinfo{author}{\bibfnamefont{S.~V.} \bibnamefont{Morozov}},
  \bibinfo{author}{\bibfnamefont{V.~I.} \bibnamefont{Fal'ko}},
  \bibinfo{author}{\bibfnamefont{M.~I.} \bibnamefont{Katsnelson}},
  \bibinfo{author}{\bibfnamefont{U.}~\bibnamefont{Zeitler}},
  \bibinfo{author}{\bibfnamefont{D.}~\bibnamefont{Jiang}},
  \bibinfo{author}{\bibfnamefont{F.}~\bibnamefont{Schedin}}, and
  \bibinfo{author}{\bibfnamefont{A.~K.} \bibnamefont{Geim}},
  \bibinfo{year}{2006}, \bibinfo{journal}{Nature Phys.}
  \textbf{\bibinfo{volume}{2}}, \bibinfo{pages}{177}.

\bibitem[{\citenamefont{Olshanetsky}
  \emph{et~al.}(2005)\citenamefont{Olshanetsky, Renard, Kvon, Portal, Woods,
  Zhang, and Harris}}]{olshanetsky:2006}
\bibinfo{author}{\bibnamefont{Olshanetsky}, \bibfnamefont{E.~B.}},
  \bibinfo{author}{\bibfnamefont{V.}~\bibnamefont{Renard}},
  \bibinfo{author}{\bibfnamefont{Z.~D.} \bibnamefont{Kvon}},
  \bibinfo{author}{\bibfnamefont{J.~C.} \bibnamefont{Portal}},
  \bibinfo{author}{\bibfnamefont{N.~J.} \bibnamefont{Woods}},
  \bibinfo{author}{\bibfnamefont{J.}~\bibnamefont{Zhang}}, and
  \bibinfo{author}{\bibfnamefont{J.~J.} \bibnamefont{Harris}},
  \bibinfo{year}{2005}, \bibinfo{journal}{Europhys. Lett.}
  \textbf{\bibinfo{volume}{71}}, \bibinfo{pages}{665}.

\bibitem[{\citenamefont{Orlita and Potemski}(2010)}]{orlita:2010}
\bibinfo{author}{\bibnamefont{Orlita}, \bibfnamefont{M.}}, and
  \bibinfo{author}{\bibfnamefont{M.}~\bibnamefont{Potemski}},
  \bibinfo{year}{2010}, \bibinfo{journal}{Semicond. Sci. Technol.}
  \textbf{\bibinfo{volume}{25}}, \bibinfo{pages}{063001}.

\bibitem[{\citenamefont{Paalanen} \emph{et~al.}(1983)\citenamefont{Paalanen,
  Tsui, and Hwang}}]{paalanen:1983}
\bibinfo{author}{\bibnamefont{Paalanen}, \bibfnamefont{M.~A.}},
  \bibinfo{author}{\bibfnamefont{D.~C.} \bibnamefont{Tsui}}, and
  \bibinfo{author}{\bibfnamefont{J.~C.~M.} \bibnamefont{Hwang}},
  \bibinfo{year}{1983}, \bibinfo{journal}{Phys. Rev. Lett.}
  \textbf{\bibinfo{volume}{51}}, \bibinfo{pages}{2226}.

\bibitem[{\citenamefont{Park}(2004)}]{park:2004}
\bibinfo{author}{\bibnamefont{Park}, \bibfnamefont{K.}}, \bibinfo{year}{2004},
  \bibinfo{journal}{Phys. Rev. B} \textbf{\bibinfo{volume}{69}},
  \bibinfo{pages}{201301(R)}.

\bibitem[{\citenamefont{Pavlovich and Epshtein}(1976)}]{pavlovich:1976}
\bibinfo{author}{\bibnamefont{Pavlovich}, \bibfnamefont{V.~V.}}, and
  \bibinfo{author}{\bibfnamefont{E.~M.} \bibnamefont{Epshtein}},
  \bibinfo{year}{1976}, \bibinfo{journal}{Sov. Phys. Semicond.}
  \textbf{\bibinfo{volume}{10}}, \bibinfo{pages}{1196}.

\bibitem[{\citenamefont{Pechenezhskii}
  \emph{et~al.}(2007)\citenamefont{Pechenezhskii, Dorozhkin, and
  Dmitriev}}]{pechenezhskii:2007}
\bibinfo{author}{\bibnamefont{Pechenezhskii}, \bibfnamefont{I.~V.}},
  \bibinfo{author}{\bibfnamefont{S.~I.} \bibnamefont{Dorozhkin}}, and
  \bibinfo{author}{\bibfnamefont{I.~A.} \bibnamefont{Dmitriev}},
  \bibinfo{year}{2007}, \bibinfo{journal}{JETP Lett.}
  \textbf{\bibinfo{volume}{85}}, \bibinfo{pages}{86}.

\bibitem[{\citenamefont{Pfeiffer and West}(2003)}]{pfeiffer:2003}
\bibinfo{author}{\bibnamefont{Pfeiffer}, \bibfnamefont{L.}}, and
  \bibinfo{author}{\bibfnamefont{K.~W.} \bibnamefont{West}},
  \bibinfo{year}{2003}, \bibinfo{journal}{Physica E}
  \textbf{\bibinfo{volume}{20}}, \bibinfo{pages}{57}.

\bibitem[{\citenamefont{Pfeiffer} \emph{et~al.}(1989)\citenamefont{Pfeiffer,
  West, Stormer, and Baldwin}}]{pfeiffer:1989}
\bibinfo{author}{\bibnamefont{Pfeiffer}, \bibfnamefont{L.}},
  \bibinfo{author}{\bibfnamefont{K.~W.} \bibnamefont{West}},
  \bibinfo{author}{\bibfnamefont{H.~L.} \bibnamefont{Stormer}}, and
  \bibinfo{author}{\bibfnamefont{K.~W.} \bibnamefont{Baldwin}},
  \bibinfo{year}{1989}, \bibinfo{journal}{Appl. Phys. Lett.}
  \textbf{\bibinfo{volume}{55}}, \bibinfo{pages}{1888}.

\bibitem[{\citenamefont{Pippard}(1989)}]{pippard:1989}
\bibinfo{author}{\bibnamefont{Pippard}, \bibfnamefont{A.~B.}},
  \bibinfo{year}{1989}, \emph{\bibinfo{title}{Magnetoresistance in Metals}}
  (\bibinfo{publisher}{Cambridge University, Cambridge}).

\bibitem[{\citenamefont{Polyakov}(1986)}]{polyakov:1986}
\bibinfo{author}{\bibnamefont{Polyakov}, \bibfnamefont{D.~G.}},
  \bibinfo{year}{1986}, \bibinfo{journal}{Sov. Phys. JETP}
  \textbf{\bibinfo{volume}{63}}, \bibinfo{pages}{317}.

\bibitem[{\citenamefont{Polyakov} \emph{et~al.}(2002)\citenamefont{Polyakov,
  Evers, and Gornyi}}]{polyakov:2002}
\bibinfo{author}{\bibnamefont{Polyakov}, \bibfnamefont{D.~G.}},
  \bibinfo{author}{\bibfnamefont{F.}~\bibnamefont{Evers}}, and
  \bibinfo{author}{\bibfnamefont{I.~V.} \bibnamefont{Gornyi}},
  \bibinfo{year}{2002}, \bibinfo{journal}{Phys. Rev. B}
  \textbf{\bibinfo{volume}{65}}, \bibinfo{pages}{125326}.

\bibitem[{\citenamefont{Polyakov} \emph{et~al.}(2001)\citenamefont{Polyakov,
  Evers, Mirlin, and W{\"o}lfle}}]{polyakov:2001}
\bibinfo{author}{\bibnamefont{Polyakov}, \bibfnamefont{D.~G.}},
  \bibinfo{author}{\bibfnamefont{F.}~\bibnamefont{Evers}},
  \bibinfo{author}{\bibfnamefont{A.~D.} \bibnamefont{Mirlin}}, and
  \bibinfo{author}{\bibfnamefont{P.}~\bibnamefont{W{\"o}lfle}},
  \bibinfo{year}{2001}, \bibinfo{journal}{Phys. Rev. B}
  \textbf{\bibinfo{volume}{64}}, \bibinfo{pages}{205306}.

\bibitem[{\citenamefont{Polyanovsky}(1988)}]{polyanovsky:1988}
\bibinfo{author}{\bibnamefont{Polyanovsky}, \bibfnamefont{V.}},
  \bibinfo{year}{1988}, \bibinfo{journal}{Sov. Phys. Semicond.}
  \textbf{\bibinfo{volume}{22}}, \bibinfo{pages}{1408}.

\bibitem[{\citenamefont{Pozhela}(1981)}]{pozhela:1981}
\bibinfo{author}{\bibnamefont{Pozhela}, \bibfnamefont{J.~K.}},
  \bibinfo{year}{1981}, \emph{\bibinfo{title}{Plasma and current instabilities
  in semiconductors}} (\bibinfo{publisher}{Pergamon, Oxford}).

\bibitem[{\citenamefont{Prange and Nee}(1968)}]{prange:1968}
\bibinfo{author}{\bibnamefont{Prange}, \bibfnamefont{R.~E.}}, and
  \bibinfo{author}{\bibfnamefont{T.-W.} \bibnamefont{Nee}},
  \bibinfo{year}{1968}, \bibinfo{journal}{Phys. Rev.}
  \textbf{\bibinfo{volume}{168}}, \bibinfo{pages}{779}.

\bibitem[{\citenamefont{Pruisken}(1990)}]{pruisken:1990}
\bibinfo{author}{\bibnamefont{Pruisken}, \bibfnamefont{A.~M.~M.}},
  \bibinfo{year}{1990}, in \emph{\bibinfo{booktitle}{The Quantum Hall Effect}},
  edited by \bibinfo{editor}{\bibfnamefont{R.~E.} \bibnamefont{Prange}} and
  \bibinfo{editor}{\bibfnamefont{S.~M.} \bibnamefont{Girvin}}
  (\bibinfo{publisher}{Springer, Berlin}).

\bibitem[{\citenamefont{Raichev}(2008)}]{raichev:2008}
\bibinfo{author}{\bibnamefont{Raichev}, \bibfnamefont{O.~E.}},
  \bibinfo{year}{2008}, \bibinfo{journal}{Phys. Rev. B}
  \textbf{\bibinfo{volume}{78}}, \bibinfo{pages}{125304}.

\bibitem[{\citenamefont{Raichev}(2009)}]{raichev:2009}
\bibinfo{author}{\bibnamefont{Raichev}, \bibfnamefont{O.~E.}},
  \bibinfo{year}{2009}, \bibinfo{journal}{Phys. Rev. B}
  \textbf{\bibinfo{volume}{80}}, \bibinfo{pages}{075318}.

\bibitem[{\citenamefont{Raichev}(2010{\natexlab{a}})}]{raichev:2010a}
\bibinfo{author}{\bibnamefont{Raichev}, \bibfnamefont{O.~E.}},
  \bibinfo{year}{2010}{\natexlab{a}}, \bibinfo{journal}{Phys. Rev. B}
  \textbf{\bibinfo{volume}{81}}, \bibinfo{pages}{195301}.

\bibitem[{\citenamefont{Raichev}(2010{\natexlab{b}})}]{raichev:2010b}
\bibinfo{author}{\bibnamefont{Raichev}, \bibfnamefont{O.~E.}},
  \bibinfo{year}{2010}{\natexlab{b}}, \bibinfo{journal}{Phys. Rev. B}
  \textbf{\bibinfo{volume}{81}}, \bibinfo{pages}{165319}.

\bibitem[{\citenamefont{Raikh and Shahbazyan}(1993)}]{raikh:1993}
\bibinfo{author}{\bibnamefont{Raikh}, \bibfnamefont{M.~E.}}, and
  \bibinfo{author}{\bibfnamefont{T.~V.} \bibnamefont{Shahbazyan}},
  \bibinfo{year}{1993}, \bibinfo{journal}{Phys. Rev. B}
  \textbf{\bibinfo{volume}{47}}, \bibinfo{pages}{1522}.

\bibitem[{\citenamefont{Raikh and Shahbazyan}(1994)}]{raikh:1994}
\bibinfo{author}{\bibnamefont{Raikh}, \bibfnamefont{M.~E.}}, and
  \bibinfo{author}{\bibfnamefont{T.~V.} \bibnamefont{Shahbazyan}},
  \bibinfo{year}{1994}, \bibinfo{journal}{Phys. Rev. B}
  \textbf{\bibinfo{volume}{49}}, \bibinfo{pages}{5531}.

\bibitem[{\citenamefont{Ridley}(1963)}]{ridley:1963}
\bibinfo{author}{\bibnamefont{Ridley}, \bibfnamefont{B.~K.}},
  \bibinfo{year}{1963}, \bibinfo{journal}{Proc. Phys. Soc.}
  \textbf{\bibinfo{volume}{82}}, \bibinfo{pages}{954}.

\bibitem[{\citenamefont{Robinson} \emph{et~al.}(2004)\citenamefont{Robinson,
  Kennett, Cooper, and Fal'ko}}]{robinson:2004}
\bibinfo{author}{\bibnamefont{Robinson}, \bibfnamefont{J.~P.}},
  \bibinfo{author}{\bibfnamefont{M.~P.} \bibnamefont{Kennett}},
  \bibinfo{author}{\bibfnamefont{N.~R.} \bibnamefont{Cooper}}, and
  \bibinfo{author}{\bibfnamefont{V.~I.} \bibnamefont{Fal'ko}},
  \bibinfo{year}{2004}, \bibinfo{journal}{Phys. Rev. Lett.}
  \textbf{\bibinfo{volume}{93}}, \bibinfo{pages}{036804}.

\bibitem[{\citenamefont{Romero} \emph{et~al.}(2008)\citenamefont{Romero,
  McHugh, Sarachik, Vitkalov, and Bykov}}]{romero:2008}
\bibinfo{author}{\bibnamefont{Romero}, \bibfnamefont{N.}},
  \bibinfo{author}{\bibfnamefont{S.}~\bibnamefont{McHugh}},
  \bibinfo{author}{\bibfnamefont{M.~P.} \bibnamefont{Sarachik}},
  \bibinfo{author}{\bibfnamefont{S.~A.} \bibnamefont{Vitkalov}}, and
  \bibinfo{author}{\bibfnamefont{A.~A.} \bibnamefont{Bykov}},
  \bibinfo{year}{2008}, \bibinfo{journal}{Phys. Rev. B}
  \textbf{\bibinfo{volume}{78}}, \bibinfo{pages}{153311}.

\bibitem[{\citenamefont{Rowe} \emph{et~al.}(2001)\citenamefont{Rowe, Nehls,
  Stradling, and Ferguson}}]{rowe:2001}
\bibinfo{author}{\bibnamefont{Rowe}, \bibfnamefont{A.~C.~H.}},
  \bibinfo{author}{\bibfnamefont{J.}~\bibnamefont{Nehls}},
  \bibinfo{author}{\bibfnamefont{R.~A.} \bibnamefont{Stradling}}, and
  \bibinfo{author}{\bibfnamefont{R.~S.} \bibnamefont{Ferguson}},
  \bibinfo{year}{2001}, \bibinfo{journal}{Phys. Rev. B}
  \textbf{\bibinfo{volume}{63}}, \bibinfo{pages}{201307(R)}.

\bibitem[{\citenamefont{Ryzhii} \emph{et~al.}(2004)\citenamefont{Ryzhii,
  Chaplik, and Suris}}]{ryzhii:2004}
\bibinfo{author}{\bibnamefont{Ryzhii}, \bibfnamefont{V.}},
  \bibinfo{author}{\bibfnamefont{A.}~\bibnamefont{Chaplik}}, and
  \bibinfo{author}{\bibfnamefont{R.}~\bibnamefont{Suris}},
  \bibinfo{year}{2004}, \bibinfo{journal}{JETP Lett.}
  \textbf{\bibinfo{volume}{80}}, \bibinfo{pages}{363}.

\bibitem[{\citenamefont{Ryzhii and Suris}(2003)}]{ryzhii:2003d}
\bibinfo{author}{\bibnamefont{Ryzhii}, \bibfnamefont{V.}}, and
  \bibinfo{author}{\bibfnamefont{R.}~\bibnamefont{Suris}},
  \bibinfo{year}{2003}, \bibinfo{journal}{J. Phys.: Condens. Matter}
  \textbf{\bibinfo{volume}{15}}, \bibinfo{pages}{6855}.

\bibitem[{\citenamefont{Ryzhii}(1970)}]{ryzhii:1970}
\bibinfo{author}{\bibnamefont{Ryzhii}, \bibfnamefont{V.~I.}},
  \bibinfo{year}{1970}, \bibinfo{journal}{Sov. Phys. Solid State}
  \textbf{\bibinfo{volume}{11}}, \bibinfo{pages}{2078}.

\bibitem[{\citenamefont{Ryzhii} \emph{et~al.}(1986)\citenamefont{Ryzhii, Suris,
  and Shchamkhalova}}]{ryzhii:1986}
\bibinfo{author}{\bibnamefont{Ryzhii}, \bibfnamefont{V.~I.}},
  \bibinfo{author}{\bibfnamefont{R.~A.} \bibnamefont{Suris}}, and
  \bibinfo{author}{\bibfnamefont{B.~S.} \bibnamefont{Shchamkhalova}},
  \bibinfo{year}{1986}, \bibinfo{journal}{Sov. Phys. Semicond.}
  \textbf{\bibinfo{volume}{20}}, \bibinfo{pages}{1299}.

\bibitem[{\citenamefont{Sadowski} \emph{et~al.}(2006)\citenamefont{Sadowski,
  Martinez, Potemski, Berger, and de~Heer}}]{sadowski:2006}
\bibinfo{author}{\bibnamefont{Sadowski}, \bibfnamefont{M.~L.}},
  \bibinfo{author}{\bibfnamefont{G.}~\bibnamefont{Martinez}},
  \bibinfo{author}{\bibfnamefont{M.}~\bibnamefont{Potemski}},
  \bibinfo{author}{\bibfnamefont{C.}~\bibnamefont{Berger}}, and
  \bibinfo{author}{\bibfnamefont{W.~A.} \bibnamefont{de~Heer}},
  \bibinfo{year}{2006}, \bibinfo{journal}{Phys. Rev. Lett.}
  \textbf{\bibinfo{volume}{97}}, \bibinfo{pages}{266405}.

\bibitem[{\citenamefont{Saku} \emph{et~al.}(1996)\citenamefont{Saku, Horikoshi,
  and Tokura}}]{saku:1996}
\bibinfo{author}{\bibnamefont{Saku}, \bibfnamefont{T.}},
  \bibinfo{author}{\bibfnamefont{Y.}~\bibnamefont{Horikoshi}}, and
  \bibinfo{author}{\bibfnamefont{Y.}~\bibnamefont{Tokura}},
  \bibinfo{year}{1996}, \bibinfo{journal}{Jpn. J. Appl. Phys.}
  \textbf{\bibinfo{volume}{35}}, \bibinfo{pages}{34}.

\bibitem[{\citenamefont{Sander} \emph{et~al.}(1998)\citenamefont{Sander,
  Holmes, Harris, Maude, and Portal}}]{sander:1998}
\bibinfo{author}{\bibnamefont{Sander}, \bibfnamefont{T.~H.}},
  \bibinfo{author}{\bibfnamefont{S.~N.} \bibnamefont{Holmes}},
  \bibinfo{author}{\bibfnamefont{J.~J.} \bibnamefont{Harris}},
  \bibinfo{author}{\bibfnamefont{D.~K.} \bibnamefont{Maude}}, and
  \bibinfo{author}{\bibfnamefont{J.~C.} \bibnamefont{Portal}},
  \bibinfo{year}{1998}, \bibinfo{journal}{Phys. Rev. B}
  \textbf{\bibinfo{volume}{58}}, \bibinfo{pages}{13856}.

\bibitem[{\citenamefont{Sch\"oll}(2001)}]{schoell:2001}
\bibinfo{author}{\bibnamefont{Sch\"oll}, \bibfnamefont{E.}},
  \bibinfo{year}{2001}, \emph{\bibinfo{title}{Nonlinear spatio-temporal
  dynamics and chaos in semiconductors}} (\bibinfo{publisher}{Cambridge
  University, Cambridge}).

\bibitem[{\citenamefont{Sedrakyan and Raikh}(2008)}]{sedrakyan:2008b}
\bibinfo{author}{\bibnamefont{Sedrakyan}, \bibfnamefont{T.~A.}}, and
  \bibinfo{author}{\bibfnamefont{M.~E.} \bibnamefont{Raikh}},
  \bibinfo{year}{2008}, \bibinfo{journal}{Phys. Rev. Lett.}
  \textbf{\bibinfo{volume}{100}}, \bibinfo{pages}{086808}.

\bibitem[{\citenamefont{Seely}(1974)}]{seely:1974}
\bibinfo{author}{\bibnamefont{Seely}, \bibfnamefont{J.~F.}},
  \bibinfo{year}{1974}, \bibinfo{journal}{Phys. Rev. A}
  \textbf{\bibinfo{volume}{10}}, \bibinfo{pages}{1863}.

\bibitem[{\citenamefont{Shayegan} \emph{et~al.}(1988)\citenamefont{Shayegan,
  Goldman, Jiang, Sajoto, and Santos}}]{shayegan:1988}
\bibinfo{author}{\bibnamefont{Shayegan}, \bibfnamefont{M.}},
  \bibinfo{author}{\bibfnamefont{V.~J.} \bibnamefont{Goldman}},
  \bibinfo{author}{\bibfnamefont{C.}~\bibnamefont{Jiang}},
  \bibinfo{author}{\bibfnamefont{T.}~\bibnamefont{Sajoto}}, and
  \bibinfo{author}{\bibfnamefont{M.}~\bibnamefont{Santos}},
  \bibinfo{year}{1988}, \bibinfo{journal}{Appl. Phys. Lett.}
  \textbf{\bibinfo{volume}{52}}, \bibinfo{pages}{1086}.

\bibitem[{\citenamefont{Shi and Xie}(2003)}]{shi:2003}
\bibinfo{author}{\bibnamefont{Shi}, \bibfnamefont{J.}}, and
  \bibinfo{author}{\bibfnamefont{X.~C.} \bibnamefont{Xie}},
  \bibinfo{year}{2003}, \bibinfo{journal}{Phys. Rev. Lett.}
  \textbf{\bibinfo{volume}{91}}, \bibinfo{pages}{086801}.

\bibitem[{\citenamefont{Shikler} \emph{et~al.}(1997)\citenamefont{Shikler,
  Heiblum, and Umansky}}]{shikler:1997}
\bibinfo{author}{\bibnamefont{Shikler}, \bibfnamefont{R.}},
  \bibinfo{author}{\bibfnamefont{M.}~\bibnamefont{Heiblum}}, and
  \bibinfo{author}{\bibfnamefont{V.}~\bibnamefont{Umansky}},
  \bibinfo{year}{1997}, \bibinfo{journal}{Phys. Rev. B}
  \textbf{\bibinfo{volume}{55}}, \bibinfo{pages}{15427}.

\bibitem[{\citenamefont{Shoenberg}(1984)}]{shoenberg:1984}
\bibinfo{author}{\bibnamefont{Shoenberg}, \bibfnamefont{D.}},
  \bibinfo{year}{1984}, \emph{\bibinfo{title}{Magnetic Oscillations in Metals}}
  (\bibinfo{publisher}{Cambridge University, Cambridge}).

\bibitem[{\citenamefont{Smet}(1997)}]{smet:1997}
\bibinfo{author}{\bibnamefont{Smet}, \bibfnamefont{J.~H.}},
  \bibinfo{year}{1997}, \bibinfo{journal}{unpublished} .

\bibitem[{\citenamefont{Smet}(1998)}]{smet:1998}
\bibinfo{author}{\bibnamefont{Smet}, \bibfnamefont{J.~H.}},
  \bibinfo{year}{1998}, in \emph{\bibinfo{booktitle}{Composite Fermions}},
  edited by \bibinfo{editor}{\bibfnamefont{O.}~\bibnamefont{Heinonen}}
  (\bibinfo{publisher}{World Scientific, Singapore}).

\bibitem[{\citenamefont{Smet} \emph{et~al.}(2005)\citenamefont{Smet, Gorshunov,
  Jiang, Pfeiffer, West, Umansky, Dressel, Meisels, Kuchar, and von
  Klitzing}}]{smet:2005}
\bibinfo{author}{\bibnamefont{Smet}, \bibfnamefont{J.~H.}},
  \bibinfo{author}{\bibfnamefont{B.}~\bibnamefont{Gorshunov}},
  \bibinfo{author}{\bibfnamefont{C.}~\bibnamefont{Jiang}},
  \bibinfo{author}{\bibfnamefont{L.}~\bibnamefont{Pfeiffer}},
  \bibinfo{author}{\bibfnamefont{K.}~\bibnamefont{West}},
  \bibinfo{author}{\bibfnamefont{V.}~\bibnamefont{Umansky}},
  \bibinfo{author}{\bibfnamefont{M.}~\bibnamefont{Dressel}},
  \bibinfo{author}{\bibfnamefont{R.}~\bibnamefont{Meisels}},
  \bibinfo{author}{\bibfnamefont{F.}~\bibnamefont{Kuchar}}, and
  \bibinfo{author}{\bibfnamefont{K.}~\bibnamefont{von Klitzing}},
  \bibinfo{year}{2005}, \bibinfo{journal}{Phys. Rev. Lett.}
  \textbf{\bibinfo{volume}{95}}, \bibinfo{pages}{116804}.

\bibitem[{\citenamefont{Sollner} \emph{et~al.}(1983)\citenamefont{Sollner,
  Goodhue, Tannenwald, Parker, and Peck}}]{sollner:1983}
\bibinfo{author}{\bibnamefont{Sollner}, \bibfnamefont{T.~C. L.~G.}},
  \bibinfo{author}{\bibfnamefont{W.~D.} \bibnamefont{Goodhue}},
  \bibinfo{author}{\bibfnamefont{P.~E.} \bibnamefont{Tannenwald}},
  \bibinfo{author}{\bibfnamefont{C.~D.} \bibnamefont{Parker}}, and
  \bibinfo{author}{\bibfnamefont{D.~D.} \bibnamefont{Peck}},
  \bibinfo{year}{1983}, \bibinfo{journal}{Appl. Phys. Lett.}
  \textbf{\bibinfo{volume}{43}}, \bibinfo{pages}{588}.

\bibitem[{\citenamefont{Stellmach} \emph{et~al.}(2007)\citenamefont{Stellmach,
  Vasile, Hirsch, Bonk, Vasilyev, Hein, Becker, and Nachtwei}}]{stellmach:2007}
\bibinfo{author}{\bibnamefont{Stellmach}, \bibfnamefont{C.}},
  \bibinfo{author}{\bibfnamefont{G.}~\bibnamefont{Vasile}},
  \bibinfo{author}{\bibfnamefont{A.}~\bibnamefont{Hirsch}},
  \bibinfo{author}{\bibfnamefont{R.}~\bibnamefont{Bonk}},
  \bibinfo{author}{\bibfnamefont{Y.~B.} \bibnamefont{Vasilyev}},
  \bibinfo{author}{\bibfnamefont{G.}~\bibnamefont{Hein}},
  \bibinfo{author}{\bibfnamefont{C.~R.} \bibnamefont{Becker}}, and
  \bibinfo{author}{\bibfnamefont{G.}~\bibnamefont{Nachtwei}},
  \bibinfo{year}{2007}, \bibinfo{journal}{Phys. Rev. B}
  \textbf{\bibinfo{volume}{76}}, \bibinfo{pages}{035341}.

\bibitem[{\citenamefont{Stern}(1983)}]{stern:1983}
\bibinfo{author}{\bibnamefont{Stern}, \bibfnamefont{F.}}, \bibinfo{year}{1983},
  \bibinfo{journal}{Appl. Phys. Lett.} \textbf{\bibinfo{volume}{43}},
  \bibinfo{pages}{974}.

\bibitem[{\citenamefont{Stone} \emph{et~al.}(2007)\citenamefont{Stone, Yang,
  Yuan, Du, Pfeiffer, and West}}]{stone:2007}
\bibinfo{author}{\bibnamefont{Stone}, \bibfnamefont{K.}},
  \bibinfo{author}{\bibfnamefont{C.~L.} \bibnamefont{Yang}},
  \bibinfo{author}{\bibfnamefont{Z.~Q.} \bibnamefont{Yuan}},
  \bibinfo{author}{\bibfnamefont{R.~R.} \bibnamefont{Du}},
  \bibinfo{author}{\bibfnamefont{L.~N.} \bibnamefont{Pfeiffer}}, and
  \bibinfo{author}{\bibfnamefont{K.~W.} \bibnamefont{West}},
  \bibinfo{year}{2007}, \bibinfo{journal}{Phys. Rev. B}
  \textbf{\bibinfo{volume}{76}}, \bibinfo{pages}{153306}.

\bibitem[{\citenamefont{Stormer}(1999)}]{stormer:1999}
\bibinfo{author}{\bibnamefont{Stormer}, \bibfnamefont{H.~L.}},
  \bibinfo{year}{1999}, \bibinfo{journal}{Rev. Mod. Phys.}
  \textbf{\bibinfo{volume}{71}}, \bibinfo{pages}{875}.

\bibitem[{\citenamefont{Studenikin}
  \emph{et~al.}(2012)\citenamefont{Studenikin, Granger, Kam, Sachrajda,
  Wasilewski, and Poole}}]{studenikin:2010}
\bibinfo{author}{\bibnamefont{Studenikin}, \bibfnamefont{S.~A.}},
  \bibinfo{author}{\bibfnamefont{G.}~\bibnamefont{Granger}},
  \bibinfo{author}{\bibfnamefont{A.}~\bibnamefont{Kam}},
  \bibinfo{author}{\bibfnamefont{A.~S.} \bibnamefont{Sachrajda}},
  \bibinfo{author}{\bibfnamefont{Z.~R.} \bibnamefont{Wasilewski}}, and
  \bibinfo{author}{\bibfnamefont{P.~J.} \bibnamefont{Poole}},
  \bibinfo{year}{2012}, \bibinfo{journal}{Phys. Rev. B}
  \textbf{\bibinfo{volume}{86}}, \bibinfo{pages}{115309}.

\bibitem[{\citenamefont{Studenikin}
  \emph{et~al.}(2004)\citenamefont{Studenikin, Potemski, Coleridge, Sachrajda,
  and Wasilewski}}]{studenikin:2004}
\bibinfo{author}{\bibnamefont{Studenikin}, \bibfnamefont{S.~A.}},
  \bibinfo{author}{\bibfnamefont{M.}~\bibnamefont{Potemski}},
  \bibinfo{author}{\bibfnamefont{P.~T.} \bibnamefont{Coleridge}},
  \bibinfo{author}{\bibfnamefont{A.~S.} \bibnamefont{Sachrajda}}, and
  \bibinfo{author}{\bibfnamefont{Z.~R.} \bibnamefont{Wasilewski}},
  \bibinfo{year}{2004}, \bibinfo{journal}{Solid State Commun.}
  \textbf{\bibinfo{volume}{129}}, \bibinfo{pages}{341}.

\bibitem[{\citenamefont{Studenikin}
  \emph{et~al.}(2005)\citenamefont{Studenikin, Potemski, Sachrajda, Hilke,
  Pfeiffer, and West}}]{studenikin:2005}
\bibinfo{author}{\bibnamefont{Studenikin}, \bibfnamefont{S.~A.}},
  \bibinfo{author}{\bibfnamefont{M.}~\bibnamefont{Potemski}},
  \bibinfo{author}{\bibfnamefont{A.}~\bibnamefont{Sachrajda}},
  \bibinfo{author}{\bibfnamefont{M.}~\bibnamefont{Hilke}},
  \bibinfo{author}{\bibfnamefont{L.~N.} \bibnamefont{Pfeiffer}}, and
  \bibinfo{author}{\bibfnamefont{K.~W.} \bibnamefont{West}},
  \bibinfo{year}{2005}, \bibinfo{journal}{Phys. Rev. B}
  \textbf{\bibinfo{volume}{71}}, \bibinfo{pages}{245313}.

\bibitem[{\citenamefont{Studenikin}
  \emph{et~al.}(2007)\citenamefont{Studenikin, Sachrajda, Gupta, Wasilewski,
  Fedorych, Byszewski, Maude, Potemski, Hilke, West, and
  Pfeiffer}}]{studenikin:2007}
\bibinfo{author}{\bibnamefont{Studenikin}, \bibfnamefont{S.~A.}},
  \bibinfo{author}{\bibfnamefont{A.~S.} \bibnamefont{Sachrajda}},
  \bibinfo{author}{\bibfnamefont{J.~A.} \bibnamefont{Gupta}},
  \bibinfo{author}{\bibfnamefont{Z.~R.} \bibnamefont{Wasilewski}},
  \bibinfo{author}{\bibfnamefont{O.~M.} \bibnamefont{Fedorych}},
  \bibinfo{author}{\bibfnamefont{M.}~\bibnamefont{Byszewski}},
  \bibinfo{author}{\bibfnamefont{D.~K.} \bibnamefont{Maude}},
  \bibinfo{author}{\bibfnamefont{M.}~\bibnamefont{Potemski}},
  \bibinfo{author}{\bibfnamefont{M.}~\bibnamefont{Hilke}},
  \bibinfo{author}{\bibfnamefont{K.~W.} \bibnamefont{West}}, and
  \bibinfo{author}{\bibfnamefont{L.~N.} \bibnamefont{Pfeiffer}},
  \bibinfo{year}{2007}, \bibinfo{journal}{Phys. Rev. B}
  \textbf{\bibinfo{volume}{76}}, \bibinfo{pages}{165321}.

\bibitem[{\citenamefont{Tan} \emph{et~al.}(2011)\citenamefont{Tan, Tan, Ma,
  Liu, Lu, and Yang}}]{tan:2011}
\bibinfo{author}{\bibnamefont{Tan}, \bibfnamefont{Z.}},
  \bibinfo{author}{\bibfnamefont{C.}~\bibnamefont{Tan}},
  \bibinfo{author}{\bibfnamefont{L.}~\bibnamefont{Ma}},
  \bibinfo{author}{\bibfnamefont{G.~T.} \bibnamefont{Liu}},
  \bibinfo{author}{\bibfnamefont{L.}~\bibnamefont{Lu}}, and
  \bibinfo{author}{\bibfnamefont{C.~L.} \bibnamefont{Yang}},
  \bibinfo{year}{2011}, \bibinfo{journal}{Phys. Rev. B}
  \textbf{\bibinfo{volume}{84}}, \bibinfo{pages}{115429}.

\bibitem[{\citenamefont{Tokura} \emph{et~al.}(1992)\citenamefont{Tokura, Saku,
  Tarucha, and Horikoshi}}]{tokura:1992}
\bibinfo{author}{\bibnamefont{Tokura}, \bibfnamefont{Y.}},
  \bibinfo{author}{\bibfnamefont{T.}~\bibnamefont{Saku}},
  \bibinfo{author}{\bibfnamefont{S.}~\bibnamefont{Tarucha}}, and
  \bibinfo{author}{\bibfnamefont{Y.}~\bibnamefont{Horikoshi}},
  \bibinfo{year}{1992}, \bibinfo{journal}{Phys. Rev. B}
  \textbf{\bibinfo{volume}{46}}, \bibinfo{pages}{15558(R)}.

\bibitem[{\citenamefont{Toner and Tu}(1995)}]{toner:1995}
\bibinfo{author}{\bibnamefont{Toner}, \bibfnamefont{J.}}, and
  \bibinfo{author}{\bibfnamefont{Y.}~\bibnamefont{Tu}}, \bibinfo{year}{1995},
  \bibinfo{journal}{Phys. Rev. Lett.} \textbf{\bibinfo{volume}{75}},
  \bibinfo{pages}{4326}.

\bibitem[{\citenamefont{Toner and Tu}(1998)}]{toner:1998}
\bibinfo{author}{\bibnamefont{Toner}, \bibfnamefont{J.}}, and
  \bibinfo{author}{\bibfnamefont{Y.}~\bibnamefont{Tu}}, \bibinfo{year}{1998},
  \bibinfo{journal}{Phys. Rev. E} \textbf{\bibinfo{volume}{58}},
  \bibinfo{pages}{4828}.

\bibitem[{\citenamefont{Torres and Kunold}(2005)}]{torres:2005}
\bibinfo{author}{\bibnamefont{Torres}, \bibfnamefont{M.}}, and
  \bibinfo{author}{\bibfnamefont{A.}~\bibnamefont{Kunold}},
  \bibinfo{year}{2005}, \bibinfo{journal}{Phys. Rev. B}
  \textbf{\bibinfo{volume}{71}}, \bibinfo{pages}{115313}.

\bibitem[{\citenamefont{Torres and Kunold}(2006)}]{torres:2006}
\bibinfo{author}{\bibnamefont{Torres}, \bibfnamefont{M.}}, and
  \bibinfo{author}{\bibfnamefont{A.}~\bibnamefont{Kunold}},
  \bibinfo{year}{2006}, \bibinfo{journal}{J. Phys.: Condens. Matter}
  \textbf{\bibinfo{volume}{18}}, \bibinfo{pages}{4029}.

\bibitem[{\citenamefont{Tsu and Esaki}(1973)}]{tsu:1973}
\bibinfo{author}{\bibnamefont{Tsu}, \bibfnamefont{R.}}, and
  \bibinfo{author}{\bibfnamefont{L.}~\bibnamefont{Esaki}},
  \bibinfo{year}{1973}, \bibinfo{journal}{Appl. Phys. Lett.}
  \textbf{\bibinfo{volume}{22}}, \bibinfo{pages}{562}.

\bibitem[{\citenamefont{Tsui}(1999)}]{tsui:1999}
\bibinfo{author}{\bibnamefont{Tsui}, \bibfnamefont{D.~C.}},
  \bibinfo{year}{1999}, \bibinfo{journal}{Rev. Mod. Phys.}
  \textbf{\bibinfo{volume}{71}}, \bibinfo{pages}{891}.

\bibitem[{\citenamefont{Tsui} \emph{et~al.}(1980)\citenamefont{Tsui, Englert,
  Cho, and Gossard}}]{tsui:1980}
\bibinfo{author}{\bibnamefont{Tsui}, \bibfnamefont{D.~C.}},
  \bibinfo{author}{\bibfnamefont{T.}~\bibnamefont{Englert}},
  \bibinfo{author}{\bibfnamefont{A.~Y.} \bibnamefont{Cho}}, and
  \bibinfo{author}{\bibfnamefont{A.~C.} \bibnamefont{Gossard}},
  \bibinfo{year}{1980}, \bibinfo{journal}{Phys. Rev. Lett.}
  \textbf{\bibinfo{volume}{44}}, \bibinfo{pages}{341}.

\bibitem[{\citenamefont{Tsui} \emph{et~al.}(1982)\citenamefont{Tsui, Stormer,
  and Gossard}}]{tsui:1982}
\bibinfo{author}{\bibnamefont{Tsui}, \bibfnamefont{D.~C.}},
  \bibinfo{author}{\bibfnamefont{H.~L.} \bibnamefont{Stormer}}, and
  \bibinfo{author}{\bibfnamefont{A.~C.} \bibnamefont{Gossard}},
  \bibinfo{year}{1982}, \bibinfo{journal}{Phys. Rev. B}
  \textbf{\bibinfo{volume}{25}}, \bibinfo{pages}{1405(R)}.

\bibitem[{\citenamefont{Tung} \emph{et~al.}(2009)\citenamefont{Tung, Yang,
  Smirnov, Pfeiffer, West, Du, and Wang}}]{tung:2009}
\bibinfo{author}{\bibnamefont{Tung}, \bibfnamefont{L.-C.}},
  \bibinfo{author}{\bibfnamefont{C.~L.} \bibnamefont{Yang}},
  \bibinfo{author}{\bibfnamefont{D.}~\bibnamefont{Smirnov}},
  \bibinfo{author}{\bibfnamefont{L.~N.} \bibnamefont{Pfeiffer}},
  \bibinfo{author}{\bibfnamefont{K.~W.} \bibnamefont{West}},
  \bibinfo{author}{\bibfnamefont{R.~R.} \bibnamefont{Du}}, and
  \bibinfo{author}{\bibfnamefont{Y.-J.} \bibnamefont{Wang}},
  \bibinfo{year}{2009}, \bibinfo{journal}{Solid State Commun.}
  \textbf{\bibinfo{volume}{149}}, \bibinfo{pages}{1531}.

\bibitem[{\citenamefont{Umansky} \emph{et~al.}(2009)\citenamefont{Umansky,
  Heiblum, Levinson, Smet, N{\"u}bler, and Dolev}}]{umansky:2009}
\bibinfo{author}{\bibnamefont{Umansky}, \bibfnamefont{V.}},
  \bibinfo{author}{\bibfnamefont{M.}~\bibnamefont{Heiblum}},
  \bibinfo{author}{\bibfnamefont{Y.}~\bibnamefont{Levinson}},
  \bibinfo{author}{\bibfnamefont{J.}~\bibnamefont{Smet}},
  \bibinfo{author}{\bibfnamefont{J.}~\bibnamefont{N{\"u}bler}}, and
  \bibinfo{author}{\bibfnamefont{M.}~\bibnamefont{Dolev}},
  \bibinfo{year}{2009}, \bibinfo{journal}{J. Cryst. Growth}
  \textbf{\bibinfo{volume}{311}}, \bibinfo{pages}{1658}.

\bibitem[{\citenamefont{Umansky} \emph{et~al.}(1997)\citenamefont{Umansky,
  de~Picciotto, and Heiblum}}]{umansky:1997}
\bibinfo{author}{\bibnamefont{Umansky}, \bibfnamefont{V.}},
  \bibinfo{author}{\bibfnamefont{R.}~\bibnamefont{de~Picciotto}}, and
  \bibinfo{author}{\bibfnamefont{M.}~\bibnamefont{Heiblum}},
  \bibinfo{year}{1997}, \bibinfo{journal}{Appl. Phys. Lett.}
  \textbf{\bibinfo{volume}{71}}, \bibinfo{pages}{683}.

\bibitem[{\citenamefont{Vavilov and Aleiner}(2004)}]{vavilov:2004}
\bibinfo{author}{\bibnamefont{Vavilov}, \bibfnamefont{M.~G.}}, and
  \bibinfo{author}{\bibfnamefont{I.~L.} \bibnamefont{Aleiner}},
  \bibinfo{year}{2004}, \bibinfo{journal}{Phys. Rev. B}
  \textbf{\bibinfo{volume}{69}}, \bibinfo{pages}{035303}.

\bibitem[{\citenamefont{Vavilov} \emph{et~al.}(2007)\citenamefont{Vavilov,
  Aleiner, and Glazman}}]{vavilov:2007}
\bibinfo{author}{\bibnamefont{Vavilov}, \bibfnamefont{M.~G.}},
  \bibinfo{author}{\bibfnamefont{I.~L.} \bibnamefont{Aleiner}}, and
  \bibinfo{author}{\bibfnamefont{L.~I.} \bibnamefont{Glazman}},
  \bibinfo{year}{2007}, \bibinfo{journal}{Phys. Rev. B}
  \textbf{\bibinfo{volume}{76}}, \bibinfo{pages}{115331}.

\bibitem[{\citenamefont{Vavilov} \emph{et~al.}(2004)\citenamefont{Vavilov,
  Dmitriev, Aleiner, Mirlin, and Polyakov}}]{vavilov:2004b}
\bibinfo{author}{\bibnamefont{Vavilov}, \bibfnamefont{M.~G.}},
  \bibinfo{author}{\bibfnamefont{I.~A.} \bibnamefont{Dmitriev}},
  \bibinfo{author}{\bibfnamefont{I.~L.} \bibnamefont{Aleiner}},
  \bibinfo{author}{\bibfnamefont{A.~D.} \bibnamefont{Mirlin}}, and
  \bibinfo{author}{\bibfnamefont{D.~G.} \bibnamefont{Polyakov}},
  \bibinfo{year}{2004}, \bibinfo{journal}{Phys. Rev. B}
  \textbf{\bibinfo{volume}{70}}, \bibinfo{pages}{161306(R)}.

\bibitem[{\citenamefont{Vicsek} \emph{et~al.}(1995)\citenamefont{Vicsek,
  Czir\'ok, Ben-Jacob, Cohen, and Shochet}}]{vicsek:1995}
\bibinfo{author}{\bibnamefont{Vicsek}, \bibfnamefont{T.}},
  \bibinfo{author}{\bibfnamefont{A.}~\bibnamefont{Czir\'ok}},
  \bibinfo{author}{\bibfnamefont{E.}~\bibnamefont{Ben-Jacob}},
  \bibinfo{author}{\bibfnamefont{I.}~\bibnamefont{Cohen}}, and
  \bibinfo{author}{\bibfnamefont{O.}~\bibnamefont{Shochet}},
  \bibinfo{year}{1995}, \bibinfo{journal}{Phys. Rev. Lett.}
  \textbf{\bibinfo{volume}{75}}, \bibinfo{pages}{1226}.

\bibitem[{\citenamefont{Vitkalov}(2009)}]{vitkalov:2009}
\bibinfo{author}{\bibnamefont{Vitkalov}, \bibfnamefont{S.}},
  \bibinfo{year}{2009}, \bibinfo{journal}{Int. J. Mod. Phys. B}
  \textbf{\bibinfo{volume}{23}}, \bibinfo{pages}{4727}.

\bibitem[{\citenamefont{Volkov and Kogan}(1967)}]{volkov:1967}
\bibinfo{author}{\bibnamefont{Volkov}, \bibfnamefont{A.~F.}}, and
  \bibinfo{author}{\bibfnamefont{S.~M.} \bibnamefont{Kogan}},
  \bibinfo{year}{1967}, \bibinfo{journal}{Sov. Phys. JETP}
  \textbf{\bibinfo{volume}{25}}, \bibinfo{pages}{1095}.

\bibitem[{\citenamefont{Volkov and Kogan}(1969)}]{volkov:1969}
\bibinfo{author}{\bibnamefont{Volkov}, \bibfnamefont{A.~F.}}, and
  \bibinfo{author}{\bibfnamefont{S.~M.} \bibnamefont{Kogan}},
  \bibinfo{year}{1969}, \bibinfo{journal}{Sov. Phys. Usp.}
  \textbf{\bibinfo{volume}{11}}, \bibinfo{pages}{881}.

\bibitem[{\citenamefont{Volkov and Pavlovskii}(2004)}]{volkov:2004}
\bibinfo{author}{\bibnamefont{Volkov}, \bibfnamefont{A.~F.}}, and
  \bibinfo{author}{\bibfnamefont{V.~V.} \bibnamefont{Pavlovskii}},
  \bibinfo{year}{2004}, \bibinfo{journal}{Phys. Rev. B}
  \textbf{\bibinfo{volume}{69}}, \bibinfo{pages}{125305}.

\bibitem[{\citenamefont{Volkov and Takhtamirov}(2007)}]{volkov:2007}
\bibinfo{author}{\bibnamefont{Volkov}, \bibfnamefont{V.~A.}}, and
  \bibinfo{author}{\bibfnamefont{E.~E.} \bibnamefont{Takhtamirov}},
  \bibinfo{year}{2007}, \bibinfo{journal}{JETP} \textbf{\bibinfo{volume}{104}},
  \bibinfo{pages}{602}.

\bibitem[{\citenamefont{Walukiewicz}
  \emph{et~al.}(1984)\citenamefont{Walukiewicz, Ruda, Lagowski, and
  Gatos}}]{walukiewicz:1984}
\bibinfo{author}{\bibnamefont{Walukiewicz}, \bibfnamefont{W.}},
  \bibinfo{author}{\bibfnamefont{H.~E.} \bibnamefont{Ruda}},
  \bibinfo{author}{\bibfnamefont{J.}~\bibnamefont{Lagowski}}, and
  \bibinfo{author}{\bibfnamefont{H.~C.} \bibnamefont{Gatos}},
  \bibinfo{year}{1984}, \bibinfo{journal}{Phys. Rev. B}
  \textbf{\bibinfo{volume}{30}}, \bibinfo{pages}{4571}.

\bibitem[{\citenamefont{Weiss} \emph{et~al.}(1989)\citenamefont{Weiss, von
  Klitzing, Ploog, and Weimann}}]{weiss:1989}
\bibinfo{author}{\bibnamefont{Weiss}, \bibfnamefont{D.}},
  \bibinfo{author}{\bibfnamefont{K.}~\bibnamefont{von Klitzing}},
  \bibinfo{author}{\bibfnamefont{K.}~\bibnamefont{Ploog}}, and
  \bibinfo{author}{\bibfnamefont{G.}~\bibnamefont{Weimann}},
  \bibinfo{year}{1989}, \bibinfo{journal}{Europhys. Lett.}
  \textbf{\bibinfo{volume}{8}}, \bibinfo{pages}{179}.

\bibitem[{\citenamefont{Weiss} \emph{et~al.}(1991)\citenamefont{Weiss, Roukes,
  Menschig, Grambow, von Klitzing, and Weimann}}]{weiss:1991b}
\bibinfo{author}{\bibnamefont{Weiss}, \bibfnamefont{D.}},
  \bibinfo{author}{\bibfnamefont{M.~L.} \bibnamefont{Roukes}},
  \bibinfo{author}{\bibfnamefont{A.}~\bibnamefont{Menschig}},
  \bibinfo{author}{\bibfnamefont{P.}~\bibnamefont{Grambow}},
  \bibinfo{author}{\bibfnamefont{K.}~\bibnamefont{von Klitzing}}, and
  \bibinfo{author}{\bibfnamefont{G.}~\bibnamefont{Weimann}},
  \bibinfo{year}{1991}, \bibinfo{journal}{Phys. Rev. Lett.}
  \textbf{\bibinfo{volume}{66}}, \bibinfo{pages}{2790}.

\bibitem[{\citenamefont{Wiedmann}
  \emph{et~al.}(2009{\natexlab{a}})\citenamefont{Wiedmann, Gusev, Raichev,
  Bakarov, and Portal}}]{wiedmann:2009a}
\bibinfo{author}{\bibnamefont{Wiedmann}, \bibfnamefont{S.}},
  \bibinfo{author}{\bibfnamefont{G.~M.} \bibnamefont{Gusev}},
  \bibinfo{author}{\bibfnamefont{O.~E.} \bibnamefont{Raichev}},
  \bibinfo{author}{\bibfnamefont{A.~K.} \bibnamefont{Bakarov}}, and
  \bibinfo{author}{\bibfnamefont{J.~C.} \bibnamefont{Portal}},
  \bibinfo{year}{2009}{\natexlab{a}}, \bibinfo{journal}{Phys. Rev. B}
  \textbf{\bibinfo{volume}{80}}, \bibinfo{pages}{035317}.

\bibitem[{\citenamefont{Wiedmann}
  \emph{et~al.}(2010{\natexlab{a}})\citenamefont{Wiedmann, Gusev, Raichev,
  Bakarov, and Portal}}]{wiedmann:2010}
\bibinfo{author}{\bibnamefont{Wiedmann}, \bibfnamefont{S.}},
  \bibinfo{author}{\bibfnamefont{G.~M.} \bibnamefont{Gusev}},
  \bibinfo{author}{\bibfnamefont{O.~E.} \bibnamefont{Raichev}},
  \bibinfo{author}{\bibfnamefont{A.~K.} \bibnamefont{Bakarov}}, and
  \bibinfo{author}{\bibfnamefont{J.~C.} \bibnamefont{Portal}},
  \bibinfo{year}{2010}{\natexlab{a}}, \bibinfo{journal}{Phys. Rev. B}
  \textbf{\bibinfo{volume}{82}}, \bibinfo{pages}{165333}.

\bibitem[{\citenamefont{Wiedmann}
  \emph{et~al.}(2010{\natexlab{b}})\citenamefont{Wiedmann, Gusev, Raichev,
  Bakarov, and Portal}}]{wiedmann:2010a}
\bibinfo{author}{\bibnamefont{Wiedmann}, \bibfnamefont{S.}},
  \bibinfo{author}{\bibfnamefont{G.~M.} \bibnamefont{Gusev}},
  \bibinfo{author}{\bibfnamefont{O.~E.} \bibnamefont{Raichev}},
  \bibinfo{author}{\bibfnamefont{A.~K.} \bibnamefont{Bakarov}}, and
  \bibinfo{author}{\bibfnamefont{J.~C.} \bibnamefont{Portal}},
  \bibinfo{year}{2010}{\natexlab{b}}, \bibinfo{journal}{Phys. Rev. B}
  \textbf{\bibinfo{volume}{81}}, \bibinfo{pages}{085311}.

\bibitem[{\citenamefont{Wiedmann}
  \emph{et~al.}(2010{\natexlab{c}})\citenamefont{Wiedmann, Gusev, Raichev,
  Bakarov, and Portal}}]{wiedmann:2010b}
\bibinfo{author}{\bibnamefont{Wiedmann}, \bibfnamefont{S.}},
  \bibinfo{author}{\bibfnamefont{G.~M.} \bibnamefont{Gusev}},
  \bibinfo{author}{\bibfnamefont{O.~E.} \bibnamefont{Raichev}},
  \bibinfo{author}{\bibfnamefont{A.~K.} \bibnamefont{Bakarov}}, and
  \bibinfo{author}{\bibfnamefont{J.~C.} \bibnamefont{Portal}},
  \bibinfo{year}{2010}{\natexlab{c}}, \bibinfo{journal}{Phys. Rev. Lett.}
  \textbf{\bibinfo{volume}{105}}, \bibinfo{pages}{026804}.

\bibitem[{\citenamefont{Wiedmann}
  \emph{et~al.}(2011{\natexlab{a}})\citenamefont{Wiedmann, Gusev, Raichev,
  Bakarov, and Portal}}]{wiedmann:2011c}
\bibinfo{author}{\bibnamefont{Wiedmann}, \bibfnamefont{S.}},
  \bibinfo{author}{\bibfnamefont{G.~M.} \bibnamefont{Gusev}},
  \bibinfo{author}{\bibfnamefont{O.~E.} \bibnamefont{Raichev}},
  \bibinfo{author}{\bibfnamefont{A.~K.} \bibnamefont{Bakarov}}, and
  \bibinfo{author}{\bibfnamefont{J.~C.} \bibnamefont{Portal}},
  \bibinfo{year}{2011}{\natexlab{a}}, \bibinfo{journal}{Phys. Rev. B}
  \textbf{\bibinfo{volume}{84}}, \bibinfo{pages}{165303}.

\bibitem[{\citenamefont{Wiedmann}
  \emph{et~al.}(2011{\natexlab{b}})\citenamefont{Wiedmann, Gusev, Raichev,
  Kr\"amer, Bakarov, and Portal}}]{wiedmann:2011b}
\bibinfo{author}{\bibnamefont{Wiedmann}, \bibfnamefont{S.}},
  \bibinfo{author}{\bibfnamefont{G.~M.} \bibnamefont{Gusev}},
  \bibinfo{author}{\bibfnamefont{O.~E.} \bibnamefont{Raichev}},
  \bibinfo{author}{\bibfnamefont{S.}~\bibnamefont{Kr\"amer}},
  \bibinfo{author}{\bibfnamefont{A.~K.} \bibnamefont{Bakarov}}, and
  \bibinfo{author}{\bibfnamefont{J.~C.} \bibnamefont{Portal}},
  \bibinfo{year}{2011}{\natexlab{b}}, \bibinfo{journal}{Phys. Rev. B}
  \textbf{\bibinfo{volume}{83}}, \bibinfo{pages}{195317}.

\bibitem[{\citenamefont{Wiedmann} \emph{et~al.}(2008)\citenamefont{Wiedmann,
  Gusev, Raichev, Lamas, Bakarov, and Portal}}]{wiedmann:2008}
\bibinfo{author}{\bibnamefont{Wiedmann}, \bibfnamefont{S.}},
  \bibinfo{author}{\bibfnamefont{G.~M.} \bibnamefont{Gusev}},
  \bibinfo{author}{\bibfnamefont{O.~E.} \bibnamefont{Raichev}},
  \bibinfo{author}{\bibfnamefont{T.~E.} \bibnamefont{Lamas}},
  \bibinfo{author}{\bibfnamefont{A.~K.} \bibnamefont{Bakarov}}, and
  \bibinfo{author}{\bibfnamefont{J.~C.} \bibnamefont{Portal}},
  \bibinfo{year}{2008}, \bibinfo{journal}{Phys. Rev. B}
  \textbf{\bibinfo{volume}{78}}, \bibinfo{pages}{121301}.

\bibitem[{\citenamefont{Wiedmann}
  \emph{et~al.}(2009{\natexlab{b}})\citenamefont{Wiedmann, Mamani, Gusev,
  Raichev, Bakarov, and Portal}}]{wiedmann:2009b}
\bibinfo{author}{\bibnamefont{Wiedmann}, \bibfnamefont{S.}},
  \bibinfo{author}{\bibfnamefont{N.~C.} \bibnamefont{Mamani}},
  \bibinfo{author}{\bibfnamefont{G.~M.} \bibnamefont{Gusev}},
  \bibinfo{author}{\bibfnamefont{O.~E.} \bibnamefont{Raichev}},
  \bibinfo{author}{\bibfnamefont{A.~K.} \bibnamefont{Bakarov}}, and
  \bibinfo{author}{\bibfnamefont{J.~C.} \bibnamefont{Portal}},
  \bibinfo{year}{2009}{\natexlab{b}}, \bibinfo{journal}{Phys. Rev. B}
  \textbf{\bibinfo{volume}{80}}, \bibinfo{pages}{245306}.

\bibitem[{\citenamefont{Willett} \emph{et~al.}(2001)\citenamefont{Willett, Hsu,
  Natelson, West, and Pfeiffer}}]{willett:2001}
\bibinfo{author}{\bibnamefont{Willett}, \bibfnamefont{R.~L.}},
  \bibinfo{author}{\bibfnamefont{J.~W.~P.} \bibnamefont{Hsu}},
  \bibinfo{author}{\bibfnamefont{D.}~\bibnamefont{Natelson}},
  \bibinfo{author}{\bibfnamefont{K.~W.} \bibnamefont{West}}, and
  \bibinfo{author}{\bibfnamefont{L.~N.} \bibnamefont{Pfeiffer}},
  \bibinfo{year}{2001}, \bibinfo{journal}{Phys. Rev. Lett.}
  \textbf{\bibinfo{volume}{87}}, \bibinfo{pages}{126803}.

\bibitem[{\citenamefont{Willett} \emph{et~al.}(2004)\citenamefont{Willett,
  Pfeiffer, and West}}]{willett:2004}
\bibinfo{author}{\bibnamefont{Willett}, \bibfnamefont{R.~L.}},
  \bibinfo{author}{\bibfnamefont{L.~N.} \bibnamefont{Pfeiffer}}, and
  \bibinfo{author}{\bibfnamefont{K.~W.} \bibnamefont{West}},
  \bibinfo{year}{2004}, \bibinfo{journal}{Phys. Rev. Lett.}
  \textbf{\bibinfo{volume}{93}}, \bibinfo{pages}{026804}.

\bibitem[{\citenamefont{Wirthmann} \emph{et~al.}(2007)\citenamefont{Wirthmann,
  McCombe, Heitmann, Holland, Friedland, and Hu}}]{wirthmann:2007}
\bibinfo{author}{\bibnamefont{Wirthmann}, \bibfnamefont{A.}},
  \bibinfo{author}{\bibfnamefont{B.~D.} \bibnamefont{McCombe}},
  \bibinfo{author}{\bibfnamefont{D.}~\bibnamefont{Heitmann}},
  \bibinfo{author}{\bibfnamefont{S.}~\bibnamefont{Holland}},
  \bibinfo{author}{\bibfnamefont{K.-J.} \bibnamefont{Friedland}}, and
  \bibinfo{author}{\bibfnamefont{C.-M.} \bibnamefont{Hu}},
  \bibinfo{year}{2007}, \bibinfo{journal}{Phys. Rev. B}
  \textbf{\bibinfo{volume}{76}}, \bibinfo{pages}{195315}.

\bibitem[{\citenamefont{Yacoby}(1968)}]{yacoby:1968}
\bibinfo{author}{\bibnamefont{Yacoby}, \bibfnamefont{Y.}},
  \bibinfo{year}{1968}, \bibinfo{journal}{Phys. Rev.}
  \textbf{\bibinfo{volume}{169}}, \bibinfo{pages}{610}.

\bibitem[{\citenamefont{Yang} \emph{et~al.}(2006)\citenamefont{Yang, Du,
  Pfeiffer, and West}}]{yang:2006}
\bibinfo{author}{\bibnamefont{Yang}, \bibfnamefont{C.~L.}},
  \bibinfo{author}{\bibfnamefont{R.~R.} \bibnamefont{Du}},
  \bibinfo{author}{\bibfnamefont{L.~N.} \bibnamefont{Pfeiffer}}, and
  \bibinfo{author}{\bibfnamefont{K.~W.} \bibnamefont{West}},
  \bibinfo{year}{2006}, \bibinfo{journal}{Phys. Rev. B}
  \textbf{\bibinfo{volume}{74}}, \bibinfo{pages}{045315}.

\bibitem[{\citenamefont{Yang}
  \emph{et~al.}(2002{\natexlab{a}})\citenamefont{Yang, Zhang, Du, Simmons, and
  Reno}}]{yang:2002}
\bibinfo{author}{\bibnamefont{Yang}, \bibfnamefont{C.~L.}},
  \bibinfo{author}{\bibfnamefont{J.}~\bibnamefont{Zhang}},
  \bibinfo{author}{\bibfnamefont{R.~R.} \bibnamefont{Du}},
  \bibinfo{author}{\bibfnamefont{J.~A.} \bibnamefont{Simmons}}, and
  \bibinfo{author}{\bibfnamefont{J.~L.} \bibnamefont{Reno}},
  \bibinfo{year}{2002}{\natexlab{a}}, \bibinfo{journal}{Phys. Rev. Lett.}
  \textbf{\bibinfo{volume}{89}}, \bibinfo{pages}{076801}.

\bibitem[{\citenamefont{Yang} \emph{et~al.}(2003)\citenamefont{Yang, Zudov,
  Knuuttila, Du, Pfeiffer, and West}}]{yang:2003}
\bibinfo{author}{\bibnamefont{Yang}, \bibfnamefont{C.~L.}},
  \bibinfo{author}{\bibfnamefont{M.~A.} \bibnamefont{Zudov}},
  \bibinfo{author}{\bibfnamefont{T.~A.} \bibnamefont{Knuuttila}},
  \bibinfo{author}{\bibfnamefont{R.~R.} \bibnamefont{Du}},
  \bibinfo{author}{\bibfnamefont{L.~N.} \bibnamefont{Pfeiffer}}, and
  \bibinfo{author}{\bibfnamefont{K.~W.} \bibnamefont{West}},
  \bibinfo{year}{2003}, \bibinfo{journal}{Phys. Rev. Lett.}
  \textbf{\bibinfo{volume}{91}}, \bibinfo{pages}{096803}.

\bibitem[{\citenamefont{Yang}
  \emph{et~al.}(2002{\natexlab{b}})\citenamefont{Yang, Zudov, Zhang, Du,
  Simmons, and Reno}}]{yang:2002b}
\bibinfo{author}{\bibnamefont{Yang}, \bibfnamefont{C.~L.}},
  \bibinfo{author}{\bibfnamefont{M.~A.} \bibnamefont{Zudov}},
  \bibinfo{author}{\bibfnamefont{J.}~\bibnamefont{Zhang}},
  \bibinfo{author}{\bibfnamefont{R.~R.} \bibnamefont{Du}},
  \bibinfo{author}{\bibfnamefont{J.~A.} \bibnamefont{Simmons}}, and
  \bibinfo{author}{\bibfnamefont{J.~L.} \bibnamefont{Reno}},
  \bibinfo{year}{2002}{\natexlab{b}}, \bibinfo{journal}{Physica E}
  \textbf{\bibinfo{volume}{12}}, \bibinfo{pages}{443}.

\bibitem[{\citenamefont{Ye} \emph{et~al.}(2001)\citenamefont{Ye, Engel, Tsui,
  Simmons, Wendt, Vawter, and Reno}}]{ye:2001}
\bibinfo{author}{\bibnamefont{Ye}, \bibfnamefont{P.~D.}},
  \bibinfo{author}{\bibfnamefont{L.~W.} \bibnamefont{Engel}},
  \bibinfo{author}{\bibfnamefont{D.~C.} \bibnamefont{Tsui}},
  \bibinfo{author}{\bibfnamefont{J.~A.} \bibnamefont{Simmons}},
  \bibinfo{author}{\bibfnamefont{J.~R.} \bibnamefont{Wendt}},
  \bibinfo{author}{\bibfnamefont{G.~A.} \bibnamefont{Vawter}}, and
  \bibinfo{author}{\bibfnamefont{J.~L.} \bibnamefont{Reno}},
  \bibinfo{year}{2001}, \bibinfo{journal}{Appl. Phys. Lett.}
  \textbf{\bibinfo{volume}{79}}, \bibinfo{pages}{2193}.

\bibitem[{\citenamefont{Yuan} \emph{et~al.}(2006)\citenamefont{Yuan, Yang, Du,
  Pfeiffer, and West}}]{yuan:2006}
\bibinfo{author}{\bibnamefont{Yuan}, \bibfnamefont{Z.~Q.}},
  \bibinfo{author}{\bibfnamefont{C.~L.} \bibnamefont{Yang}},
  \bibinfo{author}{\bibfnamefont{R.~R.} \bibnamefont{Du}},
  \bibinfo{author}{\bibfnamefont{L.~N.} \bibnamefont{Pfeiffer}}, and
  \bibinfo{author}{\bibfnamefont{K.~W.} \bibnamefont{West}},
  \bibinfo{year}{2006}, \bibinfo{journal}{Phys. Rev. B}
  \textbf{\bibinfo{volume}{74}}, \bibinfo{pages}{075313}.

\bibitem[{\citenamefont{Zakharov}(1960)}]{zakharov:1960}
\bibinfo{author}{\bibnamefont{Zakharov}, \bibfnamefont{A.~L.}},
  \bibinfo{year}{1960}, \bibinfo{journal}{Sov. Phys. JETP}
  \textbf{\bibinfo{volume}{11}}, \bibinfo{pages}{478}.

\bibitem[{\citenamefont{Zaremba}(1992)}]{zaremba:1992}
\bibinfo{author}{\bibnamefont{Zaremba}, \bibfnamefont{E.}},
  \bibinfo{year}{1992}, \bibinfo{journal}{Phys. Rev. B}
  \textbf{\bibinfo{volume}{45}}, \bibinfo{pages}{14143}.

\bibitem[{\citenamefont{Zeldovich}(1967)}]{zeldovich:1967}
\bibinfo{author}{\bibnamefont{Zeldovich}, \bibfnamefont{Y.~B.}},
  \bibinfo{year}{1967}, \bibinfo{journal}{Sov. Phys. JETP}
  \textbf{\bibinfo{volume}{24}}, \bibinfo{pages}{1006}.

\bibitem[{\citenamefont{Zhang and Gerhardts}(1990)}]{zhang:1990}
\bibinfo{author}{\bibnamefont{Zhang}, \bibfnamefont{C.}}, and
  \bibinfo{author}{\bibfnamefont{R.~R.} \bibnamefont{Gerhardts}},
  \bibinfo{year}{1990}, \bibinfo{journal}{Phys. Rev. B}
  \textbf{\bibinfo{volume}{41}}, \bibinfo{pages}{12850}.

\bibitem[{\citenamefont{Zhang} \emph{et~al.}(2004)\citenamefont{Zhang, Lyo, Du,
  Simmons, and Reno}}]{zhang:2004}
\bibinfo{author}{\bibnamefont{Zhang}, \bibfnamefont{J.}},
  \bibinfo{author}{\bibfnamefont{S.~K.} \bibnamefont{Lyo}},
  \bibinfo{author}{\bibfnamefont{R.~R.} \bibnamefont{Du}},
  \bibinfo{author}{\bibfnamefont{J.~A.} \bibnamefont{Simmons}}, and
  \bibinfo{author}{\bibfnamefont{J.~L.} \bibnamefont{Reno}},
  \bibinfo{year}{2004}, \bibinfo{journal}{Phys. Rev. Lett.}
  \textbf{\bibinfo{volume}{92}}, \bibinfo{pages}{156802}.

\bibitem[{\citenamefont{Zhang} \emph{et~al.}(2009)\citenamefont{Zhang,
  Vitkalov, and Bykov}}]{zhang:2009}
\bibinfo{author}{\bibnamefont{Zhang}, \bibfnamefont{J.~Q.}},
  \bibinfo{author}{\bibfnamefont{S.}~\bibnamefont{Vitkalov}}, and
  \bibinfo{author}{\bibfnamefont{A.~A.} \bibnamefont{Bykov}},
  \bibinfo{year}{2009}, \bibinfo{journal}{Phys. Rev. B}
  \textbf{\bibinfo{volume}{80}}, \bibinfo{pages}{045310}.

\bibitem[{\citenamefont{Zhang}
  \emph{et~al.}(2007{\natexlab{a}})\citenamefont{Zhang, Vitkalov, Bykov,
  Kalagin, and Bakarov}}]{zhang:2007b}
\bibinfo{author}{\bibnamefont{Zhang}, \bibfnamefont{J.~Q.}},
  \bibinfo{author}{\bibfnamefont{S.}~\bibnamefont{Vitkalov}},
  \bibinfo{author}{\bibfnamefont{A.~A.} \bibnamefont{Bykov}},
  \bibinfo{author}{\bibfnamefont{A.~K.} \bibnamefont{Kalagin}}, and
  \bibinfo{author}{\bibfnamefont{A.~K.} \bibnamefont{Bakarov}},
  \bibinfo{year}{2007}{\natexlab{a}}, \bibinfo{journal}{Phys. Rev. B}
  \textbf{\bibinfo{volume}{75}}, \bibinfo{pages}{081305(R)}.

\bibitem[{\citenamefont{Zhang}
  \emph{et~al.}(2007{\natexlab{b}})\citenamefont{Zhang, Chiang, Zudov,
  Pfeiffer, and West}}]{zhang:2007a}
\bibinfo{author}{\bibnamefont{Zhang}, \bibfnamefont{W.}},
  \bibinfo{author}{\bibfnamefont{H.-S.} \bibnamefont{Chiang}},
  \bibinfo{author}{\bibfnamefont{M.~A.} \bibnamefont{Zudov}},
  \bibinfo{author}{\bibfnamefont{L.~N.} \bibnamefont{Pfeiffer}}, and
  \bibinfo{author}{\bibfnamefont{K.~W.} \bibnamefont{West}},
  \bibinfo{year}{2007}{\natexlab{b}}, \bibinfo{journal}{Phys. Rev. B}
  \textbf{\bibinfo{volume}{75}}, \bibinfo{pages}{041304(R)}.

\bibitem[{\citenamefont{Zhang}
  \emph{et~al.}(2007{\natexlab{c}})\citenamefont{Zhang, Zudov, Pfeiffer, and
  West}}]{zhang:2007c}
\bibinfo{author}{\bibnamefont{Zhang}, \bibfnamefont{W.}},
  \bibinfo{author}{\bibfnamefont{M.~A.} \bibnamefont{Zudov}},
  \bibinfo{author}{\bibfnamefont{L.~N.} \bibnamefont{Pfeiffer}}, and
  \bibinfo{author}{\bibfnamefont{K.~W.} \bibnamefont{West}},
  \bibinfo{year}{2007}{\natexlab{c}}, \bibinfo{journal}{Phys. Rev. Lett.}
  \textbf{\bibinfo{volume}{98}}, \bibinfo{pages}{106804}.

\bibitem[{\citenamefont{Zhang}
  \emph{et~al.}(2008{\natexlab{a}})\citenamefont{Zhang, Zudov, Pfeiffer, and
  West}}]{zhang:2008b}
\bibinfo{author}{\bibnamefont{Zhang}, \bibfnamefont{W.}},
  \bibinfo{author}{\bibfnamefont{M.~A.} \bibnamefont{Zudov}},
  \bibinfo{author}{\bibfnamefont{L.~N.} \bibnamefont{Pfeiffer}}, and
  \bibinfo{author}{\bibfnamefont{K.~W.} \bibnamefont{West}},
  \bibinfo{year}{2008}{\natexlab{a}}, \bibinfo{journal}{Physica E}
  \textbf{\bibinfo{volume}{40}}, \bibinfo{pages}{982}.

\bibitem[{\citenamefont{Zhang}
  \emph{et~al.}(2008{\natexlab{b}})\citenamefont{Zhang, Zudov, Pfeiffer, and
  West}}]{zhang:2008}
\bibinfo{author}{\bibnamefont{Zhang}, \bibfnamefont{W.}},
  \bibinfo{author}{\bibfnamefont{M.~A.} \bibnamefont{Zudov}},
  \bibinfo{author}{\bibfnamefont{L.~N.} \bibnamefont{Pfeiffer}}, and
  \bibinfo{author}{\bibfnamefont{K.~W.} \bibnamefont{West}},
  \bibinfo{year}{2008}{\natexlab{b}}, \bibinfo{journal}{Phys. Rev. Lett.}
  \textbf{\bibinfo{volume}{100}}, \bibinfo{pages}{036805}.

\bibitem[{\citenamefont{Zhang} \emph{et~al.}(2005)\citenamefont{Zhang, Tan,
  Stormer, and Kim}}]{zhang:2005}
\bibinfo{author}{\bibnamefont{Zhang}, \bibfnamefont{Y.}},
  \bibinfo{author}{\bibfnamefont{Y.-W.} \bibnamefont{Tan}},
  \bibinfo{author}{\bibfnamefont{H.~L.} \bibnamefont{Stormer}}, and
  \bibinfo{author}{\bibfnamefont{P.}~\bibnamefont{Kim}}, \bibinfo{year}{2005},
  \bibinfo{journal}{Nature (London)} \textbf{\bibinfo{volume}{438}},
  \bibinfo{pages}{201}.

\bibitem[{\citenamefont{Zudov}(2004)}]{zudov:2004}
\bibinfo{author}{\bibnamefont{Zudov}, \bibfnamefont{M.~A.}},
  \bibinfo{year}{2004}, \bibinfo{journal}{Phys. Rev. B}
  \textbf{\bibinfo{volume}{69}}, \bibinfo{pages}{041304(R)}.

\bibitem[{\citenamefont{Zudov} \emph{et~al.}(2009)\citenamefont{Zudov, Chiang,
  Hatke, Zhang, Pfeiffer, and West}}]{zudov:2009}
\bibinfo{author}{\bibnamefont{Zudov}, \bibfnamefont{M.~A.}},
  \bibinfo{author}{\bibfnamefont{H.-S.} \bibnamefont{Chiang}},
  \bibinfo{author}{\bibfnamefont{A.~T.} \bibnamefont{Hatke}},
  \bibinfo{author}{\bibfnamefont{W.}~\bibnamefont{Zhang}},
  \bibinfo{author}{\bibfnamefont{L.~N.} \bibnamefont{Pfeiffer}}, and
  \bibinfo{author}{\bibfnamefont{K.~W.} \bibnamefont{West}},
  \bibinfo{year}{2009}, \bibinfo{journal}{Int. J. Mod. Phys. B}
  \textbf{\bibinfo{volume}{23}}, \bibinfo{pages}{2684}.

\bibitem[{\citenamefont{Zudov} \emph{et~al.}(2003)\citenamefont{Zudov, Du,
  Pfeiffer, and West}}]{zudov:2003}
\bibinfo{author}{\bibnamefont{Zudov}, \bibfnamefont{M.~A.}},
  \bibinfo{author}{\bibfnamefont{R.~R.} \bibnamefont{Du}},
  \bibinfo{author}{\bibfnamefont{L.~N.} \bibnamefont{Pfeiffer}}, and
  \bibinfo{author}{\bibfnamefont{K.~W.} \bibnamefont{West}},
  \bibinfo{year}{2003}, \bibinfo{journal}{Phys. Rev. Lett.}
  \textbf{\bibinfo{volume}{90}}, \bibinfo{pages}{046807}.

\bibitem[{\citenamefont{Zudov}
  \emph{et~al.}(2006{\natexlab{a}})\citenamefont{Zudov, Du, Pfeiffer, and
  West}}]{zudov:2006b}
\bibinfo{author}{\bibnamefont{Zudov}, \bibfnamefont{M.~A.}},
  \bibinfo{author}{\bibfnamefont{R.~R.} \bibnamefont{Du}},
  \bibinfo{author}{\bibfnamefont{L.~N.} \bibnamefont{Pfeiffer}}, and
  \bibinfo{author}{\bibfnamefont{K.~W.} \bibnamefont{West}},
  \bibinfo{year}{2006}{\natexlab{a}}, \bibinfo{journal}{Phys. Rev. Lett.}
  \textbf{\bibinfo{volume}{96}}, \bibinfo{pages}{236804}.

\bibitem[{\citenamefont{Zudov}
  \emph{et~al.}(2006{\natexlab{b}})\citenamefont{Zudov, Du, Pfeiffer, and
  West}}]{zudov:2006a}
\bibinfo{author}{\bibnamefont{Zudov}, \bibfnamefont{M.~A.}},
  \bibinfo{author}{\bibfnamefont{R.~R.} \bibnamefont{Du}},
  \bibinfo{author}{\bibfnamefont{L.~N.} \bibnamefont{Pfeiffer}}, and
  \bibinfo{author}{\bibfnamefont{K.~W.} \bibnamefont{West}},
  \bibinfo{year}{2006}{\natexlab{b}}, \bibinfo{journal}{Phys. Rev. B}
  \textbf{\bibinfo{volume}{73}}, \bibinfo{pages}{041303(R)}.

\bibitem[{\citenamefont{Zudov} \emph{et~al.}(1997)\citenamefont{Zudov, Du,
  Simmons, and Reno}}]{zudov:1997}
\bibinfo{author}{\bibnamefont{Zudov}, \bibfnamefont{M.~A.}},
  \bibinfo{author}{\bibfnamefont{R.~R.} \bibnamefont{Du}},
  \bibinfo{author}{\bibfnamefont{J.~A.} \bibnamefont{Simmons}}, and
  \bibinfo{author}{\bibfnamefont{J.~L.} \bibnamefont{Reno}},
  \bibinfo{year}{1997}, \bibinfo{journal}{arXiv:cond-mat/9711149v1} .

\bibitem[{\citenamefont{Zudov}
  \emph{et~al.}(2001{\natexlab{a}})\citenamefont{Zudov, Du, Simmons, and
  Reno}}]{zudov:2001a}
\bibinfo{author}{\bibnamefont{Zudov}, \bibfnamefont{M.~A.}},
  \bibinfo{author}{\bibfnamefont{R.~R.} \bibnamefont{Du}},
  \bibinfo{author}{\bibfnamefont{J.~A.} \bibnamefont{Simmons}}, and
  \bibinfo{author}{\bibfnamefont{J.~L.} \bibnamefont{Reno}},
  \bibinfo{year}{2001}{\natexlab{a}}, \bibinfo{journal}{Phys. Rev. B}
  \textbf{\bibinfo{volume}{64}}, \bibinfo{pages}{201311(R)}.

\bibitem[{\citenamefont{Zudov} \emph{et~al.}(2011)\citenamefont{Zudov, Hatke,
  Chiang, Pfeiffer, West, and Reno}}]{zudov:2011}
\bibinfo{author}{\bibnamefont{Zudov}, \bibfnamefont{M.~A.}},
  \bibinfo{author}{\bibfnamefont{A.~T.} \bibnamefont{Hatke}},
  \bibinfo{author}{\bibfnamefont{H.-S.} \bibnamefont{Chiang}},
  \bibinfo{author}{\bibfnamefont{L.~N.} \bibnamefont{Pfeiffer}},
  \bibinfo{author}{\bibfnamefont{K.~W.} \bibnamefont{West}}, and
  \bibinfo{author}{\bibfnamefont{J.~L.} \bibnamefont{Reno}},
  \bibinfo{year}{2011}, \bibinfo{journal}{J. Phys.: Conf. Ser.}
  \textbf{\bibinfo{volume}{334}}, \bibinfo{pages}{012007}.

\bibitem[{\citenamefont{Zudov}
  \emph{et~al.}(2001{\natexlab{b}})\citenamefont{Zudov, Ponomarev, Efros, Du,
  Simmons, and Reno}}]{zudov:2001b}
\bibinfo{author}{\bibnamefont{Zudov}, \bibfnamefont{M.~A.}},
  \bibinfo{author}{\bibfnamefont{I.~V.} \bibnamefont{Ponomarev}},
  \bibinfo{author}{\bibfnamefont{A.~L.} \bibnamefont{Efros}},
  \bibinfo{author}{\bibfnamefont{R.~R.} \bibnamefont{Du}},
  \bibinfo{author}{\bibfnamefont{J.~A.} \bibnamefont{Simmons}}, and
  \bibinfo{author}{\bibfnamefont{J.~L.} \bibnamefont{Reno}},
  \bibinfo{year}{2001}{\natexlab{b}}, \bibinfo{journal}{Phys. Rev. Lett.}
  \textbf{\bibinfo{volume}{86}}, \bibinfo{pages}{3614}.

\end{thebibliography}
\end{document}